%% file: Bepub2_arXiv.tex
\documentclass[12pt]{cernprep}
\usepackage{graphicx}
\usepackage{amssymb}
\usepackage{epstopdf}
\begin{document}
\newcommand{\dedx}{\mbox{${\rm d}E/{\rm d}x$}}
\newcommand{\EcB}{$E \! \times \! B$}
\newcommand{\omt}{$\omega \tau$}
\newcommand{\omtsq}{$(\omega \tau )^2$}
\newcommand{\rphi}{\mbox{$r \! \cdot \! \phi$}}
\newcommand{\srphi}{\mbox{$\sigma_{r \! \cdot \! \phi}$}}
\newcommand{\dg}{\mbox{`durchgriff'}}
\newcommand{\mg}{\mbox{`margaritka'}}
\newcommand{\pT}{\mbox{$p_{\rm T}$}}
\newcommand{\GeVc}{\mbox{GeV/{\it c}}}
\newcommand{\MeVc}{\mbox{MeV/{\it c}}}
\def\kr{$^{83{\rm m}}$Kr\ }
\begin{titlepage}
\docnum{CERN--PH--EP/2008--025}
\date{5 September 2008} 
\date{Revised: 9 March 2009} 
\vspace{1cm}
\title{CROSS-SECTIONS OF LARGE-ANGLE HADRON PRODUCTION \\
IN PROTON-- AND PION--NUCLEUS INTERACTIONS II: \\
BERYLLIUM NUCLEI AND BEAM MOMENTA 
FROM \mbox{\boldmath $\pm3$}~GeV/\mbox{\boldmath $c$} 
TO \mbox{\boldmath $\pm15$}~GeV/\mbox{\boldmath $c$}}

\begin{abstract}
We report on double-differential inclusive cross-sections of 
the production of secondary protons and charged pions, in the 
interactions with a 5\% $\lambda_{\rm abs}$ 
thick stationary beryllium target, of proton and pion beams with
momentum from $\pm3$~GeV/{\it c} to $\pm15$~GeV/{\it c}. Results are given for secondary particles with production angles $20^\circ < \theta < 125^\circ$. 
\end{abstract}

\vfill  \normalsize
\begin{center}
The HARP--CDP group  \\  

\vspace*{2mm} 

A.~Bolshakova$^1$, 
I.~Boyko$^1$, 
G.~Chelkov$^{1a}$, 
D.~Dedovitch$^1$, 
A.~Elagin$^{1b}$, 
M.~Gostkin$^1$,
S.~Grishin$^{1a}$,
A.~Guskov$^1$, 
Z.~Kroumchtein$^1$, 
Yu.~Nefedov$^1$, 
K.~Nikolaev$^1$, 
A.~Zhemchugov$^1$, 
F.~Dydak$^2$, 
J.~Wotschack$^{2*}$, 
A.~De~Min$^{3c}$,
V.~Ammosov$^4$, 
V.~Gapienko$^4$, 
V.~Koreshev$^4$, 
A.~Semak$^4$, 
Yu.~Sviridov$^4$, 
E.~Usenko$^{4d}$, 
V.~Zaets$^4$ 
\\
 
\vspace*{5mm} 

$^1$~{\bf Joint Institute for Nuclear Research, Dubna, Russia} \\
$^2$~{\bf CERN, Geneva, Switzerland} \\ 
$^3$~{\bf Politecnico di Milano and INFN, 
Sezione di Milano-Bicocca, Milan, Italy} \\
$^4$~{\bf Institute of High Energy Physics, Protvino, Russia} \\

\vspace*{5mm}

\submitted{(To be submitted to Eur. Phys. J. C)}
\end{center}

\vspace*{5mm}
\rule{0.9\textwidth}{0.2mm}

\begin{footnotesize}

$^a$~Also at the Moscow Institute of Physics and Technology, Moscow, Russia 

$^b$~Now at Texas A\&M University, College Station, USA 

$^c$~On leave of absence at 
Ecole Polytechnique F\'{e}d\'{e}rale, Lausanne, Switzerland 

$^d$~Now at Institute for Nuclear Research RAS, Moscow, Russia

$^*$~Corresponding author; e-mail: joerg.wotschack@cern.ch
\end{footnotesize}

\end{titlepage}


\newpage 

\section{Introduction}

The HARP experiment arose from the realization that the 
inclusive differential cross-sections of hadron production 
in the interactions of few GeV/{\it c} protons with nuclei were 
known only within a factor of two to three, while 
more precise cross-sections are in demand for several reasons. 
Consequently, the HARP detector was designed to carry 
out a programme of systematic and precise measurements of 
hadron production by protons and pions with momenta from 
1.5 to 15~GeV/{\it c}. 

The detector combined a forward spectrometer with a 
large-angle spectrometer. The latter comprised a 
cylindrical Time Projection 
Chamber (TPC) around the target and an array of 
Resistive Plate Chambers (RPCs) that surrounded the 
TPC. The purpose of the TPC was track 
reconstruction and particle identification by \dedx . The 
purpose of the RPCs was to complement the 
particle identification by time of flight.

The HARP experiment was performed at the CERN Proton Synchrotron 
in 2001 and 2002 with a set of stationary targets ranging 
from hydrogen to lead, including beryllium.

Here, we report on the large-angle production (polar angle $\theta$ in the 
range $20^\circ < \theta < 125^\circ$) 
of secondary protons and charged pions in 
the interactions with a 5\% $\lambda_{\rm abs}$ Be target
of protons and pions with beam momenta of $\pm3.0$,
$\pm5.0$, $\pm12.0$, and $\pm15.0$~GeV/{\it c}. We have reported
earlier~\cite{Beryllium1} on results from
beam momenta of $+8.9$ and $-8.0$~GeV/{\it c}.

Our work involves only the HARP large-angle spectrometer.  
The detector characteristics and our analysis algorithms have
been described in Ref.~\cite{Beryllium1}. 

The data analysis presented in this paper rests exclusively   
on the calibrations of the TPC and the RPCs that we,
the HARP--CDP group,  
published in Refs.~\cite{TPCpub} and \cite{RPCpub}.  
As discussed in Refs.~\cite{JINSTpub} and \cite{EPJCpub},
and summarized succinctly in
the Appendix of Ref.~\cite{Beryllium1},  
our calibrations 
disagree with 
those published by the `HARP 
Collaboration'~\cite{HARPTechnicalPaper,OffRPCPaper,
500pseffect,OffTPCcalibration}. 
Conclusions of independent review 
bodies on the discrepancies between our results and those from
the HARP Collaboration can be found in 
Refs.~\cite{CarliFuster,SPSCminutes}.

\section{The T9 proton and pion beams, and the target}

The protons and pions were delivered by
the T9 beam line in the East Hall of CERN's Proton Synchrotron.
This beam line supports beam momenta between 1.5 and 15~GeV/{\it c},
with a momentum bite $\Delta p/p \sim 1$\%.

Beam particle identification was provided for by two threshold  
Cherenkov counters, BCA and BCB, filled with nitrogen, and by time of 
flight over a flight path of 24.3~m. Table~\ref{beampartid} 
lists the beam instrumentation that was used at different
beam momenta for p/$\pi^+$
and for $\pi$/e separation. 

\begin{table}[h]
\caption{Beam instrumentation for p/$\pi^+$
and $\pi$/e separation}
\label{beampartid}
\begin{center}
\begin{tabular}{|c|c|c|c|}
\hline
Beam momentum [GeV/{\it c}] & p/$\pi^+$ separation & $\pi$/e separation \\
\hline
\hline
$\pm3.0$  & TOF                 & BCB (1.05 bar) \\
\hline
$\pm5.0$  & TOF                 & BCA (0.60 bar) \\
   & BCB (2.50 bar)      &                \\
\hline
$-8.0$ and $+8.9$   & BCA (1.25 bar)      &                \\
   & BCB (1.50 bar)      &                \\ 
\hline
$\pm12.0$ and $\pm15.0$ &  BCA (3.50 bar) &             \\
          &  BCB (3.50 bar) &             \\ 
\hline          
\end{tabular}
\end{center}
\end{table}

The pion beam had a contamination by muons from pion decays. 
It also had a contamination by electrons from converted
photons from $\pi^0$ decays. Only for the beam momenta of 3 and 
5~GeV/{\it c} were electrons identified by a beam Cherenkov
counter and rejected.

The fractions of muon and electron contaminations of the pion beam
were experimentally determined~\cite{T9beammuons,T9beamelectrons} 
and are listed in Table~\ref{pioncontaminations} for all beam 
momenta. For the determination of interaction cross-sections of pions, 
the muon and 
electron contaminations must be subtracted from 
the incoming flux of pion-like particles (except at the beam 
momenta of 3 and 5~GeV/{\it c}, where beam electrons were 
identified by the beam Cherenkov response and rejected).     
\begin{table}[h]
\caption{Contaminations of the pion beams by muons and electrons}
\label{pioncontaminations}
\begin{center}
\begin{tabular}{|c|c|c|c|}
\hline
Beam momentum [GeV/{\it c}] & Muon fraction & Electron fraction \\
\hline
$\pm3.0$   & $(4.1 \pm 0.4)$\% & rejected \\
$\pm5.0$   & $(5.1 \pm 0.4)$\% & rejected  \\
$-8.0$   & $(1.9 \pm 0.5)$\% & $(1.2 \pm 0.5)$\% \\
$+8.9$   & $(1.7 \pm 0.5)$\% & $(1.2 \pm 0.5)$\% \\
$\pm12$  & $(0.6 \pm 0.6)$\% & $(0.5 \pm 0.5)$\% \\
$\pm15$  & $(0.0 \pm 0.5)$\% & $(0.0 \pm 0.5)$\% \\
\hline
\end{tabular}
\end{center}
\end{table}

There is also a kaon contamination of a few per cent in the proton 
and pion beams. Because the kaon interaction cross-sections are
close to the proton and pion interaction cross-sections, this
contamination is ignored. 

The beam trajectory was determined by a set of three multiwire 
proportional chambers (MWPCs), located upstream of the target,
several metres apart. The transverse error of the 
projected impact point on the target was 0.5~mm from the 
resolution of the MWPCs, plus a
contribution from multiple scattering of the beam particles
in various materials in the beam line. Excluding the target itself, the 
latter contribution is 0.2~mm for a 8.9~GeV/{\it c} beam 
particle.

We select  
`good' beam particles by requiring the unambiguous reconstruction
of the particle trajectory with good $\chi^2$. In addition we 
require that the particle type is unambiguously identified. 
We select `good' accelerator spills by requiring minimal intensity and
a `smooth' variation of beam intensity across the 400~ms long 
spill\footnote{A smooth variation of beam intensity eases 
corrections for dynamic TPC track distortions.}.

The target was a cylinder made of 
high-purity (99.95\%) beryllium, with a density of 1.85~g/cm$^3$,
a radius of 15~mm, and a thickness of $20.5 \pm 0.1$~mm 
(5\% $\lambda_{\rm abs}$).

The finite thickness of the target leads to a
small attenuation of the number of incident beam particles. The
attenuation factor is $f_{\rm att} = 0.975$.

The size of the beam spot at the position of the target was several
millimetres in diameter, determined by the setting of the beam
optics and by multiple scattering. The nominal 
beam position\footnote{A 
right-handed Cartesian and/or spherical polar coordinate 
system is employed; the $z$ axis coincides with the beam line, with
$+z$ pointing downstream; the coordinate origin is at the 
centre of the beryllium target, 500~mm
downstream of the TPC's pad plane; 
looking downstream, the $+x$ coordinate points to
the left and the $+y$ coordinate points up; the polar angle
$\theta$ is the angle with respect to the $+z$ axis.} 
was at $x_{\rm beam} = y_{\rm beam} = 0$, however, excursions 
by several millimetres
could occur\footnote{The only relevant issue is that the trajectory
of each individual beam particle is known, whether shifted or not, 
and therefore the amount of matter to be traversed by the 
secondary hadrons.}. 
A loose fiducial cut 
$\sqrt{x^2_{\rm beam} + y^2_{\rm beam}} < 12$~mm
ensured full beam acceptance. The muon and electron 
contaminations of the 
pion beam, stated above, refer to this acceptance cut.

\section{The HARP large-angle detectors}

Our calibration work on the HARP TPC and RPCs
is described in detail in Refs.~\cite{TPCpub} and \cite{RPCpub},
and in references cited therein. In particular, we recall that 
static and dynamic TPC track distortions up to 10~mm have been 
corrected to better than 300~$\mu$m. Therefore, TPC track 
distortions do not affect the precision of our cross-section
measurements.  

The resolution $\sigma (1/p_{\rm T})$
is typically 20\% and worsens towards small relative particle
velocity $\beta$ and small polar angle $\theta$.

The absolute momentum scale is determined to be correct to 
better than 2\%, both for positively and negatively
charged particles.
 
The polar angle $\theta$ is measured in the TPC with a 
resolution of $\sim$9~mrad, for a representative 
angle of $\theta = 60^\circ$. To this a multiple scattering
error has to be added which is $\sim$7~mrad for a proton with 
$p_{\rm T} = 500$~MeV/{\it c} and $\theta = 60^\circ$, and 
$\sim$4~mrad for a pion with the same characteristics.
The polar-angle scale is correct to better than 2~mrad.     

The TPC measures \dedx\ with a resolution of 16\% for a 
track length of 300~mm.

The intrinsic efficiency of the RPCs that surround 
the TPC is better than 98\%.

The intrinsic time resolution of the RPCs is 127~ps and
the system time-of-flight resolution (that includes the
jitter of the arrival time of the beam particle at the target)
is 175~ps. 

To separate measured particles into species, we
assign on the basis of \dedx\ and $\beta$ to each particle a 
probability of being a proton,
a pion (muon), or an electron, respectively. The probabilities
add up to unity, so that the number of particles is conserved.
These probabilities are used for weighting when entering 
tracks into plots or tables.

\section{Monte Carlo simulation}

We used the Geant4 tool kit~\cite{Geant4} for the simulation 
of the HARP large-angle spectrometer.

We had expected that
Geant4 would provide us with 
reasonably realistic spectra of secondary hadrons. 
We found this expectation 
met by Geant4's so-called QGSP\_BIC physics list, but only
for the secondaries from incoming beam protons with momentum
less than 12~GeV/{\it c}.
For the secondaries from beam protons at 12 and 15~GeV/{\it c}
momentum, and from beam pions at all momenta, we found the standard 
physics lists of Geant4 unsuitable~\cite{GEANTpub}. 

To overcome this problem,
we built our own HARP\_CDP physics list
for the production of secondaries from incoming beam pions. 
It starts from Geant4's standard QBBC physics list, 
but the Quark--Gluon String Model is replaced by the 
FRITIOF string fragmentation model for
kinetic energy $E>6$~GeV; for $E<6$~GeV, the Bertini 
Cascade is used for pions, and the Binary Cascade for protons; 
elastic and quasi-elastic scattering is disabled.
Examples of the good performance of the HARP\_CDP physics list
are given in Ref.~\cite{GEANTpub}.

\section{Systematic errors}

The systematic precision of our inclusive cross-sections 
is at the few-per-cent level, from errors
in the normalization, in the momentum measurement, in
particle identification, and in the corrections applied
to the data.

The systematic error of the absolute flux normalization is 
taken as 2\%. This error arises from uncertainties in the
target thickness, in the contribution of large-angle 
scattering of beam particles, in the attenuation of beam particles in the target, and in the subtraction of
the muon and electron contaminations. Another contribution comes from the removal of events with an abnormally large number of TPC hits above threshold.

The systematic error of the track finding  
efficiency is taken as 1\% which reflects differences 
between results from different persons who conducted
eyeball scans. We also take the statistical errors of
the parameters of a fit to scan results  
as systematic error into account~\cite{Beryllium1}.
The systematic error of the correction 
for losses from the requirement of at least 10 TPC clusters 
per track is taken as 20\% of the correction which 
itself is in the range of 5 to 30\%. This estimate arose
from differences between the four TPC sectors that
were used in our analysis, and from the observed 
variations with time. 

The systematic error of the $p_{\rm T}$ scale is taken as
2\% as discussed in Ref.~\cite{TPCpub}.

The systematic errors of the proton, pion, and electron
abundances are taken as 10\%. We stress that errors on 
abundances only lead to cross-section errors in case of a strong overlap of the resolution functions
of both identification variables, \dedx\ and $\beta$. 
The systematic error of the correction for migration, absorption
of secondary protons and pions in materials, and for pion
decay into muons, is taken as 20\% of the correction, or 1\% of the cross-section, whichever is larger. These estimates reflect our experience 
with remanent differences between data and Monte Carlo 
simulations after weighting Monte Carlo events with smooth functions 
with a view to reproducing the data simultaneously in 
several variables in the best possible way.

All systematic errors are propagated into the momentum 
spectra of secondaries and then added in quadrature.

\section{Cross-section results}

In Tables~\ref{pro.probe3}--\ref{pim.pimbe15}, collated
in the Appendix of this paper, we give
the double-differential inclusive cross-sections 
${\rm d}^2 \sigma / {\rm d} p {\rm d} \Omega$
for all 36 combinations of
incoming beam particle and secondary particle, including
statistical and systematic errors. In each bin,  
the average momentum and the average polar angle are also given.

The data of Tables~\ref{pro.probe3}--\ref{pim.pimbe15} are available in ASCII format in Ref.~\cite{ASCIItables}.

Cross-sections are only given if the total error is not larger than the cross-section itself.
Since our track reconstruction algorithm is optimized for
tracks with $p_{\rm T}$ above $\sim$70~MeV/{\it c} in the
TPC volume, we do not give cross-sections from tracks with $p_{\rm T}$ 
below this value.
Because of the absorption of slow protons in the material between the
vertex and the TPC gas, and with a view to keeping the correction
for absorption losses below 30\%, cross-sections from protons are 
limited to $p > 350$~MeV/{\it c} at the interaction vertex. 
Proton cross-sections are also not given if a 10\% error on the proton energy loss in materials between the interaction vertex and the TPC volume leads to a momentum change larger than 2\%.
Pion cross-sections are not given if pions are separated from protons by less than twice the time-of-flight resolution.

The larger than usual error bars for the +15~GeV/{\it c} pion beam
are caused by scarce statistics because the beam
composition was dominated by protons.

\begin{figure}[h]
\begin{center}
\begin{tabular}{cc}
\includegraphics[height=0.30\textheight]{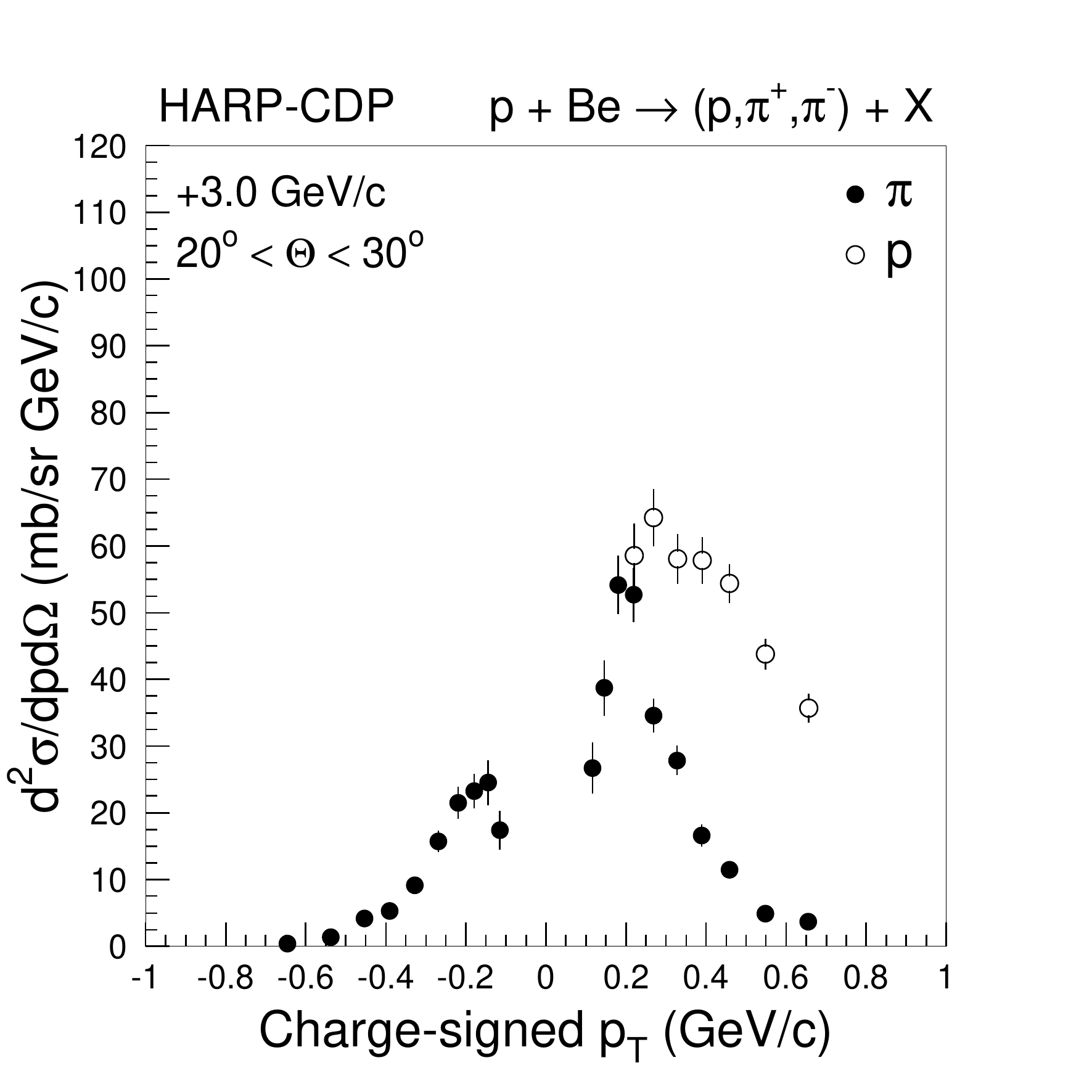} &
\includegraphics[height=0.30\textheight]{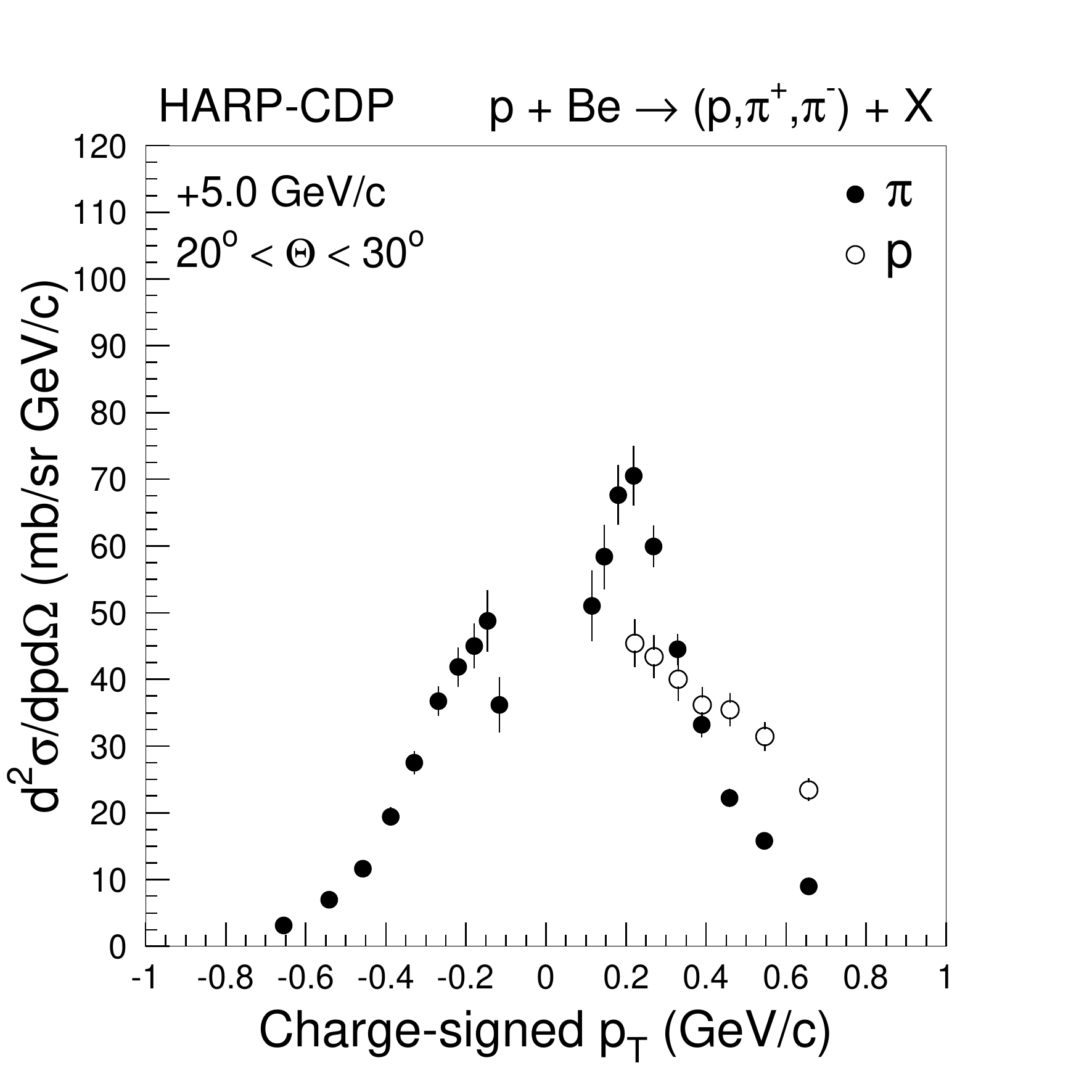} \\
\includegraphics[height=0.30\textheight]{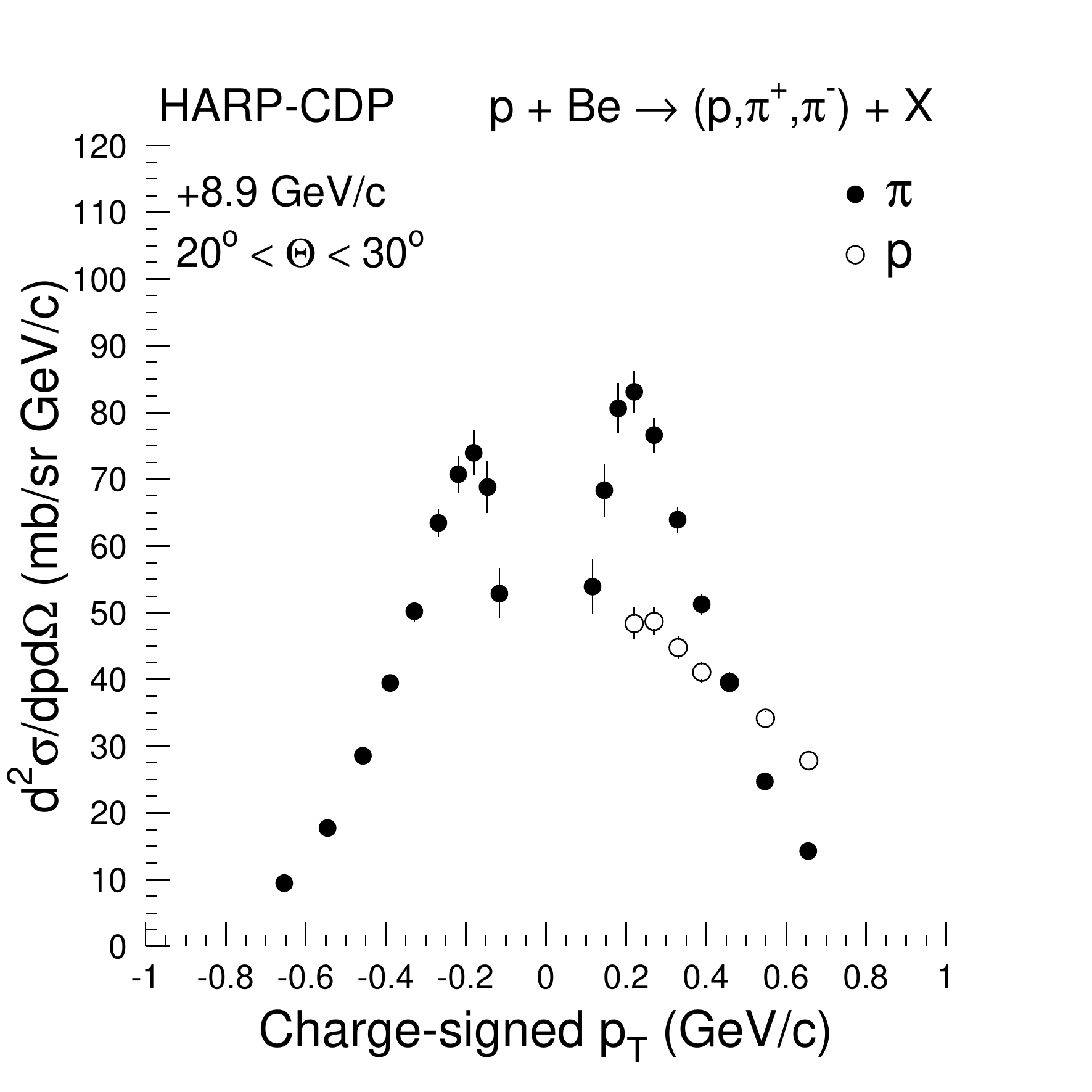} &
\includegraphics[height=0.30\textheight]{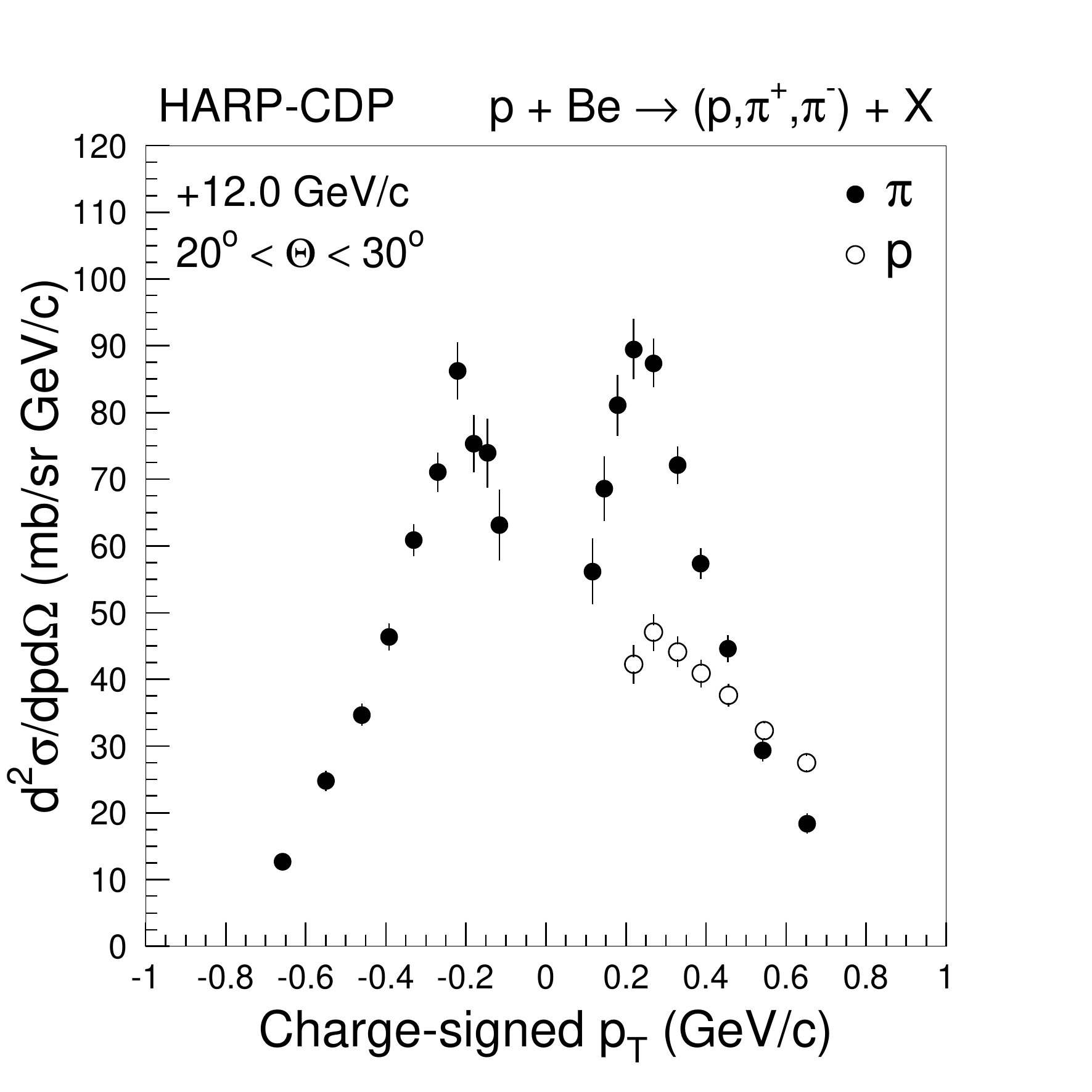} \\
\includegraphics[height=0.30\textheight]{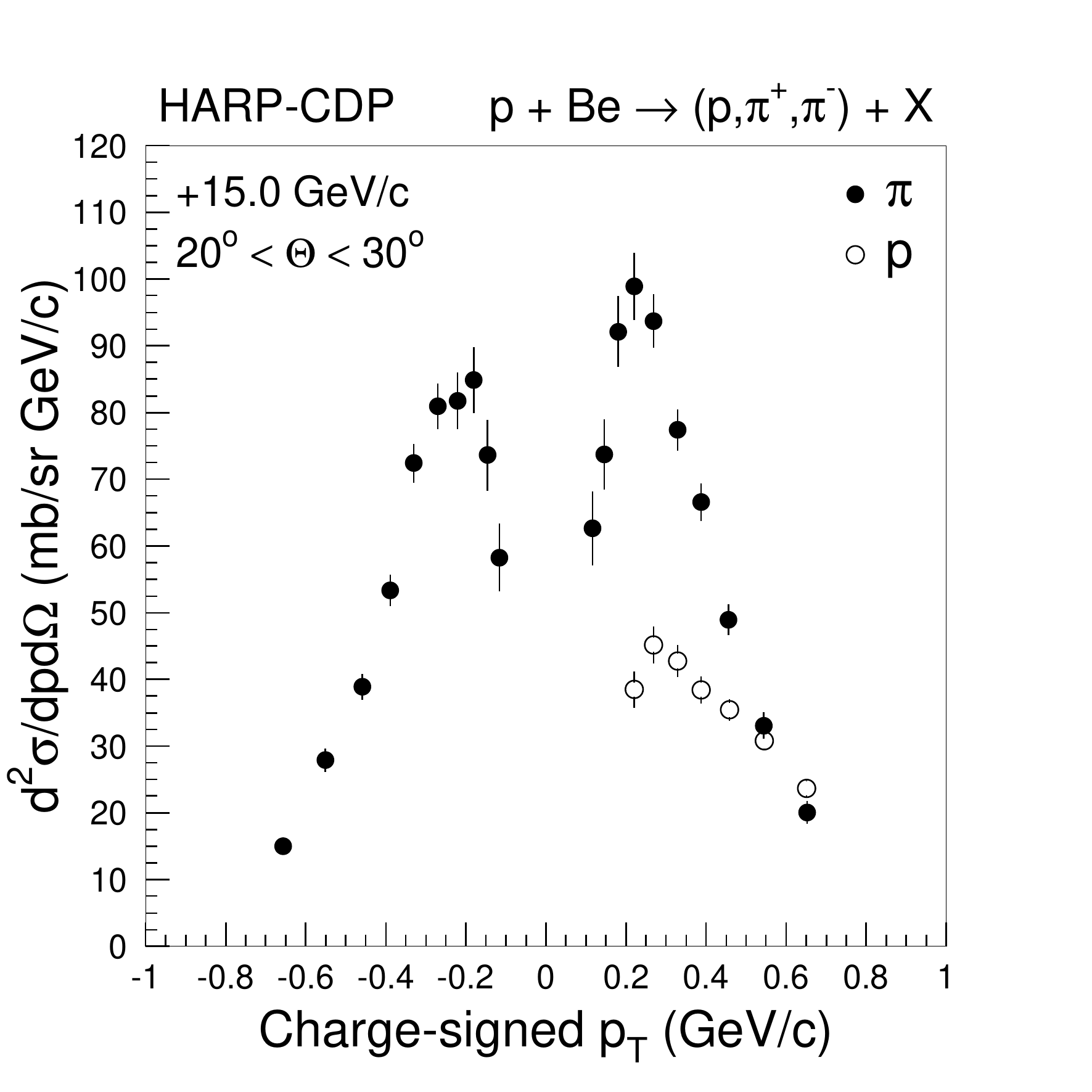} &  \\
\end{tabular}
\caption{Inclusive cross-sections of the production of secondary
protons, $\pi^+$'s, and $\pi^-$'s, by protons on beryllium nuclei, 
in the polar-angle range $20^\circ < \theta < 30^\circ$, for
different proton beam momenta, as a function of the charge-signed 
$p_{\rm T}$ of the secondaries; the shown errors are total errors.} 
\label{xsvsmompro}
\end{center}
\end{figure}

\begin{figure}[h]
\begin{center}
\begin{tabular}{cc}
\includegraphics[height=0.30\textheight]{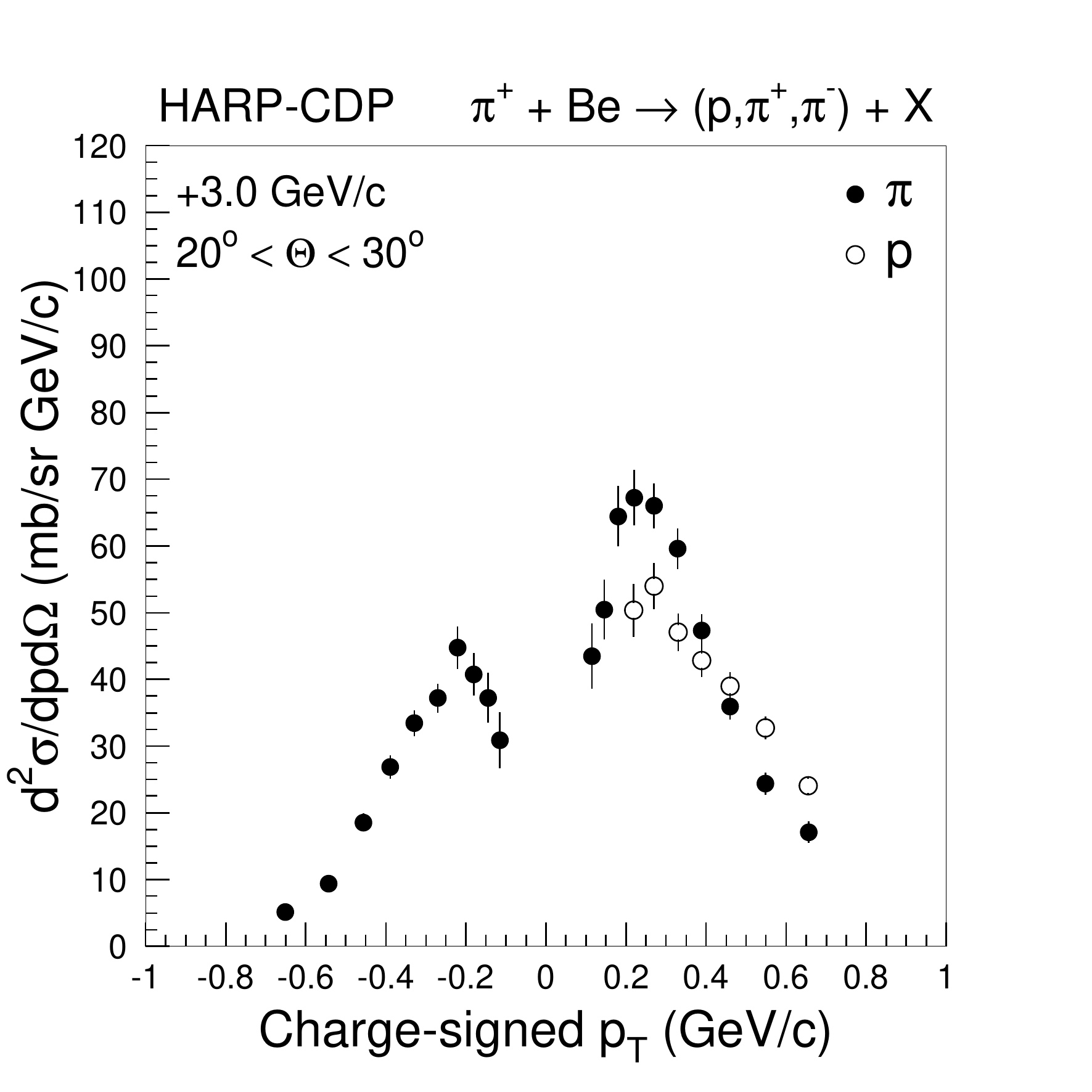} &
\includegraphics[height=0.30\textheight]{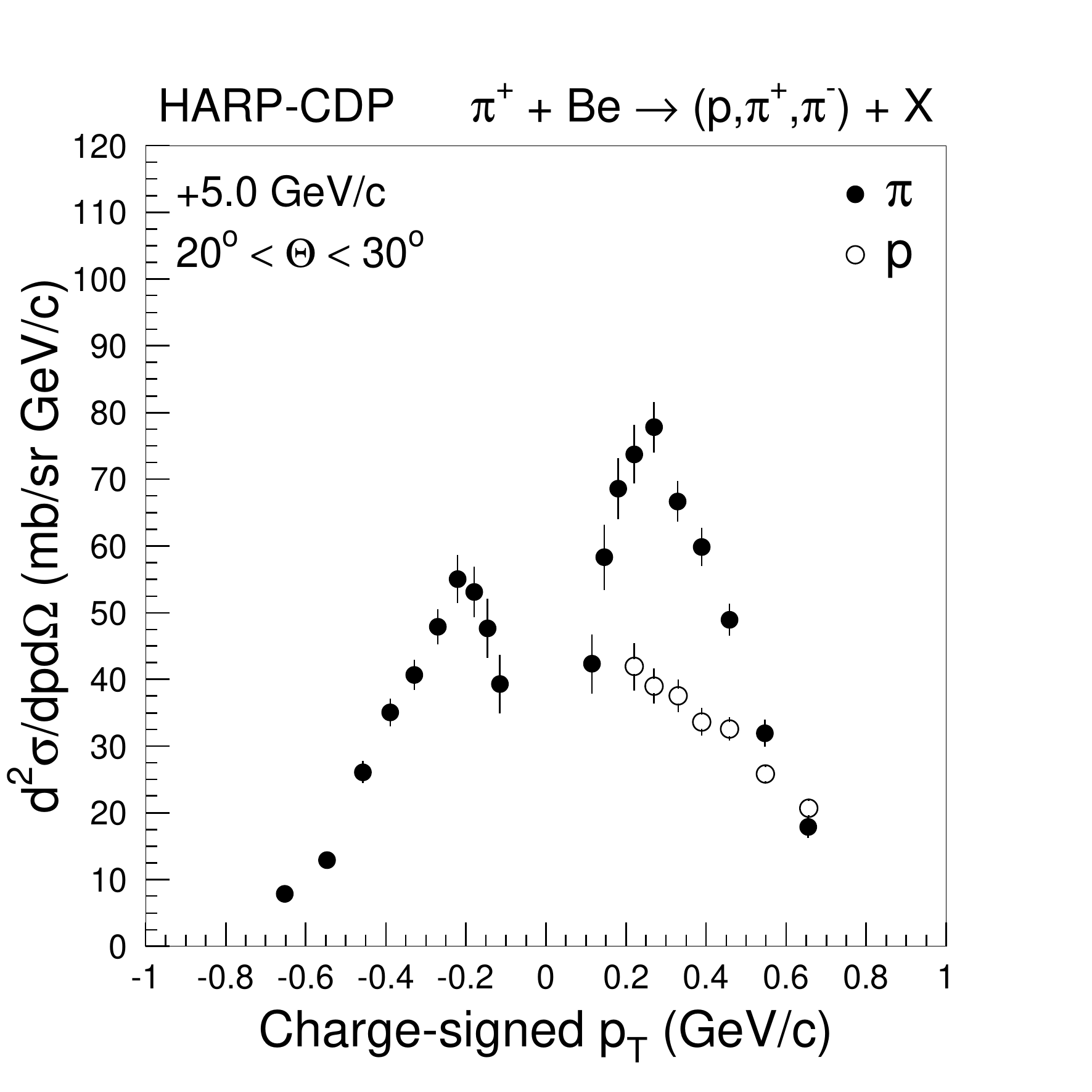} \\
\includegraphics[height=0.30\textheight]{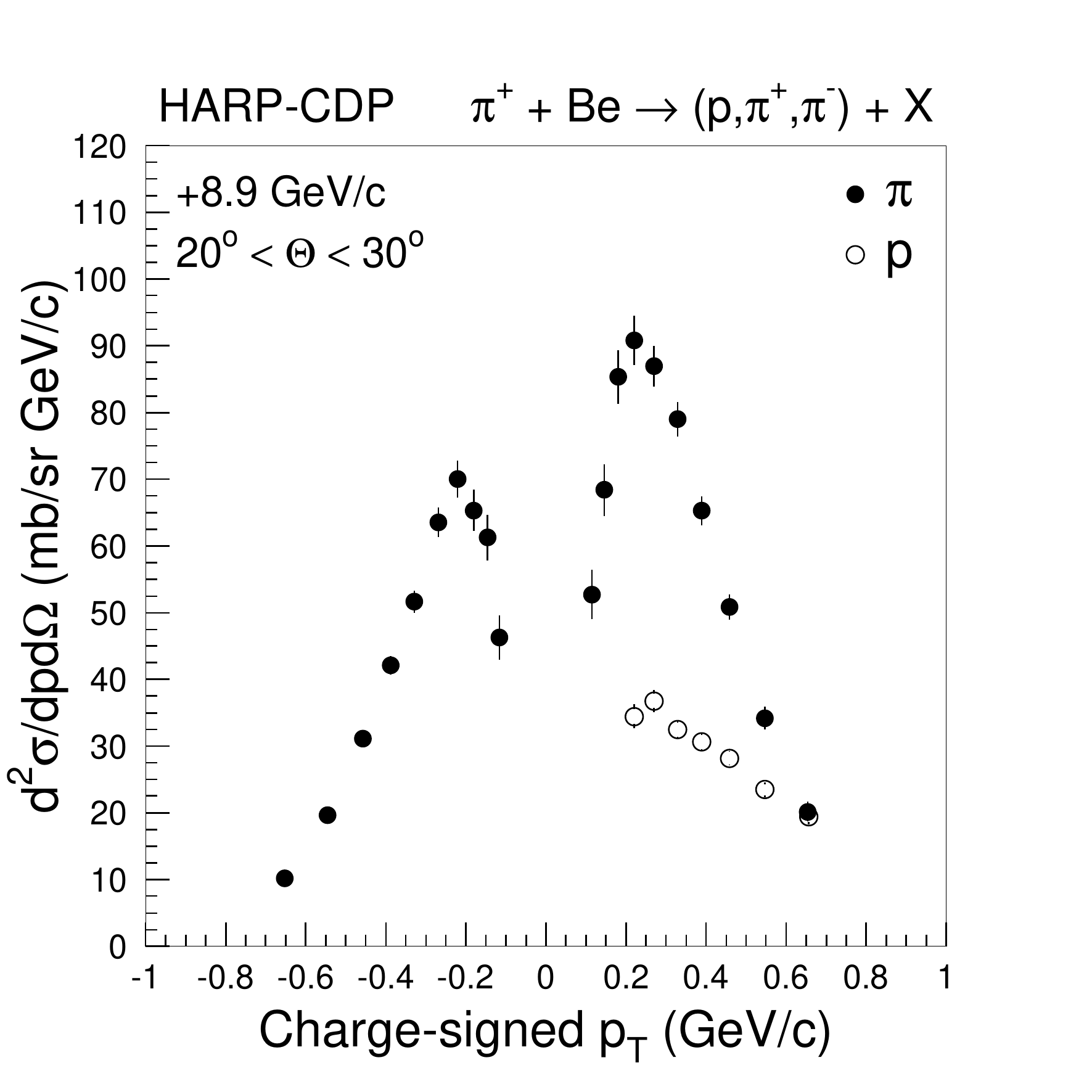} &
\includegraphics[height=0.30\textheight]{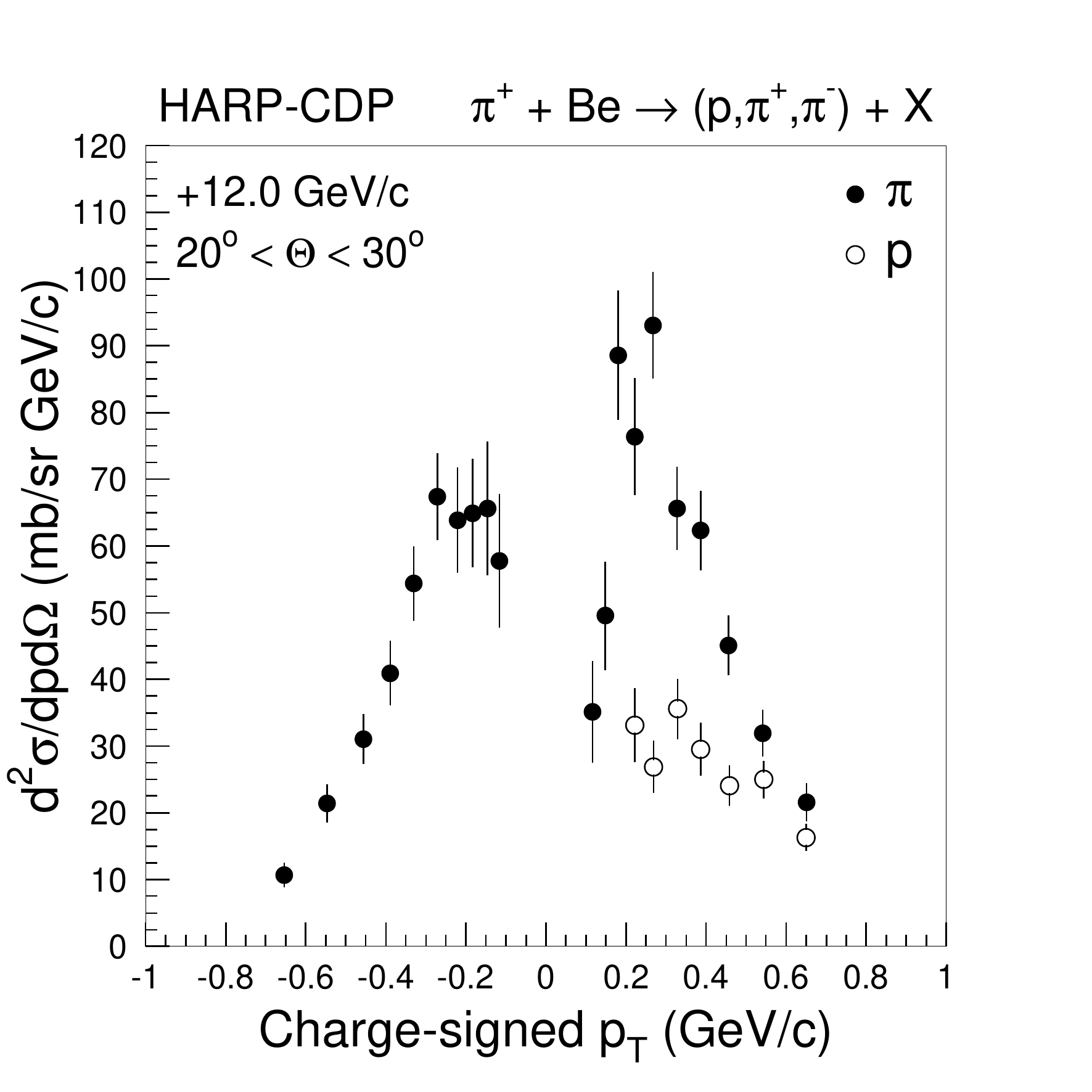} \\
\includegraphics[height=0.30\textheight]{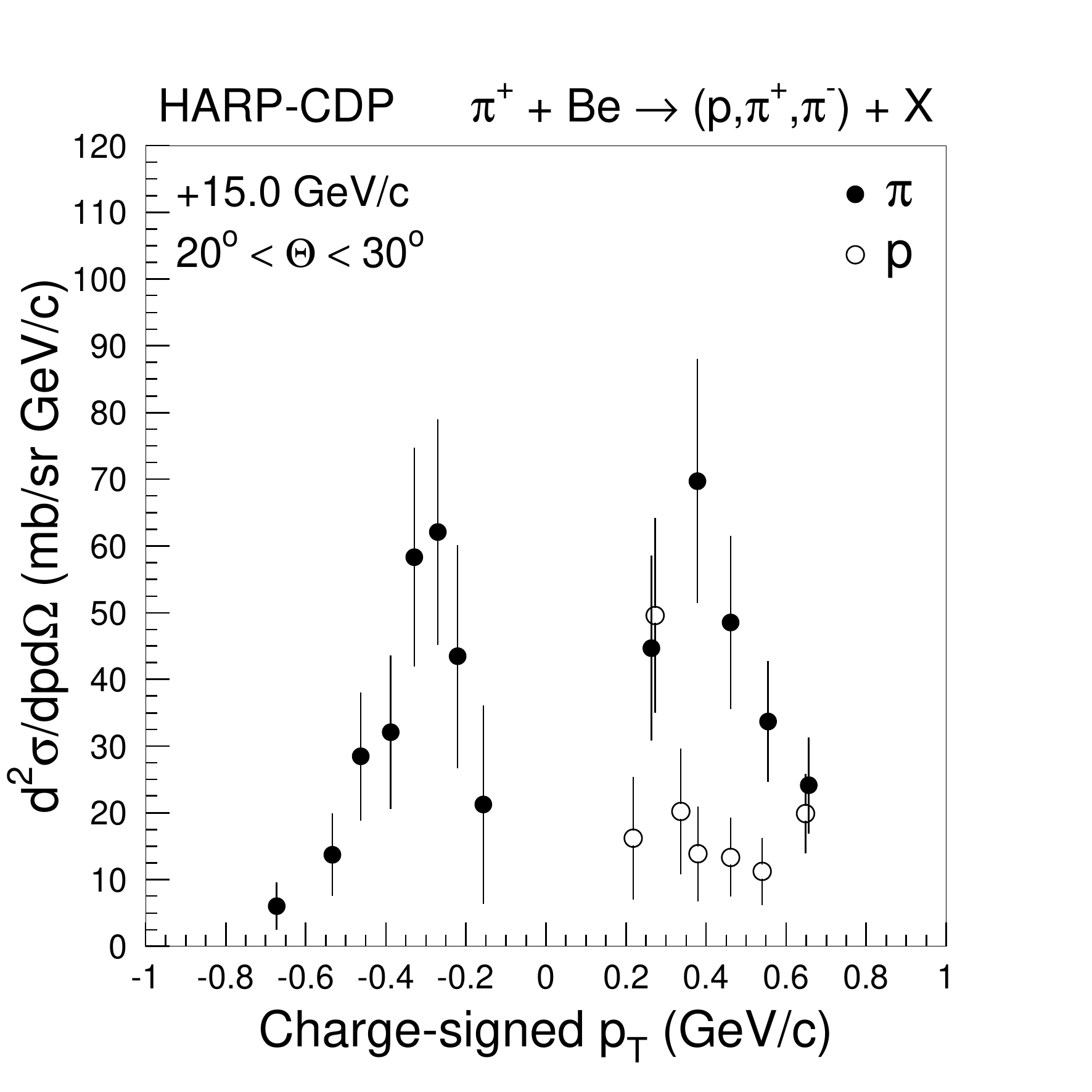} &  \\
\end{tabular}
\caption{Inclusive cross-sections of the production of secondary
protons, $\pi^+$'s, and $\pi^-$'s, by $\pi^+$'s on beryllium nuclei, 
in the polar-angle range $20^\circ < \theta < 30^\circ$, for
different proton beam momenta, as a function of the charge-signed 
$p_{\rm T}$ of the secondaries; the shown errors are total errors.}  
\label{xsvsmompip}
\end{center}
\end{figure}

\begin{figure}[h]
\begin{center}
\begin{tabular}{cc}
\includegraphics[height=0.30\textheight]{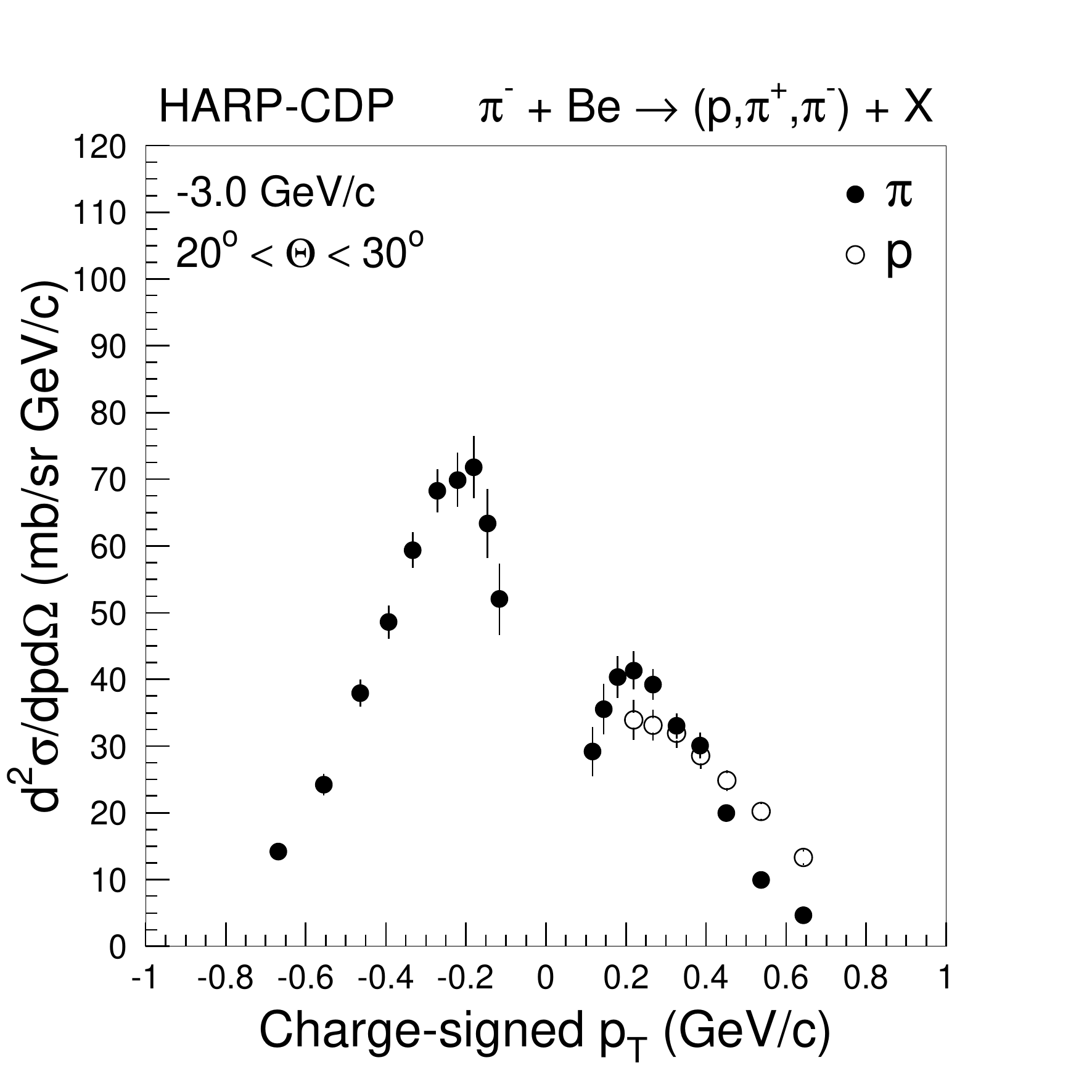} &
\includegraphics[height=0.30\textheight]{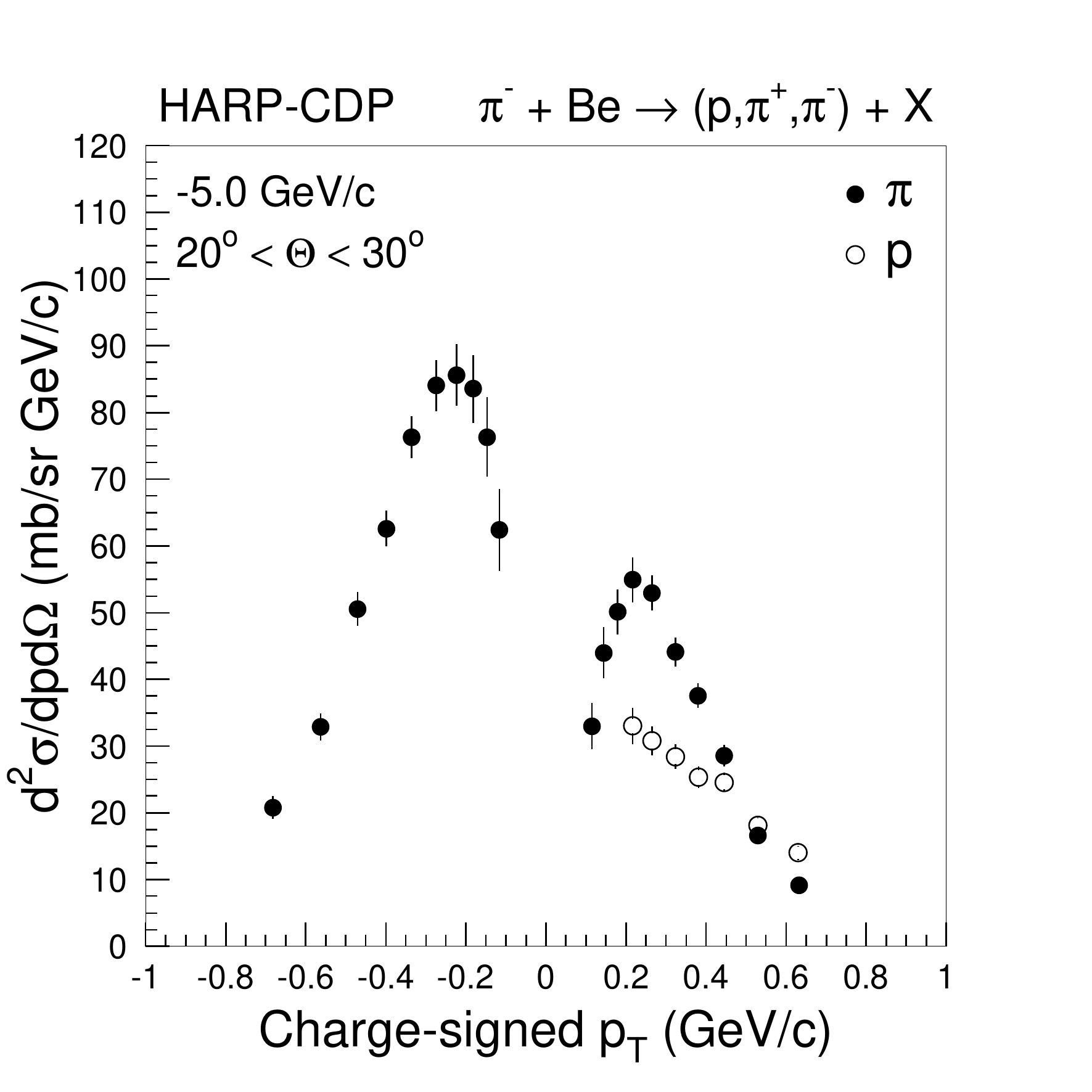} \\
\includegraphics[height=0.30\textheight]{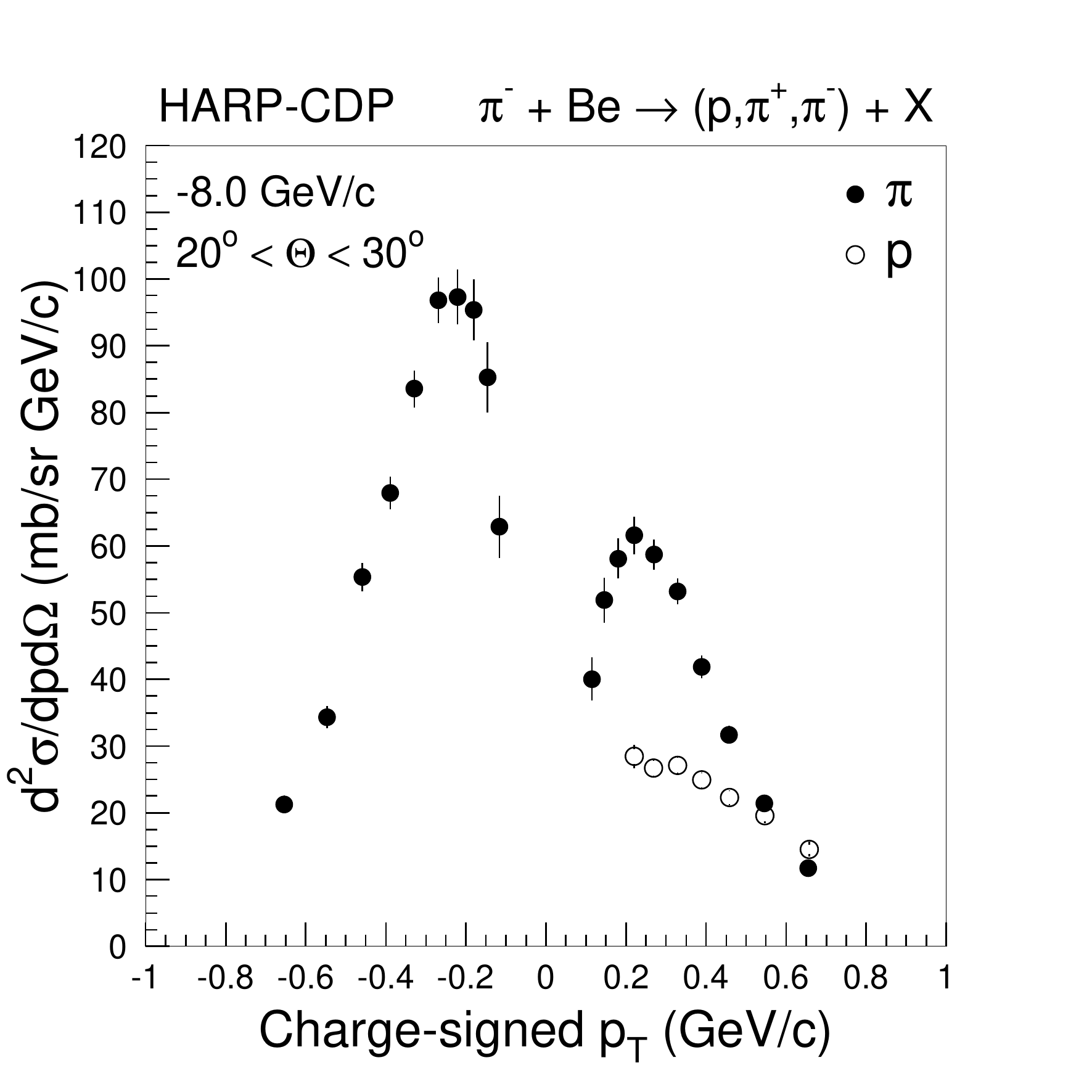} &
\includegraphics[height=0.30\textheight]{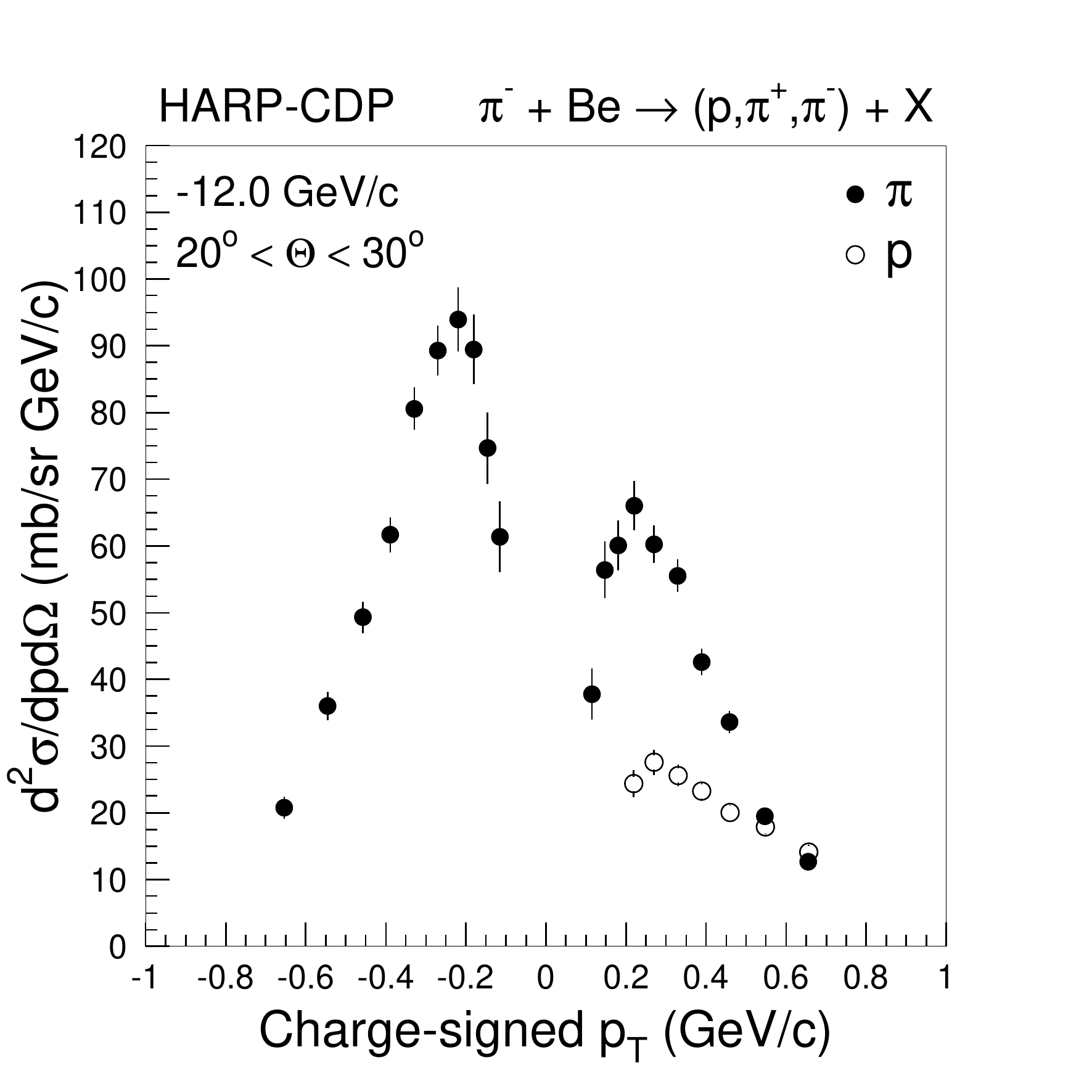} \\
\includegraphics[height=0.30\textheight]{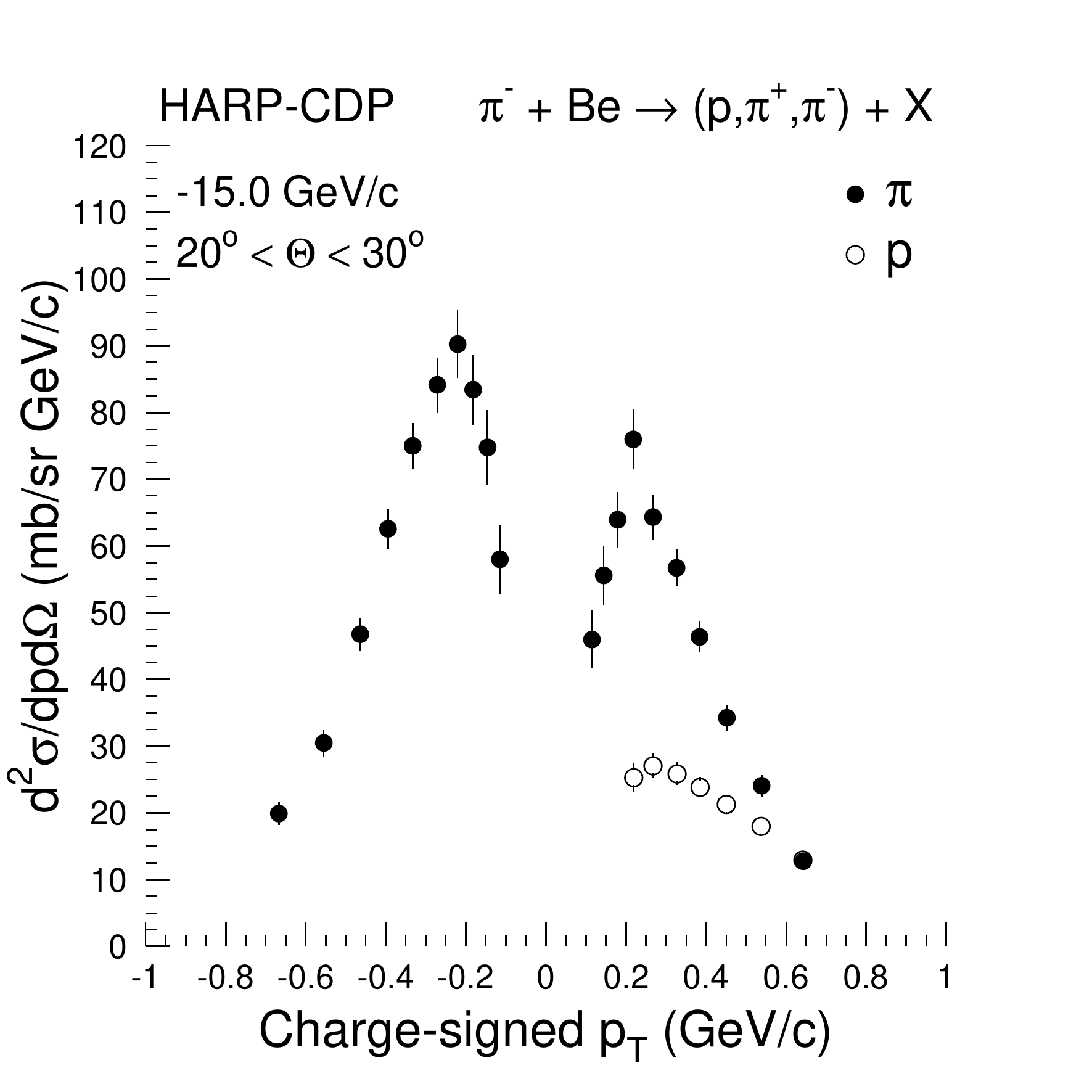} &  \\
\end{tabular}
\caption{Inclusive cross-sections of the production of secondary
protons, $\pi^+$'s, and $\pi^-$'s, by $\pi^-$'s on beryllium nuclei, 
in the polar-angle range $20^\circ < \theta < 30^\circ$, for
different proton beam momenta, as a function of the charge-signed 
$p_{\rm T}$ of the secondaries; the shown errors are total errors.} 
\label{xsvsmompim}
\end{center}
\end{figure}

\begin{figure}[h]
\begin{center}
\begin{tabular}{cc}
\includegraphics[height=0.30\textheight]{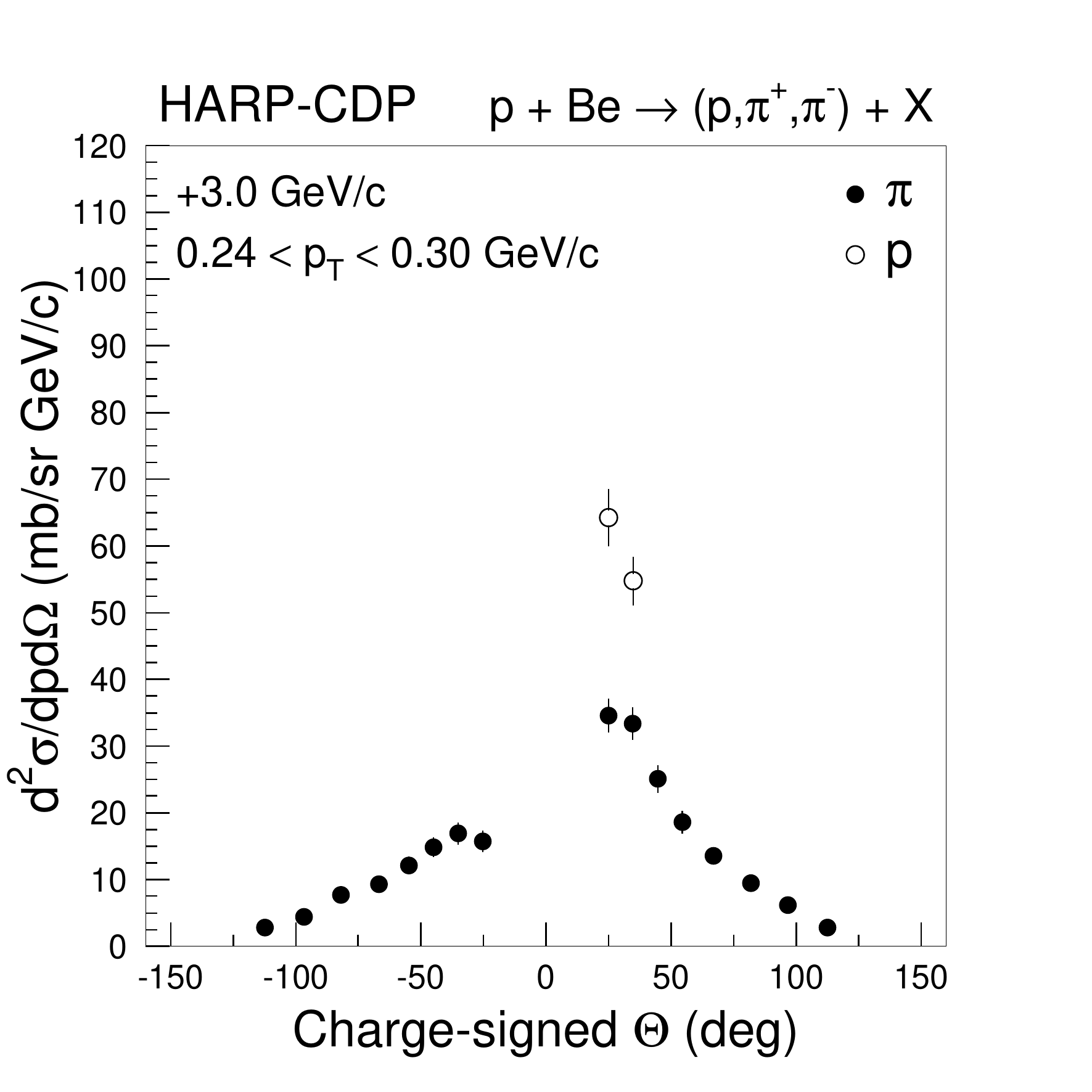} &
\includegraphics[height=0.30\textheight]{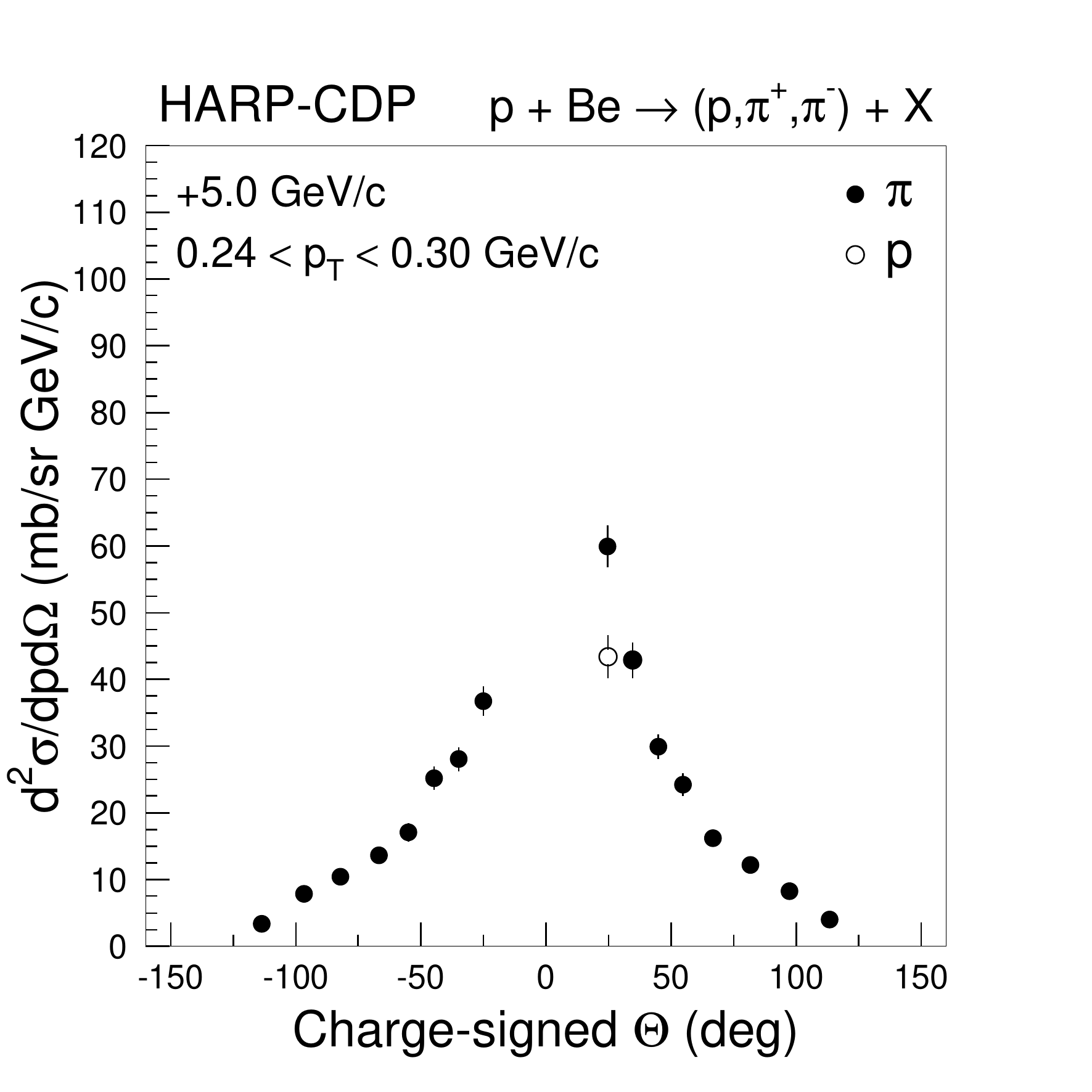} \\
\includegraphics[height=0.30\textheight]{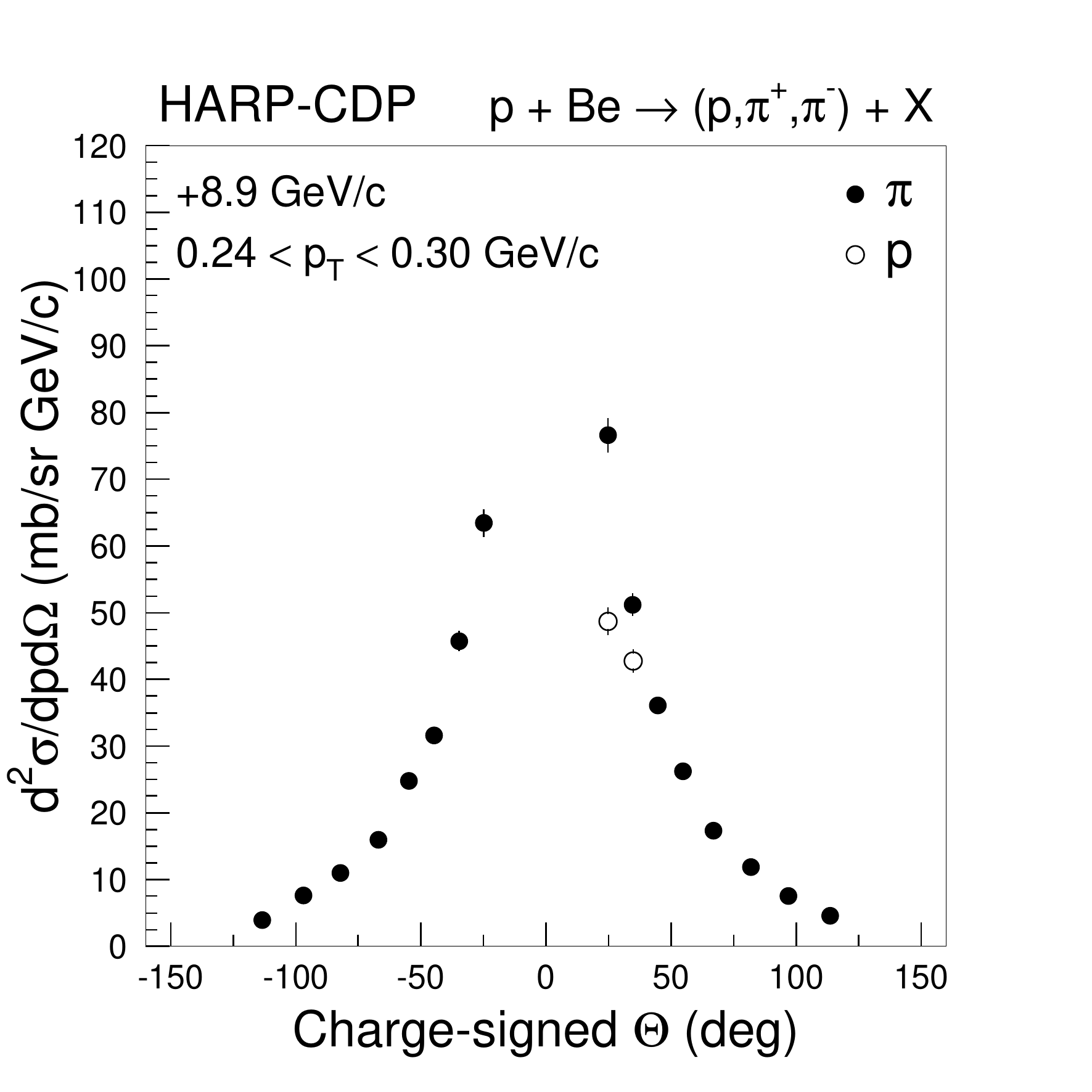} &
\includegraphics[height=0.30\textheight]{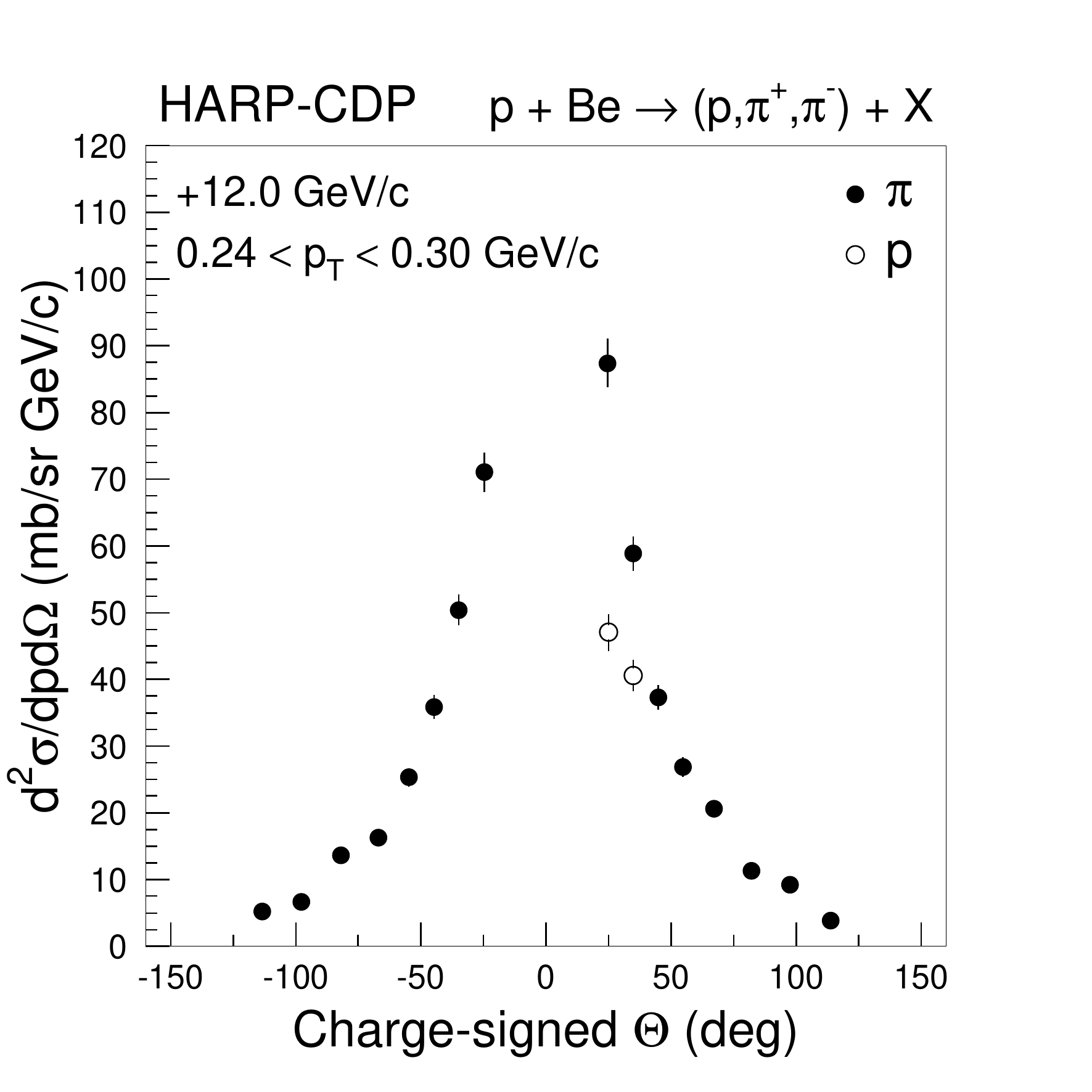} \\
\includegraphics[height=0.30\textheight]{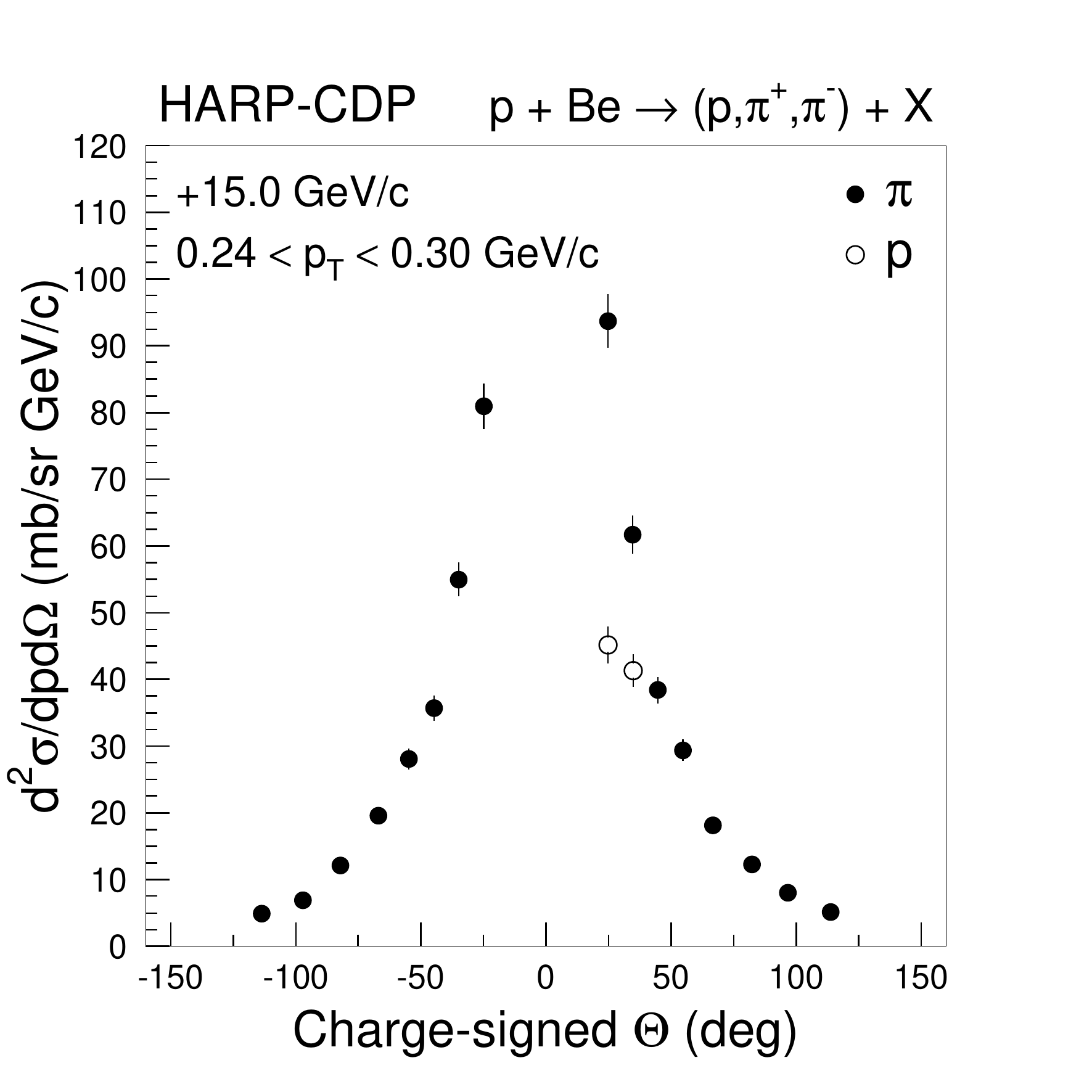} &  \\
\end{tabular}
\caption{Inclusive cross-sections of the production of secondary
protons, $\pi^+$'s, and $\pi^-$'s, with $p_{\rm T}$ in the range 
0.24--0.30~GeV/{\it c}, by protons on beryllium nuclei, for
different proton beam momenta, as a function of the charge-signed 
polar angle $\theta$ of the secondaries; the shown errors are 
total errors.} 
\label{xsvsthetapro}
\end{center}
\end{figure}

\begin{figure}[h]
\begin{center}
\begin{tabular}{cc}
\includegraphics[height=0.30\textheight]{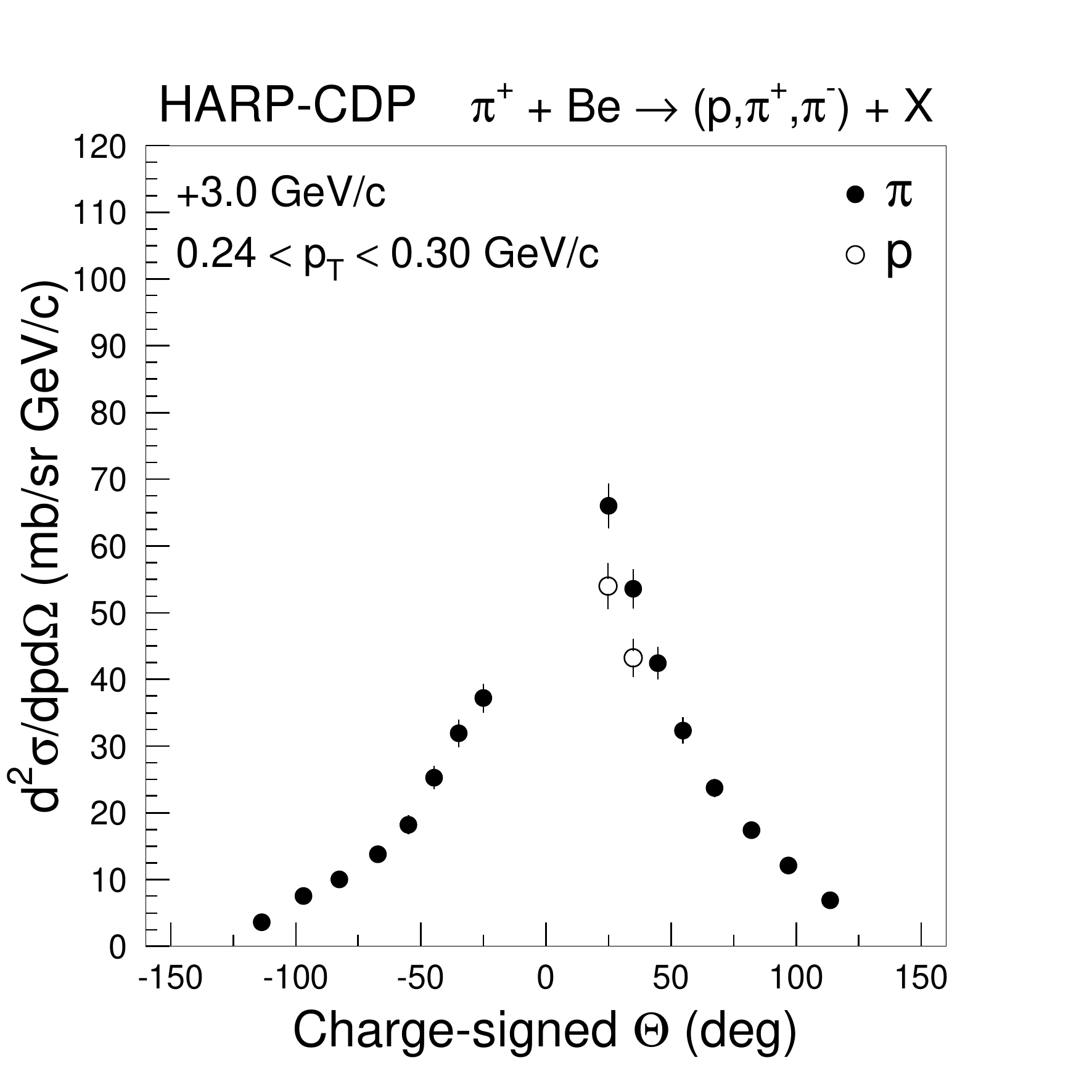} &
\includegraphics[height=0.30\textheight]{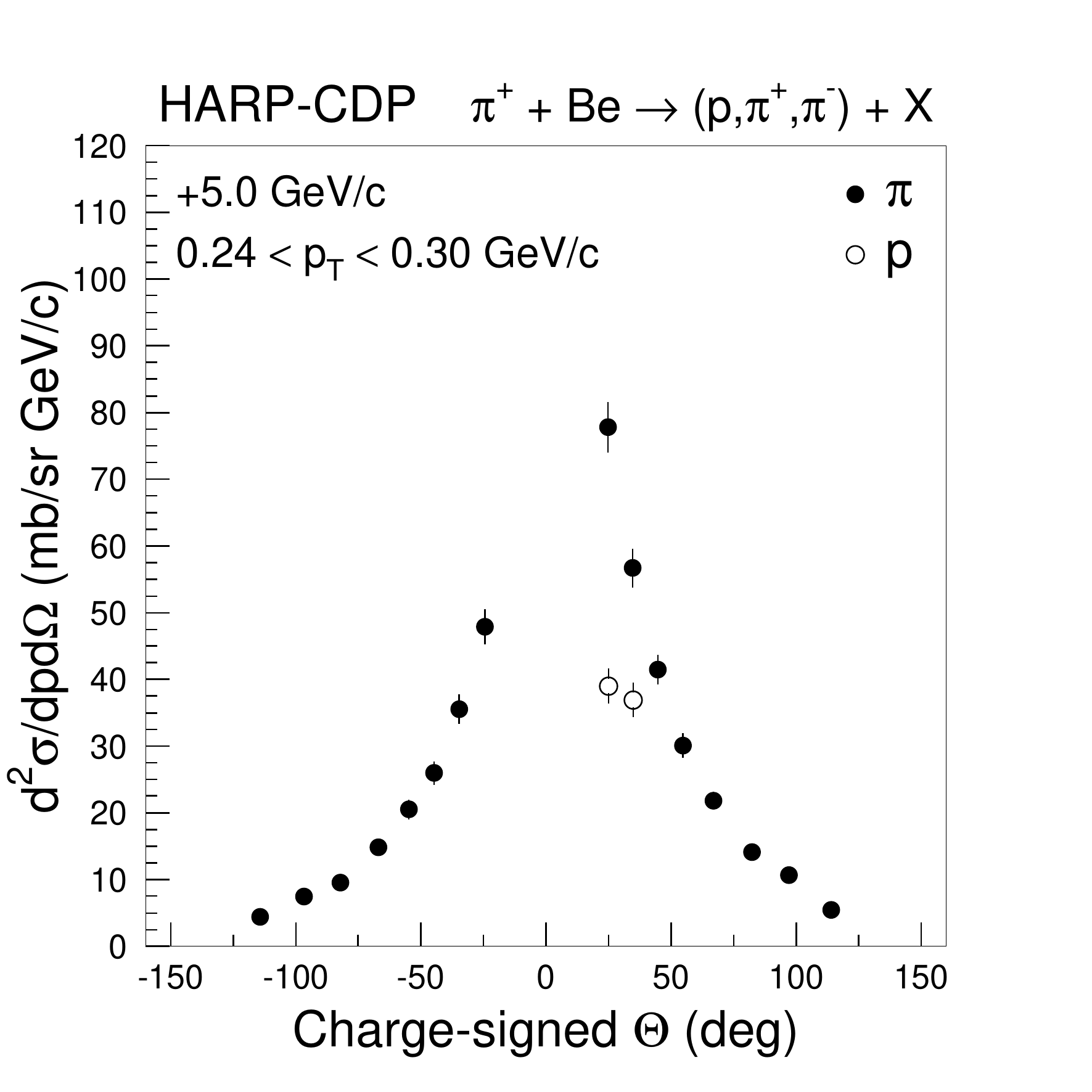} \\
\includegraphics[height=0.30\textheight]{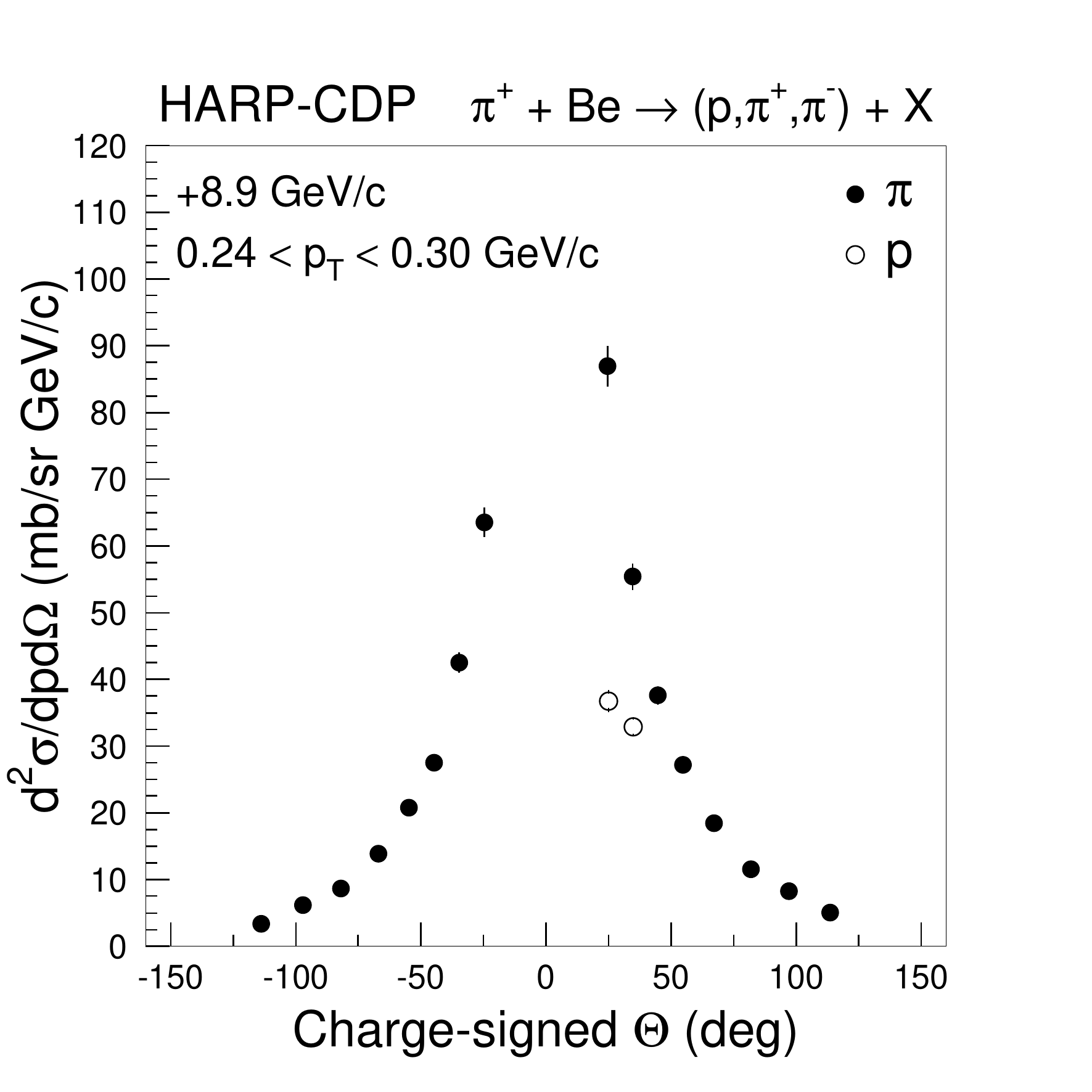} &
\includegraphics[height=0.30\textheight]{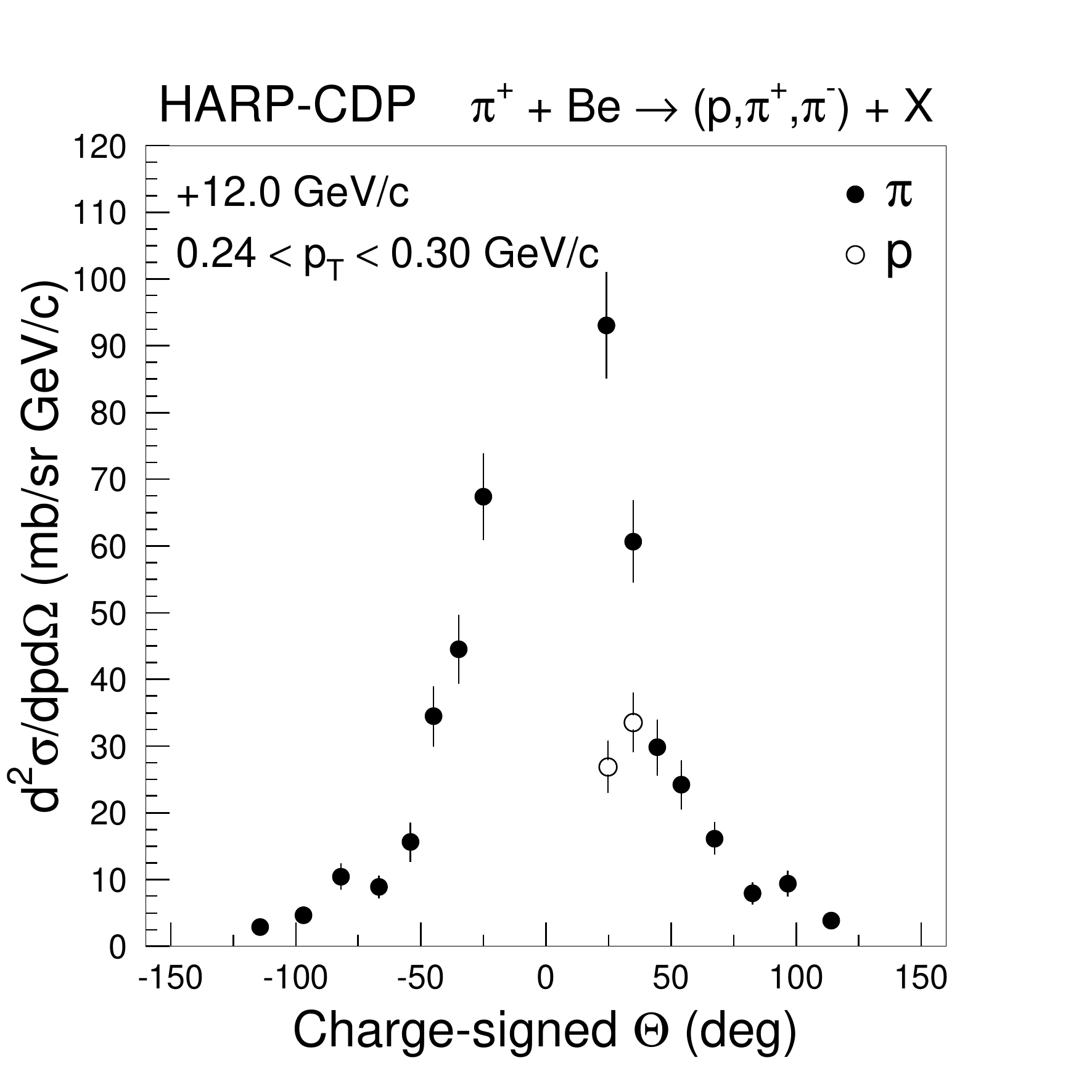} \\
\includegraphics[height=0.30\textheight]{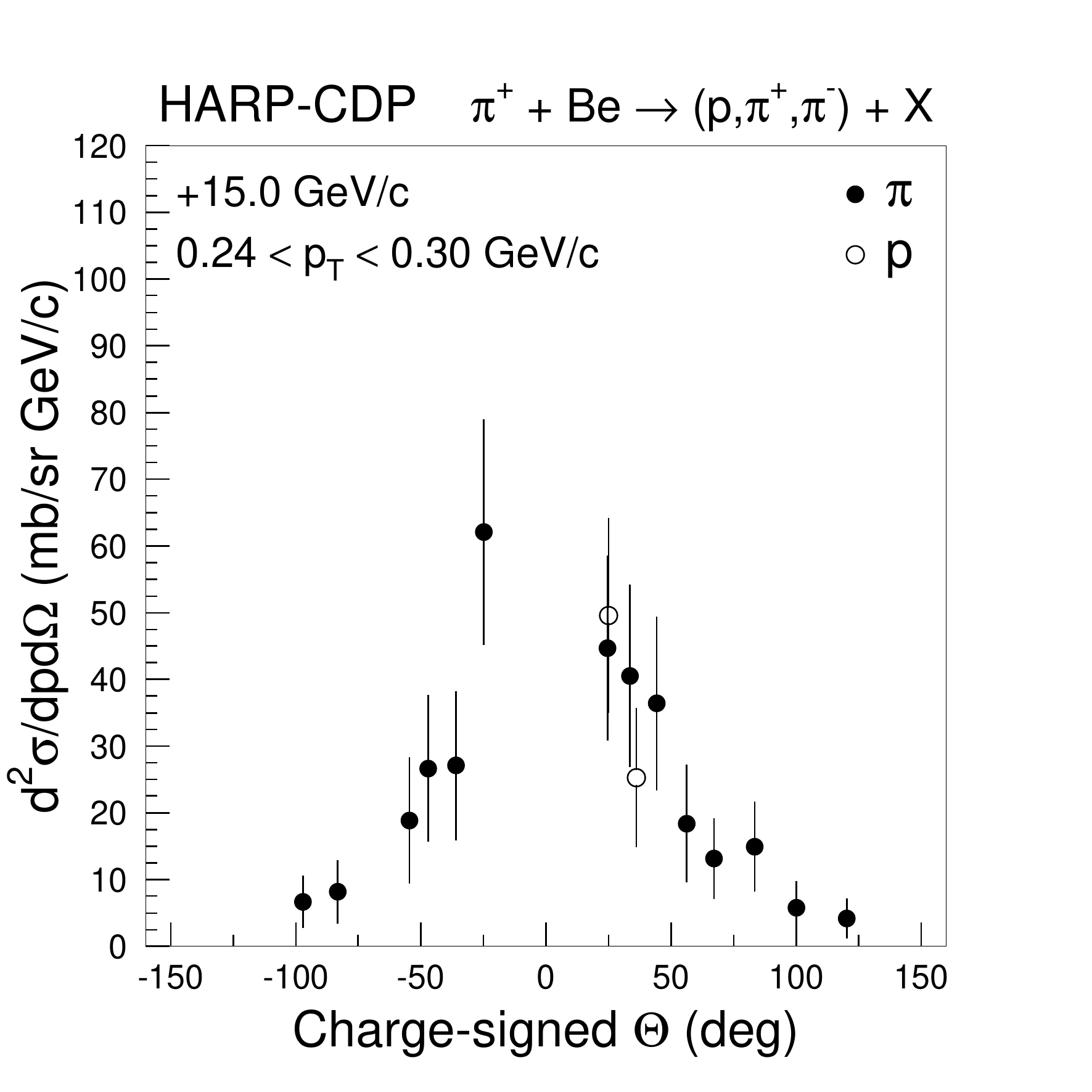} &  \\
\end{tabular}
\caption{Inclusive cross-sections of the production of secondary
protons, $\pi^+$'s, and $\pi^-$'s, with $p_{\rm T}$ in the range 
0.24--0.30~GeV/{\it c}, by $\pi^+$'s on beryllium nuclei, for
different proton beam momenta, as a function of the charge-signed 
polar angle $\theta$ of the secondaries; the shown errors are 
total errors.}  
\label{xsvsthetapip}
\end{center}
\end{figure}

\begin{figure}[h]
\begin{center}
\begin{tabular}{cc}
\includegraphics[height=0.30\textheight]{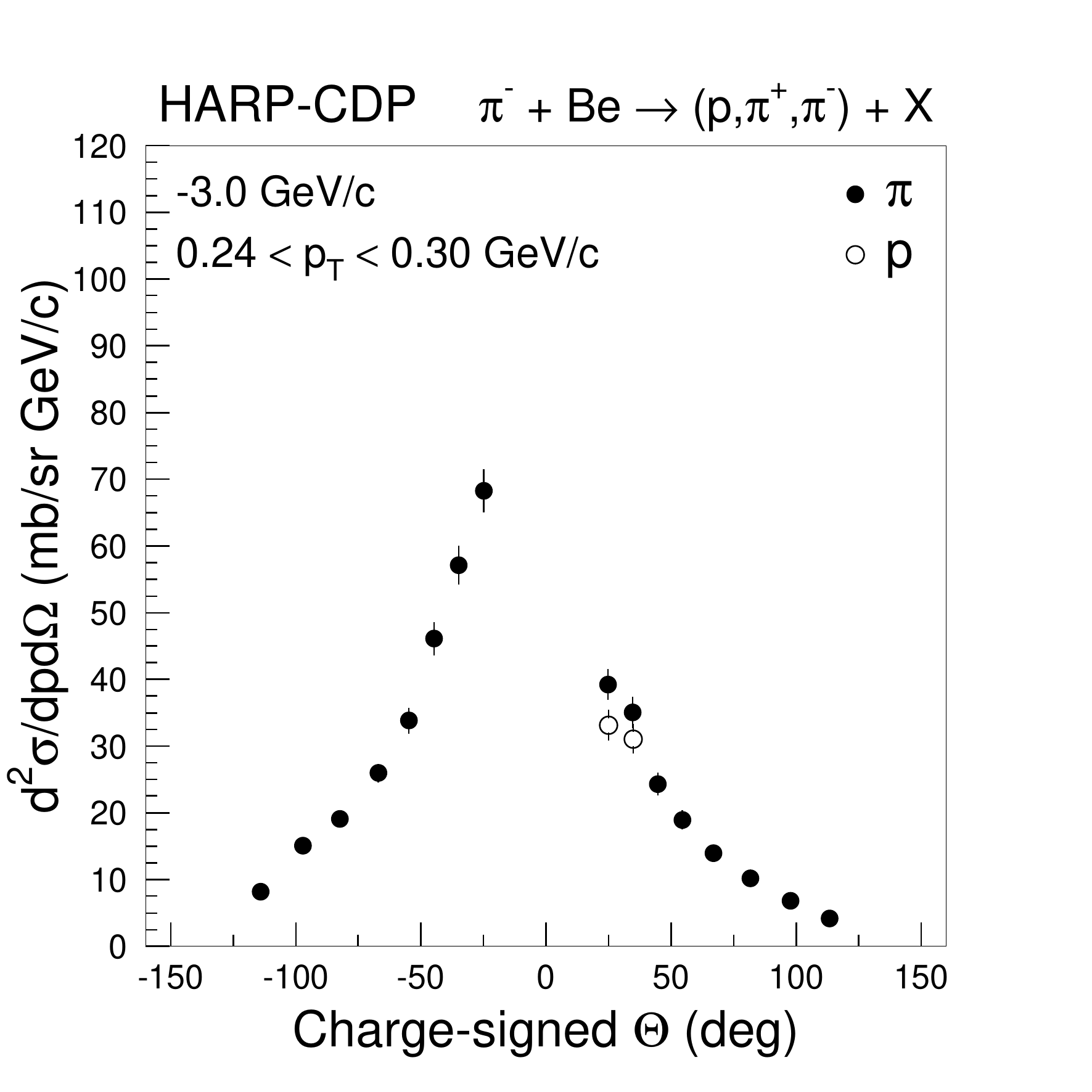} &
\includegraphics[height=0.30\textheight]{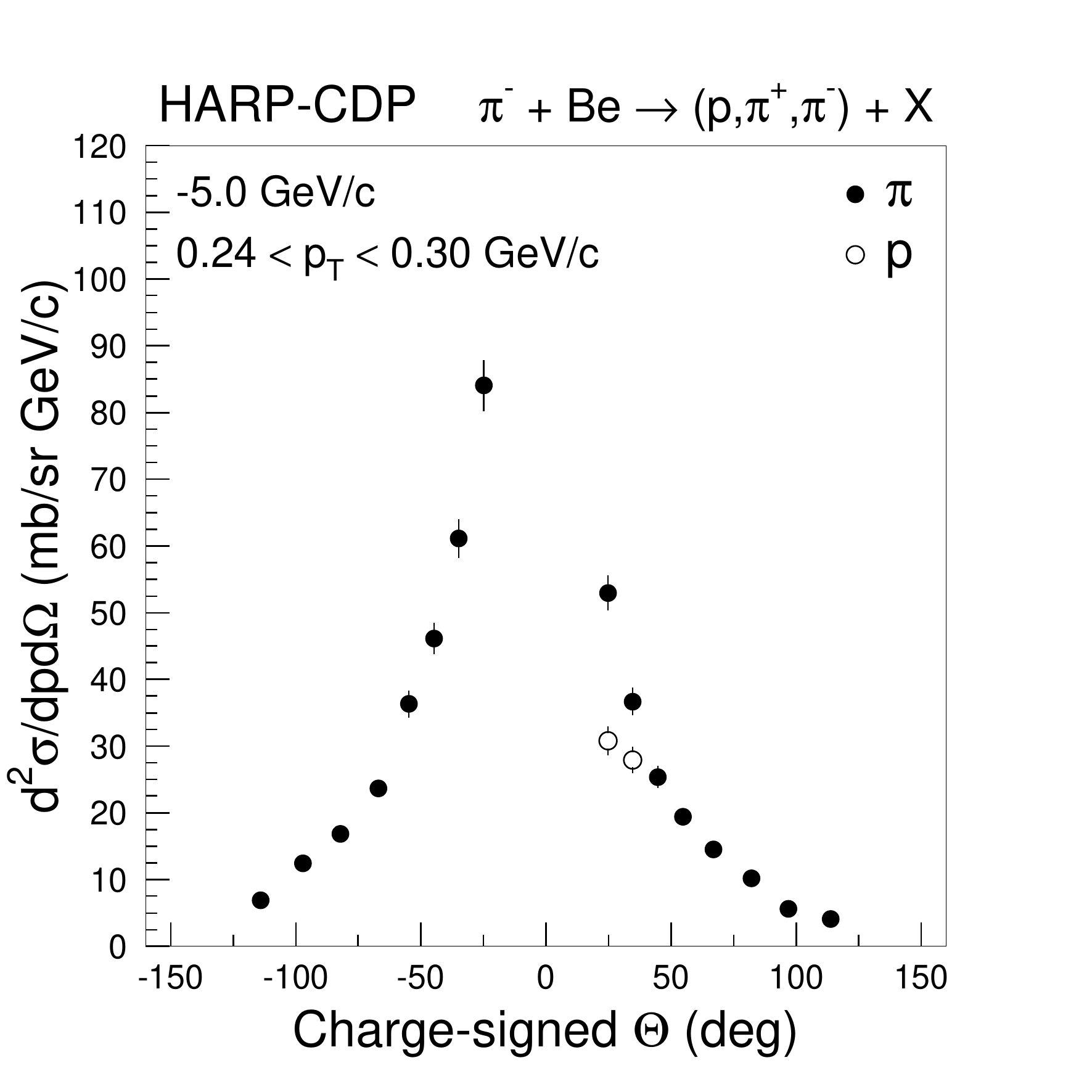} \\
\includegraphics[height=0.30\textheight]{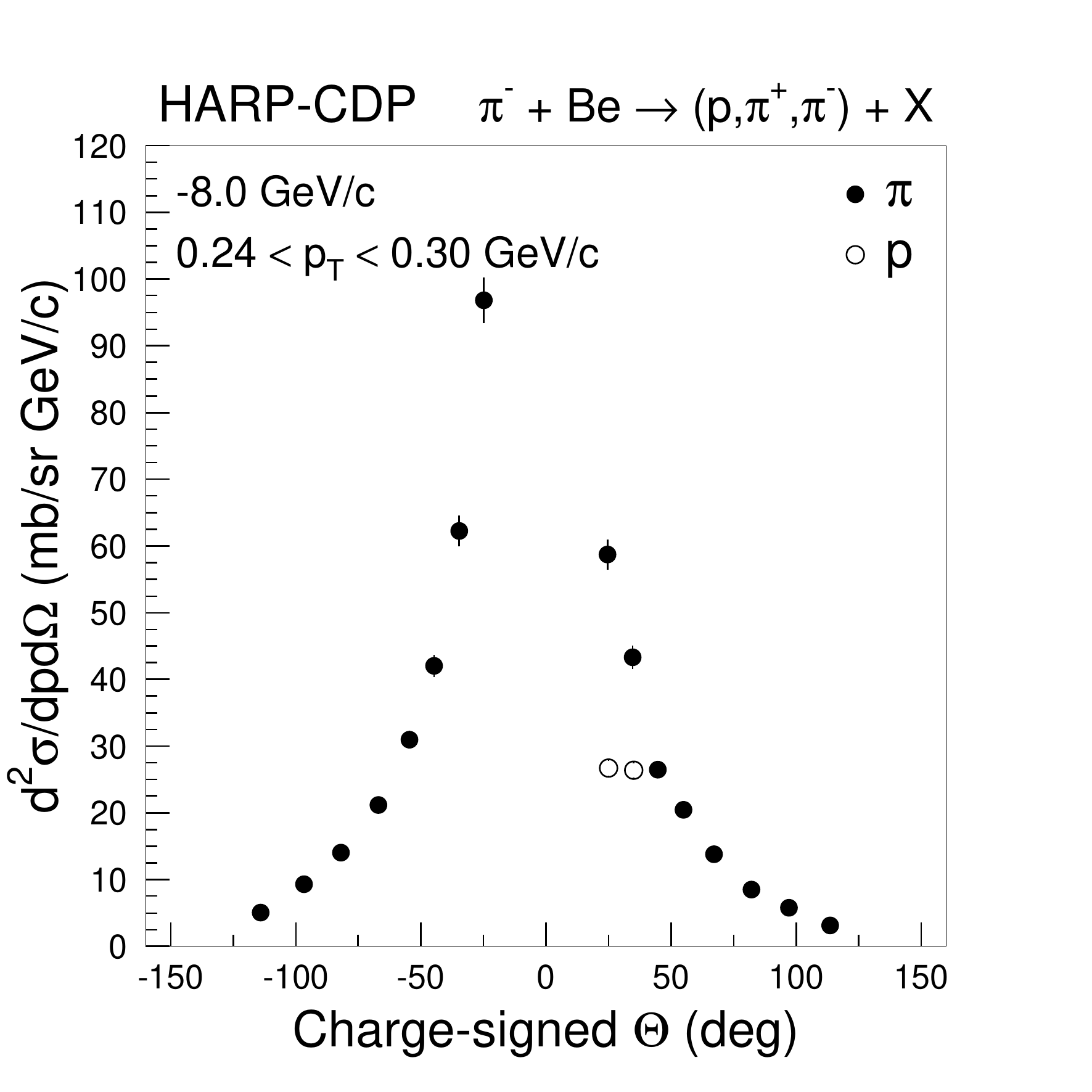} &
\includegraphics[height=0.30\textheight]{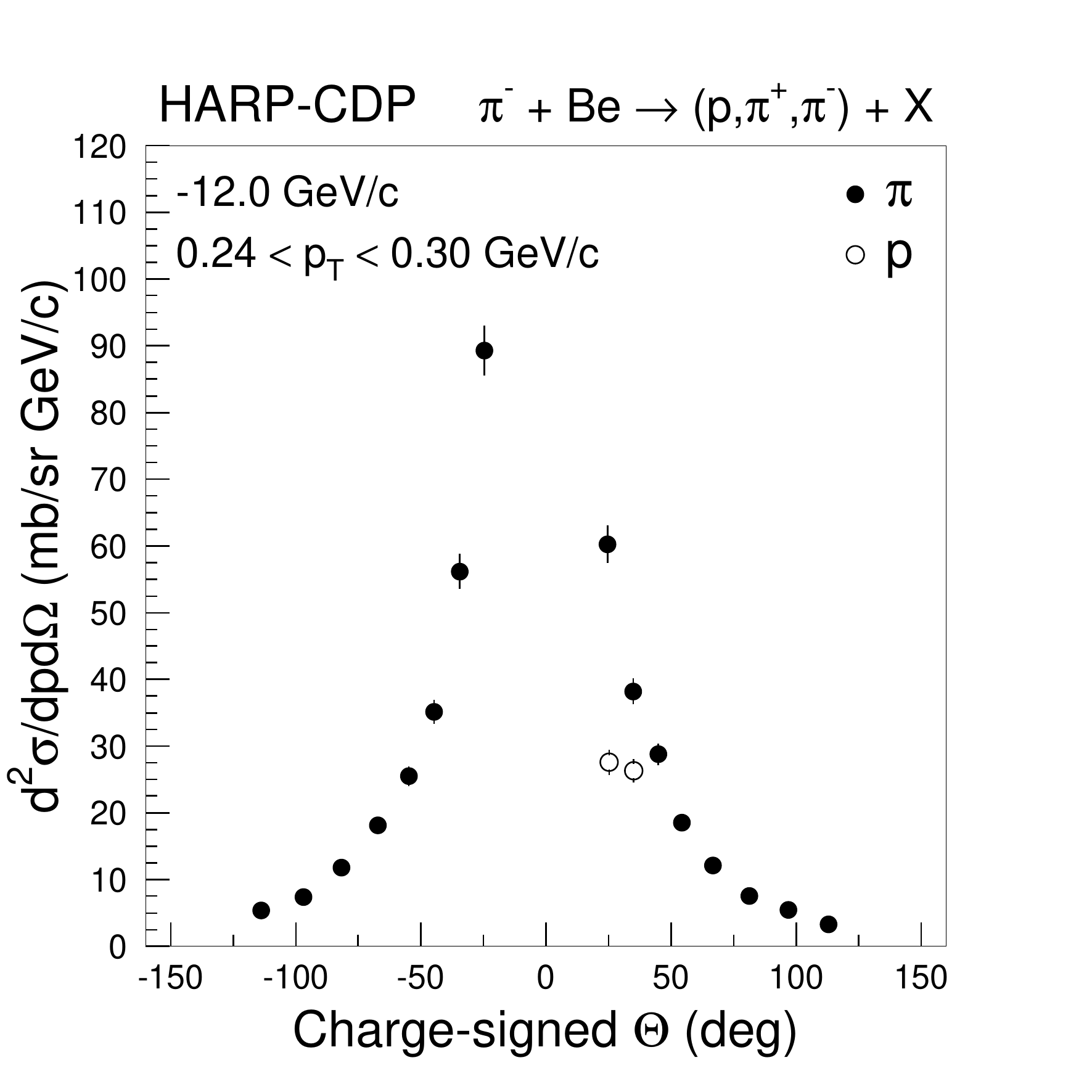} \\
\includegraphics[height=0.30\textheight]{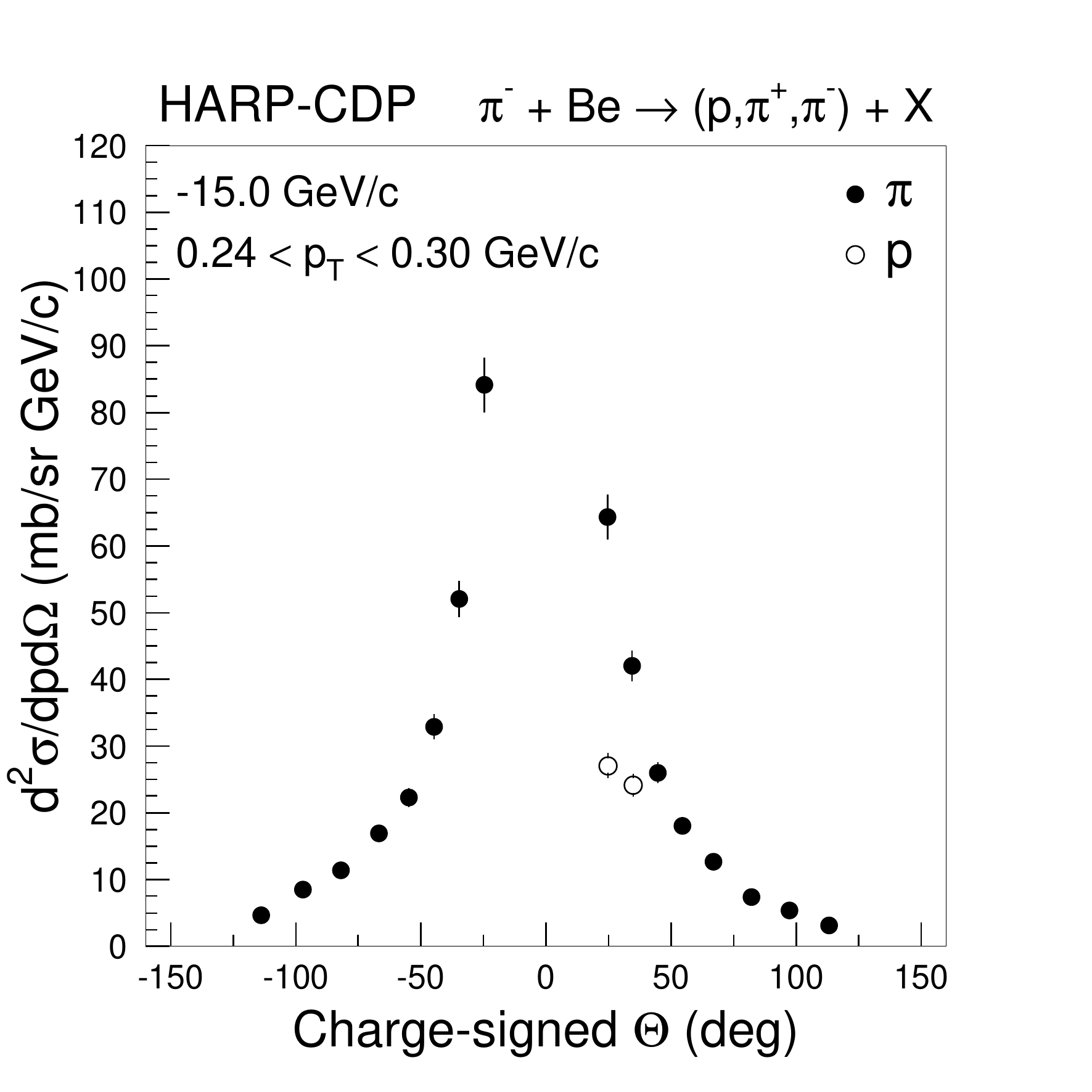} &  \\
\end{tabular}
\caption{Inclusive cross-sections of the production of secondary
protons, $\pi^+$'s, and $\pi^-$'s, with $p_{\rm T}$ in the range 
0.24--0.30~GeV/{\it c}, by $\pi^-$'s on beryllium nuclei, for
different proton beam momenta, as a function of the charge-signed 
polar angle $\theta$ of the secondaries; the shown errors are 
total errors.} 
\label{xsvsthetapim}
\end{center}
\end{figure}

\begin{figure}[h]
\begin{center}
\begin{tabular}{cc}
\includegraphics[height=0.30\textheight]{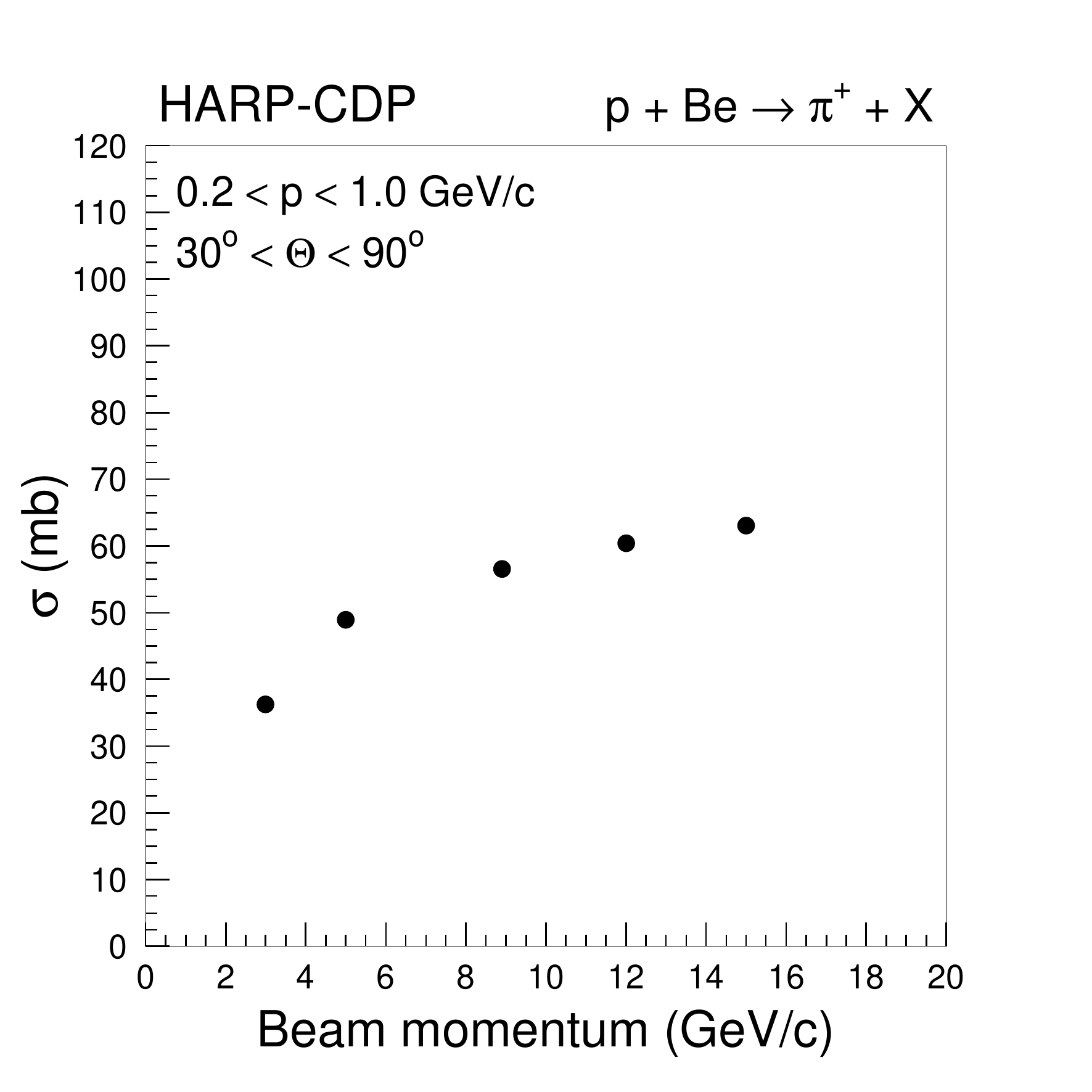} &
\includegraphics[height=0.30\textheight]{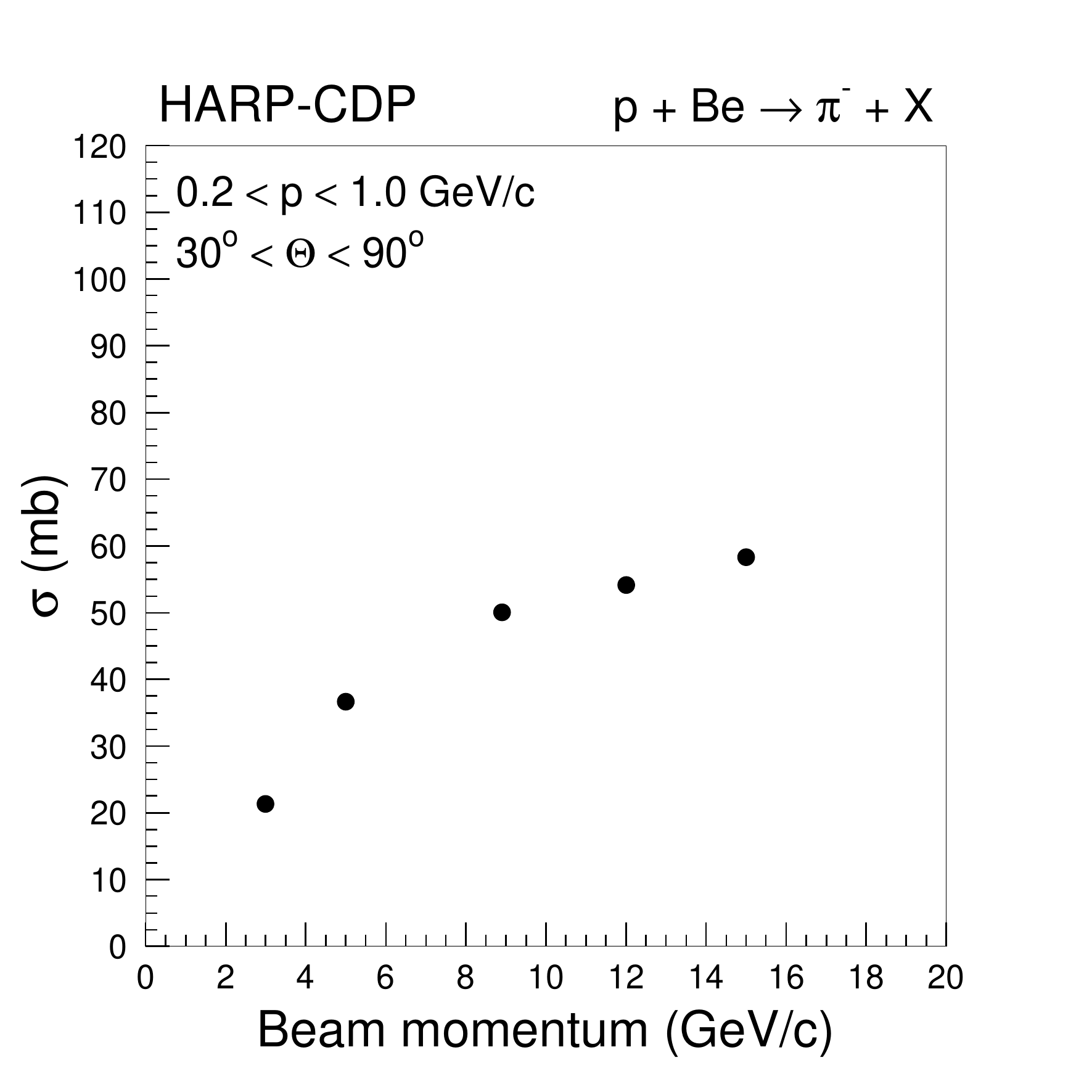} \\
\includegraphics[height=0.30\textheight]{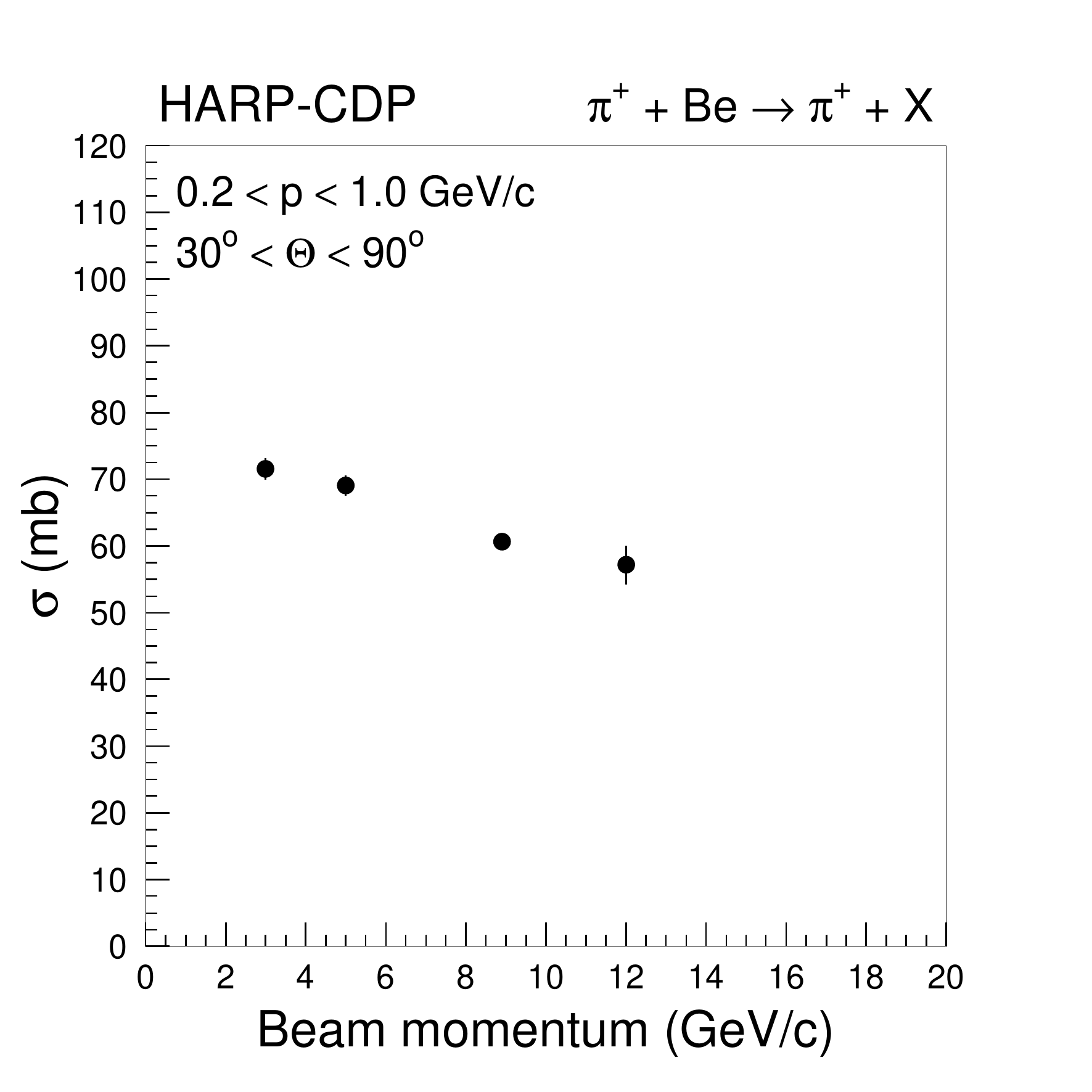} &
\includegraphics[height=0.30\textheight]{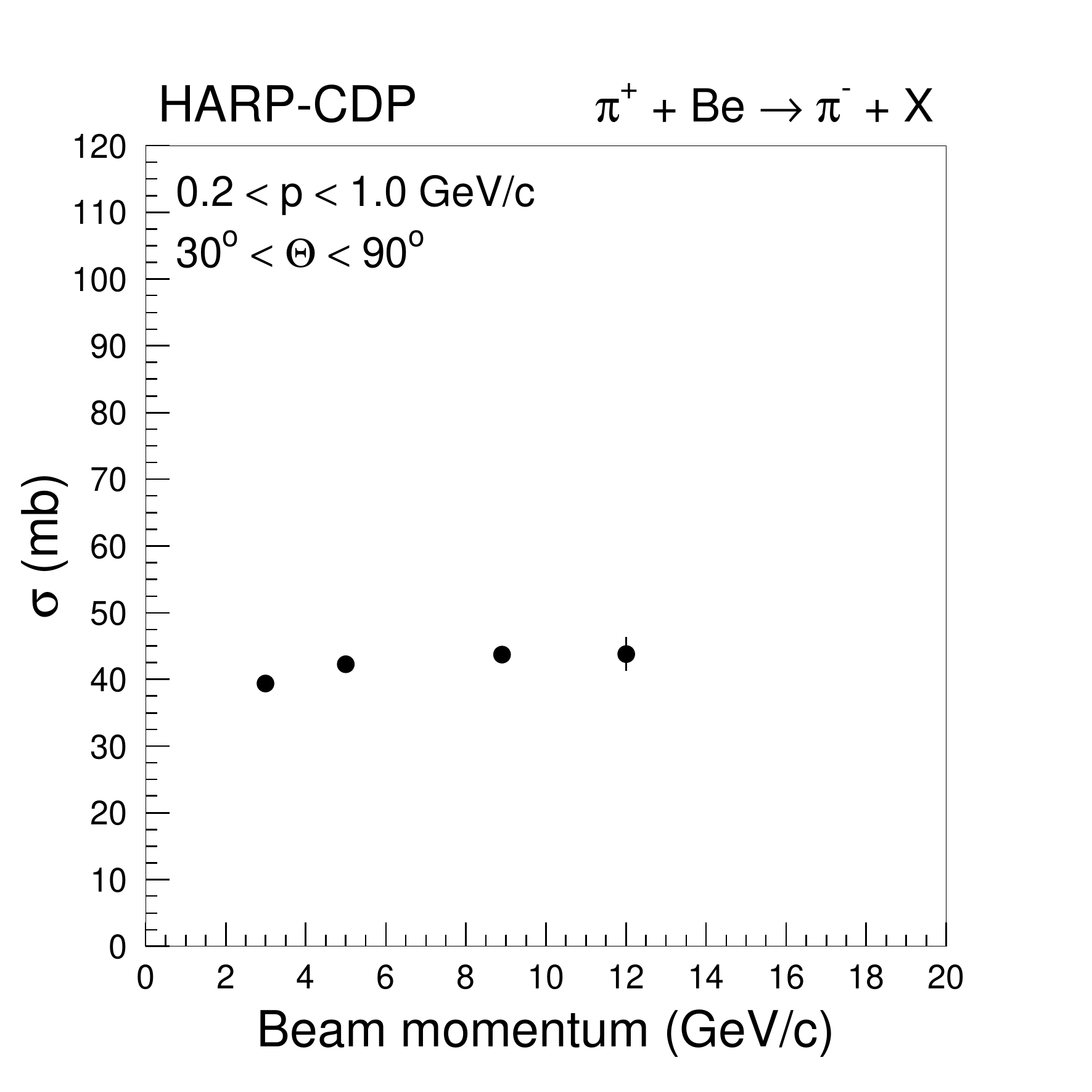} \\
\includegraphics[height=0.30\textheight]{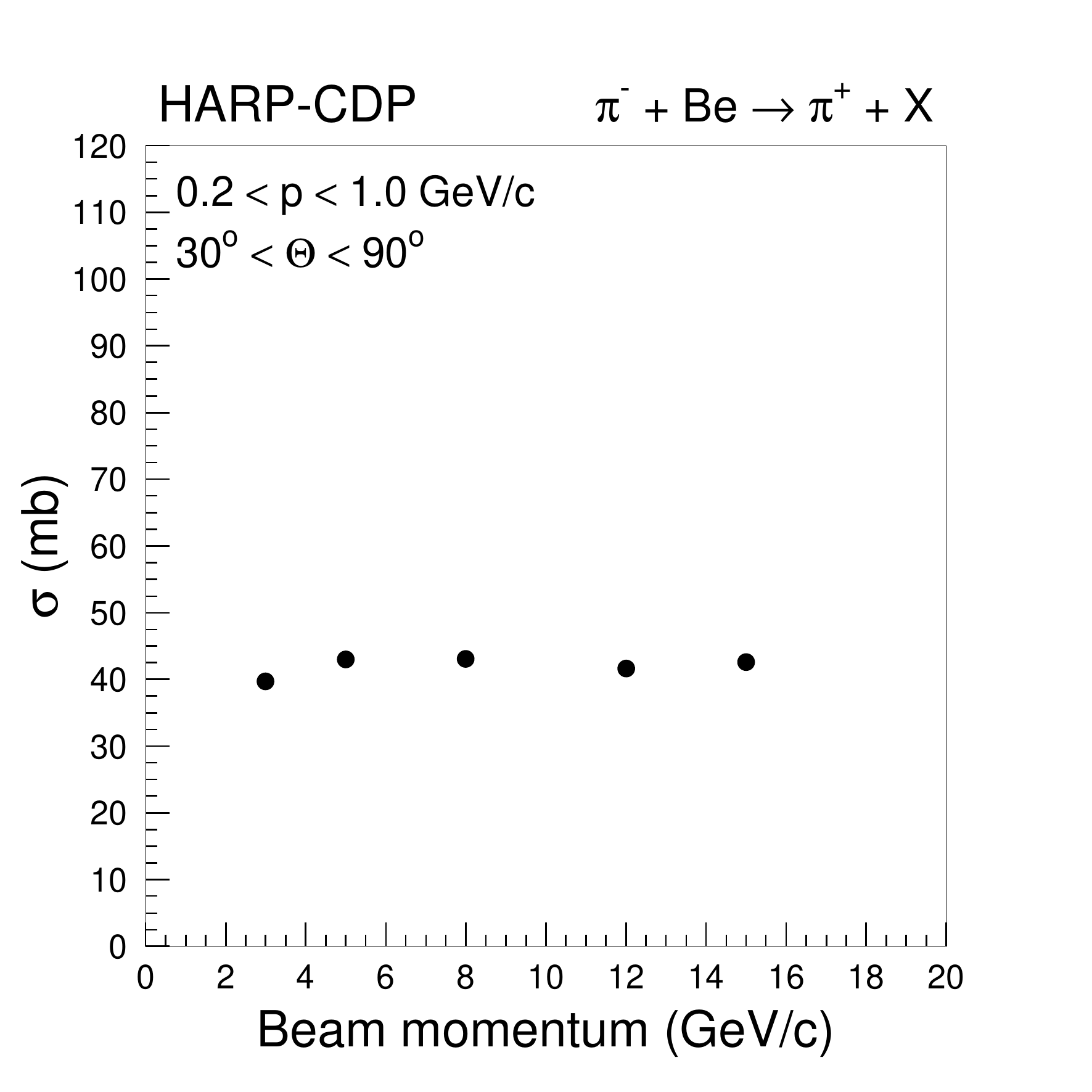} &  
\includegraphics[height=0.30\textheight]{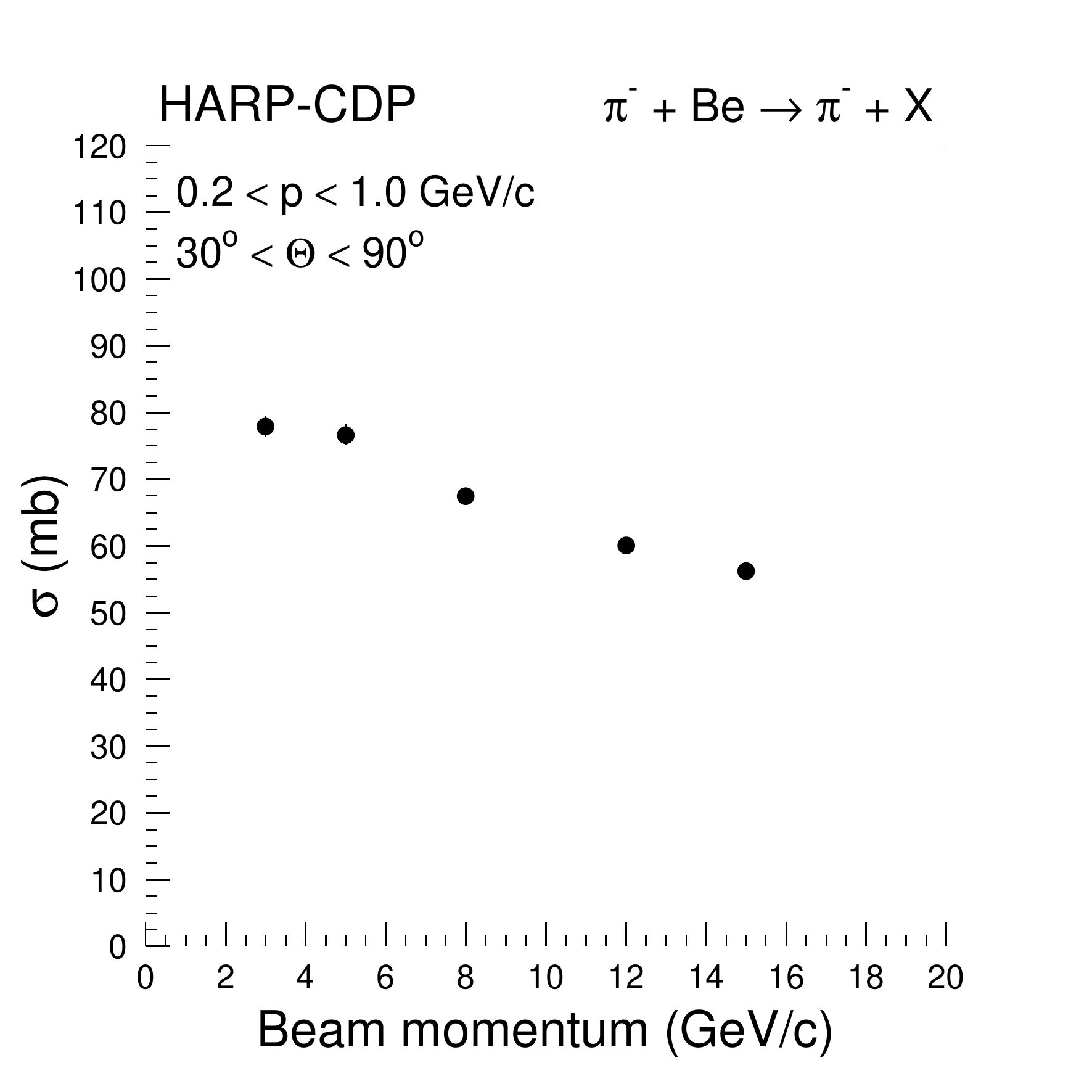} \\
\end{tabular}
\caption{Inclusive cross-sections of the production of 
secondary $\pi^+$'s and $\pi^-$'s, integrated over the momentum range 
$0.2 < p < 1.0$~GeV/{\it c} and the polar-angle range 
$30^\circ < \theta < 90^\circ$, from the interactions on beryllium nuclei
of protons (top row), $\pi^+$'s (middle row), and $\pi^-$'s (bottom row), 
as a function of the beam momentum; the shown errors are 
total errors and mostly smaller than the symbol size.} 
\label{fxsbe}
\end{center}
\end{figure}

We present in Figs.~\ref{xsvsmompro} to \ref{fxsbe} what
we consider salient features of our cross-sections.
In these figures, we also show the data from the $+8.9$~GeV/{\it c} 
and $-8.0$~GeV/c beams that we published in Ref.~\cite{Beryllium1}.

Figure~\ref{xsvsmompro} shows the inclusive cross-sections
of the production of protons, $\pi^+$'s, and $\pi^-$'s,
from incoming protons between 3~GeV/{\it c} and 15~GeV/{\it c}
momentum, as a function of their charge-signed $p_{\rm T}$.
The data refer to the polar-angle range 
$20^\circ < \theta < 30^\circ$.
Figures~\ref{xsvsmompip} and \ref{xsvsmompim} show the same
for incoming $\pi^+$'s and $\pi^-$'s.

Figure~\ref{xsvsthetapro} shows the inclusive cross-sections
of the production of protons, $\pi^+$'s, and $\pi^-$'s,
from incoming protons between 3~GeV/{\it c} and 15~GeV/{\it c}
momentum, this time as a function of their charge-signed 
polar angle $\theta$.
The data refer to the $p_{\rm T}$ range 
$0.24 < p_{\rm T} < 0.30$~GeV/{\it c}.
In this $p_{\rm T}$ range pions populate nearly all polar angles, whereas protons are absorbed at large polar angle and thus escape measurement. 
Figures~\ref{xsvsthetapip} and \ref{xsvsthetapim} show the same
for incoming $\pi^+$'s and $\pi^-$'s.

These figures highlight the rather strong differences in the 
production of proton, $\pi^+$ and $\pi^-$ secondaries for different  
beam particles and beam momenta.  

In Fig.~\ref{fxsbe}, we present the inclusive 
cross-sections of the production of 
secondary $\pi^+$'s and $\pi^-$'s, integrated over the momentum range
$0.2 < p < 1.0$~GeV/{\it c} and the polar-angle range  
$30^\circ < \theta < 90^\circ$ in the forward hemisphere, 
as a function of the beam momentum. 

\clearpage

\clearpage

\clearpage

\section{Comparison of our results with results from 
other experiments}

\subsection{Comparison with E802 results}
Experiment E802~\cite{E802} at Brookhaven National 
Laboratory measured
secondary $\pi^+$'s in the polar-angle
range $5^\circ < \theta < 58^\circ$ from the interactions of
$+14.6$~GeV/{\it c} protons with beryllium nuclei.

\begin{figure}[ht]
\begin{center}
\vspace*{8mm}
\includegraphics[width=0.7\textwidth]{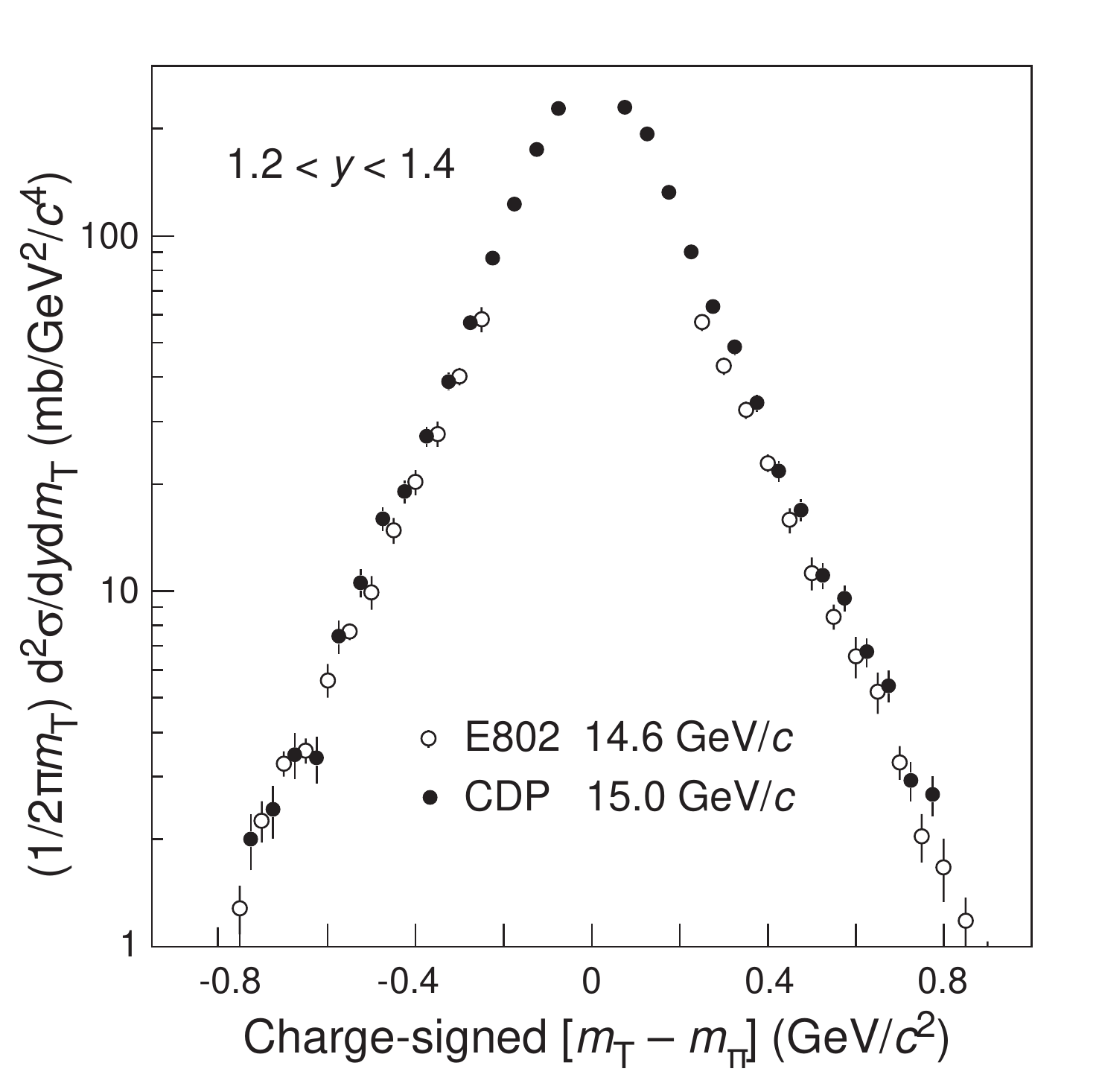} 
\caption{Comparison of our cross-sections (black circles) 
of $\pi^\pm$ production by $+15.0$~GeV/{\it c} 
protons off beryllium nuclei, 
with the cross-sections 
published by the E802 Collaboration for the proton beam 
momentum of $+14.6$~GeV/{\it c} (open circles); all errors are 
statistical only.}
\label{comparisonwithE802}
\end{center}
\end{figure}

Figure~\ref{comparisonwithE802} shows their published Lorentz-invariant 
cross-section of $\pi^+$ and $\pi^-$ production by
$+14.6$~GeV/{\it c} protons, in the rapidity range $1.2 < y < 1.4$,
as a function of $m_{\rm T} - m_{\pi}$, where $m_{\rm T}$ denotes
the pion transverse mass. Their data are compared 
with our cross-sections from the interactions of $+15.0$~GeV/{\it c} 
protons with beryllium nuclei, expressed in 
the same unit as used by E802. Since E802 quoted only statistical
errors, our data in Fig.~\ref{comparisonwithE802} are
also shown with their statistical errors.

The E802 $\pi^\pm$ cross-sections are in good agreement 
with our cross-sections measured nearly at the same proton 
beam momentum, taking into account 
the normalization uncertainty of (10--15)\% quoted by E802.   
We draw attention to the good agreement of the slopes 
of the cross-sections over two orders of magnitude.  

\subsection{Comparison with E910 results}

\begin{figure}[ht]
\begin{center}
\vspace*{8mm}
\includegraphics[width=0.7\textwidth]{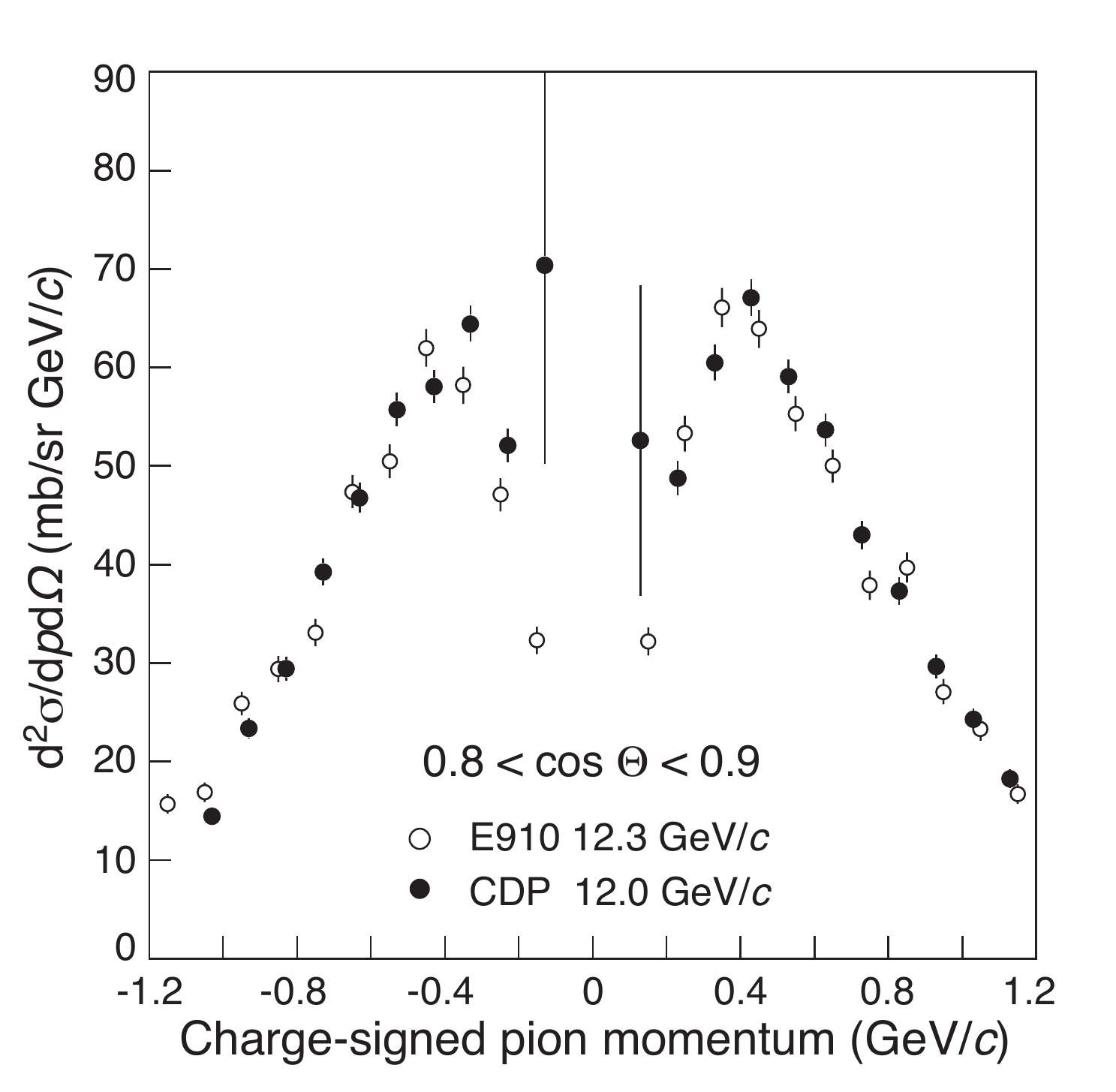} 
\caption{Comparison of our cross-sections  
of $\pi^\pm$ production by $+12.0$~GeV/{\it c} 
protons off beryllium nuclei with the cross-sections 
published by the E910 Collaboration for the proton beam 
momentum of $+12.3$~GeV/{\it c} (open circles); 
all errors are statistical only.}
\label{comparisonwithE910}
\end{center}
\end{figure}

Experiment E910~\cite{E910} at Brookhaven National Laboratory
measured secondary charged pions in the momentum
range 0.1--6~GeV/{\it c} from the interactions of $+12.3$~GeV/{\it c} 
protons with beryllium nuclei.
This experiment used a TPC for the measurement of secondaries,
with a comfortably large track length of $\sim$1.5~m. This feature,
together with a magnetic field strength of 0.5~T, 
is of particular significance, since it permits  
considerably better charge identification and proton--pion 
separation by \dedx\ than is possible in the HARP detector.
Figure~\ref{comparisonwithE910} shows their published  
cross-section ${\rm d}^2 \sigma / {\rm d}p {\rm d}\Omega$ 
of $\pi^\pm$ production by $+12.3$~GeV/{\it c} protons,
in the polar-angle range $0.8 < \cos\theta < 0.9$. 
Since E910 quoted only statistical
errors, our data in Fig.~\ref{comparisonwithE910} from the
interactions of $+12.0$~GeV/{\it c} protons are
also shown with their statistical errors. The normalization 
uncertainty quoted by E910 is $\leq$5\%. 

Also here, the E910 data are shown as published, and our data
are expressed in the same unit as used by E910.  
We draw attention to the good agreement 
in the $\pi^+ / \pi^-$ ratio 
between the cross-sections from E910 and our cross-sections. 

\subsection{Comparison with results from the HARP Collaboration}

\begin{figure}[ht]
\begin{center}
\vspace{8mm}
\begin{tabular}{cc}
\includegraphics[height=0.45\textwidth]{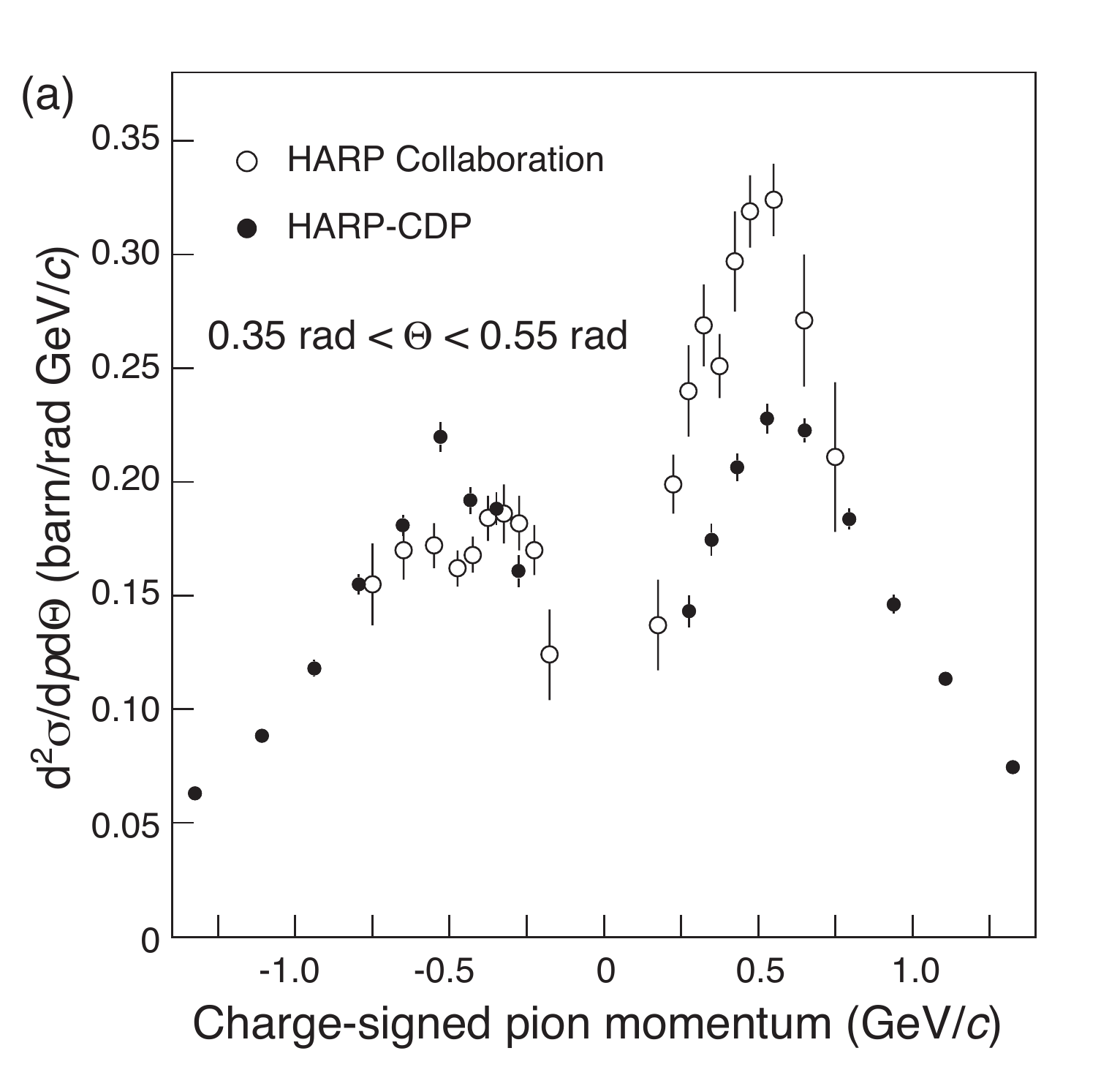} & 
\includegraphics[height=0.45\textwidth]{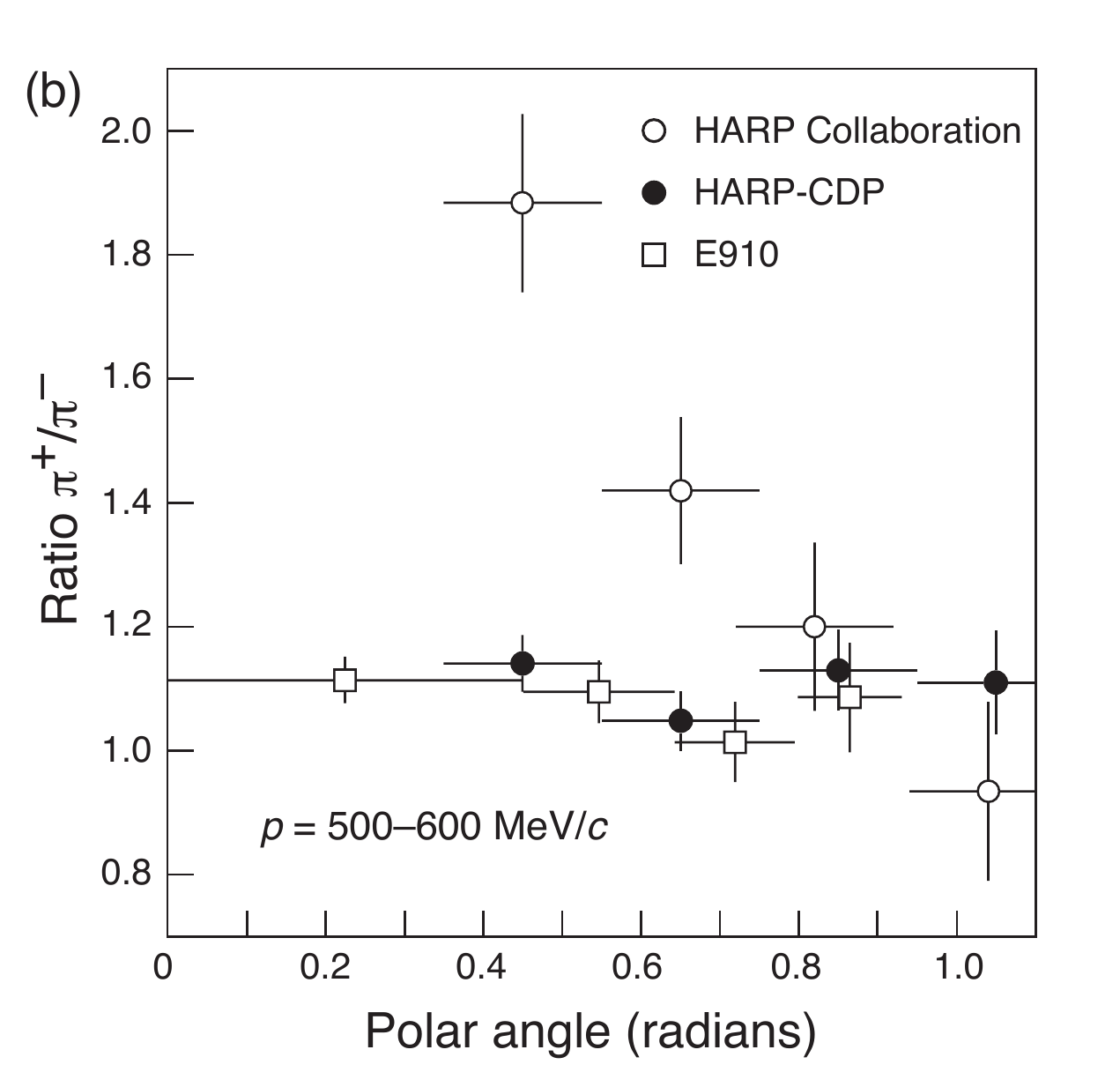}  \\
\end{tabular}
\caption{(a) Comparison of our cross-sections (black circles) 
of $\pi^\pm$ production by $+12.0$~GeV/{\it c} 
protons off beryllium nuclei with the cross-sections 
published by the HARP Collaboration (open circles); (b)
Comparison of our ratio $\pi^+/\pi^-$ at $+12.0$~GeV/{\it c} 
beam momentum as a function of the
polar angle $\theta$ with the ratios published by the 
HARP Collaboration; also shown are the ratios $\pi^+/\pi^-$
published by the 
E910 Collaboration for a $+12.3$~GeV/{\it c} beam momentum; 
for the HARP Collaboration, total errors are shown; for E910 
and our group, the shown errors are statistical only.}
\label{comparisonwithOH}
\end{center}
\end{figure}

Figure~\ref{comparisonwithOH} (a) shows the 
comparison of our
cross-sections of pion production by $+12.0$~GeV/{\it c} 
protons off beryllium nuclei with the ones 
published by the HARP Collaboration~\cite{OffLApaper},
in the polar-angle range $0.35 < \theta < 0.55$~rad.
The latter cross-sections are plotted as published, 
while we expressed our cross-sections in 
the unit used by the HARP Collaboration. 
Figure~\ref{comparisonwithOH} (b)
shows our ratio $\pi^+/\pi^-$ as a function of the
polar angle $\theta$ in comparison with the ratios published by the 
E910 Collaboration (at the slightly different proton beam 
momentum of $+12.3$~GeV/{\it c})
and by the HARP Collaboration.

The discrepancy between our results and those published by the HARP Collaboration is evident. We note the difference especially of the $\pi^+$ cross-section, and the difference in the momentum range. The discrepancy is even more serious as the same data set has been analysed by both groups.

We hold that the discrepancy is caused by problems in the HARP Collaboration's data analysis. They result primarily, but not exclusively, from a lack of understanding TPC track distortions and RPC timing signals. These problems, together with others that affect the HARP Collaboration's data analysis, are discussed in detail in Refs~\cite{JINSTpub,EPJCpub,WhiteBookseries} and in the Appendix of Ref.~\cite{Beryllium1}. 

\section{Summary}

From the analysis of data from the HARP large-angle spectrometer
(polar angle $\theta$ in the range $20^\circ < \theta < 125^\circ$), double-differential 
cross-sections ${\rm d}^2 \sigma / {\rm d}p {\rm d}\Omega$ 
of the production of secondary protons, $\pi^+$'s, and $\pi^-$'s,
have been obtained. The incoming beam particles were protons 
and pions with momenta from $\pm 3$ to $\pm 15$~GeV/{\it c}, 
impinging on a 5\% $\lambda_{\rm abs}$ thick stationary 
beryllium target. 
Our cross-sections for $\pi^+$ and $\pi^-$ production 
agree with results from other experiments but disagree with 
the results of the HARP Collaboration that were obtained 
from the same raw data.

\section*{Acknowledgements}

We are greatly indebted to many technical collaborators whose 
diligent and hard work made the HARP detector a well-functioning 
instrument. We thank all HARP colleagues who devoted time and 
effort to the design and construction of the detector, to data taking, 
and to setting up the computing and software infrastructure. 
We express our sincere gratitude to HARP's funding agencies 
for their support.

\clearpage

\appendix

\section{Cross-section Tables}


\input{table.pro.probe3.tex}
\input{table.pip.probe3.tex}
\input{table.pim.probe3.tex}
\input{table.pro.pipbe3.tex}
\input{table.pip.pipbe3.tex}
\input{table.pim.pipbe3.tex}
\input{table.pro.pimbe3.tex}
\input{table.pip.pimbe3.tex}
\input{table.pim.pimbe3.tex}
\clearpage


\input{table.pro.probe5.tex}
\input{table.pip.probe5.tex}
\input{table.pim.probe5.tex}
\input{table.pro.pipbe5.tex}
\input{table.pip.pipbe5.tex}
\input{table.pim.pipbe5.tex}
\input{table.pro.pimbe5.tex}
\input{table.pip.pimbe5.tex}
\input{table.pim.pimbe5.tex}
\clearpage


\input{table.pro.probe12.tex}
\input{table.pip.probe12.tex}
\input{table.pim.probe12.tex}
\input{table.pro.pipbe12.tex}
\input{table.pip.pipbe12.tex}
\input{table.pim.pipbe12.tex}
\input{table.pro.pimbe12.tex}
\input{table.pip.pimbe12.tex}
\input{table.pim.pimbe12.tex}
\clearpage


\input{table.pro.probe15.tex}
\input{table.pip.probe15.tex}
\input{table.pim.probe15.tex}
\input{table.pro.pipbe15.tex}
\input{table.pip.pipbe15.tex}
\input{table.pim.pipbe15.tex}
\input{table.pro.pimbe15.tex}
\input{table.pip.pimbe15.tex}
\input{table.pim.pimbe15.tex}
\clearpage

\end{document}

%% file: table.pro.probe3.tex
  
 \begin{table}[h]
 \begin{scriptsize}
 \caption{Double-differential inclusive
  cross-section ${\rm d}^2 \sigma /{\rm d}p{\rm d}\Omega$
  [mb/(GeV/{\it c} sr)] of the production of protons
  in p + Be $\rightarrow$ p + X interactions
  with $+3.0$~GeV/{\it c} beam momentum;
  the first error is statistical, the second systematic; 
 $p_{\rm T}$ in GeV/{\it c}, polar angle $\theta$ in degrees.}
 \label{pro.probe3}
 \begin{center}
 \begin{tabular}{|c||c|c|rcrcr||c|c|rcrcr|}
 \hline
   & \multicolumn{7}{c||}{$20<\theta<30$}
  & \multicolumn{7}{c|}{$30<\theta<40$} \\
 \hline
 $p_{\rm T}$ & $\langle p_{\rm T} \rangle$ & $\langle \theta \rangle$
  & \multicolumn{5}{c||}{${\rm d}^2 \sigma /{\rm d}p{\rm d}\Omega$}
  &$\langle p_{\rm T} \rangle$ & $\langle \theta \rangle$
  & \multicolumn{5}{c|}{${\rm d}^2 \sigma /{\rm d}p{\rm d}\Omega$} \\
 \hline
0.20--0.24 & 0.221 &  24.7 &  58.52 & $\!\!\pm\!\!$ &  2.79 & $\!\!\pm\!\!$ &  3.92 &  &  &  \multicolumn{5}{c|}{ } \\
0.24--0.30 & 0.269 &  25.1 &  64.24 & $\!\!\pm\!\!$ &  2.38 & $\!\!\pm\!\!$ &  3.63 & 0.271 &  34.9 &  54.76 & $\!\!\pm\!\!$ &  2.16 & $\!\!\pm\!\!$ &  2.96 \\
0.30--0.36 & 0.329 &  24.9 &  58.05 & $\!\!\pm\!\!$ &  2.24 & $\!\!\pm\!\!$ &  2.98 & 0.330 &  35.1 &  51.23 & $\!\!\pm\!\!$ &  2.05 & $\!\!\pm\!\!$ &  2.50 \\
0.36--0.42 & 0.390 &  24.9 &  57.81 & $\!\!\pm\!\!$ &  2.23 & $\!\!\pm\!\!$ &  2.75 & 0.390 &  34.9 &  52.43 & $\!\!\pm\!\!$ &  2.11 & $\!\!\pm\!\!$ &  2.44 \\
0.42--0.50 & 0.459 &  24.8 &  54.37 & $\!\!\pm\!\!$ &  1.85 & $\!\!\pm\!\!$ &  2.27 & 0.460 &  34.9 &  43.86 & $\!\!\pm\!\!$ &  1.69 & $\!\!\pm\!\!$ &  1.83 \\
0.50--0.60 & 0.548 &  24.8 &  43.78 & $\!\!\pm\!\!$ &  1.50 & $\!\!\pm\!\!$ &  1.80 & 0.549 &  34.8 &  35.55 & $\!\!\pm\!\!$ &  1.37 & $\!\!\pm\!\!$ &  1.47 \\
0.60--0.72 & 0.656 &  24.8 &  35.71 & $\!\!\pm\!\!$ &  1.28 & $\!\!\pm\!\!$ &  1.73 & 0.654 &  34.9 &  24.47 & $\!\!\pm\!\!$ &  1.03 & $\!\!\pm\!\!$ &  1.26 \\
0.72--0.90 &  &  &  \multicolumn{5}{c||}{ } & 0.800 &  35.0 &  16.01 & $\!\!\pm\!\!$ &  0.71 & $\!\!\pm\!\!$ &  1.08 \\
 \hline
 \hline
   & \multicolumn{7}{c||}{$40<\theta<50$}
  & \multicolumn{7}{c|}{$50<\theta<60$} \\
 \hline
 $p_{\rm T}$ & $\langle p_{\rm T} \rangle$ & $\langle \theta \rangle$
  & \multicolumn{5}{c||}{${\rm d}^2 \sigma /{\rm d}p{\rm d}\Omega$}
  &$\langle p_{\rm T} \rangle$ & $\langle \theta \rangle$
  & \multicolumn{5}{c|}{${\rm d}^2 \sigma /{\rm d}p{\rm d}\Omega$} \\
 \hline
0.30--0.36 & 0.332 &  45.0 &  45.49 & $\!\!\pm\!\!$ &  1.88 & $\!\!\pm\!\!$ &  2.08 &  &  &  \multicolumn{5}{c|}{ } \\
0.36--0.42 & 0.391 &  45.0 &  46.83 & $\!\!\pm\!\!$ &  1.96 & $\!\!\pm\!\!$ &  2.13 & 0.390 &  55.1 &  41.48 & $\!\!\pm\!\!$ &  1.80 & $\!\!\pm\!\!$ &  1.82 \\
0.42--0.50 & 0.461 &  45.1 &  37.82 & $\!\!\pm\!\!$ &  1.52 & $\!\!\pm\!\!$ &  1.55 & 0.461 &  55.3 &  35.87 & $\!\!\pm\!\!$ &  1.47 & $\!\!\pm\!\!$ &  1.49 \\
0.50--0.60 & 0.551 &  45.0 &  33.10 & $\!\!\pm\!\!$ &  1.32 & $\!\!\pm\!\!$ &  1.44 & 0.551 &  55.0 &  32.08 & $\!\!\pm\!\!$ &  1.27 & $\!\!\pm\!\!$ &  1.42 \\
0.60--0.72 & 0.659 &  44.9 &  27.41 & $\!\!\pm\!\!$ &  1.14 & $\!\!\pm\!\!$ &  1.42 & 0.663 &  55.2 &  24.47 & $\!\!\pm\!\!$ &  1.08 & $\!\!\pm\!\!$ &  1.41 \\
0.72--0.90 & 0.806 &  44.9 &  13.53 & $\!\!\pm\!\!$ &  0.64 & $\!\!\pm\!\!$ &  0.96 & 0.803 &  54.7 &  12.21 & $\!\!\pm\!\!$ &  0.63 & $\!\!\pm\!\!$ &  1.00 \\
0.90--1.25 & 1.037 &  44.8 &   2.97 & $\!\!\pm\!\!$ &  0.20 & $\!\!\pm\!\!$ &  0.46 & 1.030 &  54.7 &   1.69 & $\!\!\pm\!\!$ &  0.14 & $\!\!\pm\!\!$ &  0.32 \\
 \hline
 \hline
   & \multicolumn{7}{c||}{$60<\theta<75$}
  & \multicolumn{7}{c|}{$75<\theta<90$} \\
 \hline
 $p_{\rm T}$ & $\langle p_{\rm T} \rangle$ & $\langle \theta \rangle$
  & \multicolumn{5}{c||}{${\rm d}^2 \sigma /{\rm d}p{\rm d}\Omega$}
  &$\langle p_{\rm T} \rangle$ & $\langle \theta \rangle$
  & \multicolumn{5}{c|}{${\rm d}^2 \sigma /{\rm d}p{\rm d}\Omega$} \\
 \hline
0.36--0.42 & 0.391 &  67.3 &  40.49 & $\!\!\pm\!\!$ &  1.46 & $\!\!\pm\!\!$ &  1.60 &  &  &  \multicolumn{5}{c|}{ } \\
0.42--0.50 & 0.460 &  67.2 &  35.62 & $\!\!\pm\!\!$ &  1.20 & $\!\!\pm\!\!$ &  1.31 & 0.459 &  81.5 &  20.38 & $\!\!\pm\!\!$ &  0.89 & $\!\!\pm\!\!$ &  0.97 \\
0.50--0.60 & 0.549 &  67.1 &  30.01 & $\!\!\pm\!\!$ &  1.01 & $\!\!\pm\!\!$ &  1.33 & 0.547 &  81.4 &  13.19 & $\!\!\pm\!\!$ &  0.65 & $\!\!\pm\!\!$ &  0.81 \\
0.60--0.72 & 0.660 &  66.9 &  14.59 & $\!\!\pm\!\!$ &  0.65 & $\!\!\pm\!\!$ &  0.94 & 0.660 &  81.3 &   6.31 & $\!\!\pm\!\!$ &  0.44 & $\!\!\pm\!\!$ &  0.69 \\
0.72--0.90 & 0.803 &  66.4 &   5.20 & $\!\!\pm\!\!$ &  0.32 & $\!\!\pm\!\!$ &  0.57 & 0.791 &  81.4 &   1.68 & $\!\!\pm\!\!$ &  0.19 & $\!\!\pm\!\!$ &  0.26 \\
0.90--1.25 & 1.027 &  66.1 &   0.67 & $\!\!\pm\!\!$ &  0.08 & $\!\!\pm\!\!$ &  0.17 & 1.031 &  81.5 &   0.17 & $\!\!\pm\!\!$ &  0.04 & $\!\!\pm\!\!$ &  0.09 \\
 \hline
 \hline
  & \multicolumn{7}{c||}{$90<\theta<105$}
  & \multicolumn{7}{c|}{$105<\theta<125$} \\
 \hline
 $p_{\rm T}$ & $\langle p_{\rm T} \rangle$ & $\langle \theta \rangle$
  & \multicolumn{5}{c||}{${\rm d}^2 \sigma /{\rm d}p{\rm d}\Omega$}
  &$\langle p_{\rm T} \rangle$ & $\langle \theta \rangle$
  & \multicolumn{5}{c|}{${\rm d}^2 \sigma /{\rm d}p{\rm d}\Omega$} \\
 \hline
0.36--0.42 &  &  &  \multicolumn{5}{c||}{ } & 0.389 & 113.6 &   6.77 & $\!\!\pm\!\!$ &  0.48 & $\!\!\pm\!\!$ &  0.51 \\
0.42--0.50 & 0.458 &  96.5 &  10.78 & $\!\!\pm\!\!$ &  0.64 & $\!\!\pm\!\!$ &  0.79 & 0.456 & 113.0 &   4.70 & $\!\!\pm\!\!$ &  0.36 & $\!\!\pm\!\!$ &  0.40 \\
0.50--0.60 & 0.549 &  96.0 &   5.78 & $\!\!\pm\!\!$ &  0.42 & $\!\!\pm\!\!$ &  0.52 & 0.539 & 111.7 &   1.45 & $\!\!\pm\!\!$ &  0.20 & $\!\!\pm\!\!$ &  0.21 \\
0.60--0.72 & 0.659 &  95.5 &   1.58 & $\!\!\pm\!\!$ &  0.23 & $\!\!\pm\!\!$ &  0.23 & 0.659 & 112.3 &   0.23 & $\!\!\pm\!\!$ &  0.08 & $\!\!\pm\!\!$ &  0.08 \\
0.72--0.90 & 0.785 &  94.8 &   0.38 & $\!\!\pm\!\!$ &  0.11 & $\!\!\pm\!\!$ &  0.12 &  &  &  \multicolumn{5}{c|}{ } \\
 \hline
 \end{tabular}
 \end{center}
 \end{scriptsize}
 \end{table}

%% file: table.pip.probe3.tex
  
 \begin{table}[h]
 \begin{scriptsize}
 \caption{Double-differential inclusive
  cross-section ${\rm d}^2 \sigma /{\rm d}p{\rm d}\Omega$
  [mb/(GeV/{\it c} sr)] of the production of $\pi^+$'s
  in p + Be $\rightarrow$ $\pi^+$ + X interactions
  with $+3.0$~GeV/{\it c} beam momentum;
  the first error is statistical, the second systematic; 
 $p_{\rm T}$ in GeV/{\it c}, polar angle $\theta$ in degrees.}
 \label{pip.probe3}
 \begin{center}
 \begin{tabular}{|c||c|c|rcrcr||c|c|rcrcr|}
 \hline
   & \multicolumn{7}{c||}{$20<\theta<30$}
  & \multicolumn{7}{c|}{$30<\theta<40$} \\
 \hline
 $p_{\rm T}$ & $\langle p_{\rm T} \rangle$ & $\langle \theta \rangle$
  & \multicolumn{5}{c||}{${\rm d}^2 \sigma /{\rm d}p{\rm d}\Omega$}
  &$\langle p_{\rm T} \rangle$ & $\langle \theta \rangle$
  & \multicolumn{5}{c|}{${\rm d}^2 \sigma /{\rm d}p{\rm d}\Omega$} \\
 \hline
0.10--0.13 & 0.116 &  24.5 &  26.72 & $\!\!\pm\!\!$ &  2.52 & $\!\!\pm\!\!$ &  2.91 & 0.113 &  35.8 &  25.91 & $\!\!\pm\!\!$ &  2.50 & $\!\!\pm\!\!$ &  3.03 \\
0.13--0.16 & 0.146 &  24.6 &  38.71 & $\!\!\pm\!\!$ &  2.79 & $\!\!\pm\!\!$ &  3.07 & 0.146 &  34.7 &  34.51 & $\!\!\pm\!\!$ &  2.55 & $\!\!\pm\!\!$ &  2.95 \\
0.16--0.20 & 0.180 &  25.0 &  54.17 & $\!\!\pm\!\!$ &  2.77 & $\!\!\pm\!\!$ &  3.39 & 0.179 &  34.5 &  36.55 & $\!\!\pm\!\!$ &  2.23 & $\!\!\pm\!\!$ &  2.41 \\
0.20--0.24 & 0.220 &  24.9 &  52.69 & $\!\!\pm\!\!$ &  2.68 & $\!\!\pm\!\!$ &  3.04 & 0.220 &  34.8 &  39.25 & $\!\!\pm\!\!$ &  2.24 & $\!\!\pm\!\!$ &  2.35 \\
0.24--0.30 & 0.269 &  25.0 &  34.60 & $\!\!\pm\!\!$ &  1.75 & $\!\!\pm\!\!$ &  1.83 & 0.268 &  34.6 &  33.40 & $\!\!\pm\!\!$ &  1.69 & $\!\!\pm\!\!$ &  1.72 \\
0.30--0.36 & 0.328 &  25.3 &  27.87 & $\!\!\pm\!\!$ &  1.58 & $\!\!\pm\!\!$ &  1.53 & 0.329 &  35.2 &  24.06 & $\!\!\pm\!\!$ &  1.44 & $\!\!\pm\!\!$ &  1.35 \\
0.36--0.42 & 0.388 &  25.0 &  16.61 & $\!\!\pm\!\!$ &  1.17 & $\!\!\pm\!\!$ &  1.18 & 0.386 &  35.1 &  16.13 & $\!\!\pm\!\!$ &  1.16 & $\!\!\pm\!\!$ &  1.04 \\
0.42--0.50 & 0.458 &  25.1 &  11.44 & $\!\!\pm\!\!$ &  0.83 & $\!\!\pm\!\!$ &  0.77 & 0.458 &  34.8 &   9.70 & $\!\!\pm\!\!$ &  0.79 & $\!\!\pm\!\!$ &  0.65 \\
0.50--0.60 & 0.548 &  25.0 &   4.88 & $\!\!\pm\!\!$ &  0.44 & $\!\!\pm\!\!$ &  0.45 & 0.549 &  34.9 &   5.58 & $\!\!\pm\!\!$ &  0.50 & $\!\!\pm\!\!$ &  0.46 \\
0.60--0.72 & 0.656 &  24.9 &   3.72 & $\!\!\pm\!\!$ &  0.33 & $\!\!\pm\!\!$ &  0.58 & 0.654 &  35.3 &   3.21 & $\!\!\pm\!\!$ &  0.34 & $\!\!\pm\!\!$ &  0.40 \\
0.72--0.90 &  &  &  \multicolumn{5}{c||}{ } & 0.798 &  34.7 &   1.73 & $\!\!\pm\!\!$ &  0.18 & $\!\!\pm\!\!$ &  0.59 \\
 \hline
 \hline
   & \multicolumn{7}{c||}{$40<\theta<50$}
  & \multicolumn{7}{c|}{$50<\theta<60$} \\
 \hline
 $p_{\rm T}$ & $\langle p_{\rm T} \rangle$ & $\langle \theta \rangle$
  & \multicolumn{5}{c||}{${\rm d}^2 \sigma /{\rm d}p{\rm d}\Omega$}
  &$\langle p_{\rm T} \rangle$ & $\langle \theta \rangle$
  & \multicolumn{5}{c|}{${\rm d}^2 \sigma /{\rm d}p{\rm d}\Omega$} \\
 \hline
0.10--0.13 & 0.116 &  45.4 &  20.82 & $\!\!\pm\!\!$ &  2.28 & $\!\!\pm\!\!$ &  2.40 &  &  &  \multicolumn{5}{c|}{ } \\
0.13--0.16 & 0.146 &  45.0 &  27.15 & $\!\!\pm\!\!$ &  2.32 & $\!\!\pm\!\!$ &  2.26 & 0.145 &  54.4 &  27.70 & $\!\!\pm\!\!$ &  2.42 & $\!\!\pm\!\!$ &  2.54 \\
0.16--0.20 & 0.180 &  45.0 &  35.34 & $\!\!\pm\!\!$ &  2.19 & $\!\!\pm\!\!$ &  2.32 & 0.180 &  55.0 &  29.59 & $\!\!\pm\!\!$ &  2.06 & $\!\!\pm\!\!$ &  2.05 \\
0.20--0.24 & 0.219 &  44.4 &  32.30 & $\!\!\pm\!\!$ &  2.05 & $\!\!\pm\!\!$ &  2.01 & 0.220 &  54.7 &  26.36 & $\!\!\pm\!\!$ &  1.87 & $\!\!\pm\!\!$ &  1.73 \\
0.24--0.30 & 0.271 &  44.6 &  25.09 & $\!\!\pm\!\!$ &  1.47 & $\!\!\pm\!\!$ &  1.44 & 0.269 &  54.5 &  18.60 & $\!\!\pm\!\!$ &  1.27 & $\!\!\pm\!\!$ &  1.13 \\
0.30--0.36 & 0.329 &  44.6 &  18.15 & $\!\!\pm\!\!$ &  1.23 & $\!\!\pm\!\!$ &  1.10 & 0.330 &  54.6 &  13.68 & $\!\!\pm\!\!$ &  1.08 & $\!\!\pm\!\!$ &  0.92 \\
0.36--0.42 & 0.391 &  44.7 &  12.60 & $\!\!\pm\!\!$ &  1.02 & $\!\!\pm\!\!$ &  0.86 & 0.389 &  54.1 &   9.64 & $\!\!\pm\!\!$ &  0.91 & $\!\!\pm\!\!$ &  0.75 \\
0.42--0.50 & 0.457 &  44.9 &   8.79 & $\!\!\pm\!\!$ &  0.77 & $\!\!\pm\!\!$ &  0.63 & 0.460 &  54.7 &   6.56 & $\!\!\pm\!\!$ &  0.66 & $\!\!\pm\!\!$ &  0.54 \\
0.50--0.60 & 0.544 &  45.1 &   3.35 & $\!\!\pm\!\!$ &  0.41 & $\!\!\pm\!\!$ &  0.29 & 0.548 &  54.1 &   3.39 & $\!\!\pm\!\!$ &  0.41 & $\!\!\pm\!\!$ &  0.33 \\
0.60--0.72 & 0.649 &  44.4 &   2.43 & $\!\!\pm\!\!$ &  0.31 & $\!\!\pm\!\!$ &  0.33 & 0.658 &  54.2 &   2.39 & $\!\!\pm\!\!$ &  0.32 & $\!\!\pm\!\!$ &  0.36 \\
0.72--0.90 & 0.806 &  44.0 &   1.24 & $\!\!\pm\!\!$ &  0.17 & $\!\!\pm\!\!$ &  0.33 & 0.800 &  53.9 &   0.61 & $\!\!\pm\!\!$ &  0.12 & $\!\!\pm\!\!$ &  0.17 \\
0.90--1.25 &  &  &  \multicolumn{5}{c||}{ } & 1.048 &  54.2 &   0.14 & $\!\!\pm\!\!$ &  0.03 & $\!\!\pm\!\!$ &  0.04 \\
 \hline
 \hline
   & \multicolumn{7}{c||}{$60<\theta<75$}
  & \multicolumn{7}{c|}{$75<\theta<90$} \\
 \hline
 $p_{\rm T}$ & $\langle p_{\rm T} \rangle$ & $\langle \theta \rangle$
  & \multicolumn{5}{c||}{${\rm d}^2 \sigma /{\rm d}p{\rm d}\Omega$}
  &$\langle p_{\rm T} \rangle$ & $\langle \theta \rangle$
  & \multicolumn{5}{c|}{${\rm d}^2 \sigma /{\rm d}p{\rm d}\Omega$} \\
 \hline
0.13--0.16 & 0.145 &  67.5 &  19.57 & $\!\!\pm\!\!$ &  1.66 & $\!\!\pm\!\!$ &  1.71 & 0.148 &  82.6 &  19.65 & $\!\!\pm\!\!$ &  1.65 & $\!\!\pm\!\!$ &  1.89 \\
0.16--0.20 & 0.180 &  67.5 &  21.09 & $\!\!\pm\!\!$ &  1.41 & $\!\!\pm\!\!$ &  1.39 & 0.181 &  82.0 &  18.85 & $\!\!\pm\!\!$ &  1.33 & $\!\!\pm\!\!$ &  1.34 \\
0.20--0.24 & 0.221 &  67.3 &  20.59 & $\!\!\pm\!\!$ &  1.36 & $\!\!\pm\!\!$ &  1.29 & 0.218 &  81.8 &  14.36 & $\!\!\pm\!\!$ &  1.15 & $\!\!\pm\!\!$ &  0.98 \\
0.24--0.30 & 0.269 &  67.0 &  13.53 & $\!\!\pm\!\!$ &  0.88 & $\!\!\pm\!\!$ &  0.76 & 0.268 &  81.8 &   9.46 & $\!\!\pm\!\!$ &  0.75 & $\!\!\pm\!\!$ &  0.64 \\
0.30--0.36 & 0.329 &  66.6 &   8.32 & $\!\!\pm\!\!$ &  0.69 & $\!\!\pm\!\!$ &  0.53 & 0.328 &  81.6 &   7.00 & $\!\!\pm\!\!$ &  0.64 & $\!\!\pm\!\!$ &  0.57 \\
0.36--0.42 & 0.390 &  66.8 &   7.22 & $\!\!\pm\!\!$ &  0.65 & $\!\!\pm\!\!$ &  0.58 & 0.390 &  81.8 &   4.34 & $\!\!\pm\!\!$ &  0.50 & $\!\!\pm\!\!$ &  0.45 \\
0.42--0.50 & 0.461 &  66.4 &   3.92 & $\!\!\pm\!\!$ &  0.41 & $\!\!\pm\!\!$ &  0.33 & 0.460 &  81.5 &   2.48 & $\!\!\pm\!\!$ &  0.32 & $\!\!\pm\!\!$ &  0.27 \\
0.50--0.60 & 0.549 &  66.7 &   2.49 & $\!\!\pm\!\!$ &  0.29 & $\!\!\pm\!\!$ &  0.26 & 0.551 &  80.6 &   1.18 & $\!\!\pm\!\!$ &  0.20 & $\!\!\pm\!\!$ &  0.17 \\
0.60--0.72 & 0.659 &  67.0 &   1.01 & $\!\!\pm\!\!$ &  0.17 & $\!\!\pm\!\!$ &  0.16 & 0.645 &  82.4 &   0.26 & $\!\!\pm\!\!$ &  0.08 & $\!\!\pm\!\!$ &  0.08 \\
0.72--0.90 & 0.803 &  66.0 &   0.34 & $\!\!\pm\!\!$ &  0.08 & $\!\!\pm\!\!$ &  0.11 & 0.807 &  82.7 &   0.09 & $\!\!\pm\!\!$ &  0.03 & $\!\!\pm\!\!$ &  0.06 \\
 \hline
 \hline
  & \multicolumn{7}{c||}{$90<\theta<105$}
  & \multicolumn{7}{c|}{$105<\theta<125$} \\
 \hline
 $p_{\rm T}$ & $\langle p_{\rm T} \rangle$ & $\langle \theta \rangle$
  & \multicolumn{5}{c||}{${\rm d}^2 \sigma /{\rm d}p{\rm d}\Omega$}
  &$\langle p_{\rm T} \rangle$ & $\langle \theta \rangle$
  & \multicolumn{5}{c|}{${\rm d}^2 \sigma /{\rm d}p{\rm d}\Omega$} \\
 \hline
0.13--0.16 & 0.145 &  97.6 &  14.79 & $\!\!\pm\!\!$ &  1.42 & $\!\!\pm\!\!$ &  1.45 & 0.144 & 114.7 &  12.72 & $\!\!\pm\!\!$ &  1.16 & $\!\!\pm\!\!$ &  1.10 \\
0.16--0.20 & 0.183 &  97.8 &  15.13 & $\!\!\pm\!\!$ &  1.19 & $\!\!\pm\!\!$ &  1.14 & 0.180 & 113.6 &  10.23 & $\!\!\pm\!\!$ &  0.86 & $\!\!\pm\!\!$ &  0.68 \\
0.20--0.24 & 0.218 &  97.3 &  11.01 & $\!\!\pm\!\!$ &  1.01 & $\!\!\pm\!\!$ &  0.87 & 0.218 & 113.2 &   7.25 & $\!\!\pm\!\!$ &  0.71 & $\!\!\pm\!\!$ &  0.61 \\
0.24--0.30 & 0.264 &  96.7 &   6.15 & $\!\!\pm\!\!$ &  0.61 & $\!\!\pm\!\!$ &  0.49 & 0.267 & 112.5 &   2.79 & $\!\!\pm\!\!$ &  0.36 & $\!\!\pm\!\!$ &  0.28 \\
0.30--0.36 & 0.330 &  96.6 &   3.62 & $\!\!\pm\!\!$ &  0.46 & $\!\!\pm\!\!$ &  0.39 & 0.327 & 112.6 &   1.16 & $\!\!\pm\!\!$ &  0.23 & $\!\!\pm\!\!$ &  0.18 \\
0.36--0.42 & 0.386 &  96.2 &   2.62 & $\!\!\pm\!\!$ &  0.40 & $\!\!\pm\!\!$ &  0.36 & 0.400 & 112.8 &   0.83 & $\!\!\pm\!\!$ &  0.19 & $\!\!\pm\!\!$ &  0.18 \\
0.42--0.50 & 0.457 &  95.5 &   0.87 & $\!\!\pm\!\!$ &  0.19 & $\!\!\pm\!\!$ &  0.15 & 0.454 & 113.8 &   0.33 & $\!\!\pm\!\!$ &  0.10 & $\!\!\pm\!\!$ &  0.11 \\
0.50--0.60 & 0.548 &  96.3 &   0.40 & $\!\!\pm\!\!$ &  0.11 & $\!\!\pm\!\!$ &  0.11 &  &  &  \multicolumn{5}{c|}{ } \\
0.60--0.72 & 0.662 &  92.0 &   0.11 & $\!\!\pm\!\!$ &  0.05 & $\!\!\pm\!\!$ &  0.06 &  &  &  \multicolumn{5}{c|}{ } \\
 \hline
 \end{tabular}
 \end{center}
 \end{scriptsize}
 \end{table}

%% file: table.pim.probe3.tex
  
 \begin{table}[h]
 \begin{scriptsize}
 \caption{Double-differential inclusive
  cross-section ${\rm d}^2 \sigma /{\rm d}p{\rm d}\Omega$
  [mb/(GeV/{\it c} sr)] of the production of $\pi^-$'s
  in p + Be $\rightarrow$ $\pi^-$ + X interactions
  with $+3.0$~GeV/{\it c} beam momentum;
  the first error is statistical, the second systematic; 
 $p_{\rm T}$ in GeV/{\it c}, polar angle $\theta$ in degrees.}
 \label{pim.probe3}
 \begin{center}
 \begin{tabular}{|c||c|c|rcrcr||c|c|rcrcr|}
 \hline
   & \multicolumn{7}{c||}{$20<\theta<30$}
  & \multicolumn{7}{c|}{$30<\theta<40$} \\
 \hline
 $p_{\rm T}$ & $\langle p_{\rm T} \rangle$ & $\langle \theta \rangle$
  & \multicolumn{5}{c||}{${\rm d}^2 \sigma /{\rm d}p{\rm d}\Omega$}
  &$\langle p_{\rm T} \rangle$ & $\langle \theta \rangle$
  & \multicolumn{5}{c|}{${\rm d}^2 \sigma /{\rm d}p{\rm d}\Omega$} \\
 \hline
0.10--0.13 & 0.115 &  25.5 &  17.40 & $\!\!\pm\!\!$ &  2.00 & $\!\!\pm\!\!$ &  2.06 & 0.115 &  34.5 &  19.48 & $\!\!\pm\!\!$ &  2.02 & $\!\!\pm\!\!$ &  2.28 \\
0.13--0.16 & 0.145 &  24.7 &  24.57 & $\!\!\pm\!\!$ &  2.20 & $\!\!\pm\!\!$ &  2.55 & 0.146 &  34.3 &  17.52 & $\!\!\pm\!\!$ &  1.79 & $\!\!\pm\!\!$ &  1.73 \\
0.16--0.20 & 0.179 &  24.9 &  23.25 & $\!\!\pm\!\!$ &  1.76 & $\!\!\pm\!\!$ &  1.91 & 0.181 &  34.9 &  19.27 & $\!\!\pm\!\!$ &  1.55 & $\!\!\pm\!\!$ &  1.57 \\
0.20--0.24 & 0.220 &  24.9 &  21.48 & $\!\!\pm\!\!$ &  1.70 & $\!\!\pm\!\!$ &  1.69 & 0.219 &  34.7 &  22.12 & $\!\!\pm\!\!$ &  1.69 & $\!\!\pm\!\!$ &  1.65 \\
0.24--0.30 & 0.268 &  25.2 &  15.72 & $\!\!\pm\!\!$ &  1.18 & $\!\!\pm\!\!$ &  1.09 & 0.270 &  35.0 &  16.90 & $\!\!\pm\!\!$ &  1.19 & $\!\!\pm\!\!$ &  1.17 \\
0.30--0.36 & 0.327 &  25.2 &   9.18 & $\!\!\pm\!\!$ &  0.90 & $\!\!\pm\!\!$ &  0.73 & 0.330 &  34.8 &  11.79 & $\!\!\pm\!\!$ &  0.99 & $\!\!\pm\!\!$ &  0.85 \\
0.36--0.42 & 0.391 &  24.7 &   5.31 & $\!\!\pm\!\!$ &  0.68 & $\!\!\pm\!\!$ &  0.54 & 0.390 &  35.3 &   7.60 & $\!\!\pm\!\!$ &  0.81 & $\!\!\pm\!\!$ &  0.66 \\
0.42--0.50 & 0.453 &  25.4 &   4.17 & $\!\!\pm\!\!$ &  0.53 & $\!\!\pm\!\!$ &  0.47 & 0.461 &  34.8 &   3.53 & $\!\!\pm\!\!$ &  0.48 & $\!\!\pm\!\!$ &  0.37 \\
0.50--0.60 & 0.537 &  26.2 &   1.39 & $\!\!\pm\!\!$ &  0.27 & $\!\!\pm\!\!$ &  0.24 & 0.551 &  34.8 &   2.62 & $\!\!\pm\!\!$ &  0.37 & $\!\!\pm\!\!$ &  0.37 \\
0.60--0.72 & 0.645 &  24.9 &   0.38 & $\!\!\pm\!\!$ &  0.13 & $\!\!\pm\!\!$ &  0.12 & 0.647 &  35.5 &   1.07 & $\!\!\pm\!\!$ &  0.22 & $\!\!\pm\!\!$ &  0.24 \\
0.72--0.90 &  &  &  \multicolumn{5}{c||}{ } & 0.767 &  36.4 &   0.25 & $\!\!\pm\!\!$ &  0.09 & $\!\!\pm\!\!$ &  0.13 \\
 \hline
 \hline
   & \multicolumn{7}{c||}{$40<\theta<50$}
  & \multicolumn{7}{c|}{$50<\theta<60$} \\
 \hline
 $p_{\rm T}$ & $\langle p_{\rm T} \rangle$ & $\langle \theta \rangle$
  & \multicolumn{5}{c||}{${\rm d}^2 \sigma /{\rm d}p{\rm d}\Omega$}
  &$\langle p_{\rm T} \rangle$ & $\langle \theta \rangle$
  & \multicolumn{5}{c|}{${\rm d}^2 \sigma /{\rm d}p{\rm d}\Omega$} \\
 \hline
0.10--0.13 & 0.117 &  44.8 &  16.25 & $\!\!\pm\!\!$ &  1.94 & $\!\!\pm\!\!$ &  2.09 &  &  &  \multicolumn{5}{c|}{ } \\
0.13--0.16 & 0.144 &  45.1 &  16.74 & $\!\!\pm\!\!$ &  1.81 & $\!\!\pm\!\!$ &  1.69 & 0.146 &  54.6 &  18.91 & $\!\!\pm\!\!$ &  1.96 & $\!\!\pm\!\!$ &  1.99 \\
0.16--0.20 & 0.177 &  45.1 &  17.69 & $\!\!\pm\!\!$ &  1.54 & $\!\!\pm\!\!$ &  1.46 & 0.180 &  54.9 &  18.49 & $\!\!\pm\!\!$ &  1.56 & $\!\!\pm\!\!$ &  1.49 \\
0.20--0.24 & 0.219 &  44.8 &  19.00 & $\!\!\pm\!\!$ &  1.57 & $\!\!\pm\!\!$ &  1.55 & 0.221 &  55.3 &  16.17 & $\!\!\pm\!\!$ &  1.48 & $\!\!\pm\!\!$ &  1.30 \\
0.24--0.30 & 0.270 &  44.9 &  14.87 & $\!\!\pm\!\!$ &  1.12 & $\!\!\pm\!\!$ &  1.01 & 0.266 &  54.7 &  12.13 & $\!\!\pm\!\!$ &  1.02 & $\!\!\pm\!\!$ &  0.89 \\
0.30--0.36 & 0.328 &  45.0 &  11.85 & $\!\!\pm\!\!$ &  1.01 & $\!\!\pm\!\!$ &  0.94 & 0.326 &  54.8 &   9.51 & $\!\!\pm\!\!$ &  0.90 & $\!\!\pm\!\!$ &  0.76 \\
0.36--0.42 & 0.387 &  44.7 &   6.06 & $\!\!\pm\!\!$ &  0.70 & $\!\!\pm\!\!$ &  0.54 & 0.388 &  54.9 &   7.13 & $\!\!\pm\!\!$ &  0.79 & $\!\!\pm\!\!$ &  0.68 \\
0.42--0.50 & 0.455 &  44.5 &   4.13 & $\!\!\pm\!\!$ &  0.52 & $\!\!\pm\!\!$ &  0.41 & 0.454 &  55.2 &   4.28 & $\!\!\pm\!\!$ &  0.53 & $\!\!\pm\!\!$ &  0.45 \\
0.50--0.60 & 0.537 &  45.0 &   1.98 & $\!\!\pm\!\!$ &  0.32 & $\!\!\pm\!\!$ &  0.25 & 0.540 &  54.7 &   1.37 & $\!\!\pm\!\!$ &  0.27 & $\!\!\pm\!\!$ &  0.19 \\
0.60--0.72 & 0.654 &  45.0 &   0.87 & $\!\!\pm\!\!$ &  0.20 & $\!\!\pm\!\!$ &  0.19 & 0.650 &  55.0 &   1.12 & $\!\!\pm\!\!$ &  0.22 & $\!\!\pm\!\!$ &  0.26 \\
0.72--0.90 & 0.762 &  43.3 &   0.16 & $\!\!\pm\!\!$ &  0.07 & $\!\!\pm\!\!$ &  0.09 & 0.765 &  54.8 &   0.19 & $\!\!\pm\!\!$ &  0.08 & $\!\!\pm\!\!$ &  0.09 \\
 \hline
 \hline
   & \multicolumn{7}{c||}{$60<\theta<75$}
  & \multicolumn{7}{c|}{$75<\theta<90$} \\
 \hline
 $p_{\rm T}$ & $\langle p_{\rm T} \rangle$ & $\langle \theta \rangle$
  & \multicolumn{5}{c||}{${\rm d}^2 \sigma /{\rm d}p{\rm d}\Omega$}
  &$\langle p_{\rm T} \rangle$ & $\langle \theta \rangle$
  & \multicolumn{5}{c|}{${\rm d}^2 \sigma /{\rm d}p{\rm d}\Omega$} \\
 \hline
0.13--0.16 & 0.146 &  67.5 &  15.69 & $\!\!\pm\!\!$ &  1.47 & $\!\!\pm\!\!$ &  1.52 & 0.145 &  82.9 &  14.31 & $\!\!\pm\!\!$ &  1.42 & $\!\!\pm\!\!$ &  1.42 \\
0.16--0.20 & 0.178 &  67.3 &  14.06 & $\!\!\pm\!\!$ &  1.15 & $\!\!\pm\!\!$ &  1.05 & 0.179 &  82.1 &  14.32 & $\!\!\pm\!\!$ &  1.17 & $\!\!\pm\!\!$ &  1.14 \\
0.20--0.24 & 0.219 &  67.2 &  13.56 & $\!\!\pm\!\!$ &  1.10 & $\!\!\pm\!\!$ &  0.98 & 0.217 &  82.3 &  11.87 & $\!\!\pm\!\!$ &  1.03 & $\!\!\pm\!\!$ &  0.96 \\
0.24--0.30 & 0.268 &  66.8 &   9.30 & $\!\!\pm\!\!$ &  0.73 & $\!\!\pm\!\!$ &  0.63 & 0.266 &  82.0 &   7.73 & $\!\!\pm\!\!$ &  0.68 & $\!\!\pm\!\!$ &  0.61 \\
0.30--0.36 & 0.326 &  65.8 &   7.61 & $\!\!\pm\!\!$ &  0.67 & $\!\!\pm\!\!$ &  0.57 & 0.325 &  82.1 &   3.93 & $\!\!\pm\!\!$ &  0.48 & $\!\!\pm\!\!$ &  0.44 \\
0.36--0.42 & 0.387 &  67.0 &   3.96 & $\!\!\pm\!\!$ &  0.48 & $\!\!\pm\!\!$ &  0.36 & 0.382 &  81.7 &   2.65 & $\!\!\pm\!\!$ &  0.40 & $\!\!\pm\!\!$ &  0.32 \\
0.42--0.50 & 0.450 &  66.6 &   2.56 & $\!\!\pm\!\!$ &  0.33 & $\!\!\pm\!\!$ &  0.28 & 0.459 &  82.6 &   1.49 & $\!\!\pm\!\!$ &  0.25 & $\!\!\pm\!\!$ &  0.22 \\
0.50--0.60 & 0.543 &  66.6 &   1.35 & $\!\!\pm\!\!$ &  0.22 & $\!\!\pm\!\!$ &  0.18 & 0.546 &  83.1 &   0.63 & $\!\!\pm\!\!$ &  0.15 & $\!\!\pm\!\!$ &  0.13 \\
0.60--0.72 & 0.633 &  66.6 &   0.42 & $\!\!\pm\!\!$ &  0.11 & $\!\!\pm\!\!$ &  0.11 & 0.647 &  78.0 &   0.12 & $\!\!\pm\!\!$ &  0.06 & $\!\!\pm\!\!$ &  0.05 \\
 \hline
 \hline
  & \multicolumn{7}{c||}{$90<\theta<105$}
  & \multicolumn{7}{c|}{$105<\theta<125$} \\
 \hline
 $p_{\rm T}$ & $\langle p_{\rm T} \rangle$ & $\langle \theta \rangle$
  & \multicolumn{5}{c||}{${\rm d}^2 \sigma /{\rm d}p{\rm d}\Omega$}
  &$\langle p_{\rm T} \rangle$ & $\langle \theta \rangle$
  & \multicolumn{5}{c|}{${\rm d}^2 \sigma /{\rm d}p{\rm d}\Omega$} \\
 \hline
0.13--0.16 & 0.145 &  97.8 &  11.43 & $\!\!\pm\!\!$ &  1.24 & $\!\!\pm\!\!$ &  1.16 & 0.143 & 115.3 &  12.45 & $\!\!\pm\!\!$ &  1.13 & $\!\!\pm\!\!$ &  1.11 \\
0.16--0.20 & 0.178 &  97.2 &  13.43 & $\!\!\pm\!\!$ &  1.12 & $\!\!\pm\!\!$ &  1.08 & 0.178 & 113.6 &   9.85 & $\!\!\pm\!\!$ &  0.84 & $\!\!\pm\!\!$ &  0.75 \\
0.20--0.24 & 0.216 &  96.8 &   7.62 & $\!\!\pm\!\!$ &  0.85 & $\!\!\pm\!\!$ &  0.72 & 0.218 & 114.0 &   3.92 & $\!\!\pm\!\!$ &  0.52 & $\!\!\pm\!\!$ &  0.41 \\
0.24--0.30 & 0.265 &  96.7 &   4.37 & $\!\!\pm\!\!$ &  0.52 & $\!\!\pm\!\!$ &  0.45 & 0.266 & 112.3 &   2.80 & $\!\!\pm\!\!$ &  0.36 & $\!\!\pm\!\!$ &  0.35 \\
0.30--0.36 & 0.329 &  96.6 &   2.98 & $\!\!\pm\!\!$ &  0.42 & $\!\!\pm\!\!$ &  0.44 & 0.329 & 111.3 &   1.37 & $\!\!\pm\!\!$ &  0.25 & $\!\!\pm\!\!$ &  0.24 \\
0.36--0.42 & 0.393 &  96.4 &   0.98 & $\!\!\pm\!\!$ &  0.24 & $\!\!\pm\!\!$ &  0.18 & 0.388 & 112.6 &   0.48 & $\!\!\pm\!\!$ &  0.14 & $\!\!\pm\!\!$ &  0.13 \\
0.42--0.50 & 0.445 &  97.7 &   0.98 & $\!\!\pm\!\!$ &  0.21 & $\!\!\pm\!\!$ &  0.23 & 0.468 & 114.6 &   0.18 & $\!\!\pm\!\!$ &  0.07 & $\!\!\pm\!\!$ &  0.07 \\
0.50--0.60 & 0.545 &  97.1 &   0.31 & $\!\!\pm\!\!$ &  0.10 & $\!\!\pm\!\!$ &  0.16 &  &  &  \multicolumn{5}{c|}{ } \\
 \hline
 \end{tabular}
 \end{center}
 \end{scriptsize}
 \end{table}

%% file: table.pro.pipbe3.tex
  
 \begin{table}[h]
 \begin{scriptsize}
 \caption{Double-differential inclusive
  cross-section ${\rm d}^2 \sigma /{\rm d}p{\rm d}\Omega$
  [mb/(GeV/{\it c} sr)] of the production of protons
  in $\pi^+$ + Be $\rightarrow$ p + X interactions
  with $+3.0$~GeV/{\it c} beam momentum;
  the first error is statistical, the second systematic; 
 $p_{\rm T}$ in GeV/{\it c}, polar angle $\theta$ in degrees.}
 \label{pro.pipbe3}
 \begin{center}
 \begin{tabular}{|c||c|c|rcrcr||c|c|rcrcr|}
 \hline
   & \multicolumn{7}{c||}{$20<\theta<30$}
  & \multicolumn{7}{c|}{$30<\theta<40$} \\
 \hline
 $p_{\rm T}$ & $\langle p_{\rm T} \rangle$ & $\langle \theta \rangle$
  & \multicolumn{5}{c||}{${\rm d}^2 \sigma /{\rm d}p{\rm d}\Omega$}
  &$\langle p_{\rm T} \rangle$ & $\langle \theta \rangle$
  & \multicolumn{5}{c|}{${\rm d}^2 \sigma /{\rm d}p{\rm d}\Omega$} \\
 \hline
0.20--0.24 & 0.220 &  24.9 &  50.34 & $\!\!\pm\!\!$ &  1.88 & $\!\!\pm\!\!$ &  3.48 &  &  &  \multicolumn{5}{c|}{ } \\
0.24--0.30 & 0.271 &  24.9 &  53.95 & $\!\!\pm\!\!$ &  1.58 & $\!\!\pm\!\!$ &  3.07 & 0.270 &  34.9 &  43.21 & $\!\!\pm\!\!$ &  1.37 & $\!\!\pm\!\!$ &  2.53 \\
0.30--0.36 & 0.330 &  24.9 &  47.07 & $\!\!\pm\!\!$ &  1.46 & $\!\!\pm\!\!$ &  2.45 & 0.329 &  35.1 &  43.86 & $\!\!\pm\!\!$ &  1.37 & $\!\!\pm\!\!$ &  2.23 \\
0.36--0.42 & 0.390 &  25.1 &  42.80 & $\!\!\pm\!\!$ &  1.37 & $\!\!\pm\!\!$ &  2.06 & 0.389 &  34.9 &  38.22 & $\!\!\pm\!\!$ &  1.30 & $\!\!\pm\!\!$ &  1.77 \\
0.42--0.50 & 0.460 &  24.9 &  39.00 & $\!\!\pm\!\!$ &  1.11 & $\!\!\pm\!\!$ &  1.73 & 0.459 &  34.9 &  36.01 & $\!\!\pm\!\!$ &  1.10 & $\!\!\pm\!\!$ &  1.48 \\
0.50--0.60 & 0.549 &  25.0 &  32.73 & $\!\!\pm\!\!$ &  0.91 & $\!\!\pm\!\!$ &  1.38 & 0.547 &  35.0 &  31.06 & $\!\!\pm\!\!$ &  0.91 & $\!\!\pm\!\!$ &  1.31 \\
0.60--0.72 & 0.655 &  25.1 &  24.05 & $\!\!\pm\!\!$ &  0.73 & $\!\!\pm\!\!$ &  1.26 & 0.654 &  34.8 &  21.63 & $\!\!\pm\!\!$ &  0.68 & $\!\!\pm\!\!$ &  1.10 \\
0.72--0.90 &  &  &  \multicolumn{5}{c||}{ } & 0.799 &  35.0 &  14.33 & $\!\!\pm\!\!$ &  0.47 & $\!\!\pm\!\!$ &  0.97 \\
 \hline
 \hline
   & \multicolumn{7}{c||}{$40<\theta<50$}
  & \multicolumn{7}{c|}{$50<\theta<60$} \\
 \hline
 $p_{\rm T}$ & $\langle p_{\rm T} \rangle$ & $\langle \theta \rangle$
  & \multicolumn{5}{c||}{${\rm d}^2 \sigma /{\rm d}p{\rm d}\Omega$}
  &$\langle p_{\rm T} \rangle$ & $\langle \theta \rangle$
  & \multicolumn{5}{c|}{${\rm d}^2 \sigma /{\rm d}p{\rm d}\Omega$} \\
 \hline
0.30--0.36 & 0.331 &  44.9 &  40.03 & $\!\!\pm\!\!$ &  1.27 & $\!\!\pm\!\!$ &  2.00 &  &  &  \multicolumn{5}{c|}{ } \\
0.36--0.42 & 0.390 &  44.9 &  38.72 & $\!\!\pm\!\!$ &  1.29 & $\!\!\pm\!\!$ &  1.78 & 0.392 &  55.0 &  34.14 & $\!\!\pm\!\!$ &  1.18 & $\!\!\pm\!\!$ &  1.68 \\
0.42--0.50 & 0.461 &  45.2 &  30.71 & $\!\!\pm\!\!$ &  1.00 & $\!\!\pm\!\!$ &  1.31 & 0.461 &  54.9 &  28.83 & $\!\!\pm\!\!$ &  0.95 & $\!\!\pm\!\!$ &  1.26 \\
0.50--0.60 & 0.551 &  44.9 &  26.80 & $\!\!\pm\!\!$ &  0.86 & $\!\!\pm\!\!$ &  1.14 & 0.549 &  54.9 &  22.78 & $\!\!\pm\!\!$ &  0.78 & $\!\!\pm\!\!$ &  1.08 \\
0.60--0.72 & 0.659 &  45.0 &  20.70 & $\!\!\pm\!\!$ &  0.71 & $\!\!\pm\!\!$ &  1.15 & 0.658 &  55.0 &  17.89 & $\!\!\pm\!\!$ &  0.67 & $\!\!\pm\!\!$ &  1.06 \\
0.72--0.90 & 0.804 &  45.2 &  11.04 & $\!\!\pm\!\!$ &  0.41 & $\!\!\pm\!\!$ &  0.77 & 0.804 &  54.9 &   9.14 & $\!\!\pm\!\!$ &  0.39 & $\!\!\pm\!\!$ &  0.67 \\
0.90--1.25 & 1.040 &  45.0 &   2.59 & $\!\!\pm\!\!$ &  0.13 & $\!\!\pm\!\!$ &  0.30 & 1.035 &  54.6 &   1.45 & $\!\!\pm\!\!$ &  0.09 & $\!\!\pm\!\!$ &  0.21 \\
 \hline
 \hline
   & \multicolumn{7}{c||}{$60<\theta<75$}
  & \multicolumn{7}{c|}{$75<\theta<90$} \\
 \hline
 $p_{\rm T}$ & $\langle p_{\rm T} \rangle$ & $\langle \theta \rangle$
  & \multicolumn{5}{c||}{${\rm d}^2 \sigma /{\rm d}p{\rm d}\Omega$}
  &$\langle p_{\rm T} \rangle$ & $\langle \theta \rangle$
  & \multicolumn{5}{c|}{${\rm d}^2 \sigma /{\rm d}p{\rm d}\Omega$} \\
 \hline
0.36--0.42 & 0.391 &  67.4 &  32.41 & $\!\!\pm\!\!$ &  0.95 & $\!\!\pm\!\!$ &  1.35 &  &  &  \multicolumn{5}{c|}{ } \\
0.42--0.50 & 0.461 &  67.3 &  28.38 & $\!\!\pm\!\!$ &  0.78 & $\!\!\pm\!\!$ &  1.06 & 0.460 &  81.9 &  19.96 & $\!\!\pm\!\!$ &  0.65 & $\!\!\pm\!\!$ &  1.00 \\
0.50--0.60 & 0.549 &  67.0 &  22.20 & $\!\!\pm\!\!$ &  0.63 & $\!\!\pm\!\!$ &  1.01 & 0.548 &  82.0 &  12.18 & $\!\!\pm\!\!$ &  0.45 & $\!\!\pm\!\!$ &  0.77 \\
0.60--0.72 & 0.658 &  67.0 &  11.22 & $\!\!\pm\!\!$ &  0.41 & $\!\!\pm\!\!$ &  0.73 & 0.655 &  81.4 &   5.92 & $\!\!\pm\!\!$ &  0.30 & $\!\!\pm\!\!$ &  0.52 \\
0.72--0.90 & 0.803 &  66.7 &   4.69 & $\!\!\pm\!\!$ &  0.22 & $\!\!\pm\!\!$ &  0.46 & 0.804 &  82.2 &   2.17 & $\!\!\pm\!\!$ &  0.15 & $\!\!\pm\!\!$ &  0.29 \\
0.90--1.25 & 1.041 &  66.8 &   0.71 & $\!\!\pm\!\!$ &  0.06 & $\!\!\pm\!\!$ &  0.14 & 1.035 &  81.3 &   0.24 & $\!\!\pm\!\!$ &  0.04 & $\!\!\pm\!\!$ &  0.06 \\
 \hline
 \hline
  & \multicolumn{7}{c||}{$90<\theta<105$}
  & \multicolumn{7}{c|}{$105<\theta<125$} \\
 \hline
 $p_{\rm T}$ & $\langle p_{\rm T} \rangle$ & $\langle \theta \rangle$
  & \multicolumn{5}{c||}{${\rm d}^2 \sigma /{\rm d}p{\rm d}\Omega$}
  &$\langle p_{\rm T} \rangle$ & $\langle \theta \rangle$
  & \multicolumn{5}{c|}{${\rm d}^2 \sigma /{\rm d}p{\rm d}\Omega$} \\
 \hline
0.36--0.42 &  &  &  \multicolumn{5}{c||}{ } & 0.389 & 114.0 &   8.63 & $\!\!\pm\!\!$ &  0.40 & $\!\!\pm\!\!$ &  0.61 \\
0.42--0.50 & 0.459 &  97.0 &  12.05 & $\!\!\pm\!\!$ &  0.50 & $\!\!\pm\!\!$ &  0.88 & 0.459 & 113.8 &   5.58 & $\!\!\pm\!\!$ &  0.29 & $\!\!\pm\!\!$ &  0.45 \\
0.50--0.60 & 0.546 &  96.7 &   6.52 & $\!\!\pm\!\!$ &  0.32 & $\!\!\pm\!\!$ &  0.56 & 0.544 & 113.0 &   2.68 & $\!\!\pm\!\!$ &  0.19 & $\!\!\pm\!\!$ &  0.31 \\
0.60--0.72 & 0.658 &  96.4 &   2.78 & $\!\!\pm\!\!$ &  0.20 & $\!\!\pm\!\!$ &  0.34 & 0.649 & 113.3 &   0.52 & $\!\!\pm\!\!$ &  0.08 & $\!\!\pm\!\!$ &  0.11 \\
0.72--0.90 & 0.798 &  95.9 &   0.68 & $\!\!\pm\!\!$ &  0.09 & $\!\!\pm\!\!$ &  0.13 & 0.794 & 111.5 &   0.09 & $\!\!\pm\!\!$ &  0.03 & $\!\!\pm\!\!$ &  0.04 \\
0.90--1.25 & 1.018 &  95.4 &   0.05 & $\!\!\pm\!\!$ &  0.02 & $\!\!\pm\!\!$ &  0.04 &  &  &  \multicolumn{5}{c|}{ } \\
 \hline
 \end{tabular}
 \end{center}
 \end{scriptsize}
 \end{table}

%% file: table.pip.pipbe3.tex
  
 \begin{table}[h]
 \begin{scriptsize}
 \caption{Double-differential inclusive
  cross-section ${\rm d}^2 \sigma /{\rm d}p{\rm d}\Omega$
  [mb/(GeV/{\it c} sr)] of the production of $\pi^+$'s
  in $\pi^+$ + Be $\rightarrow$ $\pi^+$ + X interactions
  with $+3.0$~GeV/{\it c} beam momentum;
  the first error is statistical, the second systematic; 
 $p_{\rm T}$ in GeV/{\it c}, polar angle $\theta$ in degrees.}
 \label{pip.pipbe3}
 \begin{center}
 \begin{tabular}{|c||c|c|rcrcr||c|c|rcrcr|}
 \hline
   & \multicolumn{7}{c||}{$20<\theta<30$}
  & \multicolumn{7}{c|}{$30<\theta<40$} \\
 \hline
 $p_{\rm T}$ & $\langle p_{\rm T} \rangle$ & $\langle \theta \rangle$
  & \multicolumn{5}{c||}{${\rm d}^2 \sigma /{\rm d}p{\rm d}\Omega$}
  &$\langle p_{\rm T} \rangle$ & $\langle \theta \rangle$
  & \multicolumn{5}{c|}{${\rm d}^2 \sigma /{\rm d}p{\rm d}\Omega$} \\
 \hline
0.10--0.13 & 0.115 &  24.8 &  43.47 & $\!\!\pm\!\!$ &  2.34 & $\!\!\pm\!\!$ &  4.31 & 0.115 &  34.9 &  33.36 & $\!\!\pm\!\!$ &  2.01 & $\!\!\pm\!\!$ &  3.40 \\
0.13--0.16 & 0.145 &  24.8 &  50.46 & $\!\!\pm\!\!$ &  2.29 & $\!\!\pm\!\!$ &  3.82 & 0.146 &  34.5 &  36.70 & $\!\!\pm\!\!$ &  1.89 & $\!\!\pm\!\!$ &  2.97 \\
0.16--0.20 & 0.181 &  24.7 &  64.45 & $\!\!\pm\!\!$ &  2.17 & $\!\!\pm\!\!$ &  3.95 & 0.181 &  34.9 &  47.62 & $\!\!\pm\!\!$ &  1.84 & $\!\!\pm\!\!$ &  3.14 \\
0.20--0.24 & 0.220 &  24.8 &  67.24 & $\!\!\pm\!\!$ &  2.18 & $\!\!\pm\!\!$ &  3.57 & 0.220 &  34.7 &  55.06 & $\!\!\pm\!\!$ &  1.92 & $\!\!\pm\!\!$ &  3.08 \\
0.24--0.30 & 0.270 &  24.9 &  66.04 & $\!\!\pm\!\!$ &  1.76 & $\!\!\pm\!\!$ &  2.86 & 0.270 &  34.8 &  53.58 & $\!\!\pm\!\!$ &  1.55 & $\!\!\pm\!\!$ &  2.50 \\
0.30--0.36 & 0.329 &  24.9 &  59.60 & $\!\!\pm\!\!$ &  1.66 & $\!\!\pm\!\!$ &  2.52 & 0.329 &  34.8 &  47.63 & $\!\!\pm\!\!$ &  1.46 & $\!\!\pm\!\!$ &  2.11 \\
0.36--0.42 & 0.389 &  24.9 &  47.34 & $\!\!\pm\!\!$ &  1.46 & $\!\!\pm\!\!$ &  1.99 & 0.389 &  34.7 &  36.43 & $\!\!\pm\!\!$ &  1.26 & $\!\!\pm\!\!$ &  1.64 \\
0.42--0.50 & 0.460 &  25.0 &  35.97 & $\!\!\pm\!\!$ &  1.08 & $\!\!\pm\!\!$ &  1.66 & 0.458 &  34.9 &  32.72 & $\!\!\pm\!\!$ &  1.05 & $\!\!\pm\!\!$ &  1.51 \\
0.50--0.60 & 0.548 &  24.9 &  24.39 & $\!\!\pm\!\!$ &  0.76 & $\!\!\pm\!\!$ &  1.47 & 0.547 &  35.0 &  20.98 & $\!\!\pm\!\!$ &  0.72 & $\!\!\pm\!\!$ &  1.21 \\
0.60--0.72 & 0.657 &  25.0 &  17.10 & $\!\!\pm\!\!$ &  0.60 & $\!\!\pm\!\!$ &  1.49 & 0.654 &  34.9 &  13.91 & $\!\!\pm\!\!$ &  0.53 & $\!\!\pm\!\!$ &  1.12 \\
0.72--0.90 &  &  &  \multicolumn{5}{c||}{ } & 0.798 &  34.7 &   9.18 & $\!\!\pm\!\!$ &  0.37 & $\!\!\pm\!\!$ &  1.12 \\
 \hline
 \hline
   & \multicolumn{7}{c||}{$40<\theta<50$}
  & \multicolumn{7}{c|}{$50<\theta<60$} \\
 \hline
 $p_{\rm T}$ & $\langle p_{\rm T} \rangle$ & $\langle \theta \rangle$
  & \multicolumn{5}{c||}{${\rm d}^2 \sigma /{\rm d}p{\rm d}\Omega$}
  &$\langle p_{\rm T} \rangle$ & $\langle \theta \rangle$
  & \multicolumn{5}{c|}{${\rm d}^2 \sigma /{\rm d}p{\rm d}\Omega$} \\
 \hline
0.10--0.13 & 0.116 &  45.0 &  32.76 & $\!\!\pm\!\!$ &  2.06 & $\!\!\pm\!\!$ &  3.31 &  &  &  \multicolumn{5}{c|}{ } \\
0.13--0.16 & 0.146 &  44.7 &  38.57 & $\!\!\pm\!\!$ &  2.00 & $\!\!\pm\!\!$ &  3.15 & 0.145 &  54.9 &  32.66 & $\!\!\pm\!\!$ &  1.91 & $\!\!\pm\!\!$ &  2.70 \\
0.16--0.20 & 0.181 &  44.7 &  38.32 & $\!\!\pm\!\!$ &  1.65 & $\!\!\pm\!\!$ &  2.58 & 0.180 &  54.8 &  35.10 & $\!\!\pm\!\!$ &  1.63 & $\!\!\pm\!\!$ &  2.29 \\
0.20--0.24 & 0.220 &  44.8 &  40.72 & $\!\!\pm\!\!$ &  1.67 & $\!\!\pm\!\!$ &  2.39 & 0.220 &  54.5 &  33.05 & $\!\!\pm\!\!$ &  1.51 & $\!\!\pm\!\!$ &  2.00 \\
0.24--0.30 & 0.271 &  44.8 &  42.43 & $\!\!\pm\!\!$ &  1.38 & $\!\!\pm\!\!$ &  2.01 & 0.270 &  54.7 &  32.36 & $\!\!\pm\!\!$ &  1.22 & $\!\!\pm\!\!$ &  1.57 \\
0.30--0.36 & 0.331 &  44.6 &  33.97 & $\!\!\pm\!\!$ &  1.21 & $\!\!\pm\!\!$ &  1.55 & 0.331 &  54.7 &  27.80 & $\!\!\pm\!\!$ &  1.12 & $\!\!\pm\!\!$ &  1.28 \\
0.36--0.42 & 0.391 &  44.7 &  29.63 & $\!\!\pm\!\!$ &  1.14 & $\!\!\pm\!\!$ &  1.41 & 0.390 &  54.9 &  24.49 & $\!\!\pm\!\!$ &  1.06 & $\!\!\pm\!\!$ &  1.18 \\
0.42--0.50 & 0.460 &  44.7 &  24.72 & $\!\!\pm\!\!$ &  0.93 & $\!\!\pm\!\!$ &  1.18 & 0.461 &  54.9 &  18.68 & $\!\!\pm\!\!$ &  0.80 & $\!\!\pm\!\!$ &  1.00 \\
0.50--0.60 & 0.551 &  44.7 &  17.94 & $\!\!\pm\!\!$ &  0.69 & $\!\!\pm\!\!$ &  1.06 & 0.550 &  54.7 &  13.98 & $\!\!\pm\!\!$ &  0.61 & $\!\!\pm\!\!$ &  0.90 \\
0.60--0.72 & 0.662 &  44.6 &  12.72 & $\!\!\pm\!\!$ &  0.53 & $\!\!\pm\!\!$ &  0.97 & 0.660 &  54.6 &   8.57 & $\!\!\pm\!\!$ &  0.44 & $\!\!\pm\!\!$ &  0.72 \\
0.72--0.90 & 0.802 &  44.5 &   6.08 & $\!\!\pm\!\!$ &  0.30 & $\!\!\pm\!\!$ &  0.67 & 0.800 &  54.7 &   4.44 & $\!\!\pm\!\!$ &  0.27 & $\!\!\pm\!\!$ &  0.50 \\
0.90--1.25 &  &  &  \multicolumn{5}{c||}{ } & 1.026 &  54.3 &   0.69 & $\!\!\pm\!\!$ &  0.06 & $\!\!\pm\!\!$ &  0.14 \\
 \hline
 \hline
   & \multicolumn{7}{c||}{$60<\theta<75$}
  & \multicolumn{7}{c|}{$75<\theta<90$} \\
 \hline
 $p_{\rm T}$ & $\langle p_{\rm T} \rangle$ & $\langle \theta \rangle$
  & \multicolumn{5}{c||}{${\rm d}^2 \sigma /{\rm d}p{\rm d}\Omega$}
  &$\langle p_{\rm T} \rangle$ & $\langle \theta \rangle$
  & \multicolumn{5}{c|}{${\rm d}^2 \sigma /{\rm d}p{\rm d}\Omega$} \\
 \hline
0.13--0.16 & 0.146 &  67.2 &  27.36 & $\!\!\pm\!\!$ &  1.41 & $\!\!\pm\!\!$ &  2.13 & 0.145 &  82.3 &  23.85 & $\!\!\pm\!\!$ &  1.33 & $\!\!\pm\!\!$ &  1.95 \\
0.16--0.20 & 0.180 &  67.4 &  32.26 & $\!\!\pm\!\!$ &  1.27 & $\!\!\pm\!\!$ &  1.96 & 0.181 &  82.2 &  27.84 & $\!\!\pm\!\!$ &  1.17 & $\!\!\pm\!\!$ &  1.73 \\
0.20--0.24 & 0.220 &  67.5 &  28.37 & $\!\!\pm\!\!$ &  1.15 & $\!\!\pm\!\!$ &  1.62 & 0.219 &  82.1 &  22.35 & $\!\!\pm\!\!$ &  1.04 & $\!\!\pm\!\!$ &  1.36 \\
0.24--0.30 & 0.270 &  67.3 &  23.71 & $\!\!\pm\!\!$ &  0.86 & $\!\!\pm\!\!$ &  1.13 & 0.269 &  82.1 &  17.37 & $\!\!\pm\!\!$ &  0.74 & $\!\!\pm\!\!$ &  0.91 \\
0.30--0.36 & 0.331 &  67.2 &  20.19 & $\!\!\pm\!\!$ &  0.79 & $\!\!\pm\!\!$ &  0.93 & 0.331 &  81.5 &  12.76 & $\!\!\pm\!\!$ &  0.63 & $\!\!\pm\!\!$ &  0.71 \\
0.36--0.42 & 0.392 &  67.1 &  16.77 & $\!\!\pm\!\!$ &  0.73 & $\!\!\pm\!\!$ &  0.82 & 0.389 &  81.1 &  10.39 & $\!\!\pm\!\!$ &  0.57 & $\!\!\pm\!\!$ &  0.68 \\
0.42--0.50 & 0.461 &  66.9 &  12.41 & $\!\!\pm\!\!$ &  0.53 & $\!\!\pm\!\!$ &  0.71 & 0.460 &  82.3 &   6.55 & $\!\!\pm\!\!$ &  0.39 & $\!\!\pm\!\!$ &  0.51 \\
0.50--0.60 & 0.550 &  66.4 &   8.58 & $\!\!\pm\!\!$ &  0.39 & $\!\!\pm\!\!$ &  0.63 & 0.552 &  82.0 &   4.47 & $\!\!\pm\!\!$ &  0.29 & $\!\!\pm\!\!$ &  0.44 \\
0.60--0.72 & 0.659 &  66.6 &   4.59 & $\!\!\pm\!\!$ &  0.27 & $\!\!\pm\!\!$ &  0.47 & 0.658 &  81.5 &   2.01 & $\!\!\pm\!\!$ &  0.17 & $\!\!\pm\!\!$ &  0.27 \\
0.72--0.90 & 0.793 &  66.1 &   1.75 & $\!\!\pm\!\!$ &  0.13 & $\!\!\pm\!\!$ &  0.24 & 0.809 &  81.6 &   0.56 & $\!\!\pm\!\!$ &  0.06 & $\!\!\pm\!\!$ &  0.12 \\
0.90--1.25 & 1.029 &  66.0 &   0.20 & $\!\!\pm\!\!$ &  0.02 & $\!\!\pm\!\!$ &  0.05 & 1.035 &  82.4 &   0.05 & $\!\!\pm\!\!$ &  0.01 & $\!\!\pm\!\!$ &  0.03 \\
 \hline
 \hline
  & \multicolumn{7}{c||}{$90<\theta<105$}
  & \multicolumn{7}{c|}{$105<\theta<125$} \\
 \hline
 $p_{\rm T}$ & $\langle p_{\rm T} \rangle$ & $\langle \theta \rangle$
  & \multicolumn{5}{c||}{${\rm d}^2 \sigma /{\rm d}p{\rm d}\Omega$}
  &$\langle p_{\rm T} \rangle$ & $\langle \theta \rangle$
  & \multicolumn{5}{c|}{${\rm d}^2 \sigma /{\rm d}p{\rm d}\Omega$} \\
 \hline
0.13--0.16 & 0.146 &  97.0 &  19.93 & $\!\!\pm\!\!$ &  1.20 & $\!\!\pm\!\!$ &  1.62 & 0.145 & 114.6 &  20.18 & $\!\!\pm\!\!$ &  1.07 & $\!\!\pm\!\!$ &  1.46 \\
0.16--0.20 & 0.179 &  97.4 &  24.56 & $\!\!\pm\!\!$ &  1.12 & $\!\!\pm\!\!$ &  1.53 & 0.180 & 114.2 &  17.84 & $\!\!\pm\!\!$ &  0.83 & $\!\!\pm\!\!$ &  1.01 \\
0.20--0.24 & 0.220 &  96.8 &  19.59 & $\!\!\pm\!\!$ &  0.98 & $\!\!\pm\!\!$ &  1.16 & 0.219 & 114.1 &  11.22 & $\!\!\pm\!\!$ &  0.64 & $\!\!\pm\!\!$ &  0.69 \\
0.24--0.30 & 0.269 &  96.8 &  12.11 & $\!\!\pm\!\!$ &  0.62 & $\!\!\pm\!\!$ &  0.66 & 0.267 & 113.7 &   6.86 & $\!\!\pm\!\!$ &  0.41 & $\!\!\pm\!\!$ &  0.48 \\
0.30--0.36 & 0.329 &  96.8 &   9.74 & $\!\!\pm\!\!$ &  0.55 & $\!\!\pm\!\!$ &  0.67 & 0.329 & 114.2 &   4.83 & $\!\!\pm\!\!$ &  0.34 & $\!\!\pm\!\!$ &  0.45 \\
0.36--0.42 & 0.390 &  97.1 &   6.52 & $\!\!\pm\!\!$ &  0.46 & $\!\!\pm\!\!$ &  0.57 & 0.390 & 113.4 &   2.61 & $\!\!\pm\!\!$ &  0.24 & $\!\!\pm\!\!$ &  0.33 \\
0.42--0.50 & 0.460 &  96.3 &   3.56 & $\!\!\pm\!\!$ &  0.29 & $\!\!\pm\!\!$ &  0.38 & 0.458 & 112.8 &   1.44 & $\!\!\pm\!\!$ &  0.15 & $\!\!\pm\!\!$ &  0.23 \\
0.50--0.60 & 0.549 &  95.9 &   1.84 & $\!\!\pm\!\!$ &  0.18 & $\!\!\pm\!\!$ &  0.26 & 0.545 & 111.8 &   0.39 & $\!\!\pm\!\!$ &  0.06 & $\!\!\pm\!\!$ &  0.10 \\
0.60--0.72 & 0.657 &  96.6 &   0.69 & $\!\!\pm\!\!$ &  0.09 & $\!\!\pm\!\!$ &  0.19 & 0.640 & 111.3 &   0.23 & $\!\!\pm\!\!$ &  0.05 & $\!\!\pm\!\!$ &  0.09 \\
0.72--0.90 & 0.786 &  95.8 &   0.16 & $\!\!\pm\!\!$ &  0.03 & $\!\!\pm\!\!$ &  0.05 &  &  &  \multicolumn{5}{c|}{ } \\
 \hline
 \end{tabular}
 \end{center}
 \end{scriptsize}
 \end{table}

%% file: table.pim.pipbe3.tex
  
 \begin{table}[h]
 \begin{scriptsize}
 \caption{Double-differential inclusive
  cross-section ${\rm d}^2 \sigma /{\rm d}p{\rm d}\Omega$
  [mb/(GeV/{\it c} sr)] of the production of $\pi^-$'s
  in $\pi^+$ + Be $\rightarrow$ $\pi^-$ + X interactions
  with $+3.0$~GeV/{\it c} beam momentum;
  the first error is statistical, the second systematic; 
 $p_{\rm T}$ in GeV/{\it c}, polar angle $\theta$ in degrees.}
 \label{pim.pipbe3}
 \begin{center}
 \begin{tabular}{|c||c|c|rcrcr||c|c|rcrcr|}
 \hline
   & \multicolumn{7}{c||}{$20<\theta<30$}
  & \multicolumn{7}{c|}{$30<\theta<40$} \\
 \hline
 $p_{\rm T}$ & $\langle p_{\rm T} \rangle$ & $\langle \theta \rangle$
  & \multicolumn{5}{c||}{${\rm d}^2 \sigma /{\rm d}p{\rm d}\Omega$}
  &$\langle p_{\rm T} \rangle$ & $\langle \theta \rangle$
  & \multicolumn{5}{c|}{${\rm d}^2 \sigma /{\rm d}p{\rm d}\Omega$} \\
 \hline
0.10--0.13 & 0.116 &  24.7 &  30.88 & $\!\!\pm\!\!$ &  1.92 & $\!\!\pm\!\!$ &  3.72 & 0.116 &  34.5 &  22.19 & $\!\!\pm\!\!$ &  1.56 & $\!\!\pm\!\!$ &  2.72 \\
0.13--0.16 & 0.145 &  24.6 &  37.24 & $\!\!\pm\!\!$ &  1.96 & $\!\!\pm\!\!$ &  3.16 & 0.146 &  35.0 &  31.28 & $\!\!\pm\!\!$ &  1.75 & $\!\!\pm\!\!$ &  3.34 \\
0.16--0.20 & 0.180 &  24.9 &  40.74 & $\!\!\pm\!\!$ &  1.69 & $\!\!\pm\!\!$ &  2.75 & 0.181 &  34.8 &  29.45 & $\!\!\pm\!\!$ &  1.39 & $\!\!\pm\!\!$ &  2.40 \\
0.20--0.24 & 0.220 &  24.9 &  44.76 & $\!\!\pm\!\!$ &  1.77 & $\!\!\pm\!\!$ &  2.68 & 0.220 &  34.8 &  32.04 & $\!\!\pm\!\!$ &  1.46 & $\!\!\pm\!\!$ &  2.08 \\
0.24--0.30 & 0.269 &  25.0 &  37.20 & $\!\!\pm\!\!$ &  1.30 & $\!\!\pm\!\!$ &  1.80 & 0.269 &  34.8 &  31.91 & $\!\!\pm\!\!$ &  1.17 & $\!\!\pm\!\!$ &  1.72 \\
0.30--0.36 & 0.328 &  24.9 &  33.42 & $\!\!\pm\!\!$ &  1.24 & $\!\!\pm\!\!$ &  1.54 & 0.329 &  34.8 &  27.38 & $\!\!\pm\!\!$ &  1.09 & $\!\!\pm\!\!$ &  1.41 \\
0.36--0.42 & 0.389 &  25.1 &  26.86 & $\!\!\pm\!\!$ &  1.12 & $\!\!\pm\!\!$ &  1.35 & 0.388 &  34.8 &  21.11 & $\!\!\pm\!\!$ &  0.96 & $\!\!\pm\!\!$ &  1.16 \\
0.42--0.50 & 0.456 &  25.0 &  18.55 & $\!\!\pm\!\!$ &  0.80 & $\!\!\pm\!\!$ &  1.06 & 0.457 &  34.6 &  15.18 & $\!\!\pm\!\!$ &  0.71 & $\!\!\pm\!\!$ &  0.89 \\
0.50--0.60 & 0.543 &  25.0 &   9.36 & $\!\!\pm\!\!$ &  0.50 & $\!\!\pm\!\!$ &  0.68 & 0.543 &  34.8 &   9.29 & $\!\!\pm\!\!$ &  0.50 & $\!\!\pm\!\!$ &  0.70 \\
0.60--0.72 & 0.651 &  25.0 &   5.11 & $\!\!\pm\!\!$ &  0.35 & $\!\!\pm\!\!$ &  0.52 & 0.652 &  34.5 &   4.87 & $\!\!\pm\!\!$ &  0.34 & $\!\!\pm\!\!$ &  0.52 \\
0.72--0.90 &  &  &  \multicolumn{5}{c||}{ } & 0.784 &  35.0 &   1.98 & $\!\!\pm\!\!$ &  0.19 & $\!\!\pm\!\!$ &  0.26 \\
 \hline
 \hline
   & \multicolumn{7}{c||}{$40<\theta<50$}
  & \multicolumn{7}{c|}{$50<\theta<60$} \\
 \hline
 $p_{\rm T}$ & $\langle p_{\rm T} \rangle$ & $\langle \theta \rangle$
  & \multicolumn{5}{c||}{${\rm d}^2 \sigma /{\rm d}p{\rm d}\Omega$}
  &$\langle p_{\rm T} \rangle$ & $\langle \theta \rangle$
  & \multicolumn{5}{c|}{${\rm d}^2 \sigma /{\rm d}p{\rm d}\Omega$} \\
 \hline
0.10--0.13 & 0.115 &  44.6 &  21.93 & $\!\!\pm\!\!$ &  1.62 & $\!\!\pm\!\!$ &  2.45 &  &  &  \multicolumn{5}{c|}{ } \\
0.13--0.16 & 0.144 &  44.7 &  26.26 & $\!\!\pm\!\!$ &  1.62 & $\!\!\pm\!\!$ &  2.46 & 0.146 &  54.6 &  23.61 & $\!\!\pm\!\!$ &  1.58 & $\!\!\pm\!\!$ &  2.20 \\
0.16--0.20 & 0.180 &  44.6 &  26.11 & $\!\!\pm\!\!$ &  1.34 & $\!\!\pm\!\!$ &  1.95 & 0.180 &  54.7 &  21.65 & $\!\!\pm\!\!$ &  1.22 & $\!\!\pm\!\!$ &  1.59 \\
0.20--0.24 & 0.219 &  44.8 &  25.18 & $\!\!\pm\!\!$ &  1.30 & $\!\!\pm\!\!$ &  1.69 & 0.219 &  54.6 &  22.61 & $\!\!\pm\!\!$ &  1.26 & $\!\!\pm\!\!$ &  1.74 \\
0.24--0.30 & 0.267 &  44.8 &  25.29 & $\!\!\pm\!\!$ &  1.05 & $\!\!\pm\!\!$ &  1.36 & 0.269 &  54.9 &  18.22 & $\!\!\pm\!\!$ &  0.90 & $\!\!\pm\!\!$ &  1.12 \\
0.30--0.36 & 0.327 &  44.3 &  22.08 & $\!\!\pm\!\!$ &  1.00 & $\!\!\pm\!\!$ &  1.25 & 0.328 &  54.8 &  15.85 & $\!\!\pm\!\!$ &  0.84 & $\!\!\pm\!\!$ &  0.92 \\
0.36--0.42 & 0.387 &  44.6 &  16.81 & $\!\!\pm\!\!$ &  0.85 & $\!\!\pm\!\!$ &  1.06 & 0.388 &  54.5 &  11.86 & $\!\!\pm\!\!$ &  0.73 & $\!\!\pm\!\!$ &  0.76 \\
0.42--0.50 & 0.456 &  44.9 &  12.20 & $\!\!\pm\!\!$ &  0.65 & $\!\!\pm\!\!$ &  0.81 & 0.455 &  54.9 &   9.31 & $\!\!\pm\!\!$ &  0.57 & $\!\!\pm\!\!$ &  0.64 \\
0.50--0.60 & 0.540 &  44.8 &   6.91 & $\!\!\pm\!\!$ &  0.43 & $\!\!\pm\!\!$ &  0.56 & 0.541 &  55.0 &   5.19 & $\!\!\pm\!\!$ &  0.38 & $\!\!\pm\!\!$ &  0.46 \\
0.60--0.72 & 0.653 &  44.5 &   3.73 & $\!\!\pm\!\!$ &  0.29 & $\!\!\pm\!\!$ &  0.41 & 0.647 &  54.6 &   2.77 & $\!\!\pm\!\!$ &  0.25 & $\!\!\pm\!\!$ &  0.31 \\
0.72--0.90 & 0.788 &  44.7 &   1.56 & $\!\!\pm\!\!$ &  0.16 & $\!\!\pm\!\!$ &  0.23 & 0.789 &  54.8 &   1.06 & $\!\!\pm\!\!$ &  0.13 & $\!\!\pm\!\!$ &  0.17 \\
0.90--1.25 &  &  &  \multicolumn{5}{c||}{ } & 0.992 &  54.7 &   0.13 & $\!\!\pm\!\!$ &  0.02 & $\!\!\pm\!\!$ &  0.04 \\
 \hline
 \hline
   & \multicolumn{7}{c||}{$60<\theta<75$}
  & \multicolumn{7}{c|}{$75<\theta<90$} \\
 \hline
 $p_{\rm T}$ & $\langle p_{\rm T} \rangle$ & $\langle \theta \rangle$
  & \multicolumn{5}{c||}{${\rm d}^2 \sigma /{\rm d}p{\rm d}\Omega$}
  &$\langle p_{\rm T} \rangle$ & $\langle \theta \rangle$
  & \multicolumn{5}{c|}{${\rm d}^2 \sigma /{\rm d}p{\rm d}\Omega$} \\
 \hline
0.13--0.16 & 0.144 &  67.3 &  18.75 & $\!\!\pm\!\!$ &  1.16 & $\!\!\pm\!\!$ &  1.57 & 0.145 &  82.8 &  18.01 & $\!\!\pm\!\!$ &  1.14 & $\!\!\pm\!\!$ &  1.67 \\
0.16--0.20 & 0.179 &  67.3 &  17.85 & $\!\!\pm\!\!$ &  0.93 & $\!\!\pm\!\!$ &  1.20 & 0.179 &  82.3 &  16.80 & $\!\!\pm\!\!$ &  0.91 & $\!\!\pm\!\!$ &  1.21 \\
0.20--0.24 & 0.219 &  66.8 &  18.04 & $\!\!\pm\!\!$ &  0.91 & $\!\!\pm\!\!$ &  1.20 & 0.218 &  81.9 &  12.18 & $\!\!\pm\!\!$ &  0.75 & $\!\!\pm\!\!$ &  0.88 \\
0.24--0.30 & 0.269 &  67.1 &  13.77 & $\!\!\pm\!\!$ &  0.64 & $\!\!\pm\!\!$ &  0.77 & 0.268 &  82.5 &   9.99 & $\!\!\pm\!\!$ &  0.56 & $\!\!\pm\!\!$ &  0.65 \\
0.30--0.36 & 0.329 &  67.0 &  12.02 & $\!\!\pm\!\!$ &  0.61 & $\!\!\pm\!\!$ &  0.71 & 0.327 &  82.0 &   8.09 & $\!\!\pm\!\!$ &  0.50 & $\!\!\pm\!\!$ &  0.59 \\
0.36--0.42 & 0.386 &  66.3 &   9.40 & $\!\!\pm\!\!$ &  0.53 & $\!\!\pm\!\!$ &  0.59 & 0.388 &  82.0 &   5.41 & $\!\!\pm\!\!$ &  0.41 & $\!\!\pm\!\!$ &  0.45 \\
0.42--0.50 & 0.452 &  66.5 &   6.20 & $\!\!\pm\!\!$ &  0.37 & $\!\!\pm\!\!$ &  0.46 & 0.455 &  82.3 &   3.34 & $\!\!\pm\!\!$ &  0.28 & $\!\!\pm\!\!$ &  0.33 \\
0.50--0.60 & 0.541 &  66.4 &   4.18 & $\!\!\pm\!\!$ &  0.28 & $\!\!\pm\!\!$ &  0.39 & 0.542 &  81.5 &   1.93 & $\!\!\pm\!\!$ &  0.19 & $\!\!\pm\!\!$ &  0.24 \\
0.60--0.72 & 0.645 &  66.7 &   2.26 & $\!\!\pm\!\!$ &  0.18 & $\!\!\pm\!\!$ &  0.29 & 0.642 &  82.1 &   1.01 & $\!\!\pm\!\!$ &  0.12 & $\!\!\pm\!\!$ &  0.18 \\
0.72--0.90 & 0.784 &  66.4 &   0.52 & $\!\!\pm\!\!$ &  0.07 & $\!\!\pm\!\!$ &  0.09 & 0.785 &  82.7 &   0.18 & $\!\!\pm\!\!$ &  0.04 & $\!\!\pm\!\!$ &  0.05 \\
0.90--1.25 & 1.021 &  65.6 &   0.07 & $\!\!\pm\!\!$ &  0.02 & $\!\!\pm\!\!$ &  0.03 &  &  &  \multicolumn{5}{c|}{ } \\
 \hline
 \hline
  & \multicolumn{7}{c||}{$90<\theta<105$}
  & \multicolumn{7}{c|}{$105<\theta<125$} \\
 \hline
 $p_{\rm T}$ & $\langle p_{\rm T} \rangle$ & $\langle \theta \rangle$
  & \multicolumn{5}{c||}{${\rm d}^2 \sigma /{\rm d}p{\rm d}\Omega$}
  &$\langle p_{\rm T} \rangle$ & $\langle \theta \rangle$
  & \multicolumn{5}{c|}{${\rm d}^2 \sigma /{\rm d}p{\rm d}\Omega$} \\
 \hline
0.13--0.16 & 0.144 &  97.0 &  14.92 & $\!\!\pm\!\!$ &  1.04 & $\!\!\pm\!\!$ &  1.51 & 0.145 & 114.3 &  11.86 & $\!\!\pm\!\!$ &  0.81 & $\!\!\pm\!\!$ &  0.96 \\
0.16--0.20 & 0.180 &  97.2 &  13.45 & $\!\!\pm\!\!$ &  0.81 & $\!\!\pm\!\!$ &  1.08 & 0.178 & 114.7 &  10.27 & $\!\!\pm\!\!$ &  0.63 & $\!\!\pm\!\!$ &  0.72 \\
0.20--0.24 & 0.218 &  97.2 &  10.54 & $\!\!\pm\!\!$ &  0.72 & $\!\!\pm\!\!$ &  0.87 & 0.218 & 113.8 &   6.81 & $\!\!\pm\!\!$ &  0.50 & $\!\!\pm\!\!$ &  0.60 \\
0.24--0.30 & 0.266 &  97.0 &   7.52 & $\!\!\pm\!\!$ &  0.49 & $\!\!\pm\!\!$ &  0.58 & 0.268 & 113.6 &   3.57 & $\!\!\pm\!\!$ &  0.29 & $\!\!\pm\!\!$ &  0.34 \\
0.30--0.36 & 0.327 &  97.2 &   4.88 & $\!\!\pm\!\!$ &  0.39 & $\!\!\pm\!\!$ &  0.46 & 0.325 & 113.2 &   2.68 & $\!\!\pm\!\!$ &  0.25 & $\!\!\pm\!\!$ &  0.34 \\
0.36--0.42 & 0.385 &  97.3 &   3.56 & $\!\!\pm\!\!$ &  0.33 & $\!\!\pm\!\!$ &  0.40 & 0.389 & 113.6 &   1.45 & $\!\!\pm\!\!$ &  0.18 & $\!\!\pm\!\!$ &  0.22 \\
0.42--0.50 & 0.456 &  97.0 &   2.01 & $\!\!\pm\!\!$ &  0.22 & $\!\!\pm\!\!$ &  0.27 & 0.447 & 113.2 &   0.78 & $\!\!\pm\!\!$ &  0.11 & $\!\!\pm\!\!$ &  0.16 \\
0.50--0.60 & 0.530 &  97.9 &   0.74 & $\!\!\pm\!\!$ &  0.11 & $\!\!\pm\!\!$ &  0.14 & 0.539 & 112.2 &   0.26 & $\!\!\pm\!\!$ &  0.05 & $\!\!\pm\!\!$ &  0.09 \\
0.60--0.72 & 0.637 &  96.3 &   0.31 & $\!\!\pm\!\!$ &  0.06 & $\!\!\pm\!\!$ &  0.09 & 0.658 & 110.4 &   0.11 & $\!\!\pm\!\!$ &  0.03 & $\!\!\pm\!\!$ &  0.05 \\
0.72--0.90 & 0.801 &  97.3 &   0.08 & $\!\!\pm\!\!$ &  0.02 & $\!\!\pm\!\!$ &  0.04 &  &  &  \multicolumn{5}{c|}{ } \\
 \hline
 \end{tabular}
 \end{center}
 \end{scriptsize}
 \end{table}

%% file: table.pro.pimbe3.tex
  
 \begin{table}[h]
 \begin{scriptsize}
 \caption{Double-differential inclusive
  cross-section ${\rm d}^2 \sigma /{\rm d}p{\rm d}\Omega$
  [mb/(GeV/{\it c} sr)] of the production of protons
  in $\pi^-$ + Be $\rightarrow$ p + X interactions
  with $-3.0$~GeV/{\it c} beam momentum;
  the first error is statistical, the second systematic; 
 $p_{\rm T}$ in GeV/{\it c}, polar angle $\theta$ in degrees.}
 \label{pro.pimbe3}
 \begin{center}
 \begin{tabular}{|c||c|c|rcrcr||c|c|rcrcr|}
 \hline
   & \multicolumn{7}{c||}{$20<\theta<30$}
  & \multicolumn{7}{c|}{$30<\theta<40$} \\
 \hline
 $p_{\rm T}$ & $\langle p_{\rm T} \rangle$ & $\langle \theta \rangle$
  & \multicolumn{5}{c||}{${\rm d}^2 \sigma /{\rm d}p{\rm d}\Omega$}
  &$\langle p_{\rm T} \rangle$ & $\langle \theta \rangle$
  & \multicolumn{5}{c|}{${\rm d}^2 \sigma /{\rm d}p{\rm d}\Omega$} \\
 \hline
0.20--0.24 & 0.219 &  25.0 &  33.91 & $\!\!\pm\!\!$ &  1.47 & $\!\!\pm\!\!$ &  2.56 &  &  &  \multicolumn{5}{c|}{ } \\
0.24--0.30 & 0.267 &  25.0 &  33.11 & $\!\!\pm\!\!$ &  1.17 & $\!\!\pm\!\!$ &  2.03 & 0.268 &  34.8 &  31.08 & $\!\!\pm\!\!$ &  1.12 & $\!\!\pm\!\!$ &  1.87 \\
0.30--0.36 & 0.327 &  25.1 &  31.93 & $\!\!\pm\!\!$ &  1.16 & $\!\!\pm\!\!$ &  1.79 & 0.326 &  34.8 &  26.38 & $\!\!\pm\!\!$ &  1.01 & $\!\!\pm\!\!$ &  1.47 \\
0.36--0.42 & 0.386 &  25.1 &  28.54 & $\!\!\pm\!\!$ &  1.07 & $\!\!\pm\!\!$ &  1.56 & 0.385 &  34.9 &  23.94 & $\!\!\pm\!\!$ &  0.99 & $\!\!\pm\!\!$ &  1.30 \\
0.42--0.50 & 0.452 &  25.0 &  24.84 & $\!\!\pm\!\!$ &  0.87 & $\!\!\pm\!\!$ &  1.28 & 0.452 &  35.0 &  21.35 & $\!\!\pm\!\!$ &  0.82 & $\!\!\pm\!\!$ &  1.05 \\
0.50--0.60 & 0.538 &  25.1 &  20.22 & $\!\!\pm\!\!$ &  0.71 & $\!\!\pm\!\!$ &  0.99 & 0.541 &  34.9 &  18.89 & $\!\!\pm\!\!$ &  0.70 & $\!\!\pm\!\!$ &  0.98 \\
0.60--0.72 & 0.642 &  25.1 &  13.32 & $\!\!\pm\!\!$ &  0.51 & $\!\!\pm\!\!$ &  0.73 & 0.643 &  34.7 &  12.44 & $\!\!\pm\!\!$ &  0.51 & $\!\!\pm\!\!$ &  0.73 \\
0.72--0.90 &  &  &  \multicolumn{5}{c||}{ } & 0.781 &  35.1 &   8.20 & $\!\!\pm\!\!$ &  0.35 & $\!\!\pm\!\!$ &  0.62 \\
 \hline
 \hline
   & \multicolumn{7}{c||}{$40<\theta<50$}
  & \multicolumn{7}{c|}{$50<\theta<60$} \\
 \hline
 $p_{\rm T}$ & $\langle p_{\rm T} \rangle$ & $\langle \theta \rangle$
  & \multicolumn{5}{c||}{${\rm d}^2 \sigma /{\rm d}p{\rm d}\Omega$}
  &$\langle p_{\rm T} \rangle$ & $\langle \theta \rangle$
  & \multicolumn{5}{c|}{${\rm d}^2 \sigma /{\rm d}p{\rm d}\Omega$} \\
 \hline
0.30--0.36 & 0.330 &  45.0 &  25.49 & $\!\!\pm\!\!$ &  0.95 & $\!\!\pm\!\!$ &  1.51 &  &  &  \multicolumn{5}{c|}{ } \\
0.36--0.42 & 0.389 &  45.1 &  22.78 & $\!\!\pm\!\!$ &  0.92 & $\!\!\pm\!\!$ &  1.17 & 0.389 &  55.1 &  23.63 & $\!\!\pm\!\!$ &  0.93 & $\!\!\pm\!\!$ &  1.29 \\
0.42--0.50 & 0.458 &  45.0 &  19.31 & $\!\!\pm\!\!$ &  0.77 & $\!\!\pm\!\!$ &  1.01 & 0.459 &  55.0 &  18.58 & $\!\!\pm\!\!$ &  0.72 & $\!\!\pm\!\!$ &  1.02 \\
0.50--0.60 & 0.547 &  45.1 &  15.77 & $\!\!\pm\!\!$ &  0.64 & $\!\!\pm\!\!$ &  0.86 & 0.549 &  55.1 &  14.23 & $\!\!\pm\!\!$ &  0.61 & $\!\!\pm\!\!$ &  0.84 \\
0.60--0.72 & 0.656 &  45.0 &  11.75 & $\!\!\pm\!\!$ &  0.53 & $\!\!\pm\!\!$ &  0.81 & 0.657 &  54.8 &  10.03 & $\!\!\pm\!\!$ &  0.49 & $\!\!\pm\!\!$ &  0.72 \\
0.72--0.90 & 0.801 &  44.9 &   6.18 & $\!\!\pm\!\!$ &  0.31 & $\!\!\pm\!\!$ &  0.50 & 0.798 &  54.9 &   5.16 & $\!\!\pm\!\!$ &  0.29 & $\!\!\pm\!\!$ &  0.45 \\
0.90--1.25 & 1.034 &  44.7 &   1.94 & $\!\!\pm\!\!$ &  0.12 & $\!\!\pm\!\!$ &  0.25 & 1.028 &  54.8 &   0.84 & $\!\!\pm\!\!$ &  0.08 & $\!\!\pm\!\!$ &  0.15 \\
 \hline
 \hline
   & \multicolumn{7}{c||}{$60<\theta<75$}
  & \multicolumn{7}{c|}{$75<\theta<90$} \\
 \hline
 $p_{\rm T}$ & $\langle p_{\rm T} \rangle$ & $\langle \theta \rangle$
  & \multicolumn{5}{c||}{${\rm d}^2 \sigma /{\rm d}p{\rm d}\Omega$}
  &$\langle p_{\rm T} \rangle$ & $\langle \theta \rangle$
  & \multicolumn{5}{c|}{${\rm d}^2 \sigma /{\rm d}p{\rm d}\Omega$} \\
 \hline
0.36--0.42 & 0.385 &  67.6 &  23.08 & $\!\!\pm\!\!$ &  0.74 & $\!\!\pm\!\!$ &  1.00 &  &  &  \multicolumn{5}{c|}{ } \\
0.42--0.50 & 0.452 &  67.5 &  19.40 & $\!\!\pm\!\!$ &  0.60 & $\!\!\pm\!\!$ &  0.82 & 0.452 &  81.8 &  15.01 & $\!\!\pm\!\!$ &  0.52 & $\!\!\pm\!\!$ &  0.80 \\
0.50--0.60 & 0.538 &  67.4 &  13.83 & $\!\!\pm\!\!$ &  0.47 & $\!\!\pm\!\!$ &  0.69 & 0.535 &  81.9 &   7.82 & $\!\!\pm\!\!$ &  0.34 & $\!\!\pm\!\!$ &  0.53 \\
0.60--0.72 & 0.642 &  67.1 &   7.44 & $\!\!\pm\!\!$ &  0.33 & $\!\!\pm\!\!$ &  0.56 & 0.641 &  81.8 &   3.85 & $\!\!\pm\!\!$ &  0.24 & $\!\!\pm\!\!$ &  0.42 \\
0.72--0.90 & 0.777 &  66.5 &   2.74 & $\!\!\pm\!\!$ &  0.17 & $\!\!\pm\!\!$ &  0.33 & 0.769 &  81.7 &   1.36 & $\!\!\pm\!\!$ &  0.13 & $\!\!\pm\!\!$ &  0.23 \\
0.90--1.25 & 0.988 &  66.9 &   0.44 & $\!\!\pm\!\!$ &  0.05 & $\!\!\pm\!\!$ &  0.13 & 0.999 &  81.0 &   0.17 & $\!\!\pm\!\!$ &  0.03 & $\!\!\pm\!\!$ &  0.07 \\
 \hline
 \hline
  & \multicolumn{7}{c||}{$90<\theta<105$}
  & \multicolumn{7}{c|}{$105<\theta<125$} \\
 \hline
 $p_{\rm T}$ & $\langle p_{\rm T} \rangle$ & $\langle \theta \rangle$
  & \multicolumn{5}{c||}{${\rm d}^2 \sigma /{\rm d}p{\rm d}\Omega$}
  &$\langle p_{\rm T} \rangle$ & $\langle \theta \rangle$
  & \multicolumn{5}{c|}{${\rm d}^2 \sigma /{\rm d}p{\rm d}\Omega$} \\
 \hline
0.36--0.42 &  &  &  \multicolumn{5}{c||}{ } & 0.383 & 114.3 &   5.20 & $\!\!\pm\!\!$ &  0.29 & $\!\!\pm\!\!$ &  0.42 \\
0.42--0.50 & 0.452 &  96.7 &   7.99 & $\!\!\pm\!\!$ &  0.37 & $\!\!\pm\!\!$ &  0.66 & 0.449 & 113.5 &   3.79 & $\!\!\pm\!\!$ &  0.23 & $\!\!\pm\!\!$ &  0.31 \\
0.50--0.60 & 0.538 &  96.7 &   4.57 & $\!\!\pm\!\!$ &  0.26 & $\!\!\pm\!\!$ &  0.49 & 0.536 & 113.5 &   1.67 & $\!\!\pm\!\!$ &  0.15 & $\!\!\pm\!\!$ &  0.23 \\
0.60--0.72 & 0.640 &  97.0 &   1.57 & $\!\!\pm\!\!$ &  0.16 & $\!\!\pm\!\!$ &  0.26 & 0.638 & 112.8 &   0.40 & $\!\!\pm\!\!$ &  0.07 & $\!\!\pm\!\!$ &  0.12 \\
0.72--0.90 & 0.776 &  96.8 &   0.37 & $\!\!\pm\!\!$ &  0.07 & $\!\!\pm\!\!$ &  0.11 & 0.770 & 112.5 &   0.09 & $\!\!\pm\!\!$ &  0.03 & $\!\!\pm\!\!$ &  0.08 \\
 \hline
 \end{tabular}
 \end{center}
 \end{scriptsize}
 \end{table}

%% file: table.pip.pimbe3.tex
  
 \begin{table}[h]
 \begin{scriptsize}
 \caption{Double-differential inclusive
  cross-section ${\rm d}^2 \sigma /{\rm d}p{\rm d}\Omega$
  [mb/(GeV/{\it c} sr)] of the production of $\pi^+$'s
  in $\pi^-$ + Be $\rightarrow$ $\pi^+$ + X interactions
  with $-3.0$~GeV/{\it c} beam momentum;
  the first error is statistical, the second systematic; 
 $p_{\rm T}$ in GeV/{\it c}, polar angle $\theta$ in degrees.}
 \label{pip.pimbe3}
 \begin{center}
 \begin{tabular}{|c||c|c|rcrcr||c|c|rcrcr|}
 \hline
   & \multicolumn{7}{c||}{$20<\theta<30$}
  & \multicolumn{7}{c|}{$30<\theta<40$} \\
 \hline
 $p_{\rm T}$ & $\langle p_{\rm T} \rangle$ & $\langle \theta \rangle$
  & \multicolumn{5}{c||}{${\rm d}^2 \sigma /{\rm d}p{\rm d}\Omega$}
  &$\langle p_{\rm T} \rangle$ & $\langle \theta \rangle$
  & \multicolumn{5}{c|}{${\rm d}^2 \sigma /{\rm d}p{\rm d}\Omega$} \\
 \hline
0.10--0.13 & 0.116 &  25.0 &  29.21 & $\!\!\pm\!\!$ &  1.74 & $\!\!\pm\!\!$ &  3.26 & 0.115 &  34.9 &  23.15 & $\!\!\pm\!\!$ &  1.54 & $\!\!\pm\!\!$ &  2.62 \\
0.13--0.16 & 0.144 &  24.9 &  35.55 & $\!\!\pm\!\!$ &  1.85 & $\!\!\pm\!\!$ &  3.30 & 0.144 &  34.9 &  28.39 & $\!\!\pm\!\!$ &  1.54 & $\!\!\pm\!\!$ &  2.67 \\
0.16--0.20 & 0.180 &  24.8 &  40.36 & $\!\!\pm\!\!$ &  1.62 & $\!\!\pm\!\!$ &  2.65 & 0.180 &  34.8 &  33.48 & $\!\!\pm\!\!$ &  1.44 & $\!\!\pm\!\!$ &  2.45 \\
0.20--0.24 & 0.219 &  24.9 &  41.34 & $\!\!\pm\!\!$ &  1.60 & $\!\!\pm\!\!$ &  2.37 & 0.219 &  34.7 &  33.44 & $\!\!\pm\!\!$ &  1.43 & $\!\!\pm\!\!$ &  2.27 \\
0.24--0.30 & 0.267 &  24.9 &  39.24 & $\!\!\pm\!\!$ &  1.28 & $\!\!\pm\!\!$ &  1.87 & 0.268 &  34.7 &  35.03 & $\!\!\pm\!\!$ &  1.20 & $\!\!\pm\!\!$ &  2.00 \\
0.30--0.36 & 0.326 &  25.1 &  33.02 & $\!\!\pm\!\!$ &  1.16 & $\!\!\pm\!\!$ &  1.51 & 0.326 &  34.8 &  26.02 & $\!\!\pm\!\!$ &  1.01 & $\!\!\pm\!\!$ &  1.34 \\
0.36--0.42 & 0.385 &  24.8 &  30.08 & $\!\!\pm\!\!$ &  1.14 & $\!\!\pm\!\!$ &  1.52 & 0.384 &  34.6 &  22.82 & $\!\!\pm\!\!$ &  0.95 & $\!\!\pm\!\!$ &  1.24 \\
0.42--0.50 & 0.450 &  25.0 &  19.96 & $\!\!\pm\!\!$ &  0.77 & $\!\!\pm\!\!$ &  1.08 & 0.451 &  34.7 &  17.07 & $\!\!\pm\!\!$ &  0.73 & $\!\!\pm\!\!$ &  0.97 \\
0.50--0.60 & 0.537 &  24.9 &   9.92 & $\!\!\pm\!\!$ &  0.44 & $\!\!\pm\!\!$ &  0.71 & 0.538 &  34.8 &  10.06 & $\!\!\pm\!\!$ &  0.48 & $\!\!\pm\!\!$ &  0.70 \\
0.60--0.72 & 0.643 &  25.1 &   4.64 & $\!\!\pm\!\!$ &  0.26 & $\!\!\pm\!\!$ &  0.46 & 0.642 &  34.9 &   4.85 & $\!\!\pm\!\!$ &  0.29 & $\!\!\pm\!\!$ &  0.46 \\
0.72--0.90 &  &  &  \multicolumn{5}{c||}{ } & 0.770 &  34.9 &   2.09 & $\!\!\pm\!\!$ &  0.14 & $\!\!\pm\!\!$ &  0.29 \\
 \hline
 \hline
   & \multicolumn{7}{c||}{$40<\theta<50$}
  & \multicolumn{7}{c|}{$50<\theta<60$} \\
 \hline
 $p_{\rm T}$ & $\langle p_{\rm T} \rangle$ & $\langle \theta \rangle$
  & \multicolumn{5}{c||}{${\rm d}^2 \sigma /{\rm d}p{\rm d}\Omega$}
  &$\langle p_{\rm T} \rangle$ & $\langle \theta \rangle$
  & \multicolumn{5}{c|}{${\rm d}^2 \sigma /{\rm d}p{\rm d}\Omega$} \\
 \hline
0.10--0.13 & 0.117 &  44.9 &  21.69 & $\!\!\pm\!\!$ &  1.51 & $\!\!\pm\!\!$ &  2.58 &  &  &  \multicolumn{5}{c|}{ } \\
0.13--0.16 & 0.145 &  44.8 &  25.29 & $\!\!\pm\!\!$ &  1.56 & $\!\!\pm\!\!$ &  2.28 & 0.146 &  55.0 &  20.66 & $\!\!\pm\!\!$ &  1.37 & $\!\!\pm\!\!$ &  1.94 \\
0.16--0.20 & 0.180 &  45.0 &  25.92 & $\!\!\pm\!\!$ &  1.27 & $\!\!\pm\!\!$ &  1.97 & 0.180 &  54.9 &  25.83 & $\!\!\pm\!\!$ &  1.32 & $\!\!\pm\!\!$ &  1.83 \\
0.20--0.24 & 0.219 &  44.4 &  29.52 & $\!\!\pm\!\!$ &  1.38 & $\!\!\pm\!\!$ &  2.04 & 0.219 &  54.5 &  20.63 & $\!\!\pm\!\!$ &  1.10 & $\!\!\pm\!\!$ &  1.49 \\
0.24--0.30 & 0.269 &  44.7 &  24.30 & $\!\!\pm\!\!$ &  0.97 & $\!\!\pm\!\!$ &  1.35 & 0.270 &  54.6 &  18.92 & $\!\!\pm\!\!$ &  0.89 & $\!\!\pm\!\!$ &  1.13 \\
0.30--0.36 & 0.330 &  44.8 &  19.35 & $\!\!\pm\!\!$ &  0.88 & $\!\!\pm\!\!$ &  1.01 & 0.330 &  54.7 &  16.04 & $\!\!\pm\!\!$ &  0.81 & $\!\!\pm\!\!$ &  0.91 \\
0.36--0.42 & 0.389 &  45.0 &  15.92 & $\!\!\pm\!\!$ &  0.80 & $\!\!\pm\!\!$ &  0.92 & 0.388 &  54.8 &  13.50 & $\!\!\pm\!\!$ &  0.75 & $\!\!\pm\!\!$ &  0.81 \\
0.42--0.50 & 0.457 &  44.7 &  12.48 & $\!\!\pm\!\!$ &  0.63 & $\!\!\pm\!\!$ &  0.77 & 0.455 &  54.7 &   9.33 & $\!\!\pm\!\!$ &  0.54 & $\!\!\pm\!\!$ &  0.58 \\
0.50--0.60 & 0.547 &  44.8 &   7.93 & $\!\!\pm\!\!$ &  0.44 & $\!\!\pm\!\!$ &  0.58 & 0.544 &  54.7 &   5.37 & $\!\!\pm\!\!$ &  0.34 & $\!\!\pm\!\!$ &  0.42 \\
0.60--0.72 & 0.654 &  44.7 &   4.22 & $\!\!\pm\!\!$ &  0.28 & $\!\!\pm\!\!$ &  0.43 & 0.650 &  54.4 &   3.37 & $\!\!\pm\!\!$ &  0.26 & $\!\!\pm\!\!$ &  0.37 \\
0.72--0.90 & 0.793 &  45.2 &   1.63 & $\!\!\pm\!\!$ &  0.13 & $\!\!\pm\!\!$ &  0.22 & 0.790 &  54.1 &   1.14 & $\!\!\pm\!\!$ &  0.11 & $\!\!\pm\!\!$ &  0.18 \\
0.90--1.25 &  &  &  \multicolumn{5}{c||}{ } & 1.014 &  54.2 &   0.16 & $\!\!\pm\!\!$ &  0.02 & $\!\!\pm\!\!$ &  0.05 \\
 \hline
 \hline
   & \multicolumn{7}{c||}{$60<\theta<75$}
  & \multicolumn{7}{c|}{$75<\theta<90$} \\
 \hline
 $p_{\rm T}$ & $\langle p_{\rm T} \rangle$ & $\langle \theta \rangle$
  & \multicolumn{5}{c||}{${\rm d}^2 \sigma /{\rm d}p{\rm d}\Omega$}
  &$\langle p_{\rm T} \rangle$ & $\langle \theta \rangle$
  & \multicolumn{5}{c|}{${\rm d}^2 \sigma /{\rm d}p{\rm d}\Omega$} \\
 \hline
0.13--0.16 & 0.146 &  67.2 &  15.97 & $\!\!\pm\!\!$ &  1.01 & $\!\!\pm\!\!$ &  1.35 & 0.146 &  82.6 &  15.06 & $\!\!\pm\!\!$ &  0.99 & $\!\!\pm\!\!$ &  1.39 \\
0.16--0.20 & 0.178 &  67.4 &  17.11 & $\!\!\pm\!\!$ &  0.86 & $\!\!\pm\!\!$ &  1.14 & 0.179 &  82.5 &  14.95 & $\!\!\pm\!\!$ &  0.80 & $\!\!\pm\!\!$ &  1.06 \\
0.20--0.24 & 0.218 &  67.2 &  15.09 & $\!\!\pm\!\!$ &  0.79 & $\!\!\pm\!\!$ &  0.95 & 0.218 &  82.2 &  11.90 & $\!\!\pm\!\!$ &  0.72 & $\!\!\pm\!\!$ &  0.81 \\
0.24--0.30 & 0.266 &  66.9 &  13.94 & $\!\!\pm\!\!$ &  0.62 & $\!\!\pm\!\!$ &  0.78 & 0.266 &  81.8 &  10.16 & $\!\!\pm\!\!$ &  0.52 & $\!\!\pm\!\!$ &  0.65 \\
0.30--0.36 & 0.325 &  67.2 &  10.63 & $\!\!\pm\!\!$ &  0.55 & $\!\!\pm\!\!$ &  0.62 & 0.325 &  81.8 &   7.59 & $\!\!\pm\!\!$ &  0.46 & $\!\!\pm\!\!$ &  0.51 \\
0.36--0.42 & 0.383 &  66.4 &   7.96 & $\!\!\pm\!\!$ &  0.47 & $\!\!\pm\!\!$ &  0.50 & 0.384 &  82.3 &   5.41 & $\!\!\pm\!\!$ &  0.39 & $\!\!\pm\!\!$ &  0.43 \\
0.42--0.50 & 0.451 &  67.3 &   5.92 & $\!\!\pm\!\!$ &  0.35 & $\!\!\pm\!\!$ &  0.41 & 0.449 &  82.0 &   3.92 & $\!\!\pm\!\!$ &  0.29 & $\!\!\pm\!\!$ &  0.38 \\
0.50--0.60 & 0.539 &  67.2 &   3.52 & $\!\!\pm\!\!$ &  0.23 & $\!\!\pm\!\!$ &  0.31 & 0.536 &  82.2 &   1.94 & $\!\!\pm\!\!$ &  0.17 & $\!\!\pm\!\!$ &  0.25 \\
0.60--0.72 & 0.638 &  66.6 &   1.70 & $\!\!\pm\!\!$ &  0.14 & $\!\!\pm\!\!$ &  0.20 & 0.644 &  81.8 &   0.84 & $\!\!\pm\!\!$ &  0.09 & $\!\!\pm\!\!$ &  0.16 \\
0.72--0.90 & 0.766 &  66.2 &   0.54 & $\!\!\pm\!\!$ &  0.06 & $\!\!\pm\!\!$ &  0.11 & 0.767 &  81.5 &   0.25 & $\!\!\pm\!\!$ &  0.04 & $\!\!\pm\!\!$ &  0.06 \\
0.90--1.25 & 0.980 &  65.8 &   0.07 & $\!\!\pm\!\!$ &  0.02 & $\!\!\pm\!\!$ &  0.02 &  &  &  \multicolumn{5}{c|}{ } \\
 \hline
 \hline
  & \multicolumn{7}{c||}{$90<\theta<105$}
  & \multicolumn{7}{c|}{$105<\theta<125$} \\
 \hline
 $p_{\rm T}$ & $\langle p_{\rm T} \rangle$ & $\langle \theta \rangle$
  & \multicolumn{5}{c||}{${\rm d}^2 \sigma /{\rm d}p{\rm d}\Omega$}
  &$\langle p_{\rm T} \rangle$ & $\langle \theta \rangle$
  & \multicolumn{5}{c|}{${\rm d}^2 \sigma /{\rm d}p{\rm d}\Omega$} \\
 \hline
0.13--0.16 & 0.145 &  97.2 &  11.20 & $\!\!\pm\!\!$ &  0.84 & $\!\!\pm\!\!$ &  1.11 & 0.144 & 113.9 &  10.34 & $\!\!\pm\!\!$ &  0.72 & $\!\!\pm\!\!$ &  0.85 \\
0.16--0.20 & 0.179 &  97.4 &  12.05 & $\!\!\pm\!\!$ &  0.72 & $\!\!\pm\!\!$ &  0.92 & 0.178 & 114.4 &   8.56 & $\!\!\pm\!\!$ &  0.55 & $\!\!\pm\!\!$ &  0.61 \\
0.20--0.24 & 0.217 &  97.4 &   9.60 & $\!\!\pm\!\!$ &  0.66 & $\!\!\pm\!\!$ &  0.74 & 0.216 & 114.1 &   6.08 & $\!\!\pm\!\!$ &  0.45 & $\!\!\pm\!\!$ &  0.49 \\
0.24--0.30 & 0.266 &  97.7 &   6.84 & $\!\!\pm\!\!$ &  0.45 & $\!\!\pm\!\!$ &  0.50 & 0.265 & 113.4 &   4.15 & $\!\!\pm\!\!$ &  0.30 & $\!\!\pm\!\!$ &  0.34 \\
0.30--0.36 & 0.325 &  96.8 &   4.72 & $\!\!\pm\!\!$ &  0.36 & $\!\!\pm\!\!$ &  0.42 & 0.322 & 113.8 &   2.53 & $\!\!\pm\!\!$ &  0.23 & $\!\!\pm\!\!$ &  0.27 \\
0.36--0.42 & 0.387 &  97.0 &   2.92 & $\!\!\pm\!\!$ &  0.28 & $\!\!\pm\!\!$ &  0.31 & 0.387 & 114.2 &   1.53 & $\!\!\pm\!\!$ &  0.17 & $\!\!\pm\!\!$ &  0.22 \\
0.42--0.50 & 0.447 &  96.9 &   2.06 & $\!\!\pm\!\!$ &  0.20 & $\!\!\pm\!\!$ &  0.27 & 0.451 & 113.3 &   0.96 & $\!\!\pm\!\!$ &  0.12 & $\!\!\pm\!\!$ &  0.19 \\
0.50--0.60 & 0.531 &  95.9 &   0.90 & $\!\!\pm\!\!$ &  0.11 & $\!\!\pm\!\!$ &  0.17 & 0.525 & 113.1 &   0.31 & $\!\!\pm\!\!$ &  0.05 & $\!\!\pm\!\!$ &  0.08 \\
0.60--0.72 & 0.632 &  96.9 &   0.28 & $\!\!\pm\!\!$ &  0.05 & $\!\!\pm\!\!$ &  0.10 & 0.643 & 113.0 &   0.13 & $\!\!\pm\!\!$ &  0.03 & $\!\!\pm\!\!$ &  0.08 \\
0.72--0.90 & 0.757 &  99.2 &   0.03 & $\!\!\pm\!\!$ &  0.02 & $\!\!\pm\!\!$ &  0.01 &  &  &  \multicolumn{5}{c|}{ } \\
 \hline
 \end{tabular}
 \end{center}
 \end{scriptsize}
 \end{table}

%% file: table.pim.pimbe3.tex
  
 \begin{table}[h]
 \begin{scriptsize}
 \caption{Double-differential inclusive
  cross-section ${\rm d}^2 \sigma /{\rm d}p{\rm d}\Omega$
  [mb/(GeV/{\it c} sr)] of the production of $\pi^-$'s
  in $\pi^-$ + Be $\rightarrow$ $\pi^-$ + X interactions
  with $-3.0$~GeV/{\it c} beam momentum;
  the first error is statistical, the second systematic; 
 $p_{\rm T}$ in GeV/{\it c}, polar angle $\theta$ in degrees.}
 \label{pim.pimbe3}
 \begin{center}
 \begin{tabular}{|c||c|c|rcrcr||c|c|rcrcr|}
 \hline
   & \multicolumn{7}{c||}{$20<\theta<30$}
  & \multicolumn{7}{c|}{$30<\theta<40$} \\
 \hline
 $p_{\rm T}$ & $\langle p_{\rm T} \rangle$ & $\langle \theta \rangle$
  & \multicolumn{5}{c||}{${\rm d}^2 \sigma /{\rm d}p{\rm d}\Omega$}
  &$\langle p_{\rm T} \rangle$ & $\langle \theta \rangle$
  & \multicolumn{5}{c|}{${\rm d}^2 \sigma /{\rm d}p{\rm d}\Omega$} \\
 \hline
0.10--0.13 & 0.117 &  24.8 &  52.04 & $\!\!\pm\!\!$ &  2.38 & $\!\!\pm\!\!$ &  4.76 & 0.116 &  34.9 &  41.17 & $\!\!\pm\!\!$ &  2.09 & $\!\!\pm\!\!$ &  3.81 \\
0.13--0.16 & 0.146 &  24.8 &  63.35 & $\!\!\pm\!\!$ &  2.45 & $\!\!\pm\!\!$ &  4.58 & 0.146 &  35.0 &  49.12 & $\!\!\pm\!\!$ &  2.09 & $\!\!\pm\!\!$ &  3.69 \\
0.16--0.20 & 0.181 &  24.7 &  71.79 & $\!\!\pm\!\!$ &  2.15 & $\!\!\pm\!\!$ &  4.11 & 0.181 &  34.8 &  55.30 & $\!\!\pm\!\!$ &  1.86 & $\!\!\pm\!\!$ &  3.42 \\
0.20--0.24 & 0.221 &  24.8 &  69.90 & $\!\!\pm\!\!$ &  2.12 & $\!\!\pm\!\!$ &  3.49 & 0.221 &  34.8 &  62.10 & $\!\!\pm\!\!$ &  1.95 & $\!\!\pm\!\!$ &  3.29 \\
0.24--0.30 & 0.271 &  24.7 &  68.25 & $\!\!\pm\!\!$ &  1.67 & $\!\!\pm\!\!$ &  2.72 & 0.272 &  34.9 &  57.09 & $\!\!\pm\!\!$ &  1.50 & $\!\!\pm\!\!$ &  2.48 \\
0.30--0.36 & 0.333 &  25.0 &  59.39 & $\!\!\pm\!\!$ &  1.55 & $\!\!\pm\!\!$ &  2.20 & 0.332 &  34.8 &  50.60 & $\!\!\pm\!\!$ &  1.41 & $\!\!\pm\!\!$ &  2.03 \\
0.36--0.42 & 0.393 &  25.1 &  48.64 & $\!\!\pm\!\!$ &  1.42 & $\!\!\pm\!\!$ &  2.03 & 0.394 &  34.9 &  41.06 & $\!\!\pm\!\!$ &  1.28 & $\!\!\pm\!\!$ &  1.75 \\
0.42--0.50 & 0.464 &  24.9 &  37.94 & $\!\!\pm\!\!$ &  1.09 & $\!\!\pm\!\!$ &  1.71 & 0.464 &  35.1 &  32.43 & $\!\!\pm\!\!$ &  1.00 & $\!\!\pm\!\!$ &  1.51 \\
0.50--0.60 & 0.555 &  25.1 &  24.23 & $\!\!\pm\!\!$ &  0.77 & $\!\!\pm\!\!$ &  1.42 & 0.554 &  34.9 &  20.71 & $\!\!\pm\!\!$ &  0.69 & $\!\!\pm\!\!$ &  1.21 \\
0.60--0.72 & 0.669 &  25.1 &  14.21 & $\!\!\pm\!\!$ &  0.56 & $\!\!\pm\!\!$ &  1.12 & 0.669 &  34.8 &  13.86 & $\!\!\pm\!\!$ &  0.55 & $\!\!\pm\!\!$ &  1.09 \\
0.72--0.90 &  &  &  \multicolumn{5}{c||}{ } & 0.817 &  34.8 &   6.91 & $\!\!\pm\!\!$ &  0.33 & $\!\!\pm\!\!$ &  0.75 \\
 \hline
 \hline
   & \multicolumn{7}{c||}{$40<\theta<50$}
  & \multicolumn{7}{c|}{$50<\theta<60$} \\
 \hline
 $p_{\rm T}$ & $\langle p_{\rm T} \rangle$ & $\langle \theta \rangle$
  & \multicolumn{5}{c||}{${\rm d}^2 \sigma /{\rm d}p{\rm d}\Omega$}
  &$\langle p_{\rm T} \rangle$ & $\langle \theta \rangle$
  & \multicolumn{5}{c|}{${\rm d}^2 \sigma /{\rm d}p{\rm d}\Omega$} \\
 \hline
0.10--0.13 & 0.116 &  44.9 &  40.23 & $\!\!\pm\!\!$ &  2.14 & $\!\!\pm\!\!$ &  3.81 &  &  &  \multicolumn{5}{c|}{ } \\
0.13--0.16 & 0.145 &  44.8 &  43.31 & $\!\!\pm\!\!$ &  1.97 & $\!\!\pm\!\!$ &  3.18 & 0.145 &  54.9 &  40.62 & $\!\!\pm\!\!$ &  1.97 & $\!\!\pm\!\!$ &  3.09 \\
0.16--0.20 & 0.179 &  44.9 &  48.36 & $\!\!\pm\!\!$ &  1.74 & $\!\!\pm\!\!$ &  3.00 & 0.180 &  55.1 &  42.09 & $\!\!\pm\!\!$ &  1.68 & $\!\!\pm\!\!$ &  2.62 \\
0.20--0.24 & 0.221 &  44.8 &  50.20 & $\!\!\pm\!\!$ &  1.75 & $\!\!\pm\!\!$ &  2.76 & 0.220 &  54.9 &  40.28 & $\!\!\pm\!\!$ &  1.59 & $\!\!\pm\!\!$ &  2.17 \\
0.24--0.30 & 0.270 &  44.7 &  46.13 & $\!\!\pm\!\!$ &  1.37 & $\!\!\pm\!\!$ &  2.07 & 0.269 &  54.8 &  33.81 & $\!\!\pm\!\!$ &  1.16 & $\!\!\pm\!\!$ &  1.56 \\
0.30--0.36 & 0.329 &  44.7 &  39.36 & $\!\!\pm\!\!$ &  1.23 & $\!\!\pm\!\!$ &  1.70 & 0.329 &  54.7 &  29.67 & $\!\!\pm\!\!$ &  1.08 & $\!\!\pm\!\!$ &  1.31 \\
0.36--0.42 & 0.390 &  44.7 &  34.59 & $\!\!\pm\!\!$ &  1.17 & $\!\!\pm\!\!$ &  1.57 & 0.390 &  54.7 &  22.99 & $\!\!\pm\!\!$ &  0.96 & $\!\!\pm\!\!$ &  1.10 \\
0.42--0.50 & 0.457 &  44.7 &  25.86 & $\!\!\pm\!\!$ &  0.89 & $\!\!\pm\!\!$ &  1.30 & 0.457 &  54.7 &  18.61 & $\!\!\pm\!\!$ &  0.75 & $\!\!\pm\!\!$ &  0.98 \\
0.50--0.60 & 0.547 &  44.7 &  18.97 & $\!\!\pm\!\!$ &  0.68 & $\!\!\pm\!\!$ &  1.19 & 0.546 &  54.8 &  13.52 & $\!\!\pm\!\!$ &  0.57 & $\!\!\pm\!\!$ &  0.89 \\
0.60--0.72 & 0.653 &  44.9 &  11.24 & $\!\!\pm\!\!$ &  0.48 & $\!\!\pm\!\!$ &  0.95 & 0.654 &  54.6 &   9.31 & $\!\!\pm\!\!$ &  0.44 & $\!\!\pm\!\!$ &  0.82 \\
0.72--0.90 & 0.795 &  45.3 &   5.77 & $\!\!\pm\!\!$ &  0.30 & $\!\!\pm\!\!$ &  0.69 & 0.792 &  54.7 &   3.89 & $\!\!\pm\!\!$ &  0.24 & $\!\!\pm\!\!$ &  0.47 \\
0.90--1.25 &  &  &  \multicolumn{5}{c||}{ } & 1.015 &  54.8 &   0.55 & $\!\!\pm\!\!$ &  0.05 & $\!\!\pm\!\!$ &  0.11 \\
 \hline
 \hline
   & \multicolumn{7}{c||}{$60<\theta<75$}
  & \multicolumn{7}{c|}{$75<\theta<90$} \\
 \hline
 $p_{\rm T}$ & $\langle p_{\rm T} \rangle$ & $\langle \theta \rangle$
  & \multicolumn{5}{c||}{${\rm d}^2 \sigma /{\rm d}p{\rm d}\Omega$}
  &$\langle p_{\rm T} \rangle$ & $\langle \theta \rangle$
  & \multicolumn{5}{c|}{${\rm d}^2 \sigma /{\rm d}p{\rm d}\Omega$} \\
 \hline
0.13--0.16 & 0.146 &  67.2 &  33.59 & $\!\!\pm\!\!$ &  1.47 & $\!\!\pm\!\!$ &  2.45 & 0.147 &  82.6 &  26.96 & $\!\!\pm\!\!$ &  1.30 & $\!\!\pm\!\!$ &  2.10 \\
0.16--0.20 & 0.181 &  67.1 &  36.93 & $\!\!\pm\!\!$ &  1.28 & $\!\!\pm\!\!$ &  2.13 & 0.181 &  82.3 &  30.47 & $\!\!\pm\!\!$ &  1.17 & $\!\!\pm\!\!$ &  1.80 \\
0.20--0.24 & 0.221 &  67.1 &  31.13 & $\!\!\pm\!\!$ &  1.13 & $\!\!\pm\!\!$ &  1.64 & 0.220 &  82.0 &  26.87 & $\!\!\pm\!\!$ &  1.07 & $\!\!\pm\!\!$ &  1.47 \\
0.24--0.30 & 0.271 &  67.0 &  25.95 & $\!\!\pm\!\!$ &  0.84 & $\!\!\pm\!\!$ &  1.13 & 0.271 &  82.3 &  19.09 & $\!\!\pm\!\!$ &  0.73 & $\!\!\pm\!\!$ &  0.99 \\
0.30--0.36 & 0.333 &  67.2 &  22.56 & $\!\!\pm\!\!$ &  0.78 & $\!\!\pm\!\!$ &  1.00 & 0.330 &  81.9 &  13.08 & $\!\!\pm\!\!$ &  0.60 & $\!\!\pm\!\!$ &  0.72 \\
0.36--0.42 & 0.393 &  66.8 &  17.17 & $\!\!\pm\!\!$ &  0.69 & $\!\!\pm\!\!$ &  0.86 & 0.393 &  81.9 &  10.09 & $\!\!\pm\!\!$ &  0.52 & $\!\!\pm\!\!$ &  0.67 \\
0.42--0.50 & 0.465 &  66.5 &  13.19 & $\!\!\pm\!\!$ &  0.51 & $\!\!\pm\!\!$ &  0.76 & 0.462 &  81.7 &   8.45 & $\!\!\pm\!\!$ &  0.41 & $\!\!\pm\!\!$ &  0.65 \\
0.50--0.60 & 0.557 &  66.9 &   8.15 & $\!\!\pm\!\!$ &  0.36 & $\!\!\pm\!\!$ &  0.61 & 0.553 &  81.9 &   4.59 & $\!\!\pm\!\!$ &  0.28 & $\!\!\pm\!\!$ &  0.45 \\
0.60--0.72 & 0.667 &  66.9 &   4.73 & $\!\!\pm\!\!$ &  0.26 & $\!\!\pm\!\!$ &  0.47 & 0.664 &  81.5 &   2.18 & $\!\!\pm\!\!$ &  0.17 & $\!\!\pm\!\!$ &  0.28 \\
0.72--0.90 & 0.805 &  66.8 &   1.97 & $\!\!\pm\!\!$ &  0.13 & $\!\!\pm\!\!$ &  0.26 & 0.811 &  81.0 &   0.63 & $\!\!\pm\!\!$ &  0.06 & $\!\!\pm\!\!$ &  0.12 \\
0.90--1.25 & 1.047 &  66.1 &   0.17 & $\!\!\pm\!\!$ &  0.02 & $\!\!\pm\!\!$ &  0.04 & 1.049 &  80.5 &   0.05 & $\!\!\pm\!\!$ &  0.01 & $\!\!\pm\!\!$ &  0.02 \\
 \hline
 \hline
  & \multicolumn{7}{c||}{$90<\theta<105$}
  & \multicolumn{7}{c|}{$105<\theta<125$} \\
 \hline
 $p_{\rm T}$ & $\langle p_{\rm T} \rangle$ & $\langle \theta \rangle$
  & \multicolumn{5}{c||}{${\rm d}^2 \sigma /{\rm d}p{\rm d}\Omega$}
  &$\langle p_{\rm T} \rangle$ & $\langle \theta \rangle$
  & \multicolumn{5}{c|}{${\rm d}^2 \sigma /{\rm d}p{\rm d}\Omega$} \\
 \hline
0.13--0.16 & 0.146 &  97.1 &  27.64 & $\!\!\pm\!\!$ &  1.33 & $\!\!\pm\!\!$ &  2.12 & 0.146 & 114.8 &  25.29 & $\!\!\pm\!\!$ &  1.11 & $\!\!\pm\!\!$ &  1.59 \\
0.16--0.20 & 0.181 &  97.6 &  24.44 & $\!\!\pm\!\!$ &  1.03 & $\!\!\pm\!\!$ &  1.44 & 0.180 & 113.9 &  20.30 & $\!\!\pm\!\!$ &  0.83 & $\!\!\pm\!\!$ &  1.03 \\
0.20--0.24 & 0.221 &  97.0 &  19.49 & $\!\!\pm\!\!$ &  0.91 & $\!\!\pm\!\!$ &  1.11 & 0.219 & 114.2 &  13.21 & $\!\!\pm\!\!$ &  0.67 & $\!\!\pm\!\!$ &  0.73 \\
0.24--0.30 & 0.270 &  97.1 &  15.05 & $\!\!\pm\!\!$ &  0.66 & $\!\!\pm\!\!$ &  0.87 & 0.270 & 114.1 &   8.17 & $\!\!\pm\!\!$ &  0.41 & $\!\!\pm\!\!$ &  0.54 \\
0.30--0.36 & 0.331 &  97.0 &   9.62 & $\!\!\pm\!\!$ &  0.51 & $\!\!\pm\!\!$ &  0.68 & 0.329 & 113.3 &   5.01 & $\!\!\pm\!\!$ &  0.32 & $\!\!\pm\!\!$ &  0.46 \\
0.36--0.42 & 0.391 &  97.5 &   6.99 & $\!\!\pm\!\!$ &  0.43 & $\!\!\pm\!\!$ &  0.63 & 0.393 & 112.4 &   2.66 & $\!\!\pm\!\!$ &  0.24 & $\!\!\pm\!\!$ &  0.32 \\
0.42--0.50 & 0.466 &  96.4 &   3.84 & $\!\!\pm\!\!$ &  0.28 & $\!\!\pm\!\!$ &  0.42 & 0.459 & 111.8 &   1.28 & $\!\!\pm\!\!$ &  0.14 & $\!\!\pm\!\!$ &  0.21 \\
0.50--0.60 & 0.549 &  95.9 &   1.71 & $\!\!\pm\!\!$ &  0.16 & $\!\!\pm\!\!$ &  0.26 & 0.551 & 110.0 &   0.40 & $\!\!\pm\!\!$ &  0.06 & $\!\!\pm\!\!$ &  0.10 \\
0.60--0.72 & 0.672 &  95.8 &   0.60 & $\!\!\pm\!\!$ &  0.08 & $\!\!\pm\!\!$ &  0.13 & 0.665 & 109.4 &   0.10 & $\!\!\pm\!\!$ &  0.03 & $\!\!\pm\!\!$ &  0.04 \\
0.72--0.90 & 0.799 &  95.5 &   0.09 & $\!\!\pm\!\!$ &  0.02 & $\!\!\pm\!\!$ &  0.04 &  &  &  \multicolumn{5}{c|}{ } \\
 \hline
 \end{tabular}
 \end{center}
 \end{scriptsize}
 \end{table}

%% file: table.pro.probe5.tex
  
 \begin{table}[h]
 \begin{scriptsize}
 \caption{Double-differential inclusive
  cross-section ${\rm d}^2 \sigma /{\rm d}p{\rm d}\Omega$
  [mb/(GeV/{\it c} sr)] of the production of protons
  in p + Be $\rightarrow$ p + X interactions
  with $+5.0$~GeV/{\it c} beam momentum;
  the first error is statistical, the second systematic; 
 $p_{\rm T}$ in GeV/{\it c}, polar angle $\theta$ in degrees.}
 \label{pro.probe5}
 \begin{center}
 \begin{tabular}{|c||c|c|rcrcr||c|c|rcrcr|}
 \hline
   & \multicolumn{7}{c||}{$20<\theta<30$}
  & \multicolumn{7}{c|}{$30<\theta<40$} \\
 \hline
 $p_{\rm T}$ & $\langle p_{\rm T} \rangle$ & $\langle \theta \rangle$
  & \multicolumn{5}{c||}{${\rm d}^2 \sigma /{\rm d}p{\rm d}\Omega$}
  &$\langle p_{\rm T} \rangle$ & $\langle \theta \rangle$
  & \multicolumn{5}{c|}{${\rm d}^2 \sigma /{\rm d}p{\rm d}\Omega$} \\
 \hline
0.20--0.24 & 0.221 &  24.9 &  45.44 & $\!\!\pm\!\!$ &  1.74 & $\!\!\pm\!\!$ &  3.07 &  &  &  \multicolumn{5}{c|}{ } \\
0.24--0.30 & 0.271 &  24.8 &  43.38 & $\!\!\pm\!\!$ &  1.46 & $\!\!\pm\!\!$ &  2.86 & 0.271 &  34.7 &  42.88 & $\!\!\pm\!\!$ &  1.33 & $\!\!\pm\!\!$ &  2.32 \\
0.30--0.36 & 0.330 &  24.9 &  40.05 & $\!\!\pm\!\!$ &  1.45 & $\!\!\pm\!\!$ &  2.96 & 0.330 &  34.7 &  39.38 & $\!\!\pm\!\!$ &  1.34 & $\!\!\pm\!\!$ &  2.20 \\
0.36--0.42 & 0.390 &  24.9 &  36.14 & $\!\!\pm\!\!$ &  1.33 & $\!\!\pm\!\!$ &  2.44 & 0.389 &  34.9 &  31.72 & $\!\!\pm\!\!$ &  1.28 & $\!\!\pm\!\!$ &  2.49 \\
0.42--0.50 & 0.459 &  24.9 &  35.43 & $\!\!\pm\!\!$ &  1.13 & $\!\!\pm\!\!$ &  2.21 & 0.459 &  34.9 &  30.16 & $\!\!\pm\!\!$ &  1.13 & $\!\!\pm\!\!$ &  2.28 \\
0.50--0.60 & 0.547 &  24.8 &  31.41 & $\!\!\pm\!\!$ &  0.97 & $\!\!\pm\!\!$ &  1.94 & 0.549 &  35.1 &  22.92 & $\!\!\pm\!\!$ &  0.85 & $\!\!\pm\!\!$ &  1.82 \\
0.60--0.72 & 0.657 &  24.9 &  23.45 & $\!\!\pm\!\!$ &  0.73 & $\!\!\pm\!\!$ &  1.54 & 0.654 &  34.9 &  18.06 & $\!\!\pm\!\!$ &  0.72 & $\!\!\pm\!\!$ &  1.68 \\
0.72--0.90 &  &  &  \multicolumn{5}{c||}{ } & 0.801 &  34.7 &   9.06 & $\!\!\pm\!\!$ &  0.40 & $\!\!\pm\!\!$ &  1.11 \\
 \hline
 \hline
   & \multicolumn{7}{c||}{$40<\theta<50$}
  & \multicolumn{7}{c|}{$50<\theta<60$} \\
 \hline
 $p_{\rm T}$ & $\langle p_{\rm T} \rangle$ & $\langle \theta \rangle$
  & \multicolumn{5}{c||}{${\rm d}^2 \sigma /{\rm d}p{\rm d}\Omega$}
  &$\langle p_{\rm T} \rangle$ & $\langle \theta \rangle$
  & \multicolumn{5}{c|}{${\rm d}^2 \sigma /{\rm d}p{\rm d}\Omega$} \\
 \hline
0.30--0.36 & 0.329 &  45.0 &  38.76 & $\!\!\pm\!\!$ &  1.27 & $\!\!\pm\!\!$ &  1.79 &  &  &  \multicolumn{5}{c|}{ } \\
0.36--0.42 & 0.389 &  45.0 &  32.78 & $\!\!\pm\!\!$ &  1.18 & $\!\!\pm\!\!$ &  1.60 & 0.390 &  55.0 &  35.42 & $\!\!\pm\!\!$ &  1.19 & $\!\!\pm\!\!$ &  1.50 \\
0.42--0.50 & 0.459 &  45.0 &  24.72 & $\!\!\pm\!\!$ &  0.98 & $\!\!\pm\!\!$ &  1.84 & 0.459 &  55.1 &  30.70 & $\!\!\pm\!\!$ &  1.02 & $\!\!\pm\!\!$ &  1.41 \\
0.50--0.60 & 0.549 &  45.1 &  21.46 & $\!\!\pm\!\!$ &  0.85 & $\!\!\pm\!\!$ &  1.73 & 0.548 &  55.0 &  21.29 & $\!\!\pm\!\!$ &  0.86 & $\!\!\pm\!\!$ &  1.87 \\
0.60--0.72 & 0.657 &  44.9 &  14.46 & $\!\!\pm\!\!$ &  0.66 & $\!\!\pm\!\!$ &  1.53 & 0.656 &  54.8 &   9.95 & $\!\!\pm\!\!$ &  0.54 & $\!\!\pm\!\!$ &  1.29 \\
0.72--0.90 & 0.799 &  45.0 &   7.29 & $\!\!\pm\!\!$ &  0.41 & $\!\!\pm\!\!$ &  1.16 & 0.799 &  54.8 &   4.24 & $\!\!\pm\!\!$ &  0.32 & $\!\!\pm\!\!$ &  0.91 \\
0.90--1.25 & 1.025 &  44.5 &   1.49 & $\!\!\pm\!\!$ &  0.12 & $\!\!\pm\!\!$ &  0.55 & 1.024 &  54.5 &   0.97 & $\!\!\pm\!\!$ &  0.09 & $\!\!\pm\!\!$ &  0.58 \\
 \hline
 \hline
   & \multicolumn{7}{c||}{$60<\theta<75$}
  & \multicolumn{7}{c|}{$75<\theta<90$} \\
 \hline
 $p_{\rm T}$ & $\langle p_{\rm T} \rangle$ & $\langle \theta \rangle$
  & \multicolumn{5}{c||}{${\rm d}^2 \sigma /{\rm d}p{\rm d}\Omega$}
  &$\langle p_{\rm T} \rangle$ & $\langle \theta \rangle$
  & \multicolumn{5}{c|}{${\rm d}^2 \sigma /{\rm d}p{\rm d}\Omega$} \\
 \hline
0.36--0.42 & 0.391 &  67.6 &  33.05 & $\!\!\pm\!\!$ &  0.92 & $\!\!\pm\!\!$ &  1.28 &  &  &  \multicolumn{5}{c|}{ } \\
0.42--0.50 & 0.461 &  67.3 &  29.97 & $\!\!\pm\!\!$ &  0.78 & $\!\!\pm\!\!$ &  1.11 & 0.460 &  81.8 &  20.96 & $\!\!\pm\!\!$ &  0.64 & $\!\!\pm\!\!$ &  0.99 \\
0.50--0.60 & 0.549 &  66.9 &  19.40 & $\!\!\pm\!\!$ &  0.62 & $\!\!\pm\!\!$ &  1.29 & 0.549 &  81.5 &  10.97 & $\!\!\pm\!\!$ &  0.44 & $\!\!\pm\!\!$ &  0.77 \\
0.60--0.72 & 0.658 &  67.2 &   6.76 & $\!\!\pm\!\!$ &  0.37 & $\!\!\pm\!\!$ &  1.12 & 0.657 &  81.3 &   2.90 & $\!\!\pm\!\!$ &  0.26 & $\!\!\pm\!\!$ &  0.79 \\
0.72--0.90 & 0.798 &  66.7 &   2.75 & $\!\!\pm\!\!$ &  0.20 & $\!\!\pm\!\!$ &  0.90 & 0.800 &  81.8 &   1.42 & $\!\!\pm\!\!$ &  0.15 & $\!\!\pm\!\!$ &  0.73 \\
0.90--1.25 & 1.036 &  66.6 &   0.75 & $\!\!\pm\!\!$ &  0.08 & $\!\!\pm\!\!$ &  0.58 &  &  &  \multicolumn{5}{c|}{ } \\
 \hline
 \hline
  & \multicolumn{7}{c||}{$90<\theta<105$}
  & \multicolumn{7}{c|}{$105<\theta<125$} \\
 \hline
 $p_{\rm T}$ & $\langle p_{\rm T} \rangle$ & $\langle \theta \rangle$
  & \multicolumn{5}{c||}{${\rm d}^2 \sigma /{\rm d}p{\rm d}\Omega$}
  &$\langle p_{\rm T} \rangle$ & $\langle \theta \rangle$
  & \multicolumn{5}{c|}{${\rm d}^2 \sigma /{\rm d}p{\rm d}\Omega$} \\
 \hline
0.36--0.42 &  &  &  \multicolumn{5}{c||}{ } & 0.391 & 113.6 &   7.45 & $\!\!\pm\!\!$ &  0.38 & $\!\!\pm\!\!$ &  0.47 \\
0.42--0.50 & 0.459 &  96.7 &  10.91 & $\!\!\pm\!\!$ &  0.45 & $\!\!\pm\!\!$ &  0.79 & 0.458 & 113.1 &   3.59 & $\!\!\pm\!\!$ &  0.25 & $\!\!\pm\!\!$ &  0.31 \\
0.50--0.60 & 0.549 &  96.7 &   4.83 & $\!\!\pm\!\!$ &  0.32 & $\!\!\pm\!\!$ &  0.52 & 0.544 & 112.8 &   0.86 & $\!\!\pm\!\!$ &  0.13 & $\!\!\pm\!\!$ &  0.40 \\
0.60--0.72 & 0.661 &  96.3 &   1.07 & $\!\!\pm\!\!$ &  0.14 & $\!\!\pm\!\!$ &  0.66 &  &  &  \multicolumn{5}{c|}{ } \\
 \hline
 \end{tabular}
 \end{center}
 \end{scriptsize}
 \end{table}

%% file: table.pip.probe5.tex
  
 \begin{table}[h]
 \begin{scriptsize}
 \caption{Double-differential inclusive
  cross-section ${\rm d}^2 \sigma /{\rm d}p{\rm d}\Omega$
  [mb/(GeV/{\it c} sr)] of the production of $\pi^+$'s
  in p + Be $\rightarrow$ $\pi^+$ + X interactions
  with $+5.0$~GeV/{\it c} beam momentum;
  the first error is statistical, the second systematic; 
 $p_{\rm T}$ in GeV/{\it c}, polar angle $\theta$ in degrees.}
 \label{pip.probe5}
 \begin{center}
 \begin{tabular}{|c||c|c|rcrcr||c|c|rcrcr|}
 \hline
   & \multicolumn{7}{c||}{$20<\theta<30$}
  & \multicolumn{7}{c|}{$30<\theta<40$} \\
 \hline
 $p_{\rm T}$ & $\langle p_{\rm T} \rangle$ & $\langle \theta \rangle$
  & \multicolumn{5}{c||}{${\rm d}^2 \sigma /{\rm d}p{\rm d}\Omega$}
  &$\langle p_{\rm T} \rangle$ & $\langle \theta \rangle$
  & \multicolumn{5}{c|}{${\rm d}^2 \sigma /{\rm d}p{\rm d}\Omega$} \\
 \hline
0.10--0.13 & 0.115 &  25.1 &  51.01 & $\!\!\pm\!\!$ &  2.51 & $\!\!\pm\!\!$ &  4.66 & 0.115 &  34.8 &  36.25 & $\!\!\pm\!\!$ &  2.02 & $\!\!\pm\!\!$ &  3.39 \\
0.13--0.16 & 0.146 &  25.0 &  58.36 & $\!\!\pm\!\!$ &  2.44 & $\!\!\pm\!\!$ &  4.22 & 0.146 &  34.7 &  40.18 & $\!\!\pm\!\!$ &  1.97 & $\!\!\pm\!\!$ &  3.02 \\
0.16--0.20 & 0.180 &  24.7 &  67.65 & $\!\!\pm\!\!$ &  2.16 & $\!\!\pm\!\!$ &  3.93 & 0.180 &  34.6 &  49.15 & $\!\!\pm\!\!$ &  1.87 & $\!\!\pm\!\!$ &  3.07 \\
0.20--0.24 & 0.220 &  24.6 &  70.51 & $\!\!\pm\!\!$ &  2.21 & $\!\!\pm\!\!$ &  3.90 & 0.220 &  34.5 &  44.32 & $\!\!\pm\!\!$ &  1.66 & $\!\!\pm\!\!$ &  2.50 \\
0.24--0.30 & 0.269 &  24.7 &  59.93 & $\!\!\pm\!\!$ &  1.63 & $\!\!\pm\!\!$ &  2.68 & 0.269 &  34.6 &  42.81 & $\!\!\pm\!\!$ &  1.36 & $\!\!\pm\!\!$ &  2.00 \\
0.30--0.36 & 0.328 &  24.9 &  44.50 & $\!\!\pm\!\!$ &  1.38 & $\!\!\pm\!\!$ &  1.93 & 0.328 &  34.6 &  32.99 & $\!\!\pm\!\!$ &  1.18 & $\!\!\pm\!\!$ &  1.51 \\
0.36--0.42 & 0.389 &  24.8 &  33.20 & $\!\!\pm\!\!$ &  1.18 & $\!\!\pm\!\!$ &  1.56 & 0.389 &  34.6 &  28.65 & $\!\!\pm\!\!$ &  1.12 & $\!\!\pm\!\!$ &  1.47 \\
0.42--0.50 & 0.459 &  24.8 &  22.23 & $\!\!\pm\!\!$ &  0.81 & $\!\!\pm\!\!$ &  1.14 & 0.456 &  34.8 &  21.63 & $\!\!\pm\!\!$ &  0.82 & $\!\!\pm\!\!$ &  1.13 \\
0.50--0.60 & 0.546 &  24.6 &  15.78 & $\!\!\pm\!\!$ &  0.60 & $\!\!\pm\!\!$ &  1.02 & 0.545 &  34.6 &  12.17 & $\!\!\pm\!\!$ &  0.53 & $\!\!\pm\!\!$ &  0.77 \\
0.60--0.72 & 0.657 &  24.9 &   8.99 & $\!\!\pm\!\!$ &  0.37 & $\!\!\pm\!\!$ &  0.81 & 0.655 &  34.6 &   6.93 & $\!\!\pm\!\!$ &  0.34 & $\!\!\pm\!\!$ &  0.61 \\
0.72--0.90 &  &  &  \multicolumn{5}{c||}{ } & 0.799 &  34.7 &   3.92 & $\!\!\pm\!\!$ &  0.21 & $\!\!\pm\!\!$ &  0.51 \\
 \hline
 \hline
   & \multicolumn{7}{c||}{$40<\theta<50$}
  & \multicolumn{7}{c|}{$50<\theta<60$} \\
 \hline
 $p_{\rm T}$ & $\langle p_{\rm T} \rangle$ & $\langle \theta \rangle$
  & \multicolumn{5}{c||}{${\rm d}^2 \sigma /{\rm d}p{\rm d}\Omega$}
  &$\langle p_{\rm T} \rangle$ & $\langle \theta \rangle$
  & \multicolumn{5}{c|}{${\rm d}^2 \sigma /{\rm d}p{\rm d}\Omega$} \\
 \hline
0.10--0.13 & 0.116 &  44.8 &  23.50 & $\!\!\pm\!\!$ &  1.57 & $\!\!\pm\!\!$ &  2.46 &  &  &  \multicolumn{5}{c|}{ } \\
0.13--0.16 & 0.146 &  44.9 &  33.42 & $\!\!\pm\!\!$ &  1.81 & $\!\!\pm\!\!$ &  2.64 & 0.145 &  54.9 &  25.86 & $\!\!\pm\!\!$ &  1.63 & $\!\!\pm\!\!$ &  2.26 \\
0.16--0.20 & 0.180 &  44.8 &  38.70 & $\!\!\pm\!\!$ &  1.66 & $\!\!\pm\!\!$ &  2.60 & 0.181 &  54.7 &  27.80 & $\!\!\pm\!\!$ &  1.39 & $\!\!\pm\!\!$ &  1.90 \\
0.20--0.24 & 0.219 &  44.7 &  35.19 & $\!\!\pm\!\!$ &  1.56 & $\!\!\pm\!\!$ &  2.17 & 0.219 &  55.0 &  27.15 & $\!\!\pm\!\!$ &  1.39 & $\!\!\pm\!\!$ &  1.78 \\
0.24--0.30 & 0.270 &  44.8 &  29.92 & $\!\!\pm\!\!$ &  1.14 & $\!\!\pm\!\!$ &  1.50 & 0.269 &  54.8 &  24.24 & $\!\!\pm\!\!$ &  1.02 & $\!\!\pm\!\!$ &  1.33 \\
0.30--0.36 & 0.329 &  44.7 &  25.71 & $\!\!\pm\!\!$ &  1.06 & $\!\!\pm\!\!$ &  1.30 & 0.329 &  54.8 &  16.87 & $\!\!\pm\!\!$ &  0.83 & $\!\!\pm\!\!$ &  0.96 \\
0.36--0.42 & 0.389 &  44.8 &  19.30 & $\!\!\pm\!\!$ &  0.92 & $\!\!\pm\!\!$ &  1.07 & 0.388 &  54.8 &  13.97 & $\!\!\pm\!\!$ &  0.78 & $\!\!\pm\!\!$ &  0.84 \\
0.42--0.50 & 0.458 &  44.7 &  13.41 & $\!\!\pm\!\!$ &  0.62 & $\!\!\pm\!\!$ &  0.79 & 0.458 &  54.7 &  11.10 & $\!\!\pm\!\!$ &  0.60 & $\!\!\pm\!\!$ &  0.71 \\
0.50--0.60 & 0.547 &  44.5 &   8.76 & $\!\!\pm\!\!$ &  0.48 & $\!\!\pm\!\!$ &  0.64 & 0.545 &  54.7 &   7.02 & $\!\!\pm\!\!$ &  0.42 & $\!\!\pm\!\!$ &  0.54 \\
0.60--0.72 & 0.653 &  44.5 &   4.99 & $\!\!\pm\!\!$ &  0.31 & $\!\!\pm\!\!$ &  0.46 & 0.659 &  54.5 &   3.19 & $\!\!\pm\!\!$ &  0.25 & $\!\!\pm\!\!$ &  0.32 \\
0.72--0.90 & 0.793 &  44.4 &   2.10 & $\!\!\pm\!\!$ &  0.15 & $\!\!\pm\!\!$ &  0.28 & 0.795 &  54.2 &   1.58 & $\!\!\pm\!\!$ &  0.14 & $\!\!\pm\!\!$ &  0.23 \\
0.90--1.25 &  &  &  \multicolumn{5}{c||}{ } & 1.027 &  54.5 &   0.22 & $\!\!\pm\!\!$ &  0.03 & $\!\!\pm\!\!$ &  0.06 \\
 \hline
 \hline
   & \multicolumn{7}{c||}{$60<\theta<75$}
  & \multicolumn{7}{c|}{$75<\theta<90$} \\
 \hline
 $p_{\rm T}$ & $\langle p_{\rm T} \rangle$ & $\langle \theta \rangle$
  & \multicolumn{5}{c||}{${\rm d}^2 \sigma /{\rm d}p{\rm d}\Omega$}
  &$\langle p_{\rm T} \rangle$ & $\langle \theta \rangle$
  & \multicolumn{5}{c|}{${\rm d}^2 \sigma /{\rm d}p{\rm d}\Omega$} \\
 \hline
0.13--0.16 & 0.145 &  67.3 &  18.92 & $\!\!\pm\!\!$ &  1.13 & $\!\!\pm\!\!$ &  1.60 & 0.145 &  82.7 &  15.63 & $\!\!\pm\!\!$ &  1.00 & $\!\!\pm\!\!$ &  1.44 \\
0.16--0.20 & 0.180 &  67.1 &  22.04 & $\!\!\pm\!\!$ &  1.01 & $\!\!\pm\!\!$ &  1.45 & 0.181 &  82.5 &  16.42 & $\!\!\pm\!\!$ &  0.85 & $\!\!\pm\!\!$ &  1.13 \\
0.20--0.24 & 0.220 &  66.7 &  22.90 & $\!\!\pm\!\!$ &  1.01 & $\!\!\pm\!\!$ &  1.36 & 0.220 &  82.2 &  16.89 & $\!\!\pm\!\!$ &  0.89 & $\!\!\pm\!\!$ &  1.08 \\
0.24--0.30 & 0.269 &  66.8 &  16.22 & $\!\!\pm\!\!$ &  0.70 & $\!\!\pm\!\!$ &  0.88 & 0.268 &  81.8 &  12.19 & $\!\!\pm\!\!$ &  0.60 & $\!\!\pm\!\!$ &  0.72 \\
0.30--0.36 & 0.329 &  66.9 &  12.06 & $\!\!\pm\!\!$ &  0.60 & $\!\!\pm\!\!$ &  0.72 & 0.330 &  82.1 &   8.26 & $\!\!\pm\!\!$ &  0.49 & $\!\!\pm\!\!$ &  0.60 \\
0.36--0.42 & 0.391 &  66.6 &   8.23 & $\!\!\pm\!\!$ &  0.46 & $\!\!\pm\!\!$ &  0.54 & 0.389 &  81.7 &   6.39 & $\!\!\pm\!\!$ &  0.45 & $\!\!\pm\!\!$ &  0.59 \\
0.42--0.50 & 0.461 &  66.5 &   6.45 & $\!\!\pm\!\!$ &  0.36 & $\!\!\pm\!\!$ &  0.45 & 0.457 &  82.5 &   2.81 & $\!\!\pm\!\!$ &  0.24 & $\!\!\pm\!\!$ &  0.27 \\
0.50--0.60 & 0.547 &  66.5 &   4.04 & $\!\!\pm\!\!$ &  0.27 & $\!\!\pm\!\!$ &  0.35 & 0.547 &  81.9 &   2.09 & $\!\!\pm\!\!$ &  0.20 & $\!\!\pm\!\!$ &  0.25 \\
0.60--0.72 & 0.657 &  67.1 &   2.29 & $\!\!\pm\!\!$ &  0.18 & $\!\!\pm\!\!$ &  0.27 & 0.661 &  81.6 &   1.04 & $\!\!\pm\!\!$ &  0.13 & $\!\!\pm\!\!$ &  0.17 \\
0.72--0.90 & 0.800 &  66.5 &   0.72 & $\!\!\pm\!\!$ &  0.08 & $\!\!\pm\!\!$ &  0.12 & 0.800 &  81.2 &   0.17 & $\!\!\pm\!\!$ &  0.03 & $\!\!\pm\!\!$ &  0.06 \\
0.90--1.25 & 1.019 &  65.3 &   0.09 & $\!\!\pm\!\!$ &  0.02 & $\!\!\pm\!\!$ &  0.03 &  &  &  \multicolumn{5}{c|}{ } \\
 \hline
 \hline
  & \multicolumn{7}{c||}{$90<\theta<105$}
  & \multicolumn{7}{c|}{$105<\theta<125$} \\
 \hline
 $p_{\rm T}$ & $\langle p_{\rm T} \rangle$ & $\langle \theta \rangle$
  & \multicolumn{5}{c||}{${\rm d}^2 \sigma /{\rm d}p{\rm d}\Omega$}
  &$\langle p_{\rm T} \rangle$ & $\langle \theta \rangle$
  & \multicolumn{5}{c|}{${\rm d}^2 \sigma /{\rm d}p{\rm d}\Omega$} \\
 \hline
0.13--0.16 & 0.146 &  97.5 &  14.62 & $\!\!\pm\!\!$ &  0.95 & $\!\!\pm\!\!$ &  1.43 & 0.145 & 114.7 &  14.61 & $\!\!\pm\!\!$ &  0.82 & $\!\!\pm\!\!$ &  1.21 \\
0.16--0.20 & 0.181 &  97.0 &  16.88 & $\!\!\pm\!\!$ &  0.91 & $\!\!\pm\!\!$ &  1.19 & 0.178 & 114.6 &  11.77 & $\!\!\pm\!\!$ &  0.65 & $\!\!\pm\!\!$ &  0.76 \\
0.20--0.24 & 0.218 &  97.3 &  13.26 & $\!\!\pm\!\!$ &  0.77 & $\!\!\pm\!\!$ &  0.94 & 0.220 & 113.9 &   7.47 & $\!\!\pm\!\!$ &  0.50 & $\!\!\pm\!\!$ &  0.59 \\
0.24--0.30 & 0.269 &  97.3 &   8.23 & $\!\!\pm\!\!$ &  0.50 & $\!\!\pm\!\!$ &  0.58 & 0.268 & 113.3 &   4.04 & $\!\!\pm\!\!$ &  0.31 & $\!\!\pm\!\!$ &  0.37 \\
0.30--0.36 & 0.327 &  97.1 &   4.11 & $\!\!\pm\!\!$ &  0.33 & $\!\!\pm\!\!$ &  0.39 & 0.331 & 112.9 &   1.85 & $\!\!\pm\!\!$ &  0.19 & $\!\!\pm\!\!$ &  0.24 \\
0.36--0.42 & 0.387 &  96.9 &   2.70 & $\!\!\pm\!\!$ &  0.27 & $\!\!\pm\!\!$ &  0.31 & 0.391 & 113.7 &   1.22 & $\!\!\pm\!\!$ &  0.17 & $\!\!\pm\!\!$ &  0.21 \\
0.42--0.50 & 0.454 &  96.8 &   1.63 & $\!\!\pm\!\!$ &  0.17 & $\!\!\pm\!\!$ &  0.23 & 0.458 & 112.6 &   0.37 & $\!\!\pm\!\!$ &  0.07 & $\!\!\pm\!\!$ &  0.08 \\
0.50--0.60 & 0.539 &  97.1 &   0.72 & $\!\!\pm\!\!$ &  0.12 & $\!\!\pm\!\!$ &  0.13 & 0.539 & 113.2 &   0.12 & $\!\!\pm\!\!$ &  0.03 & $\!\!\pm\!\!$ &  0.05 \\
0.60--0.72 & 0.659 &  96.6 &   0.17 & $\!\!\pm\!\!$ &  0.04 & $\!\!\pm\!\!$ &  0.06 &  &  &  \multicolumn{5}{c|}{ } \\
 \hline
 \end{tabular}
 \end{center}
 \end{scriptsize}
 \end{table}

%% file: table.pim.probe5.tex
  
 \begin{table}[h]
 \begin{scriptsize}
 \caption{Double-differential inclusive
  cross-section ${\rm d}^2 \sigma /{\rm d}p{\rm d}\Omega$
  [mb/(GeV/{\it c} sr)] of the production of $\pi^-$'s
  in p + Be $\rightarrow$ $\pi^-$ + X interactions
  with $+5.0$~GeV/{\it c} beam momentum;
  the first error is statistical, the second systematic; 
 $p_{\rm T}$ in GeV/{\it c}, polar angle $\theta$ in degrees.}
 \label{pim.probe5}
 \begin{center}
 \begin{tabular}{|c||c|c|rcrcr||c|c|rcrcr|}
 \hline
   & \multicolumn{7}{c||}{$20<\theta<30$}
  & \multicolumn{7}{c|}{$30<\theta<40$} \\
 \hline
 $p_{\rm T}$ & $\langle p_{\rm T} \rangle$ & $\langle \theta \rangle$
  & \multicolumn{5}{c||}{${\rm d}^2 \sigma /{\rm d}p{\rm d}\Omega$}
  &$\langle p_{\rm T} \rangle$ & $\langle \theta \rangle$
  & \multicolumn{5}{c|}{${\rm d}^2 \sigma /{\rm d}p{\rm d}\Omega$} \\
 \hline
0.10--0.13 & 0.116 &  24.7 &  36.19 & $\!\!\pm\!\!$ &  2.03 & $\!\!\pm\!\!$ &  3.60 & 0.115 &  34.9 &  29.24 & $\!\!\pm\!\!$ &  1.80 & $\!\!\pm\!\!$ &  2.99 \\
0.13--0.16 & 0.145 &  24.8 &  48.74 & $\!\!\pm\!\!$ &  2.25 & $\!\!\pm\!\!$ &  4.00 & 0.145 &  34.9 &  32.40 & $\!\!\pm\!\!$ &  1.69 & $\!\!\pm\!\!$ &  2.71 \\
0.16--0.20 & 0.180 &  24.9 &  44.98 & $\!\!\pm\!\!$ &  1.71 & $\!\!\pm\!\!$ &  2.89 & 0.180 &  34.7 &  37.10 & $\!\!\pm\!\!$ &  1.63 & $\!\!\pm\!\!$ &  2.56 \\
0.20--0.24 & 0.219 &  24.8 &  41.84 & $\!\!\pm\!\!$ &  1.69 & $\!\!\pm\!\!$ &  2.42 & 0.220 &  34.9 &  35.98 & $\!\!\pm\!\!$ &  1.54 & $\!\!\pm\!\!$ &  2.21 \\
0.24--0.30 & 0.269 &  25.0 &  36.76 & $\!\!\pm\!\!$ &  1.27 & $\!\!\pm\!\!$ &  1.78 & 0.268 &  34.9 &  28.04 & $\!\!\pm\!\!$ &  1.08 & $\!\!\pm\!\!$ &  1.45 \\
0.30--0.36 & 0.329 &  25.1 &  27.54 & $\!\!\pm\!\!$ &  1.11 & $\!\!\pm\!\!$ &  1.38 & 0.329 &  34.7 &  23.88 & $\!\!\pm\!\!$ &  1.01 & $\!\!\pm\!\!$ &  1.24 \\
0.36--0.42 & 0.388 &  24.9 &  19.44 & $\!\!\pm\!\!$ &  0.91 & $\!\!\pm\!\!$ &  1.10 & 0.388 &  34.8 &  15.08 & $\!\!\pm\!\!$ &  0.75 & $\!\!\pm\!\!$ &  0.91 \\
0.42--0.50 & 0.458 &  25.1 &  11.60 & $\!\!\pm\!\!$ &  0.62 & $\!\!\pm\!\!$ &  0.74 & 0.457 &  34.9 &  14.42 & $\!\!\pm\!\!$ &  0.72 & $\!\!\pm\!\!$ &  0.99 \\
0.50--0.60 & 0.542 &  25.0 &   6.97 & $\!\!\pm\!\!$ &  0.43 & $\!\!\pm\!\!$ &  0.56 & 0.549 &  35.0 &   7.28 & $\!\!\pm\!\!$ &  0.43 & $\!\!\pm\!\!$ &  0.58 \\
0.60--0.72 & 0.656 &  25.4 &   3.14 & $\!\!\pm\!\!$ &  0.25 & $\!\!\pm\!\!$ &  0.35 & 0.657 &  34.9 &   3.57 & $\!\!\pm\!\!$ &  0.28 & $\!\!\pm\!\!$ &  0.38 \\
0.72--0.90 &  &  &  \multicolumn{5}{c||}{ } & 0.782 &  34.6 &   1.13 & $\!\!\pm\!\!$ &  0.12 & $\!\!\pm\!\!$ &  0.16 \\
 \hline
 \hline
   & \multicolumn{7}{c||}{$40<\theta<50$}
  & \multicolumn{7}{c|}{$50<\theta<60$} \\
 \hline
 $p_{\rm T}$ & $\langle p_{\rm T} \rangle$ & $\langle \theta \rangle$
  & \multicolumn{5}{c||}{${\rm d}^2 \sigma /{\rm d}p{\rm d}\Omega$}
  &$\langle p_{\rm T} \rangle$ & $\langle \theta \rangle$
  & \multicolumn{5}{c|}{${\rm d}^2 \sigma /{\rm d}p{\rm d}\Omega$} \\
 \hline
0.10--0.13 & 0.115 &  44.8 &  23.15 & $\!\!\pm\!\!$ &  1.60 & $\!\!\pm\!\!$ &  2.53 &  &  &  \multicolumn{5}{c|}{ } \\
0.13--0.16 & 0.146 &  44.9 &  31.22 & $\!\!\pm\!\!$ &  1.78 & $\!\!\pm\!\!$ &  2.67 & 0.146 &  54.8 &  21.23 & $\!\!\pm\!\!$ &  1.40 & $\!\!\pm\!\!$ &  1.93 \\
0.16--0.20 & 0.180 &  45.1 &  28.87 & $\!\!\pm\!\!$ &  1.43 & $\!\!\pm\!\!$ &  2.02 & 0.179 &  54.9 &  23.44 & $\!\!\pm\!\!$ &  1.26 & $\!\!\pm\!\!$ &  1.68 \\
0.20--0.24 & 0.219 &  44.9 &  27.86 & $\!\!\pm\!\!$ &  1.36 & $\!\!\pm\!\!$ &  1.88 & 0.220 &  54.7 &  23.77 & $\!\!\pm\!\!$ &  1.30 & $\!\!\pm\!\!$ &  1.67 \\
0.24--0.30 & 0.269 &  44.8 &  25.22 & $\!\!\pm\!\!$ &  1.07 & $\!\!\pm\!\!$ &  1.40 & 0.270 &  54.9 &  17.08 & $\!\!\pm\!\!$ &  0.85 & $\!\!\pm\!\!$ &  1.08 \\
0.30--0.36 & 0.330 &  44.9 &  16.84 & $\!\!\pm\!\!$ &  0.83 & $\!\!\pm\!\!$ &  0.94 & 0.330 &  54.7 &  12.20 & $\!\!\pm\!\!$ &  0.71 & $\!\!\pm\!\!$ &  0.79 \\
0.36--0.42 & 0.388 &  44.7 &  16.05 & $\!\!\pm\!\!$ &  0.86 & $\!\!\pm\!\!$ &  1.05 & 0.388 &  54.8 &  12.25 & $\!\!\pm\!\!$ &  0.72 & $\!\!\pm\!\!$ &  0.88 \\
0.42--0.50 & 0.459 &  44.8 &   8.95 & $\!\!\pm\!\!$ &  0.50 & $\!\!\pm\!\!$ &  0.63 & 0.456 &  54.7 &   7.76 & $\!\!\pm\!\!$ &  0.50 & $\!\!\pm\!\!$ &  0.57 \\
0.50--0.60 & 0.544 &  44.9 &   6.37 & $\!\!\pm\!\!$ &  0.42 & $\!\!\pm\!\!$ &  0.54 & 0.546 &  54.7 &   4.02 & $\!\!\pm\!\!$ &  0.31 & $\!\!\pm\!\!$ &  0.36 \\
0.60--0.72 & 0.653 &  44.6 &   2.79 & $\!\!\pm\!\!$ &  0.23 & $\!\!\pm\!\!$ &  0.31 & 0.649 &  54.8 &   2.20 & $\!\!\pm\!\!$ &  0.22 & $\!\!\pm\!\!$ &  0.27 \\
0.72--0.90 & 0.790 &  44.1 &   1.05 & $\!\!\pm\!\!$ &  0.12 & $\!\!\pm\!\!$ &  0.16 & 0.797 &  54.9 &   0.63 & $\!\!\pm\!\!$ &  0.08 & $\!\!\pm\!\!$ &  0.13 \\
0.90--1.25 &  &  &  \multicolumn{5}{c||}{ } & 1.015 &  54.9 &   0.15 & $\!\!\pm\!\!$ &  0.03 & $\!\!\pm\!\!$ &  0.05 \\
 \hline
 \hline
   & \multicolumn{7}{c||}{$60<\theta<75$}
  & \multicolumn{7}{c|}{$75<\theta<90$} \\
 \hline
 $p_{\rm T}$ & $\langle p_{\rm T} \rangle$ & $\langle \theta \rangle$
  & \multicolumn{5}{c||}{${\rm d}^2 \sigma /{\rm d}p{\rm d}\Omega$}
  &$\langle p_{\rm T} \rangle$ & $\langle \theta \rangle$
  & \multicolumn{5}{c|}{${\rm d}^2 \sigma /{\rm d}p{\rm d}\Omega$} \\
 \hline
0.13--0.16 & 0.145 &  67.4 &  18.89 & $\!\!\pm\!\!$ &  1.11 & $\!\!\pm\!\!$ &  1.65 & 0.146 &  82.6 &  16.53 & $\!\!\pm\!\!$ &  1.03 & $\!\!\pm\!\!$ &  1.51 \\
0.16--0.20 & 0.180 &  67.4 &  20.36 & $\!\!\pm\!\!$ &  0.97 & $\!\!\pm\!\!$ &  1.35 & 0.180 &  82.2 &  16.12 & $\!\!\pm\!\!$ &  0.84 & $\!\!\pm\!\!$ &  1.12 \\
0.20--0.24 & 0.220 &  67.0 &  17.89 & $\!\!\pm\!\!$ &  0.91 & $\!\!\pm\!\!$ &  1.07 & 0.220 &  81.8 &  15.50 & $\!\!\pm\!\!$ &  0.84 & $\!\!\pm\!\!$ &  1.01 \\
0.24--0.30 & 0.267 &  66.7 &  13.63 & $\!\!\pm\!\!$ &  0.64 & $\!\!\pm\!\!$ &  0.77 & 0.267 &  82.1 &  10.46 & $\!\!\pm\!\!$ &  0.55 & $\!\!\pm\!\!$ &  0.67 \\
0.30--0.36 & 0.329 &  66.8 &   9.42 & $\!\!\pm\!\!$ &  0.51 & $\!\!\pm\!\!$ &  0.61 & 0.328 &  82.4 &   5.98 & $\!\!\pm\!\!$ &  0.41 & $\!\!\pm\!\!$ &  0.47 \\
0.36--0.42 & 0.386 &  66.7 &   7.66 & $\!\!\pm\!\!$ &  0.48 & $\!\!\pm\!\!$ &  0.55 & 0.386 &  82.0 &   4.54 & $\!\!\pm\!\!$ &  0.37 & $\!\!\pm\!\!$ &  0.43 \\
0.42--0.50 & 0.454 &  67.1 &   5.19 & $\!\!\pm\!\!$ &  0.33 & $\!\!\pm\!\!$ &  0.41 & 0.453 &  81.5 &   3.26 & $\!\!\pm\!\!$ &  0.26 & $\!\!\pm\!\!$ &  0.35 \\
0.50--0.60 & 0.540 &  66.6 &   2.69 & $\!\!\pm\!\!$ &  0.22 & $\!\!\pm\!\!$ &  0.25 & 0.542 &  81.2 &   1.53 & $\!\!\pm\!\!$ &  0.16 & $\!\!\pm\!\!$ &  0.21 \\
0.60--0.72 & 0.646 &  66.1 &   1.24 & $\!\!\pm\!\!$ &  0.13 & $\!\!\pm\!\!$ &  0.16 & 0.657 &  80.6 &   0.44 & $\!\!\pm\!\!$ &  0.07 & $\!\!\pm\!\!$ &  0.10 \\
0.72--0.90 & 0.782 &  66.8 &   0.28 & $\!\!\pm\!\!$ &  0.05 & $\!\!\pm\!\!$ &  0.06 & 0.786 &  82.9 &   0.04 & $\!\!\pm\!\!$ &  0.02 & $\!\!\pm\!\!$ &  0.02 \\
0.90--1.25 & 1.076 &  66.4 &   0.04 & $\!\!\pm\!\!$ &  0.01 & $\!\!\pm\!\!$ &  0.02 &  &  &  \multicolumn{5}{c|}{ } \\
 \hline
 \hline
  & \multicolumn{7}{c||}{$90<\theta<105$}
  & \multicolumn{7}{c|}{$105<\theta<125$} \\
 \hline
 $p_{\rm T}$ & $\langle p_{\rm T} \rangle$ & $\langle \theta \rangle$
  & \multicolumn{5}{c||}{${\rm d}^2 \sigma /{\rm d}p{\rm d}\Omega$}
  &$\langle p_{\rm T} \rangle$ & $\langle \theta \rangle$
  & \multicolumn{5}{c|}{${\rm d}^2 \sigma /{\rm d}p{\rm d}\Omega$} \\
 \hline
0.13--0.16 & 0.145 &  97.5 &  15.36 & $\!\!\pm\!\!$ &  0.99 & $\!\!\pm\!\!$ &  1.44 & 0.144 & 114.8 &  15.54 & $\!\!\pm\!\!$ &  0.88 & $\!\!\pm\!\!$ &  1.21 \\
0.16--0.20 & 0.179 &  97.6 &  14.54 & $\!\!\pm\!\!$ &  0.80 & $\!\!\pm\!\!$ &  1.03 & 0.179 & 114.0 &  12.75 & $\!\!\pm\!\!$ &  0.69 & $\!\!\pm\!\!$ &  0.86 \\
0.20--0.24 & 0.219 &  97.2 &  12.63 & $\!\!\pm\!\!$ &  0.77 & $\!\!\pm\!\!$ &  0.89 & 0.218 & 114.1 &   6.18 & $\!\!\pm\!\!$ &  0.46 & $\!\!\pm\!\!$ &  0.49 \\
0.24--0.30 & 0.268 &  96.7 &   7.83 & $\!\!\pm\!\!$ &  0.48 & $\!\!\pm\!\!$ &  0.58 & 0.266 & 113.6 &   3.39 & $\!\!\pm\!\!$ &  0.28 & $\!\!\pm\!\!$ &  0.33 \\
0.30--0.36 & 0.326 &  96.7 &   4.32 & $\!\!\pm\!\!$ &  0.36 & $\!\!\pm\!\!$ &  0.46 & 0.323 & 112.7 &   2.20 & $\!\!\pm\!\!$ &  0.22 & $\!\!\pm\!\!$ &  0.30 \\
0.36--0.42 & 0.386 &  96.8 &   2.26 & $\!\!\pm\!\!$ &  0.25 & $\!\!\pm\!\!$ &  0.29 & 0.385 & 113.2 &   1.22 & $\!\!\pm\!\!$ &  0.16 & $\!\!\pm\!\!$ &  0.22 \\
0.42--0.50 & 0.448 &  96.1 &   1.37 & $\!\!\pm\!\!$ &  0.17 & $\!\!\pm\!\!$ &  0.20 & 0.444 & 111.8 &   0.53 & $\!\!\pm\!\!$ &  0.10 & $\!\!\pm\!\!$ &  0.12 \\
0.50--0.60 & 0.538 &  96.8 &   0.49 & $\!\!\pm\!\!$ &  0.10 & $\!\!\pm\!\!$ &  0.10 & 0.538 & 111.8 &   0.13 & $\!\!\pm\!\!$ &  0.04 & $\!\!\pm\!\!$ &  0.05 \\
0.60--0.72 & 0.635 &  96.9 &   0.13 & $\!\!\pm\!\!$ &  0.04 & $\!\!\pm\!\!$ &  0.05 &  &  &  \multicolumn{5}{c|}{ } \\
 \hline
 \end{tabular}
 \end{center}
 \end{scriptsize}
 \end{table}

%% file: table.pro.pipbe5.tex
  
 \begin{table}[h]
 \begin{scriptsize}
 \caption{Double-differential inclusive
  cross-section ${\rm d}^2 \sigma /{\rm d}p{\rm d}\Omega$
  [mb/(GeV/{\it c} sr)] of the production of protons
  in $\pi^+$ + Be $\rightarrow$ p + X interactions
  with $+5.0$~GeV/{\it c} beam momentum;
  the first error is statistical, the second systematic; 
 $p_{\rm T}$ in GeV/{\it c}, polar angle $\theta$ in degrees.}
 \label{pro.pipbe5}
 \begin{center}
 \begin{tabular}{|c||c|c|rcrcr||c|c|rcrcr|}
 \hline
   & \multicolumn{7}{c||}{$20<\theta<30$}
  & \multicolumn{7}{c|}{$30<\theta<40$} \\
 \hline
 $p_{\rm T}$ & $\langle p_{\rm T} \rangle$ & $\langle \theta \rangle$
  & \multicolumn{5}{c||}{${\rm d}^2 \sigma /{\rm d}p{\rm d}\Omega$}
  &$\langle p_{\rm T} \rangle$ & $\langle \theta \rangle$
  & \multicolumn{5}{c|}{${\rm d}^2 \sigma /{\rm d}p{\rm d}\Omega$} \\
 \hline
0.20--0.24 & 0.221 &  24.9 &  41.92 & $\!\!\pm\!\!$ &  1.44 & $\!\!\pm\!\!$ &  3.27 &  &  &  \multicolumn{5}{c|}{ } \\
0.24--0.30 & 0.269 &  25.1 &  39.00 & $\!\!\pm\!\!$ &  1.13 & $\!\!\pm\!\!$ &  2.33 & 0.271 &  34.8 &  36.94 & $\!\!\pm\!\!$ &  1.07 & $\!\!\pm\!\!$ &  2.36 \\
0.30--0.36 & 0.330 &  25.1 &  37.51 & $\!\!\pm\!\!$ &  1.10 & $\!\!\pm\!\!$ &  2.14 & 0.330 &  34.9 &  33.98 & $\!\!\pm\!\!$ &  1.03 & $\!\!\pm\!\!$ &  1.76 \\
0.36--0.42 & 0.389 &  25.3 &  33.64 & $\!\!\pm\!\!$ &  1.04 & $\!\!\pm\!\!$ &  1.74 & 0.389 &  34.9 &  33.01 & $\!\!\pm\!\!$ &  1.02 & $\!\!\pm\!\!$ &  1.68 \\
0.42--0.50 & 0.459 &  25.0 &  32.55 & $\!\!\pm\!\!$ &  0.87 & $\!\!\pm\!\!$ &  1.53 & 0.458 &  35.0 &  28.64 & $\!\!\pm\!\!$ &  0.83 & $\!\!\pm\!\!$ &  1.30 \\
0.50--0.60 & 0.548 &  24.7 &  25.81 & $\!\!\pm\!\!$ &  0.67 & $\!\!\pm\!\!$ &  1.19 & 0.548 &  34.9 &  23.33 & $\!\!\pm\!\!$ &  0.66 & $\!\!\pm\!\!$ &  1.06 \\
0.60--0.72 & 0.656 &  25.2 &  20.73 & $\!\!\pm\!\!$ &  0.54 & $\!\!\pm\!\!$ &  1.07 & 0.656 &  34.8 &  18.88 & $\!\!\pm\!\!$ &  0.54 & $\!\!\pm\!\!$ &  0.99 \\
0.72--0.90 &  &  &  \multicolumn{5}{c||}{ } & 0.802 &  34.9 &  12.09 & $\!\!\pm\!\!$ &  0.36 & $\!\!\pm\!\!$ &  0.80 \\
 \hline
 \hline
   & \multicolumn{7}{c||}{$40<\theta<50$}
  & \multicolumn{7}{c|}{$50<\theta<60$} \\
 \hline
 $p_{\rm T}$ & $\langle p_{\rm T} \rangle$ & $\langle \theta \rangle$
  & \multicolumn{5}{c||}{${\rm d}^2 \sigma /{\rm d}p{\rm d}\Omega$}
  &$\langle p_{\rm T} \rangle$ & $\langle \theta \rangle$
  & \multicolumn{5}{c|}{${\rm d}^2 \sigma /{\rm d}p{\rm d}\Omega$} \\
 \hline
0.30--0.36 & 0.329 &  45.0 &  32.30 & $\!\!\pm\!\!$ &  0.99 & $\!\!\pm\!\!$ &  1.69 &  &  &  \multicolumn{5}{c|}{ } \\
0.36--0.42 & 0.389 &  45.1 &  31.21 & $\!\!\pm\!\!$ &  0.97 & $\!\!\pm\!\!$ &  1.65 & 0.389 &  55.0 &  29.76 & $\!\!\pm\!\!$ &  0.95 & $\!\!\pm\!\!$ &  1.61 \\
0.42--0.50 & 0.458 &  44.9 &  26.21 & $\!\!\pm\!\!$ &  0.79 & $\!\!\pm\!\!$ &  1.22 & 0.459 &  54.9 &  24.78 & $\!\!\pm\!\!$ &  0.76 & $\!\!\pm\!\!$ &  1.30 \\
0.50--0.60 & 0.547 &  44.9 &  19.55 & $\!\!\pm\!\!$ &  0.61 & $\!\!\pm\!\!$ &  0.98 & 0.545 &  55.0 &  18.18 & $\!\!\pm\!\!$ &  0.59 & $\!\!\pm\!\!$ &  0.94 \\
0.60--0.72 & 0.656 &  44.8 &  14.89 & $\!\!\pm\!\!$ &  0.49 & $\!\!\pm\!\!$ &  0.87 & 0.654 &  55.0 &  13.12 & $\!\!\pm\!\!$ &  0.48 & $\!\!\pm\!\!$ &  0.83 \\
0.72--0.90 & 0.799 &  44.8 &   9.60 & $\!\!\pm\!\!$ &  0.33 & $\!\!\pm\!\!$ &  0.68 & 0.796 &  54.7 &   6.78 & $\!\!\pm\!\!$ &  0.28 & $\!\!\pm\!\!$ &  0.55 \\
0.90--1.25 & 1.028 &  44.7 &   3.10 & $\!\!\pm\!\!$ &  0.14 & $\!\!\pm\!\!$ &  0.33 & 1.025 &  54.6 &   1.72 & $\!\!\pm\!\!$ &  0.10 & $\!\!\pm\!\!$ &  0.22 \\
 \hline
 \hline
   & \multicolumn{7}{c||}{$60<\theta<75$}
  & \multicolumn{7}{c|}{$75<\theta<90$} \\
 \hline
 $p_{\rm T}$ & $\langle p_{\rm T} \rangle$ & $\langle \theta \rangle$
  & \multicolumn{5}{c||}{${\rm d}^2 \sigma /{\rm d}p{\rm d}\Omega$}
  &$\langle p_{\rm T} \rangle$ & $\langle \theta \rangle$
  & \multicolumn{5}{c|}{${\rm d}^2 \sigma /{\rm d}p{\rm d}\Omega$} \\
 \hline
0.36--0.42 & 0.391 &  67.2 &  28.20 & $\!\!\pm\!\!$ &  0.74 & $\!\!\pm\!\!$ &  1.21 &  &  &  \multicolumn{5}{c|}{ } \\
0.42--0.50 & 0.461 &  67.3 &  24.13 & $\!\!\pm\!\!$ &  0.62 & $\!\!\pm\!\!$ &  1.06 & 0.459 &  81.9 &  16.51 & $\!\!\pm\!\!$ &  0.50 & $\!\!\pm\!\!$ &  0.85 \\
0.50--0.60 & 0.549 &  67.2 &  17.32 & $\!\!\pm\!\!$ &  0.47 & $\!\!\pm\!\!$ &  0.83 & 0.548 &  81.9 &  10.75 & $\!\!\pm\!\!$ &  0.37 & $\!\!\pm\!\!$ &  0.72 \\
0.60--0.72 & 0.657 &  67.2 &  10.56 & $\!\!\pm\!\!$ &  0.35 & $\!\!\pm\!\!$ &  0.70 & 0.659 &  81.9 &   4.87 & $\!\!\pm\!\!$ &  0.24 & $\!\!\pm\!\!$ &  0.48 \\
0.72--0.90 & 0.801 &  66.8 &   3.97 & $\!\!\pm\!\!$ &  0.17 & $\!\!\pm\!\!$ &  0.40 & 0.803 &  81.3 &   1.78 & $\!\!\pm\!\!$ &  0.12 & $\!\!\pm\!\!$ &  0.24 \\
0.90--1.25 & 1.039 &  66.9 &   0.80 & $\!\!\pm\!\!$ &  0.06 & $\!\!\pm\!\!$ &  0.14 & 1.031 &  81.4 &   0.33 & $\!\!\pm\!\!$ &  0.04 & $\!\!\pm\!\!$ &  0.07 \\
 \hline
 \hline
  & \multicolumn{7}{c||}{$90<\theta<105$}
  & \multicolumn{7}{c|}{$105<\theta<125$} \\
 \hline
 $p_{\rm T}$ & $\langle p_{\rm T} \rangle$ & $\langle \theta \rangle$
  & \multicolumn{5}{c||}{${\rm d}^2 \sigma /{\rm d}p{\rm d}\Omega$}
  &$\langle p_{\rm T} \rangle$ & $\langle \theta \rangle$
  & \multicolumn{5}{c|}{${\rm d}^2 \sigma /{\rm d}p{\rm d}\Omega$} \\
 \hline
0.36--0.42 &  &  &  \multicolumn{5}{c||}{ } & 0.388 & 113.8 &   7.25 & $\!\!\pm\!\!$ &  0.32 & $\!\!\pm\!\!$ &  0.54 \\
0.42--0.50 & 0.459 &  96.9 &   9.72 & $\!\!\pm\!\!$ &  0.37 & $\!\!\pm\!\!$ &  0.80 & 0.458 & 113.3 &   4.57 & $\!\!\pm\!\!$ &  0.22 & $\!\!\pm\!\!$ &  0.42 \\
0.50--0.60 & 0.546 &  97.0 &   5.52 & $\!\!\pm\!\!$ &  0.26 & $\!\!\pm\!\!$ &  0.56 & 0.546 & 113.0 &   2.00 & $\!\!\pm\!\!$ &  0.14 & $\!\!\pm\!\!$ &  0.27 \\
0.60--0.72 & 0.659 &  96.6 &   2.20 & $\!\!\pm\!\!$ &  0.16 & $\!\!\pm\!\!$ &  0.29 & 0.657 & 112.9 &   0.55 & $\!\!\pm\!\!$ &  0.07 & $\!\!\pm\!\!$ &  0.14 \\
0.72--0.90 & 0.803 &  96.6 &   0.63 & $\!\!\pm\!\!$ &  0.08 & $\!\!\pm\!\!$ &  0.13 & 0.794 & 111.9 &   0.12 & $\!\!\pm\!\!$ &  0.03 & $\!\!\pm\!\!$ &  0.05 \\
0.90--1.25 & 1.032 &  96.2 &   0.08 & $\!\!\pm\!\!$ &  0.02 & $\!\!\pm\!\!$ &  0.03 &  &  &  \multicolumn{5}{c|}{ } \\
 \hline
 \end{tabular}
 \end{center}
 \end{scriptsize}
 \end{table}

%% file: table.pip.pipbe5.tex
  
 \begin{table}[h]
 \begin{scriptsize}
 \caption{Double-differential inclusive
  cross-section ${\rm d}^2 \sigma /{\rm d}p{\rm d}\Omega$
  [mb/(GeV/{\it c} sr)] of the production of $\pi^+$'s
  in $\pi^+$ + Be $\rightarrow$ $\pi^+$ + X interactions
  with $+5.0$~GeV/{\it c} beam momentum;
  the first error is statistical, the second systematic; 
 $p_{\rm T}$ in GeV/{\it c}, polar angle $\theta$ in degrees.}
 \label{pip.pipbe5}
 \begin{center}
 \begin{tabular}{|c||c|c|rcrcr||c|c|rcrcr|}
 \hline
   & \multicolumn{7}{c||}{$20<\theta<30$}
  & \multicolumn{7}{c|}{$30<\theta<40$} \\
 \hline
 $p_{\rm T}$ & $\langle p_{\rm T} \rangle$ & $\langle \theta \rangle$
  & \multicolumn{5}{c||}{${\rm d}^2 \sigma /{\rm d}p{\rm d}\Omega$}
  &$\langle p_{\rm T} \rangle$ & $\langle \theta \rangle$
  & \multicolumn{5}{c|}{${\rm d}^2 \sigma /{\rm d}p{\rm d}\Omega$} \\
 \hline
0.10--0.13 & 0.115 &  25.0 &  42.31 & $\!\!\pm\!\!$ &  1.93 & $\!\!\pm\!\!$ &  4.05 & 0.115 &  34.8 &  36.97 & $\!\!\pm\!\!$ &  1.85 & $\!\!\pm\!\!$ &  3.45 \\
0.13--0.16 & 0.146 &  24.9 &  58.31 & $\!\!\pm\!\!$ &  2.13 & $\!\!\pm\!\!$ &  4.42 & 0.145 &  35.0 &  48.46 & $\!\!\pm\!\!$ &  1.93 & $\!\!\pm\!\!$ &  3.70 \\
0.16--0.20 & 0.181 &  24.8 &  68.61 & $\!\!\pm\!\!$ &  1.91 & $\!\!\pm\!\!$ &  4.17 & 0.180 &  34.7 &  52.84 & $\!\!\pm\!\!$ &  1.67 & $\!\!\pm\!\!$ &  3.25 \\
0.20--0.24 & 0.220 &  24.7 &  73.75 & $\!\!\pm\!\!$ &  1.89 & $\!\!\pm\!\!$ &  3.96 & 0.221 &  34.8 &  55.36 & $\!\!\pm\!\!$ &  1.67 & $\!\!\pm\!\!$ &  3.06 \\
0.24--0.30 & 0.269 &  24.7 &  77.78 & $\!\!\pm\!\!$ &  1.61 & $\!\!\pm\!\!$ &  3.42 & 0.270 &  34.7 &  56.69 & $\!\!\pm\!\!$ &  1.37 & $\!\!\pm\!\!$ &  2.60 \\
0.30--0.36 & 0.329 &  24.9 &  66.69 & $\!\!\pm\!\!$ &  1.48 & $\!\!\pm\!\!$ &  2.70 & 0.329 &  34.8 &  49.78 & $\!\!\pm\!\!$ &  1.27 & $\!\!\pm\!\!$ &  2.19 \\
0.36--0.42 & 0.389 &  24.7 &  59.86 & $\!\!\pm\!\!$ &  1.42 & $\!\!\pm\!\!$ &  2.49 & 0.389 &  34.8 &  46.53 & $\!\!\pm\!\!$ &  1.24 & $\!\!\pm\!\!$ &  2.06 \\
0.42--0.50 & 0.458 &  24.8 &  48.91 & $\!\!\pm\!\!$ &  1.11 & $\!\!\pm\!\!$ &  2.14 & 0.459 &  34.8 &  35.48 & $\!\!\pm\!\!$ &  0.96 & $\!\!\pm\!\!$ &  1.59 \\
0.50--0.60 & 0.547 &  24.8 &  31.94 & $\!\!\pm\!\!$ &  0.76 & $\!\!\pm\!\!$ &  1.90 & 0.546 &  34.6 &  23.41 & $\!\!\pm\!\!$ &  0.66 & $\!\!\pm\!\!$ &  1.40 \\
0.60--0.72 & 0.655 &  24.9 &  17.89 & $\!\!\pm\!\!$ &  0.49 & $\!\!\pm\!\!$ &  1.53 & 0.655 &  34.4 &  12.85 & $\!\!\pm\!\!$ &  0.41 & $\!\!\pm\!\!$ &  1.09 \\
0.72--0.90 &  &  &  \multicolumn{5}{c||}{ } & 0.798 &  34.7 &   6.48 & $\!\!\pm\!\!$ &  0.24 & $\!\!\pm\!\!$ &  0.79 \\
 \hline
 \hline
   & \multicolumn{7}{c||}{$40<\theta<50$}
  & \multicolumn{7}{c|}{$50<\theta<60$} \\
 \hline
 $p_{\rm T}$ & $\langle p_{\rm T} \rangle$ & $\langle \theta \rangle$
  & \multicolumn{5}{c||}{${\rm d}^2 \sigma /{\rm d}p{\rm d}\Omega$}
  &$\langle p_{\rm T} \rangle$ & $\langle \theta \rangle$
  & \multicolumn{5}{c|}{${\rm d}^2 \sigma /{\rm d}p{\rm d}\Omega$} \\
 \hline
0.10--0.13 & 0.116 &  45.1 &  32.76 & $\!\!\pm\!\!$ &  1.81 & $\!\!\pm\!\!$ &  3.26 &  &  &  \multicolumn{5}{c|}{ } \\
0.13--0.16 & 0.145 &  44.8 &  36.22 & $\!\!\pm\!\!$ &  1.69 & $\!\!\pm\!\!$ &  2.79 & 0.146 &  54.7 &  27.92 & $\!\!\pm\!\!$ &  1.52 & $\!\!\pm\!\!$ &  2.32 \\
0.16--0.20 & 0.180 &  44.8 &  40.05 & $\!\!\pm\!\!$ &  1.49 & $\!\!\pm\!\!$ &  2.59 & 0.180 &  54.9 &  34.30 & $\!\!\pm\!\!$ &  1.37 & $\!\!\pm\!\!$ &  2.34 \\
0.20--0.24 & 0.220 &  44.9 &  45.24 & $\!\!\pm\!\!$ &  1.54 & $\!\!\pm\!\!$ &  2.63 & 0.220 &  54.9 &  34.14 & $\!\!\pm\!\!$ &  1.35 & $\!\!\pm\!\!$ &  2.23 \\
0.24--0.30 & 0.271 &  44.7 &  41.47 & $\!\!\pm\!\!$ &  1.19 & $\!\!\pm\!\!$ &  1.95 & 0.269 &  54.8 &  30.07 & $\!\!\pm\!\!$ &  1.02 & $\!\!\pm\!\!$ &  1.57 \\
0.30--0.36 & 0.330 &  44.7 &  36.64 & $\!\!\pm\!\!$ &  1.12 & $\!\!\pm\!\!$ &  1.60 & 0.328 &  54.6 &  26.84 & $\!\!\pm\!\!$ &  0.96 & $\!\!\pm\!\!$ &  1.28 \\
0.36--0.42 & 0.388 &  44.6 &  33.64 & $\!\!\pm\!\!$ &  1.08 & $\!\!\pm\!\!$ &  1.50 & 0.388 &  54.8 &  23.40 & $\!\!\pm\!\!$ &  0.92 & $\!\!\pm\!\!$ &  1.24 \\
0.42--0.50 & 0.458 &  44.7 &  24.32 & $\!\!\pm\!\!$ &  0.79 & $\!\!\pm\!\!$ &  1.14 & 0.458 &  54.7 &  15.27 & $\!\!\pm\!\!$ &  0.61 & $\!\!\pm\!\!$ &  0.90 \\
0.50--0.60 & 0.546 &  44.5 &  14.69 & $\!\!\pm\!\!$ &  0.52 & $\!\!\pm\!\!$ &  0.91 & 0.547 &  54.7 &  11.27 & $\!\!\pm\!\!$ &  0.46 & $\!\!\pm\!\!$ &  0.83 \\
0.60--0.72 & 0.655 &  44.5 &   8.97 & $\!\!\pm\!\!$ &  0.37 & $\!\!\pm\!\!$ &  0.74 & 0.656 &  54.6 &   5.92 & $\!\!\pm\!\!$ &  0.30 & $\!\!\pm\!\!$ &  0.56 \\
0.72--0.90 & 0.792 &  44.3 &   3.99 & $\!\!\pm\!\!$ &  0.19 & $\!\!\pm\!\!$ &  0.46 & 0.803 &  54.3 &   3.18 & $\!\!\pm\!\!$ &  0.19 & $\!\!\pm\!\!$ &  0.39 \\
0.90--1.25 &  &  &  \multicolumn{5}{c||}{ } & 1.026 &  54.2 &   0.69 & $\!\!\pm\!\!$ &  0.06 & $\!\!\pm\!\!$ &  0.13 \\
 \hline
 \hline
   & \multicolumn{7}{c||}{$60<\theta<75$}
  & \multicolumn{7}{c|}{$75<\theta<90$} \\
 \hline
 $p_{\rm T}$ & $\langle p_{\rm T} \rangle$ & $\langle \theta \rangle$
  & \multicolumn{5}{c||}{${\rm d}^2 \sigma /{\rm d}p{\rm d}\Omega$}
  &$\langle p_{\rm T} \rangle$ & $\langle \theta \rangle$
  & \multicolumn{5}{c|}{${\rm d}^2 \sigma /{\rm d}p{\rm d}\Omega$} \\
 \hline
0.13--0.16 & 0.146 &  67.1 &  23.55 & $\!\!\pm\!\!$ &  1.12 & $\!\!\pm\!\!$ &  1.93 & 0.146 &  82.6 &  20.77 & $\!\!\pm\!\!$ &  1.07 & $\!\!\pm\!\!$ &  1.80 \\
0.16--0.20 & 0.181 &  67.2 &  28.89 & $\!\!\pm\!\!$ &  1.02 & $\!\!\pm\!\!$ &  1.94 & 0.181 &  82.4 &  20.75 & $\!\!\pm\!\!$ &  0.87 & $\!\!\pm\!\!$ &  1.35 \\
0.20--0.24 & 0.221 &  67.1 &  29.33 & $\!\!\pm\!\!$ &  1.03 & $\!\!\pm\!\!$ &  1.87 & 0.220 &  82.4 &  18.97 & $\!\!\pm\!\!$ &  0.83 & $\!\!\pm\!\!$ &  1.17 \\
0.24--0.30 & 0.269 &  67.0 &  21.83 & $\!\!\pm\!\!$ &  0.72 & $\!\!\pm\!\!$ &  1.11 & 0.269 &  82.5 &  14.12 & $\!\!\pm\!\!$ &  0.59 & $\!\!\pm\!\!$ &  0.78 \\
0.30--0.36 & 0.330 &  66.9 &  17.07 & $\!\!\pm\!\!$ &  0.63 & $\!\!\pm\!\!$ &  0.83 & 0.330 &  82.1 &  11.05 & $\!\!\pm\!\!$ &  0.50 & $\!\!\pm\!\!$ &  0.71 \\
0.36--0.42 & 0.391 &  66.6 &  14.13 & $\!\!\pm\!\!$ &  0.56 & $\!\!\pm\!\!$ &  0.81 & 0.391 &  82.1 &   7.59 & $\!\!\pm\!\!$ &  0.41 & $\!\!\pm\!\!$ &  0.58 \\
0.42--0.50 & 0.458 &  66.8 &  10.28 & $\!\!\pm\!\!$ &  0.41 & $\!\!\pm\!\!$ &  0.69 & 0.460 &  81.9 &   5.40 & $\!\!\pm\!\!$ &  0.30 & $\!\!\pm\!\!$ &  0.46 \\
0.50--0.60 & 0.549 &  66.4 &   6.46 & $\!\!\pm\!\!$ &  0.29 & $\!\!\pm\!\!$ &  0.53 & 0.549 &  81.7 &   3.37 & $\!\!\pm\!\!$ &  0.21 & $\!\!\pm\!\!$ &  0.35 \\
0.60--0.72 & 0.658 &  66.5 &   3.71 & $\!\!\pm\!\!$ &  0.20 & $\!\!\pm\!\!$ &  0.40 & 0.657 &  81.2 &   1.88 & $\!\!\pm\!\!$ &  0.15 & $\!\!\pm\!\!$ &  0.25 \\
0.72--0.90 & 0.799 &  66.1 &   1.54 & $\!\!\pm\!\!$ &  0.11 & $\!\!\pm\!\!$ &  0.21 & 0.791 &  81.8 &   0.50 & $\!\!\pm\!\!$ &  0.06 & $\!\!\pm\!\!$ &  0.10 \\
0.90--1.25 & 1.040 &  66.3 &   0.23 & $\!\!\pm\!\!$ &  0.02 & $\!\!\pm\!\!$ &  0.05 & 1.008 &  80.8 &   0.06 & $\!\!\pm\!\!$ &  0.01 & $\!\!\pm\!\!$ &  0.02 \\
 \hline
 \hline
  & \multicolumn{7}{c||}{$90<\theta<105$}
  & \multicolumn{7}{c|}{$105<\theta<125$} \\
 \hline
 $p_{\rm T}$ & $\langle p_{\rm T} \rangle$ & $\langle \theta \rangle$
  & \multicolumn{5}{c||}{${\rm d}^2 \sigma /{\rm d}p{\rm d}\Omega$}
  &$\langle p_{\rm T} \rangle$ & $\langle \theta \rangle$
  & \multicolumn{5}{c|}{${\rm d}^2 \sigma /{\rm d}p{\rm d}\Omega$} \\
 \hline
0.13--0.16 & 0.146 &  97.3 &  19.16 & $\!\!\pm\!\!$ &  1.04 & $\!\!\pm\!\!$ &  1.63 & 0.145 & 114.8 &  16.24 & $\!\!\pm\!\!$ &  0.82 & $\!\!\pm\!\!$ &  1.32 \\
0.16--0.20 & 0.180 &  96.9 &  19.87 & $\!\!\pm\!\!$ &  0.87 & $\!\!\pm\!\!$ &  1.24 & 0.180 & 113.9 &  13.80 & $\!\!\pm\!\!$ &  0.62 & $\!\!\pm\!\!$ &  0.93 \\
0.20--0.24 & 0.219 &  97.0 &  15.37 & $\!\!\pm\!\!$ &  0.75 & $\!\!\pm\!\!$ &  0.98 & 0.218 & 114.4 &   9.21 & $\!\!\pm\!\!$ &  0.50 & $\!\!\pm\!\!$ &  0.76 \\
0.24--0.30 & 0.269 &  97.1 &  10.66 & $\!\!\pm\!\!$ &  0.50 & $\!\!\pm\!\!$ &  0.69 & 0.269 & 113.9 &   5.49 & $\!\!\pm\!\!$ &  0.31 & $\!\!\pm\!\!$ &  0.47 \\
0.30--0.36 & 0.329 &  96.8 &   6.28 & $\!\!\pm\!\!$ &  0.38 & $\!\!\pm\!\!$ &  0.50 & 0.327 & 113.7 &   3.74 & $\!\!\pm\!\!$ &  0.26 & $\!\!\pm\!\!$ &  0.38 \\
0.36--0.42 & 0.390 &  97.0 &   4.88 & $\!\!\pm\!\!$ &  0.33 & $\!\!\pm\!\!$ &  0.48 & 0.386 & 113.5 &   1.95 & $\!\!\pm\!\!$ &  0.18 & $\!\!\pm\!\!$ &  0.27 \\
0.42--0.50 & 0.460 &  96.7 &   3.43 & $\!\!\pm\!\!$ &  0.24 & $\!\!\pm\!\!$ &  0.39 & 0.460 & 112.7 &   1.19 & $\!\!\pm\!\!$ &  0.12 & $\!\!\pm\!\!$ &  0.20 \\
0.50--0.60 & 0.544 &  96.8 &   1.59 & $\!\!\pm\!\!$ &  0.15 & $\!\!\pm\!\!$ &  0.23 & 0.542 & 112.1 &   0.28 & $\!\!\pm\!\!$ &  0.05 & $\!\!\pm\!\!$ &  0.09 \\
0.60--0.72 & 0.657 &  96.4 &   0.46 & $\!\!\pm\!\!$ &  0.07 & $\!\!\pm\!\!$ &  0.10 & 0.656 & 110.7 &   0.14 & $\!\!\pm\!\!$ &  0.03 & $\!\!\pm\!\!$ &  0.07 \\
0.72--0.90 & 0.792 &  95.9 &   0.10 & $\!\!\pm\!\!$ &  0.02 & $\!\!\pm\!\!$ &  0.03 &  &  &  \multicolumn{5}{c|}{ } \\
 \hline
 \end{tabular}
 \end{center}
 \end{scriptsize}
 \end{table}

%% file: table.pim.pipbe5.tex
  
 \begin{table}[h]
 \begin{scriptsize}
 \caption{Double-differential inclusive
  cross-section ${\rm d}^2 \sigma /{\rm d}p{\rm d}\Omega$
  [mb/(GeV/{\it c} sr)] of the production of $\pi^-$'s
  in $\pi^+$ + Be $\rightarrow$ $\pi^-$ + X interactions
  with $+5.0$~GeV/{\it c} beam momentum;
  the first error is statistical, the second systematic; 
 $p_{\rm T}$ in GeV/{\it c}, polar angle $\theta$ in degrees.}
 \label{pim.pipbe5}
 \begin{center}
 \begin{tabular}{|c||c|c|rcrcr||c|c|rcrcr|}
 \hline
   & \multicolumn{7}{c||}{$20<\theta<30$}
  & \multicolumn{7}{c|}{$30<\theta<40$} \\
 \hline
 $p_{\rm T}$ & $\langle p_{\rm T} \rangle$ & $\langle \theta \rangle$
  & \multicolumn{5}{c||}{${\rm d}^2 \sigma /{\rm d}p{\rm d}\Omega$}
  &$\langle p_{\rm T} \rangle$ & $\langle \theta \rangle$
  & \multicolumn{5}{c|}{${\rm d}^2 \sigma /{\rm d}p{\rm d}\Omega$} \\
 \hline
0.10--0.13 & 0.115 &  24.8 &  39.31 & $\!\!\pm\!\!$ &  1.81 & $\!\!\pm\!\!$ &  4.02 & 0.115 &  34.5 &  33.52 & $\!\!\pm\!\!$ &  1.71 & $\!\!\pm\!\!$ &  3.42 \\
0.13--0.16 & 0.145 &  24.8 &  47.64 & $\!\!\pm\!\!$ &  1.88 & $\!\!\pm\!\!$ &  4.01 & 0.145 &  34.8 &  36.33 & $\!\!\pm\!\!$ &  1.66 & $\!\!\pm\!\!$ &  3.22 \\
0.16--0.20 & 0.180 &  24.6 &  53.08 & $\!\!\pm\!\!$ &  1.62 & $\!\!\pm\!\!$ &  3.40 & 0.180 &  34.7 &  36.96 & $\!\!\pm\!\!$ &  1.38 & $\!\!\pm\!\!$ &  2.60 \\
0.20--0.24 & 0.221 &  24.6 &  55.02 & $\!\!\pm\!\!$ &  1.64 & $\!\!\pm\!\!$ &  3.20 & 0.220 &  34.7 &  40.08 & $\!\!\pm\!\!$ &  1.40 & $\!\!\pm\!\!$ &  2.55 \\
0.24--0.30 & 0.270 &  24.4 &  47.90 & $\!\!\pm\!\!$ &  1.23 & $\!\!\pm\!\!$ &  2.37 & 0.270 &  34.5 &  35.56 & $\!\!\pm\!\!$ &  1.06 & $\!\!\pm\!\!$ &  1.92 \\
0.30--0.36 & 0.329 &  24.8 &  40.67 & $\!\!\pm\!\!$ &  1.15 & $\!\!\pm\!\!$ &  1.91 & 0.329 &  34.7 &  29.35 & $\!\!\pm\!\!$ &  0.96 & $\!\!\pm\!\!$ &  1.62 \\
0.36--0.42 & 0.389 &  24.9 &  35.08 & $\!\!\pm\!\!$ &  1.08 & $\!\!\pm\!\!$ &  1.77 & 0.389 &  34.6 &  22.84 & $\!\!\pm\!\!$ &  0.85 & $\!\!\pm\!\!$ &  1.23 \\
0.42--0.50 & 0.457 &  24.7 &  26.10 & $\!\!\pm\!\!$ &  0.83 & $\!\!\pm\!\!$ &  1.44 & 0.456 &  34.8 &  17.25 & $\!\!\pm\!\!$ &  0.66 & $\!\!\pm\!\!$ &  1.03 \\
0.50--0.60 & 0.547 &  25.0 &  12.95 & $\!\!\pm\!\!$ &  0.50 & $\!\!\pm\!\!$ &  0.89 & 0.545 &  34.6 &  10.78 & $\!\!\pm\!\!$ &  0.45 & $\!\!\pm\!\!$ &  0.79 \\
0.60--0.72 & 0.652 &  24.8 &   7.87 & $\!\!\pm\!\!$ &  0.35 & $\!\!\pm\!\!$ &  0.73 & 0.651 &  34.7 &   5.75 & $\!\!\pm\!\!$ &  0.30 & $\!\!\pm\!\!$ &  0.57 \\
0.72--0.90 &  &  &  \multicolumn{5}{c||}{ } & 0.796 &  34.9 &   2.74 & $\!\!\pm\!\!$ &  0.18 & $\!\!\pm\!\!$ &  0.34 \\
 \hline
 \hline
   & \multicolumn{7}{c||}{$40<\theta<50$}
  & \multicolumn{7}{c|}{$50<\theta<60$} \\
 \hline
 $p_{\rm T}$ & $\langle p_{\rm T} \rangle$ & $\langle \theta \rangle$
  & \multicolumn{5}{c||}{${\rm d}^2 \sigma /{\rm d}p{\rm d}\Omega$}
  &$\langle p_{\rm T} \rangle$ & $\langle \theta \rangle$
  & \multicolumn{5}{c|}{${\rm d}^2 \sigma /{\rm d}p{\rm d}\Omega$} \\
 \hline
0.10--0.13 & 0.116 &  44.7 &  24.87 & $\!\!\pm\!\!$ &  1.51 & $\!\!\pm\!\!$ &  2.68 &  &  &  \multicolumn{5}{c|}{ } \\
0.13--0.16 & 0.145 &  44.8 &  27.53 & $\!\!\pm\!\!$ &  1.46 & $\!\!\pm\!\!$ &  2.34 & 0.145 &  54.7 &  22.78 & $\!\!\pm\!\!$ &  1.31 & $\!\!\pm\!\!$ &  2.11 \\
0.16--0.20 & 0.180 &  44.6 &  27.67 & $\!\!\pm\!\!$ &  1.21 & $\!\!\pm\!\!$ &  2.02 & 0.180 &  54.5 &  23.64 & $\!\!\pm\!\!$ &  1.13 & $\!\!\pm\!\!$ &  1.87 \\
0.20--0.24 & 0.220 &  44.7 &  26.37 & $\!\!\pm\!\!$ &  1.14 & $\!\!\pm\!\!$ &  1.82 & 0.220 &  54.9 &  22.43 & $\!\!\pm\!\!$ &  1.06 & $\!\!\pm\!\!$ &  1.75 \\
0.24--0.30 & 0.269 &  44.8 &  25.96 & $\!\!\pm\!\!$ &  0.94 & $\!\!\pm\!\!$ &  1.43 & 0.271 &  54.7 &  20.51 & $\!\!\pm\!\!$ &  0.83 & $\!\!\pm\!\!$ &  1.24 \\
0.30--0.36 & 0.329 &  44.7 &  21.55 & $\!\!\pm\!\!$ &  0.83 & $\!\!\pm\!\!$ &  1.17 & 0.330 &  54.7 &  16.40 & $\!\!\pm\!\!$ &  0.74 & $\!\!\pm\!\!$ &  1.03 \\
0.36--0.42 & 0.389 &  44.8 &  17.95 & $\!\!\pm\!\!$ &  0.78 & $\!\!\pm\!\!$ &  1.01 & 0.389 &  54.6 &  14.18 & $\!\!\pm\!\!$ &  0.70 & $\!\!\pm\!\!$ &  0.93 \\
0.42--0.50 & 0.456 &  44.6 &  13.69 & $\!\!\pm\!\!$ &  0.59 & $\!\!\pm\!\!$ &  0.83 & 0.456 &  54.9 &   9.04 & $\!\!\pm\!\!$ &  0.47 & $\!\!\pm\!\!$ &  0.66 \\
0.50--0.60 & 0.546 &  44.6 &   8.04 & $\!\!\pm\!\!$ &  0.39 & $\!\!\pm\!\!$ &  0.62 & 0.544 &  54.7 &   6.05 & $\!\!\pm\!\!$ &  0.34 & $\!\!\pm\!\!$ &  0.60 \\
0.60--0.72 & 0.653 &  44.7 &   4.49 & $\!\!\pm\!\!$ &  0.27 & $\!\!\pm\!\!$ &  0.47 & 0.654 &  54.5 &   3.34 & $\!\!\pm\!\!$ &  0.24 & $\!\!\pm\!\!$ &  0.40 \\
0.72--0.90 & 0.792 &  44.7 &   2.05 & $\!\!\pm\!\!$ &  0.16 & $\!\!\pm\!\!$ &  0.27 & 0.799 &  54.6 &   1.48 & $\!\!\pm\!\!$ &  0.13 & $\!\!\pm\!\!$ &  0.22 \\
0.90--1.25 &  &  &  \multicolumn{5}{c||}{ } & 1.015 &  54.6 &   0.36 & $\!\!\pm\!\!$ &  0.04 & $\!\!\pm\!\!$ &  0.08 \\
 \hline
 \hline
   & \multicolumn{7}{c||}{$60<\theta<75$}
  & \multicolumn{7}{c|}{$75<\theta<90$} \\
 \hline
 $p_{\rm T}$ & $\langle p_{\rm T} \rangle$ & $\langle \theta \rangle$
  & \multicolumn{5}{c||}{${\rm d}^2 \sigma /{\rm d}p{\rm d}\Omega$}
  &$\langle p_{\rm T} \rangle$ & $\langle \theta \rangle$
  & \multicolumn{5}{c|}{${\rm d}^2 \sigma /{\rm d}p{\rm d}\Omega$} \\
 \hline
0.13--0.16 & 0.145 &  67.0 &  20.23 & $\!\!\pm\!\!$ &  1.03 & $\!\!\pm\!\!$ &  1.78 & 0.146 &  82.5 &  15.76 & $\!\!\pm\!\!$ &  0.91 & $\!\!\pm\!\!$ &  1.44 \\
0.16--0.20 & 0.179 &  67.0 &  19.37 & $\!\!\pm\!\!$ &  0.83 & $\!\!\pm\!\!$ &  1.39 & 0.179 &  82.4 &  16.64 & $\!\!\pm\!\!$ &  0.77 & $\!\!\pm\!\!$ &  1.18 \\
0.20--0.24 & 0.219 &  67.3 &  16.20 & $\!\!\pm\!\!$ &  0.74 & $\!\!\pm\!\!$ &  1.17 & 0.218 &  82.2 &  13.44 & $\!\!\pm\!\!$ &  0.69 & $\!\!\pm\!\!$ &  0.98 \\
0.24--0.30 & 0.267 &  66.9 &  14.82 & $\!\!\pm\!\!$ &  0.57 & $\!\!\pm\!\!$ &  0.86 & 0.269 &  82.1 &   9.55 & $\!\!\pm\!\!$ &  0.47 & $\!\!\pm\!\!$ &  0.61 \\
0.30--0.36 & 0.329 &  67.0 &  11.16 & $\!\!\pm\!\!$ &  0.50 & $\!\!\pm\!\!$ &  0.67 & 0.329 &  82.4 &   7.24 & $\!\!\pm\!\!$ &  0.40 & $\!\!\pm\!\!$ &  0.56 \\
0.36--0.42 & 0.388 &  67.1 &   8.44 & $\!\!\pm\!\!$ &  0.44 & $\!\!\pm\!\!$ &  0.61 & 0.387 &  82.1 &   4.66 & $\!\!\pm\!\!$ &  0.32 & $\!\!\pm\!\!$ &  0.46 \\
0.42--0.50 & 0.454 &  66.8 &   6.11 & $\!\!\pm\!\!$ &  0.32 & $\!\!\pm\!\!$ &  0.50 & 0.456 &  81.9 &   3.37 & $\!\!\pm\!\!$ &  0.23 & $\!\!\pm\!\!$ &  0.36 \\
0.50--0.60 & 0.539 &  67.1 &   3.67 & $\!\!\pm\!\!$ &  0.22 & $\!\!\pm\!\!$ &  0.36 & 0.540 &  81.6 &   2.13 & $\!\!\pm\!\!$ &  0.17 & $\!\!\pm\!\!$ &  0.27 \\
0.60--0.72 & 0.646 &  66.6 &   2.23 & $\!\!\pm\!\!$ &  0.16 & $\!\!\pm\!\!$ &  0.27 & 0.653 &  80.8 &   1.05 & $\!\!\pm\!\!$ &  0.12 & $\!\!\pm\!\!$ &  0.17 \\
0.72--0.90 & 0.784 &  66.3 &   0.94 & $\!\!\pm\!\!$ &  0.09 & $\!\!\pm\!\!$ &  0.15 & 0.784 &  81.9 &   0.37 & $\!\!\pm\!\!$ &  0.05 & $\!\!\pm\!\!$ &  0.09 \\
0.90--1.25 & 1.011 &  65.9 &   0.15 & $\!\!\pm\!\!$ &  0.02 & $\!\!\pm\!\!$ &  0.05 & 1.015 &  82.7 &   0.03 & $\!\!\pm\!\!$ &  0.01 & $\!\!\pm\!\!$ &  0.02 \\
 \hline
 \hline
  & \multicolumn{7}{c||}{$90<\theta<105$}
  & \multicolumn{7}{c|}{$105<\theta<125$} \\
 \hline
 $p_{\rm T}$ & $\langle p_{\rm T} \rangle$ & $\langle \theta \rangle$
  & \multicolumn{5}{c||}{${\rm d}^2 \sigma /{\rm d}p{\rm d}\Omega$}
  &$\langle p_{\rm T} \rangle$ & $\langle \theta \rangle$
  & \multicolumn{5}{c|}{${\rm d}^2 \sigma /{\rm d}p{\rm d}\Omega$} \\
 \hline
0.13--0.16 & 0.145 &  96.7 &  15.55 & $\!\!\pm\!\!$ &  0.92 & $\!\!\pm\!\!$ &  1.48 & 0.145 & 114.5 &  12.24 & $\!\!\pm\!\!$ &  0.70 & $\!\!\pm\!\!$ &  1.11 \\
0.16--0.20 & 0.179 &  97.3 &  13.18 & $\!\!\pm\!\!$ &  0.70 & $\!\!\pm\!\!$ &  0.96 & 0.178 & 114.0 &   9.88 & $\!\!\pm\!\!$ &  0.52 & $\!\!\pm\!\!$ &  0.77 \\
0.20--0.24 & 0.217 &  97.2 &   9.96 & $\!\!\pm\!\!$ &  0.60 & $\!\!\pm\!\!$ &  0.80 & 0.218 & 114.1 &   7.30 & $\!\!\pm\!\!$ &  0.44 & $\!\!\pm\!\!$ &  0.70 \\
0.24--0.30 & 0.268 &  96.6 &   7.43 & $\!\!\pm\!\!$ &  0.41 & $\!\!\pm\!\!$ &  0.57 & 0.267 & 114.2 &   4.38 & $\!\!\pm\!\!$ &  0.28 & $\!\!\pm\!\!$ &  0.44 \\
0.30--0.36 & 0.328 &  96.2 &   4.13 & $\!\!\pm\!\!$ &  0.30 & $\!\!\pm\!\!$ &  0.42 & 0.326 & 113.5 &   2.46 & $\!\!\pm\!\!$ &  0.21 & $\!\!\pm\!\!$ &  0.30 \\
0.36--0.42 & 0.387 &  97.2 &   2.84 & $\!\!\pm\!\!$ &  0.25 & $\!\!\pm\!\!$ &  0.35 & 0.381 & 112.6 &   1.57 & $\!\!\pm\!\!$ &  0.17 & $\!\!\pm\!\!$ &  0.24 \\
0.42--0.50 & 0.456 &  96.9 &   2.18 & $\!\!\pm\!\!$ &  0.19 & $\!\!\pm\!\!$ &  0.30 & 0.453 & 113.1 &   0.87 & $\!\!\pm\!\!$ &  0.10 & $\!\!\pm\!\!$ &  0.18 \\
0.50--0.60 & 0.539 &  97.4 &   1.08 & $\!\!\pm\!\!$ &  0.12 & $\!\!\pm\!\!$ &  0.18 & 0.538 & 112.2 &   0.40 & $\!\!\pm\!\!$ &  0.06 & $\!\!\pm\!\!$ &  0.14 \\
0.60--0.72 & 0.653 &  94.8 &   0.37 & $\!\!\pm\!\!$ &  0.06 & $\!\!\pm\!\!$ &  0.10 & 0.658 & 111.9 &   0.06 & $\!\!\pm\!\!$ &  0.02 & $\!\!\pm\!\!$ &  0.04 \\
0.72--0.90 & 0.791 &  95.7 &   0.04 & $\!\!\pm\!\!$ &  0.02 & $\!\!\pm\!\!$ &  0.03 &  &  &  \multicolumn{5}{c|}{ } \\
 \hline
 \end{tabular}
 \end{center}
 \end{scriptsize}
 \end{table}

%% file: table.pro.pimbe5.tex
  
 \begin{table}[h]
 \begin{scriptsize}
 \caption{Double-differential inclusive
  cross-section ${\rm d}^2 \sigma /{\rm d}p{\rm d}\Omega$
  [mb/(GeV/{\it c} sr)] of the production of protons
  in $\pi^-$ + Be $\rightarrow$ P + X interactions
  with $-5.0$~GeV/{\it c} beam momentum;
  the first error is statistical, the second systematic; 
 $p_{\rm T}$ in GeV/{\it c}, polar angle $\theta$ in degrees.}
 \label{pro.pimbe5}
 \begin{center}
 \begin{tabular}{|c||c|c|rcrcr||c|c|rcrcr|}
 \hline
   & \multicolumn{7}{c||}{$20<\theta<30$}
  & \multicolumn{7}{c|}{$30<\theta<40$} \\
 \hline
 $p_{\rm T}$ & $\langle p_{\rm T} \rangle$ & $\langle \theta \rangle$
  & \multicolumn{5}{c||}{${\rm d}^2 \sigma /{\rm d}p{\rm d}\Omega$}
  &$\langle p_{\rm T} \rangle$ & $\langle \theta \rangle$
  & \multicolumn{5}{c|}{${\rm d}^2 \sigma /{\rm d}p{\rm d}\Omega$} \\
 \hline
0.20--0.24 & 0.217 &  25.0 &  33.01 & $\!\!\pm\!\!$ &  1.18 & $\!\!\pm\!\!$ &  2.52 &  &  &  \multicolumn{5}{c|}{ } \\
0.24--0.30 & 0.265 &  24.9 &  30.80 & $\!\!\pm\!\!$ &  0.92 & $\!\!\pm\!\!$ &  1.93 & 0.265 &  34.7 &  27.93 & $\!\!\pm\!\!$ &  0.85 & $\!\!\pm\!\!$ &  1.81 \\
0.30--0.36 & 0.323 &  25.1 &  28.41 & $\!\!\pm\!\!$ &  0.89 & $\!\!\pm\!\!$ &  1.65 & 0.323 &  34.9 &  27.26 & $\!\!\pm\!\!$ &  0.85 & $\!\!\pm\!\!$ &  1.50 \\
0.36--0.42 & 0.381 &  25.1 &  25.33 & $\!\!\pm\!\!$ &  0.83 & $\!\!\pm\!\!$ &  1.39 & 0.381 &  34.8 &  22.35 & $\!\!\pm\!\!$ &  0.78 & $\!\!\pm\!\!$ &  1.25 \\
0.42--0.50 & 0.445 &  24.9 &  24.57 & $\!\!\pm\!\!$ &  0.70 & $\!\!\pm\!\!$ &  1.20 & 0.447 &  34.9 &  19.39 & $\!\!\pm\!\!$ &  0.62 & $\!\!\pm\!\!$ &  1.01 \\
0.50--0.60 & 0.530 &  24.9 &  18.14 & $\!\!\pm\!\!$ &  0.53 & $\!\!\pm\!\!$ &  0.90 & 0.529 &  35.0 &  15.80 & $\!\!\pm\!\!$ &  0.51 & $\!\!\pm\!\!$ &  0.82 \\
0.60--0.72 & 0.630 &  25.0 &  14.01 & $\!\!\pm\!\!$ &  0.42 & $\!\!\pm\!\!$ &  0.80 & 0.633 &  34.9 &  12.53 & $\!\!\pm\!\!$ &  0.42 & $\!\!\pm\!\!$ &  0.73 \\
0.72--0.90 &  &  &  \multicolumn{5}{c||}{ } & 0.764 &  35.0 &   7.80 & $\!\!\pm\!\!$ &  0.27 & $\!\!\pm\!\!$ &  0.61 \\
 \hline
 \hline
   & \multicolumn{7}{c||}{$40<\theta<50$}
  & \multicolumn{7}{c|}{$50<\theta<60$} \\
 \hline
 $p_{\rm T}$ & $\langle p_{\rm T} \rangle$ & $\langle \theta \rangle$
  & \multicolumn{5}{c||}{${\rm d}^2 \sigma /{\rm d}p{\rm d}\Omega$}
  &$\langle p_{\rm T} \rangle$ & $\langle \theta \rangle$
  & \multicolumn{5}{c|}{${\rm d}^2 \sigma /{\rm d}p{\rm d}\Omega$} \\
 \hline
0.30--0.36 & 0.330 &  45.1 &  22.87 & $\!\!\pm\!\!$ &  0.75 & $\!\!\pm\!\!$ &  1.28 &  &  &  \multicolumn{5}{c|}{ } \\
0.36--0.42 & 0.388 &  45.2 &  22.59 & $\!\!\pm\!\!$ &  0.76 & $\!\!\pm\!\!$ &  1.28 & 0.389 &  54.9 &  21.50 & $\!\!\pm\!\!$ &  0.72 & $\!\!\pm\!\!$ &  1.13 \\
0.42--0.50 & 0.458 &  45.2 &  18.42 & $\!\!\pm\!\!$ &  0.61 & $\!\!\pm\!\!$ &  0.97 & 0.459 &  54.9 &  18.94 & $\!\!\pm\!\!$ &  0.61 & $\!\!\pm\!\!$ &  1.01 \\
0.50--0.60 & 0.548 &  44.9 &  14.52 & $\!\!\pm\!\!$ &  0.50 & $\!\!\pm\!\!$ &  0.79 & 0.548 &  55.0 &  12.04 & $\!\!\pm\!\!$ &  0.45 & $\!\!\pm\!\!$ &  0.72 \\
0.60--0.72 & 0.655 &  45.0 &  10.32 & $\!\!\pm\!\!$ &  0.39 & $\!\!\pm\!\!$ &  0.66 & 0.653 &  54.9 &   8.46 & $\!\!\pm\!\!$ &  0.36 & $\!\!\pm\!\!$ &  0.62 \\
0.72--0.90 & 0.800 &  45.0 &   5.60 & $\!\!\pm\!\!$ &  0.24 & $\!\!\pm\!\!$ &  0.47 & 0.798 &  54.8 &   4.78 & $\!\!\pm\!\!$ &  0.23 & $\!\!\pm\!\!$ &  0.44 \\
0.90--1.25 & 1.029 &  44.9 &   1.92 & $\!\!\pm\!\!$ &  0.10 & $\!\!\pm\!\!$ &  0.23 & 1.036 &  54.7 &   1.19 & $\!\!\pm\!\!$ &  0.08 & $\!\!\pm\!\!$ &  0.18 \\
 \hline
 \hline
   & \multicolumn{7}{c||}{$60<\theta<75$}
  & \multicolumn{7}{c|}{$75<\theta<90$} \\
 \hline
 $p_{\rm T}$ & $\langle p_{\rm T} \rangle$ & $\langle \theta \rangle$
  & \multicolumn{5}{c||}{${\rm d}^2 \sigma /{\rm d}p{\rm d}\Omega$}
  &$\langle p_{\rm T} \rangle$ & $\langle \theta \rangle$
  & \multicolumn{5}{c|}{${\rm d}^2 \sigma /{\rm d}p{\rm d}\Omega$} \\
 \hline
0.36--0.42 & 0.389 &  67.7 &  20.57 & $\!\!\pm\!\!$ &  0.85 & $\!\!\pm\!\!$ &  1.12 &  &  &  \multicolumn{5}{c|}{ } \\
0.42--0.50 & 0.458 &  67.5 &  17.79 & $\!\!\pm\!\!$ &  0.48 & $\!\!\pm\!\!$ &  0.74 & 0.458 &  81.9 &  13.29 & $\!\!\pm\!\!$ &  0.40 & $\!\!\pm\!\!$ &  0.74 \\
0.50--0.60 & 0.546 &  67.2 &  12.81 & $\!\!\pm\!\!$ &  0.38 & $\!\!\pm\!\!$ &  0.65 & 0.547 &  81.7 &   7.92 & $\!\!\pm\!\!$ &  0.28 & $\!\!\pm\!\!$ &  0.56 \\
0.60--0.72 & 0.654 &  67.0 &   7.26 & $\!\!\pm\!\!$ &  0.26 & $\!\!\pm\!\!$ &  0.55 & 0.654 &  81.8 &   3.20 & $\!\!\pm\!\!$ &  0.18 & $\!\!\pm\!\!$ &  0.38 \\
0.72--0.90 & 0.794 &  67.0 &   2.99 & $\!\!\pm\!\!$ &  0.15 & $\!\!\pm\!\!$ &  0.34 & 0.799 &  81.7 &   1.38 & $\!\!\pm\!\!$ &  0.11 & $\!\!\pm\!\!$ &  0.23 \\
0.90--1.25 & 1.035 &  66.7 &   0.71 & $\!\!\pm\!\!$ &  0.06 & $\!\!\pm\!\!$ &  0.14 & 1.033 &  81.3 &   0.25 & $\!\!\pm\!\!$ &  0.03 & $\!\!\pm\!\!$ &  0.08 \\
 \hline
 \hline
  & \multicolumn{7}{c||}{$90<\theta<105$}
  & \multicolumn{7}{c|}{$105<\theta<125$} \\
 \hline
 $p_{\rm T}$ & $\langle p_{\rm T} \rangle$ & $\langle \theta \rangle$
  & \multicolumn{5}{c||}{${\rm d}^2 \sigma /{\rm d}p{\rm d}\Omega$}
  &$\langle p_{\rm T} \rangle$ & $\langle \theta \rangle$
  & \multicolumn{5}{c|}{${\rm d}^2 \sigma /{\rm d}p{\rm d}\Omega$} \\
 \hline
0.36--0.42 &  &  &  \multicolumn{5}{c||}{ } & 0.388 & 113.8 &   5.82 & $\!\!\pm\!\!$ &  0.26 & $\!\!\pm\!\!$ &  0.47 \\
0.42--0.50 & 0.456 &  96.9 &   7.12 & $\!\!\pm\!\!$ &  0.29 & $\!\!\pm\!\!$ &  0.55 & 0.455 & 113.5 &   3.35 & $\!\!\pm\!\!$ &  0.17 & $\!\!\pm\!\!$ &  0.29 \\
0.50--0.60 & 0.547 &  96.1 &   3.69 & $\!\!\pm\!\!$ &  0.19 & $\!\!\pm\!\!$ &  0.38 & 0.545 & 112.7 &   1.35 & $\!\!\pm\!\!$ &  0.11 & $\!\!\pm\!\!$ &  0.19 \\
0.60--0.72 & 0.652 &  95.8 &   1.24 & $\!\!\pm\!\!$ &  0.12 & $\!\!\pm\!\!$ &  0.21 & 0.642 & 111.6 &   0.20 & $\!\!\pm\!\!$ &  0.04 & $\!\!\pm\!\!$ &  0.08 \\
0.72--0.90 & 0.790 &  95.5 &   0.39 & $\!\!\pm\!\!$ &  0.06 & $\!\!\pm\!\!$ &  0.12 & 0.790 & 113.9 &   0.07 & $\!\!\pm\!\!$ &  0.02 & $\!\!\pm\!\!$ &  0.06 \\
0.90--1.25 & 1.027 &  94.2 &   0.07 & $\!\!\pm\!\!$ &  0.02 & $\!\!\pm\!\!$ &  0.06 &  &  &  \multicolumn{5}{c|}{ } \\
 \hline
 \end{tabular}
 \end{center}
 \end{scriptsize}
 \end{table}

%% file: table.pip.pimbe5.tex
  
 \begin{table}[h]
 \begin{scriptsize}
 \caption{Double-differential inclusive
  cross-section ${\rm d}^2 \sigma /{\rm d}p{\rm d}\Omega$
  [mb/(GeV/{\it c} sr)] of the production of $\pi^+$'s
  in $\pi^-$ + Be $\rightarrow$ $\pi^+$ + X interactions
  with $-5.0$~GeV/{\it c} beam momentum;
  the first error is statistical, the second systematic; 
 $p_{\rm T}$ in GeV/{\it c}, polar angle $\theta$ in degrees.}
 \label{pip.pimbe5}
 \begin{center}
 \begin{tabular}{|c||c|c|rcrcr||c|c|rcrcr|}
 \hline
   & \multicolumn{7}{c||}{$20<\theta<30$}
  & \multicolumn{7}{c|}{$30<\theta<40$} \\
 \hline
 $p_{\rm T}$ & $\langle p_{\rm T} \rangle$ & $\langle \theta \rangle$
  & \multicolumn{5}{c||}{${\rm d}^2 \sigma /{\rm d}p{\rm d}\Omega$}
  &$\langle p_{\rm T} \rangle$ & $\langle \theta \rangle$
  & \multicolumn{5}{c|}{${\rm d}^2 \sigma /{\rm d}p{\rm d}\Omega$} \\
 \hline
0.10--0.13 & 0.115 &  24.9 &  33.00 & $\!\!\pm\!\!$ &  1.51 & $\!\!\pm\!\!$ &  3.18 & 0.115 &  34.7 &  27.01 & $\!\!\pm\!\!$ &  1.43 & $\!\!\pm\!\!$ &  2.80 \\
0.13--0.16 & 0.145 &  24.8 &  43.98 & $\!\!\pm\!\!$ &  1.64 & $\!\!\pm\!\!$ &  3.42 & 0.143 &  34.9 &  32.52 & $\!\!\pm\!\!$ &  1.40 & $\!\!\pm\!\!$ &  2.65 \\
0.16--0.20 & 0.179 &  24.5 &  50.11 & $\!\!\pm\!\!$ &  1.43 & $\!\!\pm\!\!$ &  3.04 & 0.178 &  34.6 &  36.54 & $\!\!\pm\!\!$ &  1.26 & $\!\!\pm\!\!$ &  2.44 \\
0.20--0.24 & 0.217 &  24.7 &  54.94 & $\!\!\pm\!\!$ &  1.48 & $\!\!\pm\!\!$ &  3.04 & 0.217 &  34.6 &  36.76 & $\!\!\pm\!\!$ &  1.19 & $\!\!\pm\!\!$ &  2.15 \\
0.24--0.30 & 0.265 &  24.7 &  52.93 & $\!\!\pm\!\!$ &  1.20 & $\!\!\pm\!\!$ &  2.40 & 0.265 &  34.6 &  36.68 & $\!\!\pm\!\!$ &  0.98 & $\!\!\pm\!\!$ &  1.79 \\
0.30--0.36 & 0.323 &  24.6 &  44.13 & $\!\!\pm\!\!$ &  1.07 & $\!\!\pm\!\!$ &  1.87 & 0.323 &  34.6 &  30.76 & $\!\!\pm\!\!$ &  0.89 & $\!\!\pm\!\!$ &  1.42 \\
0.36--0.42 & 0.380 &  24.7 &  37.58 & $\!\!\pm\!\!$ &  0.99 & $\!\!\pm\!\!$ &  1.57 & 0.380 &  34.6 &  24.16 & $\!\!\pm\!\!$ &  0.78 & $\!\!\pm\!\!$ &  1.20 \\
0.42--0.50 & 0.445 &  24.7 &  28.54 & $\!\!\pm\!\!$ &  0.76 & $\!\!\pm\!\!$ &  1.39 & 0.446 &  34.8 &  23.02 & $\!\!\pm\!\!$ &  0.71 & $\!\!\pm\!\!$ &  1.24 \\
0.50--0.60 & 0.530 &  24.7 &  16.58 & $\!\!\pm\!\!$ &  0.49 & $\!\!\pm\!\!$ &  1.09 & 0.529 &  34.8 &  13.34 & $\!\!\pm\!\!$ &  0.45 & $\!\!\pm\!\!$ &  0.93 \\
0.60--0.72 & 0.632 &  24.9 &   9.14 & $\!\!\pm\!\!$ &  0.31 & $\!\!\pm\!\!$ &  0.85 & 0.629 &  34.8 &   7.58 & $\!\!\pm\!\!$ &  0.29 & $\!\!\pm\!\!$ &  0.74 \\
0.72--0.90 &  &  &  \multicolumn{5}{c||}{ } & 0.763 &  34.8 &   3.37 & $\!\!\pm\!\!$ &  0.16 & $\!\!\pm\!\!$ &  0.46 \\
 \hline
 \hline
   & \multicolumn{7}{c||}{$40<\theta<50$}
  & \multicolumn{7}{c|}{$50<\theta<60$} \\
 \hline
 $p_{\rm T}$ & $\langle p_{\rm T} \rangle$ & $\langle \theta \rangle$
  & \multicolumn{5}{c||}{${\rm d}^2 \sigma /{\rm d}p{\rm d}\Omega$}
  &$\langle p_{\rm T} \rangle$ & $\langle \theta \rangle$
  & \multicolumn{5}{c|}{${\rm d}^2 \sigma /{\rm d}p{\rm d}\Omega$} \\
 \hline
0.10--0.13 & 0.117 &  45.1 &  23.14 & $\!\!\pm\!\!$ &  1.34 & $\!\!\pm\!\!$ &  2.64 &  &  &  \multicolumn{5}{c|}{ } \\
0.13--0.16 & 0.145 &  44.8 &  25.22 & $\!\!\pm\!\!$ &  1.27 & $\!\!\pm\!\!$ &  2.17 & 0.146 &  54.3 &  20.21 & $\!\!\pm\!\!$ &  1.12 & $\!\!\pm\!\!$ &  1.88 \\
0.16--0.20 & 0.180 &  44.7 &  27.78 & $\!\!\pm\!\!$ &  1.09 & $\!\!\pm\!\!$ &  1.98 & 0.179 &  54.8 &  23.74 & $\!\!\pm\!\!$ &  1.04 & $\!\!\pm\!\!$ &  1.85 \\
0.20--0.24 & 0.219 &  44.8 &  27.37 & $\!\!\pm\!\!$ &  1.05 & $\!\!\pm\!\!$ &  1.97 & 0.219 &  54.6 &  22.27 & $\!\!\pm\!\!$ &  0.97 & $\!\!\pm\!\!$ &  1.63 \\
0.24--0.30 & 0.270 &  44.7 &  25.38 & $\!\!\pm\!\!$ &  0.83 & $\!\!\pm\!\!$ &  1.40 & 0.270 &  54.7 &  19.38 & $\!\!\pm\!\!$ &  0.72 & $\!\!\pm\!\!$ &  1.14 \\
0.30--0.36 & 0.329 &  44.7 &  22.78 & $\!\!\pm\!\!$ &  0.77 & $\!\!\pm\!\!$ &  1.16 & 0.329 &  54.7 &  17.75 & $\!\!\pm\!\!$ &  0.70 & $\!\!\pm\!\!$ &  0.94 \\
0.36--0.42 & 0.388 &  44.9 &  19.22 & $\!\!\pm\!\!$ &  0.73 & $\!\!\pm\!\!$ &  0.98 & 0.390 &  54.7 &  13.10 & $\!\!\pm\!\!$ &  0.61 & $\!\!\pm\!\!$ &  0.74 \\
0.42--0.50 & 0.457 &  44.8 &  14.34 & $\!\!\pm\!\!$ &  0.55 & $\!\!\pm\!\!$ &  0.75 & 0.458 &  54.9 &  10.81 & $\!\!\pm\!\!$ &  0.47 & $\!\!\pm\!\!$ &  0.68 \\
0.50--0.60 & 0.544 &  44.9 &   8.70 & $\!\!\pm\!\!$ &  0.37 & $\!\!\pm\!\!$ &  0.59 & 0.546 &  54.6 &   6.19 & $\!\!\pm\!\!$ &  0.31 & $\!\!\pm\!\!$ &  0.49 \\
0.60--0.72 & 0.652 &  44.6 &   4.70 & $\!\!\pm\!\!$ &  0.24 & $\!\!\pm\!\!$ &  0.44 & 0.655 &  55.0 &   3.56 & $\!\!\pm\!\!$ &  0.22 & $\!\!\pm\!\!$ &  0.38 \\
0.72--0.90 & 0.799 &  44.3 &   2.36 & $\!\!\pm\!\!$ &  0.13 & $\!\!\pm\!\!$ &  0.31 & 0.797 &  54.4 &   1.34 & $\!\!\pm\!\!$ &  0.10 & $\!\!\pm\!\!$ &  0.18 \\
0.90--1.25 &  &  &  \multicolumn{5}{c||}{ } & 1.022 &  54.5 &   0.39 & $\!\!\pm\!\!$ &  0.03 & $\!\!\pm\!\!$ &  0.09 \\
 \hline
 \hline
   & \multicolumn{7}{c||}{$60<\theta<75$}
  & \multicolumn{7}{c|}{$75<\theta<90$} \\
 \hline
 $p_{\rm T}$ & $\langle p_{\rm T} \rangle$ & $\langle \theta \rangle$
  & \multicolumn{5}{c||}{${\rm d}^2 \sigma /{\rm d}p{\rm d}\Omega$}
  &$\langle p_{\rm T} \rangle$ & $\langle \theta \rangle$
  & \multicolumn{5}{c|}{${\rm d}^2 \sigma /{\rm d}p{\rm d}\Omega$} \\
 \hline
0.13--0.16 & 0.145 &  67.4 &  16.41 & $\!\!\pm\!\!$ &  0.84 & $\!\!\pm\!\!$ &  1.43 & 0.145 &  82.6 &  14.39 & $\!\!\pm\!\!$ &  0.78 & $\!\!\pm\!\!$ &  1.41 \\
0.16--0.20 & 0.180 &  67.4 &  18.12 & $\!\!\pm\!\!$ &  0.72 & $\!\!\pm\!\!$ &  1.30 & 0.181 &  82.1 &  14.48 & $\!\!\pm\!\!$ &  0.65 & $\!\!\pm\!\!$ &  1.09 \\
0.20--0.24 & 0.220 &  67.0 &  15.53 & $\!\!\pm\!\!$ &  0.65 & $\!\!\pm\!\!$ &  1.08 & 0.220 &  82.6 &  12.97 & $\!\!\pm\!\!$ &  0.61 & $\!\!\pm\!\!$ &  0.92 \\
0.24--0.30 & 0.269 &  67.0 &  14.54 & $\!\!\pm\!\!$ &  0.53 & $\!\!\pm\!\!$ &  0.80 & 0.269 &  82.2 &  10.19 & $\!\!\pm\!\!$ &  0.45 & $\!\!\pm\!\!$ &  0.66 \\
0.30--0.36 & 0.329 &  66.9 &  10.50 & $\!\!\pm\!\!$ &  0.45 & $\!\!\pm\!\!$ &  0.58 & 0.329 &  82.0 &   6.22 & $\!\!\pm\!\!$ &  0.34 & $\!\!\pm\!\!$ &  0.45 \\
0.36--0.42 & 0.388 &  66.8 &   8.80 & $\!\!\pm\!\!$ &  0.40 & $\!\!\pm\!\!$ &  0.55 & 0.388 &  82.2 &   4.95 & $\!\!\pm\!\!$ &  0.30 & $\!\!\pm\!\!$ &  0.42 \\
0.42--0.50 & 0.458 &  67.0 &   6.64 & $\!\!\pm\!\!$ &  0.30 & $\!\!\pm\!\!$ &  0.46 & 0.457 &  82.1 &   3.35 & $\!\!\pm\!\!$ &  0.21 & $\!\!\pm\!\!$ &  0.31 \\
0.50--0.60 & 0.544 &  66.6 &   4.12 & $\!\!\pm\!\!$ &  0.21 & $\!\!\pm\!\!$ &  0.36 & 0.545 &  81.5 &   2.13 & $\!\!\pm\!\!$ &  0.15 & $\!\!\pm\!\!$ &  0.25 \\
0.60--0.72 & 0.652 &  66.8 &   1.94 & $\!\!\pm\!\!$ &  0.13 & $\!\!\pm\!\!$ &  0.23 & 0.652 &  81.6 &   1.12 & $\!\!\pm\!\!$ &  0.10 & $\!\!\pm\!\!$ &  0.17 \\
0.72--0.90 & 0.786 &  66.5 &   0.71 & $\!\!\pm\!\!$ &  0.06 & $\!\!\pm\!\!$ &  0.11 & 0.792 &  80.7 &   0.20 & $\!\!\pm\!\!$ &  0.03 & $\!\!\pm\!\!$ &  0.05 \\
0.90--1.25 & 1.026 &  65.7 &   0.12 & $\!\!\pm\!\!$ &  0.02 & $\!\!\pm\!\!$ &  0.04 & 1.000 &  79.8 &   0.04 & $\!\!\pm\!\!$ &  0.01 & $\!\!\pm\!\!$ &  0.03 \\
 \hline
 \hline
  & \multicolumn{7}{c||}{$90<\theta<105$}
  & \multicolumn{7}{c|}{$105<\theta<125$} \\
 \hline
 $p_{\rm T}$ & $\langle p_{\rm T} \rangle$ & $\langle \theta \rangle$
  & \multicolumn{5}{c||}{${\rm d}^2 \sigma /{\rm d}p{\rm d}\Omega$}
  &$\langle p_{\rm T} \rangle$ & $\langle \theta \rangle$
  & \multicolumn{5}{c|}{${\rm d}^2 \sigma /{\rm d}p{\rm d}\Omega$} \\
 \hline
0.13--0.16 & 0.145 &  97.9 &  11.04 & $\!\!\pm\!\!$ &  0.69 & $\!\!\pm\!\!$ &  1.04 & 0.145 & 114.8 &  11.76 & $\!\!\pm\!\!$ &  0.62 & $\!\!\pm\!\!$ &  1.01 \\
0.16--0.20 & 0.179 &  97.6 &  12.33 & $\!\!\pm\!\!$ &  0.62 & $\!\!\pm\!\!$ &  0.92 & 0.179 & 114.7 &   8.52 & $\!\!\pm\!\!$ &  0.43 & $\!\!\pm\!\!$ &  0.61 \\
0.20--0.24 & 0.218 &  97.3 &  11.46 & $\!\!\pm\!\!$ &  0.58 & $\!\!\pm\!\!$ &  0.88 & 0.218 & 114.5 &   6.71 & $\!\!\pm\!\!$ &  0.38 & $\!\!\pm\!\!$ &  0.58 \\
0.24--0.30 & 0.269 &  97.0 &   5.59 & $\!\!\pm\!\!$ &  0.32 & $\!\!\pm\!\!$ &  0.45 & 0.269 & 113.9 &   4.11 & $\!\!\pm\!\!$ &  0.25 & $\!\!\pm\!\!$ &  0.34 \\
0.30--0.36 & 0.329 &  97.1 &   3.98 & $\!\!\pm\!\!$ &  0.28 & $\!\!\pm\!\!$ &  0.37 & 0.328 & 113.6 &   2.55 & $\!\!\pm\!\!$ &  0.19 & $\!\!\pm\!\!$ &  0.30 \\
0.36--0.42 & 0.387 &  96.4 &   2.91 & $\!\!\pm\!\!$ &  0.22 & $\!\!\pm\!\!$ &  0.32 & 0.387 & 113.2 &   1.51 & $\!\!\pm\!\!$ &  0.15 & $\!\!\pm\!\!$ &  0.23 \\
0.42--0.50 & 0.456 &  96.5 &   2.12 & $\!\!\pm\!\!$ &  0.18 & $\!\!\pm\!\!$ &  0.28 & 0.453 & 113.7 &   0.50 & $\!\!\pm\!\!$ &  0.07 & $\!\!\pm\!\!$ &  0.09 \\
0.50--0.60 & 0.547 &  96.5 &   0.95 & $\!\!\pm\!\!$ &  0.10 & $\!\!\pm\!\!$ &  0.17 & 0.528 & 112.1 &   0.18 & $\!\!\pm\!\!$ &  0.04 & $\!\!\pm\!\!$ &  0.07 \\
0.60--0.72 & 0.650 &  96.6 &   0.33 & $\!\!\pm\!\!$ &  0.05 & $\!\!\pm\!\!$ &  0.11 & 0.646 & 109.7 &   0.04 & $\!\!\pm\!\!$ &  0.02 & $\!\!\pm\!\!$ &  0.02 \\
0.72--0.90 & 0.775 &  95.5 &   0.07 & $\!\!\pm\!\!$ &  0.02 & $\!\!\pm\!\!$ &  0.03 &  &  &  \multicolumn{5}{c|}{ } \\
 \hline
 \end{tabular}
 \end{center}
 \end{scriptsize}
 \end{table}

%% file: table.pim.pimbe5.tex
  
 \begin{table}[h]
 \begin{scriptsize}
 \caption{Double-differential inclusive
  cross-section ${\rm d}^2 \sigma /{\rm d}p{\rm d}\Omega$
  [mb/(GeV/{\it c} sr)] of the production of $\pi^-$'s
  in $\pi^-$ + Be $\rightarrow$ $\pi^-$ + X interactions
  with $-5.0$~GeV/{\it c} beam momentum;
  the first error is statistical, the second systematic; 
 $p_{\rm T}$ in GeV/{\it c}, polar angle $\theta$ in degrees.}
 \label{pim.pimbe5}
 \begin{center}
 \begin{tabular}{|c||c|c|rcrcr||c|c|rcrcr|}
 \hline
   & \multicolumn{7}{c||}{$20<\theta<30$}
  & \multicolumn{7}{c|}{$30<\theta<40$} \\
 \hline
 $p_{\rm T}$ & $\langle p_{\rm T} \rangle$ & $\langle \theta \rangle$
  & \multicolumn{5}{c||}{${\rm d}^2 \sigma /{\rm d}p{\rm d}\Omega$}
  &$\langle p_{\rm T} \rangle$ & $\langle \theta \rangle$
  & \multicolumn{5}{c|}{${\rm d}^2 \sigma /{\rm d}p{\rm d}\Omega$} \\
 \hline
0.10--0.13 & 0.117 &  24.8 &  62.37 & $\!\!\pm\!\!$ &  2.16 & $\!\!\pm\!\!$ &  5.74 & 0.117 &  34.8 &  48.21 & $\!\!\pm\!\!$ &  1.91 & $\!\!\pm\!\!$ &  4.36 \\
0.13--0.16 & 0.147 &  24.8 &  76.32 & $\!\!\pm\!\!$ &  2.20 & $\!\!\pm\!\!$ &  5.52 & 0.146 &  34.9 &  61.93 & $\!\!\pm\!\!$ &  1.99 & $\!\!\pm\!\!$ &  4.39 \\
0.16--0.20 & 0.182 &  24.7 &  83.55 & $\!\!\pm\!\!$ &  1.87 & $\!\!\pm\!\!$ &  4.73 & 0.182 &  34.9 &  62.67 & $\!\!\pm\!\!$ &  1.65 & $\!\!\pm\!\!$ &  3.65 \\
0.20--0.24 & 0.223 &  24.8 &  85.60 & $\!\!\pm\!\!$ &  1.86 & $\!\!\pm\!\!$ &  4.23 & 0.223 &  34.8 &  65.86 & $\!\!\pm\!\!$ &  1.66 & $\!\!\pm\!\!$ &  3.39 \\
0.24--0.30 & 0.274 &  24.8 &  84.03 & $\!\!\pm\!\!$ &  1.51 & $\!\!\pm\!\!$ &  3.48 & 0.274 &  34.8 &  61.12 & $\!\!\pm\!\!$ &  1.25 & $\!\!\pm\!\!$ &  2.60 \\
0.30--0.36 & 0.335 &  24.7 &  76.28 & $\!\!\pm\!\!$ &  1.44 & $\!\!\pm\!\!$ &  2.75 & 0.336 &  34.7 &  56.80 & $\!\!\pm\!\!$ &  1.22 & $\!\!\pm\!\!$ &  2.27 \\
0.36--0.42 & 0.398 &  24.8 &  62.57 & $\!\!\pm\!\!$ &  1.32 & $\!\!\pm\!\!$ &  2.34 & 0.400 &  34.8 &  46.82 & $\!\!\pm\!\!$ &  1.12 & $\!\!\pm\!\!$ &  1.91 \\
0.42--0.50 & 0.471 &  24.8 &  50.57 & $\!\!\pm\!\!$ &  1.04 & $\!\!\pm\!\!$ &  2.31 & 0.472 &  34.8 &  37.67 & $\!\!\pm\!\!$ &  0.88 & $\!\!\pm\!\!$ &  1.73 \\
0.50--0.60 & 0.563 &  24.7 &  32.88 & $\!\!\pm\!\!$ &  0.72 & $\!\!\pm\!\!$ &  1.87 & 0.564 &  34.7 &  23.45 & $\!\!\pm\!\!$ &  0.61 & $\!\!\pm\!\!$ &  1.39 \\
0.60--0.72 & 0.682 &  24.8 &  20.81 & $\!\!\pm\!\!$ &  0.52 & $\!\!\pm\!\!$ &  1.62 & 0.681 &  34.6 &  13.46 & $\!\!\pm\!\!$ &  0.41 & $\!\!\pm\!\!$ &  1.09 \\
0.72--0.90 &  &  &  \multicolumn{5}{c||}{ } & 0.833 &  34.4 &   6.88 & $\!\!\pm\!\!$ &  0.25 & $\!\!\pm\!\!$ &  0.73 \\
 \hline
 \hline
   & \multicolumn{7}{c||}{$40<\theta<50$}
  & \multicolumn{7}{c|}{$50<\theta<60$} \\
 \hline
 $p_{\rm T}$ & $\langle p_{\rm T} \rangle$ & $\langle \theta \rangle$
  & \multicolumn{5}{c||}{${\rm d}^2 \sigma /{\rm d}p{\rm d}\Omega$}
  &$\langle p_{\rm T} \rangle$ & $\langle \theta \rangle$
  & \multicolumn{5}{c|}{${\rm d}^2 \sigma /{\rm d}p{\rm d}\Omega$} \\
 \hline
0.10--0.13 & 0.115 &  44.9 &  40.66 & $\!\!\pm\!\!$ &  1.85 & $\!\!\pm\!\!$ &  3.91 &  &  &  \multicolumn{5}{c|}{ } \\
0.13--0.16 & 0.145 &  45.0 &  46.69 & $\!\!\pm\!\!$ &  1.72 & $\!\!\pm\!\!$ &  3.41 & 0.146 &  55.0 &  39.53 & $\!\!\pm\!\!$ &  1.61 & $\!\!\pm\!\!$ &  3.02 \\
0.16--0.20 & 0.180 &  44.9 &  48.40 & $\!\!\pm\!\!$ &  1.47 & $\!\!\pm\!\!$ &  2.96 & 0.180 &  54.7 &  39.50 & $\!\!\pm\!\!$ &  1.32 & $\!\!\pm\!\!$ &  2.52 \\
0.20--0.24 & 0.220 &  44.8 &  51.14 & $\!\!\pm\!\!$ &  1.49 & $\!\!\pm\!\!$ &  2.77 & 0.220 &  54.8 &  39.04 & $\!\!\pm\!\!$ &  1.29 & $\!\!\pm\!\!$ &  2.29 \\
0.24--0.30 & 0.270 &  44.7 &  46.15 & $\!\!\pm\!\!$ &  1.14 & $\!\!\pm\!\!$ &  2.03 & 0.269 &  54.7 &  36.30 & $\!\!\pm\!\!$ &  1.01 & $\!\!\pm\!\!$ &  1.76 \\
0.30--0.36 & 0.329 &  44.8 &  41.24 & $\!\!\pm\!\!$ &  1.06 & $\!\!\pm\!\!$ &  1.68 & 0.330 &  54.7 &  28.89 & $\!\!\pm\!\!$ &  0.89 & $\!\!\pm\!\!$ &  1.25 \\
0.36--0.42 & 0.388 &  44.7 &  31.58 & $\!\!\pm\!\!$ &  0.92 & $\!\!\pm\!\!$ &  1.32 & 0.389 &  54.8 &  24.62 & $\!\!\pm\!\!$ &  0.84 & $\!\!\pm\!\!$ &  1.18 \\
0.42--0.50 & 0.457 &  44.7 &  26.31 & $\!\!\pm\!\!$ &  0.74 & $\!\!\pm\!\!$ &  1.28 & 0.458 &  54.7 &  16.87 & $\!\!\pm\!\!$ &  0.58 & $\!\!\pm\!\!$ &  0.98 \\
0.50--0.60 & 0.546 &  44.7 &  17.44 & $\!\!\pm\!\!$ &  0.53 & $\!\!\pm\!\!$ &  1.12 & 0.546 &  54.6 &  10.96 & $\!\!\pm\!\!$ &  0.41 & $\!\!\pm\!\!$ &  0.81 \\
0.60--0.72 & 0.652 &  44.7 &   9.12 & $\!\!\pm\!\!$ &  0.34 & $\!\!\pm\!\!$ &  0.80 & 0.652 &  54.7 &   6.94 & $\!\!\pm\!\!$ &  0.31 & $\!\!\pm\!\!$ &  0.66 \\
0.72--0.90 & 0.798 &  44.5 &   4.34 & $\!\!\pm\!\!$ &  0.20 & $\!\!\pm\!\!$ &  0.50 & 0.790 &  54.6 &   2.73 & $\!\!\pm\!\!$ &  0.16 & $\!\!\pm\!\!$ &  0.33 \\
0.90--1.25 &  &  &  \multicolumn{5}{c||}{ } & 1.015 &  54.7 &   0.67 & $\!\!\pm\!\!$ &  0.06 & $\!\!\pm\!\!$ &  0.12 \\
 \hline
 \hline
   & \multicolumn{7}{c||}{$60<\theta<75$}
  & \multicolumn{7}{c|}{$75<\theta<90$} \\
 \hline
 $p_{\rm T}$ & $\langle p_{\rm T} \rangle$ & $\langle \theta \rangle$
  & \multicolumn{5}{c||}{${\rm d}^2 \sigma /{\rm d}p{\rm d}\Omega$}
  &$\langle p_{\rm T} \rangle$ & $\langle \theta \rangle$
  & \multicolumn{5}{c|}{${\rm d}^2 \sigma /{\rm d}p{\rm d}\Omega$} \\
 \hline
0.13--0.16 & 0.146 &  67.2 &  30.13 & $\!\!\pm\!\!$ &  1.14 & $\!\!\pm\!\!$ &  2.19 & 0.145 &  82.5 &  26.15 & $\!\!\pm\!\!$ &  1.05 & $\!\!\pm\!\!$ &  2.03 \\
0.16--0.20 & 0.180 &  67.4 &  32.26 & $\!\!\pm\!\!$ &  0.98 & $\!\!\pm\!\!$ &  1.86 & 0.179 &  82.4 &  28.40 & $\!\!\pm\!\!$ &  0.91 & $\!\!\pm\!\!$ &  1.66 \\
0.20--0.24 & 0.220 &  67.1 &  30.28 & $\!\!\pm\!\!$ &  0.95 & $\!\!\pm\!\!$ &  1.67 & 0.220 &  82.2 &  24.92 & $\!\!\pm\!\!$ &  0.86 & $\!\!\pm\!\!$ &  1.34 \\
0.24--0.30 & 0.268 &  67.0 &  23.68 & $\!\!\pm\!\!$ &  0.67 & $\!\!\pm\!\!$ &  1.02 & 0.269 &  82.1 &  16.88 & $\!\!\pm\!\!$ &  0.57 & $\!\!\pm\!\!$ &  0.81 \\
0.30--0.36 & 0.329 &  66.8 &  19.13 & $\!\!\pm\!\!$ &  0.60 & $\!\!\pm\!\!$ &  0.84 & 0.328 &  81.8 &  12.14 & $\!\!\pm\!\!$ &  0.48 & $\!\!\pm\!\!$ &  0.70 \\
0.36--0.42 & 0.389 &  67.1 &  15.73 & $\!\!\pm\!\!$ &  0.54 & $\!\!\pm\!\!$ &  0.84 & 0.389 &  81.7 &   8.58 & $\!\!\pm\!\!$ &  0.39 & $\!\!\pm\!\!$ &  0.64 \\
0.42--0.50 & 0.459 &  66.9 &  10.41 & $\!\!\pm\!\!$ &  0.37 & $\!\!\pm\!\!$ &  0.67 & 0.458 &  81.6 &   6.55 & $\!\!\pm\!\!$ &  0.30 & $\!\!\pm\!\!$ &  0.56 \\
0.50--0.60 & 0.544 &  66.5 &   7.04 & $\!\!\pm\!\!$ &  0.27 & $\!\!\pm\!\!$ &  0.58 & 0.547 &  81.5 &   3.33 & $\!\!\pm\!\!$ &  0.19 & $\!\!\pm\!\!$ &  0.35 \\
0.60--0.72 & 0.652 &  66.8 &   3.77 & $\!\!\pm\!\!$ &  0.19 & $\!\!\pm\!\!$ &  0.40 & 0.650 &  80.8 &   1.82 & $\!\!\pm\!\!$ &  0.13 & $\!\!\pm\!\!$ &  0.25 \\
0.72--0.90 & 0.790 &  66.4 &   1.60 & $\!\!\pm\!\!$ &  0.10 & $\!\!\pm\!\!$ &  0.22 & 0.787 &  81.2 &   0.39 & $\!\!\pm\!\!$ &  0.05 & $\!\!\pm\!\!$ &  0.08 \\
0.90--1.25 & 1.032 &  67.0 &   0.18 & $\!\!\pm\!\!$ &  0.02 & $\!\!\pm\!\!$ &  0.04 & 0.992 &  79.4 &   0.03 & $\!\!\pm\!\!$ &  0.01 & $\!\!\pm\!\!$ &  0.02 \\
 \hline
 \hline
  & \multicolumn{7}{c||}{$90<\theta<105$}
  & \multicolumn{7}{c|}{$105<\theta<125$} \\
 \hline
 $p_{\rm T}$ & $\langle p_{\rm T} \rangle$ & $\langle \theta \rangle$
  & \multicolumn{5}{c||}{${\rm d}^2 \sigma /{\rm d}p{\rm d}\Omega$}
  &$\langle p_{\rm T} \rangle$ & $\langle \theta \rangle$
  & \multicolumn{5}{c|}{${\rm d}^2 \sigma /{\rm d}p{\rm d}\Omega$} \\
 \hline
0.13--0.16 & 0.146 &  97.0 &  24.14 & $\!\!\pm\!\!$ &  1.05 & $\!\!\pm\!\!$ &  1.88 & 0.145 & 114.3 &  20.22 & $\!\!\pm\!\!$ &  0.81 & $\!\!\pm\!\!$ &  1.33 \\
0.16--0.20 & 0.179 &  97.3 &  24.91 & $\!\!\pm\!\!$ &  0.89 & $\!\!\pm\!\!$ &  1.44 & 0.179 & 114.3 &  17.32 & $\!\!\pm\!\!$ &  0.63 & $\!\!\pm\!\!$ &  0.95 \\
0.20--0.24 & 0.219 &  97.3 &  17.32 & $\!\!\pm\!\!$ &  0.71 & $\!\!\pm\!\!$ &  1.00 & 0.218 & 114.3 &  11.14 & $\!\!\pm\!\!$ &  0.49 & $\!\!\pm\!\!$ &  0.75 \\
0.24--0.30 & 0.267 &  97.2 &  12.43 & $\!\!\pm\!\!$ &  0.48 & $\!\!\pm\!\!$ &  0.77 & 0.267 & 114.0 &   6.86 & $\!\!\pm\!\!$ &  0.31 & $\!\!\pm\!\!$ &  0.51 \\
0.30--0.36 & 0.328 &  96.9 &   7.52 & $\!\!\pm\!\!$ &  0.37 & $\!\!\pm\!\!$ &  0.58 & 0.329 & 113.6 &   3.70 & $\!\!\pm\!\!$ &  0.23 & $\!\!\pm\!\!$ &  0.36 \\
0.36--0.42 & 0.388 &  96.3 &   4.89 & $\!\!\pm\!\!$ &  0.30 & $\!\!\pm\!\!$ &  0.46 & 0.385 & 114.0 &   2.56 & $\!\!\pm\!\!$ &  0.19 & $\!\!\pm\!\!$ &  0.31 \\
0.42--0.50 & 0.459 &  96.8 &   3.47 & $\!\!\pm\!\!$ &  0.22 & $\!\!\pm\!\!$ &  0.40 & 0.458 & 113.0 &   1.39 & $\!\!\pm\!\!$ &  0.12 & $\!\!\pm\!\!$ &  0.22 \\
0.50--0.60 & 0.545 &  96.6 &   1.74 & $\!\!\pm\!\!$ &  0.14 & $\!\!\pm\!\!$ &  0.25 & 0.535 & 111.3 &   0.56 & $\!\!\pm\!\!$ &  0.06 & $\!\!\pm\!\!$ &  0.14 \\
0.60--0.72 & 0.640 &  96.4 &   0.56 & $\!\!\pm\!\!$ &  0.07 & $\!\!\pm\!\!$ &  0.12 & 0.653 & 110.5 &   0.18 & $\!\!\pm\!\!$ &  0.03 & $\!\!\pm\!\!$ &  0.08 \\
0.72--0.90 & 0.789 &  97.1 &   0.15 & $\!\!\pm\!\!$ &  0.03 & $\!\!\pm\!\!$ &  0.06 &  &  &  \multicolumn{5}{c|}{ } \\
 \hline
 \end{tabular}
 \end{center}
 \end{scriptsize}
 \end{table}

%% file: table.pro.probe12.tex
  
 \begin{table}[h]
 \begin{scriptsize}
 \caption{Double-differential inclusive
  cross-section ${\rm d}^2 \sigma /{\rm d}p{\rm d}\Omega$
  [mb/(GeV/{\it c} sr)] of the production of protons
  in p + Be $\rightarrow$ p + X interactions
  with $+12.0$~GeV/{\it c} beam momentum;
  the first error is statistical, the second systematic; 
 $p_{\rm T}$ in GeV/{\it c}, polar angle $\theta$ in degrees.}
 \label{pro.probe12}
 \begin{center}
 \begin{tabular}{|c||c|c|rcrcr||c|c|rcrcr|}
 \hline
   & \multicolumn{7}{c||}{$20<\theta<30$}
  & \multicolumn{7}{c|}{$30<\theta<40$} \\
 \hline
 $p_{\rm T}$ & $\langle p_{\rm T} \rangle$ & $\langle \theta \rangle$
  & \multicolumn{5}{c||}{${\rm d}^2 \sigma /{\rm d}p{\rm d}\Omega$}
  &$\langle p_{\rm T} \rangle$ & $\langle \theta \rangle$
  & \multicolumn{5}{c|}{${\rm d}^2 \sigma /{\rm d}p{\rm d}\Omega$} \\
 \hline
0.20--0.24 & 0.220 &  24.8 &  42.26 & $\!\!\pm\!\!$ &  1.72 & $\!\!\pm\!\!$ &  2.32 &  &  &  \multicolumn{5}{c|}{ } \\
0.24--0.30 & 0.269 &  25.0 &  47.06 & $\!\!\pm\!\!$ &  1.45 & $\!\!\pm\!\!$ &  2.34 & 0.270 &  34.9 &  40.58 & $\!\!\pm\!\!$ &  1.33 & $\!\!\pm\!\!$ &  1.92 \\
0.30--0.36 & 0.329 &  25.2 &  44.12 & $\!\!\pm\!\!$ &  1.43 & $\!\!\pm\!\!$ &  1.87 & 0.328 &  35.1 &  39.35 & $\!\!\pm\!\!$ &  1.31 & $\!\!\pm\!\!$ &  1.55 \\
0.36--0.42 & 0.388 &  25.2 &  40.91 & $\!\!\pm\!\!$ &  1.35 & $\!\!\pm\!\!$ &  1.53 & 0.387 &  35.0 &  37.43 & $\!\!\pm\!\!$ &  1.31 & $\!\!\pm\!\!$ &  1.27 \\
0.42--0.50 & 0.456 &  24.8 &  37.62 & $\!\!\pm\!\!$ &  1.11 & $\!\!\pm\!\!$ &  1.23 & 0.457 &  34.9 &  30.10 & $\!\!\pm\!\!$ &  1.00 & $\!\!\pm\!\!$ &  1.02 \\
0.50--0.60 & 0.545 &  24.9 &  32.32 & $\!\!\pm\!\!$ &  0.91 & $\!\!\pm\!\!$ &  1.06 & 0.544 &  34.8 &  28.13 & $\!\!\pm\!\!$ &  0.89 & $\!\!\pm\!\!$ &  0.90 \\
0.60--0.72 & 0.651 &  24.9 &  27.50 & $\!\!\pm\!\!$ &  0.78 & $\!\!\pm\!\!$ &  1.08 & 0.652 &  34.8 &  20.54 & $\!\!\pm\!\!$ &  0.68 & $\!\!\pm\!\!$ &  0.85 \\
0.72--0.90 &  &  &  \multicolumn{5}{c||}{ } & 0.796 &  34.8 &  14.17 & $\!\!\pm\!\!$ &  0.47 & $\!\!\pm\!\!$ &  0.82 \\
 \hline
 \hline
   & \multicolumn{7}{c||}{$40<\theta<50$}
  & \multicolumn{7}{c|}{$50<\theta<60$} \\
 \hline
 $p_{\rm T}$ & $\langle p_{\rm T} \rangle$ & $\langle \theta \rangle$
  & \multicolumn{5}{c||}{${\rm d}^2 \sigma /{\rm d}p{\rm d}\Omega$}
  &$\langle p_{\rm T} \rangle$ & $\langle \theta \rangle$
  & \multicolumn{5}{c|}{${\rm d}^2 \sigma /{\rm d}p{\rm d}\Omega$} \\
 \hline
0.30--0.36 & 0.328 &  44.9 &  37.76 & $\!\!\pm\!\!$ &  1.30 & $\!\!\pm\!\!$ &  1.37 &  &  &  \multicolumn{5}{c|}{ } \\
0.36--0.42 & 0.389 &  45.3 &  36.16 & $\!\!\pm\!\!$ &  1.28 & $\!\!\pm\!\!$ &  1.11 & 0.389 &  55.0 &  29.71 & $\!\!\pm\!\!$ &  1.12 & $\!\!\pm\!\!$ &  0.91 \\
0.42--0.50 & 0.459 &  44.8 &  29.11 & $\!\!\pm\!\!$ &  0.99 & $\!\!\pm\!\!$ &  0.88 & 0.459 &  54.8 &  27.99 & $\!\!\pm\!\!$ &  0.95 & $\!\!\pm\!\!$ &  0.85 \\
0.50--0.60 & 0.549 &  44.8 &  22.87 & $\!\!\pm\!\!$ &  0.80 & $\!\!\pm\!\!$ &  0.77 & 0.548 &  55.0 &  18.48 & $\!\!\pm\!\!$ &  0.71 & $\!\!\pm\!\!$ &  0.70 \\
0.60--0.72 & 0.655 &  44.9 &  17.01 & $\!\!\pm\!\!$ &  0.64 & $\!\!\pm\!\!$ &  0.74 & 0.658 &  55.0 &  14.18 & $\!\!\pm\!\!$ &  0.59 & $\!\!\pm\!\!$ &  0.70 \\
0.72--0.90 & 0.798 &  44.7 &  10.26 & $\!\!\pm\!\!$ &  0.41 & $\!\!\pm\!\!$ &  0.63 & 0.798 &  54.8 &   8.30 & $\!\!\pm\!\!$ &  0.38 & $\!\!\pm\!\!$ &  0.56 \\
0.90--1.25 & 1.040 &  44.7 &   4.00 & $\!\!\pm\!\!$ &  0.19 & $\!\!\pm\!\!$ &  0.38 & 1.031 &  54.9 &   2.21 & $\!\!\pm\!\!$ &  0.14 & $\!\!\pm\!\!$ &  0.24 \\
 \hline
 \hline
   & \multicolumn{7}{c||}{$60<\theta<75$}
  & \multicolumn{7}{c|}{$75<\theta<90$} \\
 \hline
 $p_{\rm T}$ & $\langle p_{\rm T} \rangle$ & $\langle \theta \rangle$
  & \multicolumn{5}{c||}{${\rm d}^2 \sigma /{\rm d}p{\rm d}\Omega$}
  &$\langle p_{\rm T} \rangle$ & $\langle \theta \rangle$
  & \multicolumn{5}{c|}{${\rm d}^2 \sigma /{\rm d}p{\rm d}\Omega$} \\
 \hline
0.36--0.42 & 0.393 &  67.1 &  26.96 & $\!\!\pm\!\!$ &  0.83 & $\!\!\pm\!\!$ &  1.02 &  &  &  \multicolumn{5}{c|}{ } \\
0.42--0.50 & 0.466 &  67.1 &  23.83 & $\!\!\pm\!\!$ &  0.71 & $\!\!\pm\!\!$ &  0.73 & 0.465 &  82.0 &  17.60 & $\!\!\pm\!\!$ &  0.60 & $\!\!\pm\!\!$ &  0.71 \\
0.50--0.60 & 0.556 &  67.2 &  18.53 & $\!\!\pm\!\!$ &  0.58 & $\!\!\pm\!\!$ &  0.71 & 0.555 &  81.7 &  12.56 & $\!\!\pm\!\!$ &  0.48 & $\!\!\pm\!\!$ &  0.65 \\
0.60--0.72 & 0.666 &  67.5 &  11.57 & $\!\!\pm\!\!$ &  0.44 & $\!\!\pm\!\!$ &  0.69 & 0.668 &  81.4 &   6.10 & $\!\!\pm\!\!$ &  0.31 & $\!\!\pm\!\!$ &  0.48 \\
0.72--0.90 & 0.814 &  66.9 &   5.29 & $\!\!\pm\!\!$ &  0.25 & $\!\!\pm\!\!$ &  0.49 & 0.814 &  81.6 &   2.56 & $\!\!\pm\!\!$ &  0.18 & $\!\!\pm\!\!$ &  0.27 \\
0.90--1.25 & 1.059 &  66.9 &   1.32 & $\!\!\pm\!\!$ &  0.09 & $\!\!\pm\!\!$ &  0.19 & 1.052 &  81.0 &   0.62 & $\!\!\pm\!\!$ &  0.06 & $\!\!\pm\!\!$ &  0.10 \\
 \hline
 \hline
  & \multicolumn{7}{c||}{$90<\theta<105$}
  & \multicolumn{7}{c|}{$105<\theta<125$} \\
 \hline
 $p_{\rm T}$ & $\langle p_{\rm T} \rangle$ & $\langle \theta \rangle$
  & \multicolumn{5}{c||}{${\rm d}^2 \sigma /{\rm d}p{\rm d}\Omega$}
  &$\langle p_{\rm T} \rangle$ & $\langle \theta \rangle$
  & \multicolumn{5}{c|}{${\rm d}^2 \sigma /{\rm d}p{\rm d}\Omega$} \\
 \hline
0.36--0.42 &  &  &  \multicolumn{5}{c||}{ } & 0.393 & 113.4 &   7.70 & $\!\!\pm\!\!$ &  0.40 & $\!\!\pm\!\!$ &  0.32 \\
0.42--0.50 & 0.462 &  97.0 &  10.79 & $\!\!\pm\!\!$ &  0.49 & $\!\!\pm\!\!$ &  0.62 & 0.462 & 113.2 &   4.28 & $\!\!\pm\!\!$ &  0.26 & $\!\!\pm\!\!$ &  0.23 \\
0.50--0.60 & 0.553 &  96.6 &   5.52 & $\!\!\pm\!\!$ &  0.31 & $\!\!\pm\!\!$ &  0.40 & 0.551 & 112.6 &   2.52 & $\!\!\pm\!\!$ &  0.20 & $\!\!\pm\!\!$ &  0.24 \\
0.60--0.72 & 0.660 &  96.8 &   2.69 & $\!\!\pm\!\!$ &  0.22 & $\!\!\pm\!\!$ &  0.26 & 0.664 & 111.9 &   0.68 & $\!\!\pm\!\!$ &  0.10 & $\!\!\pm\!\!$ &  0.10 \\
0.72--0.90 & 0.814 &  95.8 &   1.00 & $\!\!\pm\!\!$ &  0.11 & $\!\!\pm\!\!$ &  0.12 & 0.804 & 112.0 &   0.21 & $\!\!\pm\!\!$ &  0.04 & $\!\!\pm\!\!$ &  0.05 \\
0.90--1.25 & 1.013 &  95.6 &   0.30 & $\!\!\pm\!\!$ &  0.05 & $\!\!\pm\!\!$ &  0.05 & 1.070 & 112.1 &   0.05 & $\!\!\pm\!\!$ &  0.02 & $\!\!\pm\!\!$ &  0.02 \\
 \hline
 \end{tabular}
 \end{center}
 \end{scriptsize}
 \end{table}

%% file: table.pip.probe12.tex
  
 \begin{table}[h]
 \begin{scriptsize}
 \caption{Double-differential inclusive
  cross-section ${\rm d}^2 \sigma /{\rm d}p{\rm d}\Omega$
  [mb/(GeV/{\it c} sr)] of the production of $\pi^+$'s
  in p + Be $\rightarrow$ $\pi^+$ + X interactions
  with $+12.0$~GeV/{\it c} beam momentum;
  the first error is statistical, the second systematic; 
 $p_{\rm T}$ in GeV/{\it c}, polar angle $\theta$ in degrees.}
 \label{pip.probe12}
 \begin{center}
 \begin{tabular}{|c||c|c|rcrcr||c|c|rcrcr|}
 \hline
   & \multicolumn{7}{c||}{$20<\theta<30$}
  & \multicolumn{7}{c|}{$30<\theta<40$} \\
 \hline
 $p_{\rm T}$ & $\langle p_{\rm T} \rangle$ & $\langle \theta \rangle$
  & \multicolumn{5}{c||}{${\rm d}^2 \sigma /{\rm d}p{\rm d}\Omega$}
  &$\langle p_{\rm T} \rangle$ & $\langle \theta \rangle$
  & \multicolumn{5}{c|}{${\rm d}^2 \sigma /{\rm d}p{\rm d}\Omega$} \\
 \hline
0.10--0.13 & 0.116 &  24.5 &  56.17 & $\!\!\pm\!\!$ &  2.74 & $\!\!\pm\!\!$ &  4.10 & 0.116 &  34.9 &  37.64 & $\!\!\pm\!\!$ &  2.12 & $\!\!\pm\!\!$ &  2.76 \\
0.13--0.16 & 0.145 &  24.6 &  68.54 & $\!\!\pm\!\!$ &  2.74 & $\!\!\pm\!\!$ &  3.99 & 0.146 &  34.7 &  49.10 & $\!\!\pm\!\!$ &  2.30 & $\!\!\pm\!\!$ &  2.82 \\
0.16--0.20 & 0.180 &  24.4 &  81.06 & $\!\!\pm\!\!$ &  2.45 & $\!\!\pm\!\!$ &  3.93 & 0.180 &  34.7 &  59.23 & $\!\!\pm\!\!$ &  2.18 & $\!\!\pm\!\!$ &  2.84 \\
0.20--0.24 & 0.220 &  24.6 &  89.45 & $\!\!\pm\!\!$ &  2.60 & $\!\!\pm\!\!$ &  3.70 & 0.219 &  34.7 &  57.42 & $\!\!\pm\!\!$ &  2.05 & $\!\!\pm\!\!$ &  2.36 \\
0.24--0.30 & 0.269 &  24.6 &  87.39 & $\!\!\pm\!\!$ &  2.08 & $\!\!\pm\!\!$ &  3.02 & 0.269 &  34.8 &  58.85 & $\!\!\pm\!\!$ &  1.70 & $\!\!\pm\!\!$ &  1.99 \\
0.30--0.36 & 0.328 &  24.7 &  72.09 & $\!\!\pm\!\!$ &  1.84 & $\!\!\pm\!\!$ &  2.14 & 0.329 &  34.6 &  44.45 & $\!\!\pm\!\!$ &  1.44 & $\!\!\pm\!\!$ &  1.29 \\
0.36--0.42 & 0.387 &  24.6 &  57.37 & $\!\!\pm\!\!$ &  1.63 & $\!\!\pm\!\!$ &  1.67 & 0.388 &  34.7 &  40.10 & $\!\!\pm\!\!$ &  1.39 & $\!\!\pm\!\!$ &  1.13 \\
0.42--0.50 & 0.455 &  24.7 &  44.57 & $\!\!\pm\!\!$ &  1.24 & $\!\!\pm\!\!$ &  1.58 & 0.456 &  34.7 &  29.50 & $\!\!\pm\!\!$ &  0.99 & $\!\!\pm\!\!$ &  0.97 \\
0.50--0.60 & 0.541 &  24.8 &  29.37 & $\!\!\pm\!\!$ &  0.88 & $\!\!\pm\!\!$ &  1.50 & 0.544 &  34.7 &  20.26 & $\!\!\pm\!\!$ &  0.74 & $\!\!\pm\!\!$ &  0.95 \\
0.60--0.72 & 0.653 &  24.5 &  18.38 & $\!\!\pm\!\!$ &  0.62 & $\!\!\pm\!\!$ &  1.42 & 0.650 &  34.7 &  11.91 & $\!\!\pm\!\!$ &  0.48 & $\!\!\pm\!\!$ &  0.84 \\
0.72--0.90 &  &  &  \multicolumn{5}{c||}{ } & 0.793 &  34.4 &   5.37 & $\!\!\pm\!\!$ &  0.24 & $\!\!\pm\!\!$ &  0.60 \\
 \hline
 \hline
   & \multicolumn{7}{c||}{$40<\theta<50$}
  & \multicolumn{7}{c|}{$50<\theta<60$} \\
 \hline
 $p_{\rm T}$ & $\langle p_{\rm T} \rangle$ & $\langle \theta \rangle$
  & \multicolumn{5}{c||}{${\rm d}^2 \sigma /{\rm d}p{\rm d}\Omega$}
  &$\langle p_{\rm T} \rangle$ & $\langle \theta \rangle$
  & \multicolumn{5}{c|}{${\rm d}^2 \sigma /{\rm d}p{\rm d}\Omega$} \\
 \hline
0.10--0.13 & 0.116 &  44.9 &  29.54 & $\!\!\pm\!\!$ &  1.89 & $\!\!\pm\!\!$ &  2.23 &  &  &  \multicolumn{5}{c|}{ } \\
0.13--0.16 & 0.145 &  45.1 &  31.65 & $\!\!\pm\!\!$ &  1.79 & $\!\!\pm\!\!$ &  1.86 & 0.146 &  55.0 &  30.54 & $\!\!\pm\!\!$ &  1.73 & $\!\!\pm\!\!$ &  1.91 \\
0.16--0.20 & 0.180 &  44.7 &  39.88 & $\!\!\pm\!\!$ &  1.74 & $\!\!\pm\!\!$ &  1.95 & 0.180 &  54.7 &  32.99 & $\!\!\pm\!\!$ &  1.60 & $\!\!\pm\!\!$ &  1.61 \\
0.20--0.24 & 0.220 &  45.1 &  36.17 & $\!\!\pm\!\!$ &  1.59 & $\!\!\pm\!\!$ &  1.51 & 0.220 &  55.0 &  32.11 & $\!\!\pm\!\!$ &  1.57 & $\!\!\pm\!\!$ &  1.32 \\
0.24--0.30 & 0.269 &  44.9 &  37.30 & $\!\!\pm\!\!$ &  1.34 & $\!\!\pm\!\!$ &  1.27 & 0.269 &  54.8 &  26.84 & $\!\!\pm\!\!$ &  1.13 & $\!\!\pm\!\!$ &  0.90 \\
0.30--0.36 & 0.329 &  44.6 &  27.91 & $\!\!\pm\!\!$ &  1.13 & $\!\!\pm\!\!$ &  0.83 & 0.329 &  54.7 &  21.59 & $\!\!\pm\!\!$ &  1.01 & $\!\!\pm\!\!$ &  0.64 \\
0.36--0.42 & 0.387 &  44.6 &  27.97 & $\!\!\pm\!\!$ &  1.17 & $\!\!\pm\!\!$ &  0.84 & 0.389 &  54.7 &  17.68 & $\!\!\pm\!\!$ &  0.89 & $\!\!\pm\!\!$ &  0.56 \\
0.42--0.50 & 0.458 &  44.6 &  19.97 & $\!\!\pm\!\!$ &  0.85 & $\!\!\pm\!\!$ &  0.69 & 0.457 &  54.7 &  12.98 & $\!\!\pm\!\!$ &  0.64 & $\!\!\pm\!\!$ &  0.49 \\
0.50--0.60 & 0.545 &  44.6 &  12.52 & $\!\!\pm\!\!$ &  0.56 & $\!\!\pm\!\!$ &  0.58 & 0.547 &  54.6 &   8.98 & $\!\!\pm\!\!$ &  0.51 & $\!\!\pm\!\!$ &  0.46 \\
0.60--0.72 & 0.655 &  44.3 &   8.16 & $\!\!\pm\!\!$ &  0.42 & $\!\!\pm\!\!$ &  0.54 & 0.654 &  54.4 &   4.23 & $\!\!\pm\!\!$ &  0.29 & $\!\!\pm\!\!$ &  0.30 \\
0.72--0.90 & 0.792 &  44.4 &   3.41 & $\!\!\pm\!\!$ &  0.21 & $\!\!\pm\!\!$ &  0.35 & 0.793 &  54.8 &   2.24 & $\!\!\pm\!\!$ &  0.17 & $\!\!\pm\!\!$ &  0.23 \\
0.90--1.25 &  &  &  \multicolumn{5}{c||}{ } & 1.012 &  54.0 &   0.43 & $\!\!\pm\!\!$ &  0.04 & $\!\!\pm\!\!$ &  0.08 \\
 \hline
 \hline
   & \multicolumn{7}{c||}{$60<\theta<75$}
  & \multicolumn{7}{c|}{$75<\theta<90$} \\
 \hline
 $p_{\rm T}$ & $\langle p_{\rm T} \rangle$ & $\langle \theta \rangle$
  & \multicolumn{5}{c||}{${\rm d}^2 \sigma /{\rm d}p{\rm d}\Omega$}
  &$\langle p_{\rm T} \rangle$ & $\langle \theta \rangle$
  & \multicolumn{5}{c|}{${\rm d}^2 \sigma /{\rm d}p{\rm d}\Omega$} \\
 \hline
0.13--0.16 & 0.146 &  67.2 &  18.07 & $\!\!\pm\!\!$ &  1.05 & $\!\!\pm\!\!$ &  1.18 & 0.146 &  82.6 &  14.65 & $\!\!\pm\!\!$ &  0.92 & $\!\!\pm\!\!$ &  1.00 \\
0.16--0.20 & 0.181 &  67.2 &  23.29 & $\!\!\pm\!\!$ &  1.06 & $\!\!\pm\!\!$ &  1.14 & 0.181 &  82.3 &  17.94 & $\!\!\pm\!\!$ &  0.93 & $\!\!\pm\!\!$ &  0.96 \\
0.20--0.24 & 0.221 &  67.2 &  22.23 & $\!\!\pm\!\!$ &  1.04 & $\!\!\pm\!\!$ &  0.89 & 0.221 &  81.9 &  15.36 & $\!\!\pm\!\!$ &  0.80 & $\!\!\pm\!\!$ &  0.59 \\
0.24--0.30 & 0.272 &  67.3 &  20.62 & $\!\!\pm\!\!$ &  0.80 & $\!\!\pm\!\!$ &  0.67 & 0.271 &  82.0 &  11.32 & $\!\!\pm\!\!$ &  0.58 & $\!\!\pm\!\!$ &  0.36 \\
0.30--0.36 & 0.333 &  66.7 &  15.24 & $\!\!\pm\!\!$ &  0.71 & $\!\!\pm\!\!$ &  0.47 & 0.332 &  81.9 &   8.92 & $\!\!\pm\!\!$ &  0.52 & $\!\!\pm\!\!$ &  0.30 \\
0.36--0.42 & 0.396 &  66.5 &  10.73 & $\!\!\pm\!\!$ &  0.56 & $\!\!\pm\!\!$ &  0.37 & 0.395 &  81.8 &   6.43 & $\!\!\pm\!\!$ &  0.44 & $\!\!\pm\!\!$ &  0.28 \\
0.42--0.50 & 0.464 &  67.0 &   7.43 & $\!\!\pm\!\!$ &  0.40 & $\!\!\pm\!\!$ &  0.33 & 0.465 &  82.3 &   4.08 & $\!\!\pm\!\!$ &  0.30 & $\!\!\pm\!\!$ &  0.23 \\
0.50--0.60 & 0.555 &  66.3 &   4.91 & $\!\!\pm\!\!$ &  0.29 & $\!\!\pm\!\!$ &  0.30 & 0.553 &  81.8 &   2.81 & $\!\!\pm\!\!$ &  0.23 & $\!\!\pm\!\!$ &  0.22 \\
0.60--0.72 & 0.666 &  66.4 &   3.29 & $\!\!\pm\!\!$ &  0.23 & $\!\!\pm\!\!$ &  0.28 & 0.661 &  81.3 &   1.49 & $\!\!\pm\!\!$ &  0.16 & $\!\!\pm\!\!$ &  0.16 \\
0.72--0.90 & 0.806 &  66.1 &   0.96 & $\!\!\pm\!\!$ &  0.09 & $\!\!\pm\!\!$ &  0.12 & 0.813 &  82.8 &   0.33 & $\!\!\pm\!\!$ &  0.05 & $\!\!\pm\!\!$ &  0.05 \\
0.90--1.25 & 1.043 &  65.3 &   0.17 & $\!\!\pm\!\!$ &  0.02 & $\!\!\pm\!\!$ &  0.04 & 1.030 &  81.1 &   0.09 & $\!\!\pm\!\!$ &  0.02 & $\!\!\pm\!\!$ &  0.02 \\
 \hline
 \hline
  & \multicolumn{7}{c||}{$90<\theta<105$}
  & \multicolumn{7}{c|}{$105<\theta<125$} \\
 \hline
 $p_{\rm T}$ & $\langle p_{\rm T} \rangle$ & $\langle \theta \rangle$
  & \multicolumn{5}{c||}{${\rm d}^2 \sigma /{\rm d}p{\rm d}\Omega$}
  &$\langle p_{\rm T} \rangle$ & $\langle \theta \rangle$
  & \multicolumn{5}{c|}{${\rm d}^2 \sigma /{\rm d}p{\rm d}\Omega$} \\
 \hline
0.13--0.16 & 0.147 &  97.4 &  12.17 & $\!\!\pm\!\!$ &  0.87 & $\!\!\pm\!\!$ &  0.80 & 0.146 & 115.1 &  11.86 & $\!\!\pm\!\!$ &  0.72 & $\!\!\pm\!\!$ &  0.76 \\
0.16--0.20 & 0.181 &  97.4 &  13.20 & $\!\!\pm\!\!$ &  0.74 & $\!\!\pm\!\!$ &  0.71 & 0.181 & 113.9 &  10.37 & $\!\!\pm\!\!$ &  0.58 & $\!\!\pm\!\!$ &  0.55 \\
0.20--0.24 & 0.220 &  97.3 &  11.44 & $\!\!\pm\!\!$ &  0.71 & $\!\!\pm\!\!$ &  0.45 & 0.219 & 114.0 &   8.53 & $\!\!\pm\!\!$ &  0.56 & $\!\!\pm\!\!$ &  0.36 \\
0.24--0.30 & 0.270 &  97.6 &   9.23 & $\!\!\pm\!\!$ &  0.55 & $\!\!\pm\!\!$ &  0.34 & 0.269 & 113.7 &   3.86 & $\!\!\pm\!\!$ &  0.29 & $\!\!\pm\!\!$ &  0.17 \\
0.30--0.36 & 0.331 &  97.2 &   4.67 & $\!\!\pm\!\!$ &  0.38 & $\!\!\pm\!\!$ &  0.21 & 0.330 & 113.6 &   2.57 & $\!\!\pm\!\!$ &  0.24 & $\!\!\pm\!\!$ &  0.18 \\
0.36--0.42 & 0.393 &  97.2 &   3.51 & $\!\!\pm\!\!$ &  0.33 & $\!\!\pm\!\!$ &  0.22 & 0.393 & 113.6 &   1.53 & $\!\!\pm\!\!$ &  0.17 & $\!\!\pm\!\!$ &  0.13 \\
0.42--0.50 & 0.466 &  96.3 &   2.68 & $\!\!\pm\!\!$ &  0.25 & $\!\!\pm\!\!$ &  0.22 & 0.459 & 113.7 &   0.67 & $\!\!\pm\!\!$ &  0.10 & $\!\!\pm\!\!$ &  0.08 \\
0.50--0.60 & 0.550 &  96.1 &   0.95 & $\!\!\pm\!\!$ &  0.12 & $\!\!\pm\!\!$ &  0.11 & 0.563 & 112.5 &   0.26 & $\!\!\pm\!\!$ &  0.06 & $\!\!\pm\!\!$ &  0.04 \\
0.60--0.72 & 0.662 &  96.1 &   0.30 & $\!\!\pm\!\!$ &  0.06 & $\!\!\pm\!\!$ &  0.05 & 0.669 & 108.6 &   0.08 & $\!\!\pm\!\!$ &  0.03 & $\!\!\pm\!\!$ &  0.02 \\
0.72--0.90 & 0.812 &  97.0 &   0.11 & $\!\!\pm\!\!$ &  0.03 & $\!\!\pm\!\!$ &  0.02 &  &  &  \multicolumn{5}{c|}{ } \\
 \hline
 \end{tabular}
 \end{center}
 \end{scriptsize}
 \end{table}

%% file: table.pim.probe12.tex
  
 \begin{table}[h]
 \begin{scriptsize}
 \caption{Double-differential inclusive
  cross-section ${\rm d}^2 \sigma /{\rm d}p{\rm d}\Omega$
  [mb/(GeV/{\it c} sr)] of the production of $\pi^-$'s
  in p + Be $\rightarrow$ $\pi^-$ + X interactions
  with $+12.0$~GeV/{\it c} beam momentum;
  the first error is statistical, the second systematic; 
 $p_{\rm T}$ in GeV/{\it c}, polar angle $\theta$ in degrees.}
 \label{pim.probe12}
 \begin{center}
 \begin{tabular}{|c||c|c|rcrcr||c|c|rcrcr|}
 \hline
   & \multicolumn{7}{c||}{$20<\theta<30$}
  & \multicolumn{7}{c|}{$30<\theta<40$} \\
 \hline
 $p_{\rm T}$ & $\langle p_{\rm T} \rangle$ & $\langle \theta \rangle$
  & \multicolumn{5}{c||}{${\rm d}^2 \sigma /{\rm d}p{\rm d}\Omega$}
  &$\langle p_{\rm T} \rangle$ & $\langle \theta \rangle$
  & \multicolumn{5}{c|}{${\rm d}^2 \sigma /{\rm d}p{\rm d}\Omega$} \\
 \hline
0.10--0.13 & 0.116 &  24.7 &  63.15 & $\!\!\pm\!\!$ &  2.80 & $\!\!\pm\!\!$ &  4.48 & 0.115 &  34.6 &  36.61 & $\!\!\pm\!\!$ &  1.99 & $\!\!\pm\!\!$ &  2.85 \\
0.13--0.16 & 0.146 &  24.8 &  73.93 & $\!\!\pm\!\!$ &  2.87 & $\!\!\pm\!\!$ &  4.28 & 0.146 &  34.8 &  50.78 & $\!\!\pm\!\!$ &  2.28 & $\!\!\pm\!\!$ &  3.02 \\
0.16--0.20 & 0.180 &  24.6 &  75.32 & $\!\!\pm\!\!$ &  2.35 & $\!\!\pm\!\!$ &  3.65 & 0.181 &  34.5 &  54.11 & $\!\!\pm\!\!$ &  1.94 & $\!\!\pm\!\!$ &  2.67 \\
0.20--0.24 & 0.220 &  24.6 &  86.26 & $\!\!\pm\!\!$ &  2.56 & $\!\!\pm\!\!$ &  3.50 & 0.220 &  34.8 &  57.19 & $\!\!\pm\!\!$ &  2.04 & $\!\!\pm\!\!$ &  2.36 \\
0.24--0.30 & 0.270 &  24.6 &  71.03 & $\!\!\pm\!\!$ &  1.84 & $\!\!\pm\!\!$ &  2.33 & 0.270 &  34.8 &  50.41 & $\!\!\pm\!\!$ &  1.55 & $\!\!\pm\!\!$ &  1.67 \\
0.30--0.36 & 0.330 &  24.8 &  60.88 & $\!\!\pm\!\!$ &  1.71 & $\!\!\pm\!\!$ &  1.70 & 0.331 &  34.8 &  42.13 & $\!\!\pm\!\!$ &  1.42 & $\!\!\pm\!\!$ &  1.19 \\
0.36--0.42 & 0.391 &  24.9 &  46.35 & $\!\!\pm\!\!$ &  1.50 & $\!\!\pm\!\!$ &  1.33 & 0.390 &  34.8 &  35.09 & $\!\!\pm\!\!$ &  1.30 & $\!\!\pm\!\!$ &  1.01 \\
0.42--0.50 & 0.460 &  24.6 &  34.68 & $\!\!\pm\!\!$ &  1.11 & $\!\!\pm\!\!$ &  1.25 & 0.458 &  34.7 &  22.92 & $\!\!\pm\!\!$ &  0.86 & $\!\!\pm\!\!$ &  0.81 \\
0.50--0.60 & 0.550 &  24.8 &  24.77 & $\!\!\pm\!\!$ &  0.85 & $\!\!\pm\!\!$ &  1.25 & 0.548 &  35.0 &  16.44 & $\!\!\pm\!\!$ &  0.70 & $\!\!\pm\!\!$ &  0.81 \\
0.60--0.72 & 0.657 &  24.7 &  12.69 & $\!\!\pm\!\!$ &  0.54 & $\!\!\pm\!\!$ &  0.91 & 0.658 &  34.9 &   7.61 & $\!\!\pm\!\!$ &  0.39 & $\!\!\pm\!\!$ &  0.53 \\
0.72--0.90 &  &  &  \multicolumn{5}{c||}{ } & 0.805 &  35.1 &   3.67 & $\!\!\pm\!\!$ &  0.23 & $\!\!\pm\!\!$ &  0.37 \\
 \hline
 \hline
   & \multicolumn{7}{c||}{$40<\theta<50$}
  & \multicolumn{7}{c|}{$50<\theta<60$} \\
 \hline
 $p_{\rm T}$ & $\langle p_{\rm T} \rangle$ & $\langle \theta \rangle$
  & \multicolumn{5}{c||}{${\rm d}^2 \sigma /{\rm d}p{\rm d}\Omega$}
  &$\langle p_{\rm T} \rangle$ & $\langle \theta \rangle$
  & \multicolumn{5}{c|}{${\rm d}^2 \sigma /{\rm d}p{\rm d}\Omega$} \\
 \hline
0.10--0.13 & 0.116 &  44.8 &  26.29 & $\!\!\pm\!\!$ &  1.63 & $\!\!\pm\!\!$ &  2.21 &  &  &  \multicolumn{5}{c|}{ } \\
0.13--0.16 & 0.145 &  44.8 &  31.91 & $\!\!\pm\!\!$ &  1.69 & $\!\!\pm\!\!$ &  2.00 & 0.144 &  54.9 &  29.42 & $\!\!\pm\!\!$ &  1.74 & $\!\!\pm\!\!$ &  2.00 \\
0.16--0.20 & 0.180 &  44.9 &  42.89 & $\!\!\pm\!\!$ &  1.85 & $\!\!\pm\!\!$ &  2.15 & 0.179 &  55.0 &  28.79 & $\!\!\pm\!\!$ &  1.41 & $\!\!\pm\!\!$ &  1.44 \\
0.20--0.24 & 0.220 &  45.1 &  36.82 & $\!\!\pm\!\!$ &  1.59 & $\!\!\pm\!\!$ &  1.55 & 0.219 &  54.7 &  30.71 & $\!\!\pm\!\!$ &  1.48 & $\!\!\pm\!\!$ &  1.27 \\
0.24--0.30 & 0.269 &  44.7 &  35.90 & $\!\!\pm\!\!$ &  1.32 & $\!\!\pm\!\!$ &  1.21 & 0.269 &  54.8 &  25.32 & $\!\!\pm\!\!$ &  1.10 & $\!\!\pm\!\!$ &  0.84 \\
0.30--0.36 & 0.329 &  44.5 &  28.51 & $\!\!\pm\!\!$ &  1.12 & $\!\!\pm\!\!$ &  0.82 & 0.329 &  54.7 &  17.52 & $\!\!\pm\!\!$ &  0.84 & $\!\!\pm\!\!$ &  0.52 \\
0.36--0.42 & 0.390 &  44.7 &  23.91 & $\!\!\pm\!\!$ &  1.06 & $\!\!\pm\!\!$ &  0.73 & 0.388 &  54.6 &  16.15 & $\!\!\pm\!\!$ &  0.86 & $\!\!\pm\!\!$ &  0.53 \\
0.42--0.50 & 0.458 &  44.6 &  15.10 & $\!\!\pm\!\!$ &  0.67 & $\!\!\pm\!\!$ &  0.57 & 0.455 &  54.8 &   9.68 & $\!\!\pm\!\!$ &  0.55 & $\!\!\pm\!\!$ &  0.41 \\
0.50--0.60 & 0.545 &  44.7 &   9.36 & $\!\!\pm\!\!$ &  0.47 & $\!\!\pm\!\!$ &  0.50 & 0.547 &  54.8 &   7.63 & $\!\!\pm\!\!$ &  0.46 & $\!\!\pm\!\!$ &  0.44 \\
0.60--0.72 & 0.653 &  45.0 &   5.50 & $\!\!\pm\!\!$ &  0.36 & $\!\!\pm\!\!$ &  0.41 & 0.653 &  54.5 &   3.63 & $\!\!\pm\!\!$ &  0.27 & $\!\!\pm\!\!$ &  0.29 \\
0.72--0.90 & 0.793 &  44.8 &   2.16 & $\!\!\pm\!\!$ &  0.17 & $\!\!\pm\!\!$ &  0.23 & 0.785 &  54.5 &   1.51 & $\!\!\pm\!\!$ &  0.14 & $\!\!\pm\!\!$ &  0.16 \\
0.90--1.25 &  &  &  \multicolumn{5}{c||}{ } & 1.021 &  54.3 &   0.21 & $\!\!\pm\!\!$ &  0.03 & $\!\!\pm\!\!$ &  0.04 \\
 \hline
 \hline
   & \multicolumn{7}{c||}{$60<\theta<75$}
  & \multicolumn{7}{c|}{$75<\theta<90$} \\
 \hline
 $p_{\rm T}$ & $\langle p_{\rm T} \rangle$ & $\langle \theta \rangle$
  & \multicolumn{5}{c||}{${\rm d}^2 \sigma /{\rm d}p{\rm d}\Omega$}
  &$\langle p_{\rm T} \rangle$ & $\langle \theta \rangle$
  & \multicolumn{5}{c|}{${\rm d}^2 \sigma /{\rm d}p{\rm d}\Omega$} \\
 \hline
0.13--0.16 & 0.145 &  67.4 &  19.97 & $\!\!\pm\!\!$ &  1.09 & $\!\!\pm\!\!$ &  1.39 & 0.145 &  82.4 &  16.93 & $\!\!\pm\!\!$ &  0.99 & $\!\!\pm\!\!$ &  1.21 \\
0.16--0.20 & 0.180 &  67.5 &  22.77 & $\!\!\pm\!\!$ &  0.99 & $\!\!\pm\!\!$ &  1.15 & 0.178 &  82.3 &  18.92 & $\!\!\pm\!\!$ &  0.90 & $\!\!\pm\!\!$ &  1.00 \\
0.20--0.24 & 0.218 &  67.2 &  21.03 & $\!\!\pm\!\!$ &  0.97 & $\!\!\pm\!\!$ &  0.81 & 0.217 &  82.3 &  16.53 & $\!\!\pm\!\!$ &  0.83 & $\!\!\pm\!\!$ &  0.63 \\
0.24--0.30 & 0.267 &  66.9 &  16.28 & $\!\!\pm\!\!$ &  0.68 & $\!\!\pm\!\!$ &  0.51 & 0.265 &  81.9 &  13.64 & $\!\!\pm\!\!$ &  0.65 & $\!\!\pm\!\!$ &  0.43 \\
0.30--0.36 & 0.327 &  66.8 &  13.78 & $\!\!\pm\!\!$ &  0.64 & $\!\!\pm\!\!$ &  0.42 & 0.324 &  81.8 &   7.88 & $\!\!\pm\!\!$ &  0.47 & $\!\!\pm\!\!$ &  0.30 \\
0.36--0.42 & 0.384 &  66.7 &   9.72 & $\!\!\pm\!\!$ &  0.52 & $\!\!\pm\!\!$ &  0.35 & 0.383 &  82.0 &   5.92 & $\!\!\pm\!\!$ &  0.42 & $\!\!\pm\!\!$ &  0.30 \\
0.42--0.50 & 0.452 &  66.2 &   5.78 & $\!\!\pm\!\!$ &  0.33 & $\!\!\pm\!\!$ &  0.29 & 0.451 &  81.4 &   3.52 & $\!\!\pm\!\!$ &  0.27 & $\!\!\pm\!\!$ &  0.22 \\
0.50--0.60 & 0.535 &  66.4 &   4.40 & $\!\!\pm\!\!$ &  0.28 & $\!\!\pm\!\!$ &  0.29 & 0.534 &  82.4 &   2.34 & $\!\!\pm\!\!$ &  0.20 & $\!\!\pm\!\!$ &  0.20 \\
0.60--0.72 & 0.638 &  66.3 &   1.97 & $\!\!\pm\!\!$ &  0.16 & $\!\!\pm\!\!$ &  0.17 & 0.636 &  81.4 &   0.74 & $\!\!\pm\!\!$ &  0.11 & $\!\!\pm\!\!$ &  0.09 \\
0.72--0.90 & 0.773 &  66.4 &   0.76 & $\!\!\pm\!\!$ &  0.08 & $\!\!\pm\!\!$ &  0.10 & 0.752 &  83.0 &   0.18 & $\!\!\pm\!\!$ &  0.04 & $\!\!\pm\!\!$ &  0.03 \\
0.90--1.25 & 0.988 &  66.5 &   0.11 & $\!\!\pm\!\!$ &  0.02 & $\!\!\pm\!\!$ &  0.02 & 1.041 &  78.8 &   0.02 & $\!\!\pm\!\!$ &  0.01 & $\!\!\pm\!\!$ &  0.01 \\
 \hline
 \hline
  & \multicolumn{7}{c||}{$90<\theta<105$}
  & \multicolumn{7}{c|}{$105<\theta<125$} \\
 \hline
 $p_{\rm T}$ & $\langle p_{\rm T} \rangle$ & $\langle \theta \rangle$
  & \multicolumn{5}{c||}{${\rm d}^2 \sigma /{\rm d}p{\rm d}\Omega$}
  &$\langle p_{\rm T} \rangle$ & $\langle \theta \rangle$
  & \multicolumn{5}{c|}{${\rm d}^2 \sigma /{\rm d}p{\rm d}\Omega$} \\
 \hline
0.13--0.16 & 0.144 &  97.3 &  15.64 & $\!\!\pm\!\!$ &  0.95 & $\!\!\pm\!\!$ &  1.21 & 0.145 & 114.2 &  13.59 & $\!\!\pm\!\!$ &  0.73 & $\!\!\pm\!\!$ &  1.01 \\
0.16--0.20 & 0.179 &  96.9 &  13.02 & $\!\!\pm\!\!$ &  0.69 & $\!\!\pm\!\!$ &  0.78 & 0.178 & 114.4 &   8.83 & $\!\!\pm\!\!$ &  0.47 & $\!\!\pm\!\!$ &  0.53 \\
0.20--0.24 & 0.218 &  96.6 &  10.87 & $\!\!\pm\!\!$ &  0.66 & $\!\!\pm\!\!$ &  0.47 & 0.218 & 113.7 &   6.54 & $\!\!\pm\!\!$ &  0.43 & $\!\!\pm\!\!$ &  0.35 \\
0.24--0.30 & 0.265 &  97.7 &   6.63 & $\!\!\pm\!\!$ &  0.40 & $\!\!\pm\!\!$ &  0.27 & 0.265 & 113.4 &   5.25 & $\!\!\pm\!\!$ &  0.34 & $\!\!\pm\!\!$ &  0.28 \\
0.30--0.36 & 0.324 &  97.2 &   5.36 & $\!\!\pm\!\!$ &  0.38 & $\!\!\pm\!\!$ &  0.27 & 0.325 & 113.5 &   2.20 & $\!\!\pm\!\!$ &  0.21 & $\!\!\pm\!\!$ &  0.16 \\
0.36--0.42 & 0.387 &  96.6 &   2.69 & $\!\!\pm\!\!$ &  0.27 & $\!\!\pm\!\!$ &  0.18 & 0.380 & 112.3 &   1.38 & $\!\!\pm\!\!$ &  0.18 & $\!\!\pm\!\!$ &  0.14 \\
0.42--0.50 & 0.453 &  96.4 &   1.47 & $\!\!\pm\!\!$ &  0.15 & $\!\!\pm\!\!$ &  0.13 & 0.444 & 111.2 &   0.91 & $\!\!\pm\!\!$ &  0.14 & $\!\!\pm\!\!$ &  0.12 \\
0.50--0.60 & 0.529 &  96.6 &   0.95 & $\!\!\pm\!\!$ &  0.13 & $\!\!\pm\!\!$ &  0.12 & 0.526 & 111.0 &   0.25 & $\!\!\pm\!\!$ &  0.05 & $\!\!\pm\!\!$ &  0.04 \\
0.60--0.72 & 0.643 &  97.4 &   0.20 & $\!\!\pm\!\!$ &  0.05 & $\!\!\pm\!\!$ &  0.03 & 0.623 & 115.0 &   0.04 & $\!\!\pm\!\!$ &  0.03 & $\!\!\pm\!\!$ &  0.02 \\
0.72--0.90 & 0.762 &  94.4 &   0.11 & $\!\!\pm\!\!$ &  0.03 & $\!\!\pm\!\!$ &  0.02 &  &  &  \multicolumn{5}{c|}{ } \\
 \hline
 \end{tabular}
 \end{center}
 \end{scriptsize}
 \end{table}

%% file: table.pro.pipbe12.tex
  
 \begin{table}[h]
 \begin{scriptsize}
 \caption{Double-differential inclusive
  cross-section ${\rm d}^2 \sigma /{\rm d}p{\rm d}\Omega$
  [mb/(GeV/{\it c} sr)] of the production of protons
  in $\pi^+$ + Be $\rightarrow$ p + X interactions
  with $+12.0$~GeV/{\it c} beam momentum;
  the first error is statistical, the second systematic; 
 $p_{\rm T}$ in GeV/{\it c}, polar angle $\theta$ in degrees.}
 \label{pro.pipbe12}
 \begin{center}
 \begin{tabular}{|c||c|c|rcrcr||c|c|rcrcr|}
 \hline
   & \multicolumn{7}{c||}{$20<\theta<30$}
  & \multicolumn{7}{c|}{$30<\theta<40$} \\
 \hline
 $p_{\rm T}$ & $\langle p_{\rm T} \rangle$ & $\langle \theta \rangle$
  & \multicolumn{5}{c||}{${\rm d}^2 \sigma /{\rm d}p{\rm d}\Omega$}
  &$\langle p_{\rm T} \rangle$ & $\langle \theta \rangle$
  & \multicolumn{5}{c|}{${\rm d}^2 \sigma /{\rm d}p{\rm d}\Omega$} \\
 \hline
0.20--0.24 & 0.222 &  24.8 &  33.15 & $\!\!\pm\!\!$ &  5.14 & $\!\!\pm\!\!$ &  2.10 &  &  &  \multicolumn{5}{c|}{ } \\
0.24--0.30 & 0.269 &  24.9 &  26.85 & $\!\!\pm\!\!$ &  3.64 & $\!\!\pm\!\!$ &  1.49 & 0.271 &  34.9 &  33.55 & $\!\!\pm\!\!$ &  4.04 & $\!\!\pm\!\!$ &  1.83 \\
0.30--0.36 & 0.329 &  24.6 &  35.59 & $\!\!\pm\!\!$ &  4.18 & $\!\!\pm\!\!$ &  1.77 & 0.330 &  34.9 &  34.02 & $\!\!\pm\!\!$ &  4.00 & $\!\!\pm\!\!$ &  1.65 \\
0.36--0.42 & 0.386 &  25.0 &  29.52 & $\!\!\pm\!\!$ &  3.74 & $\!\!\pm\!\!$ &  1.33 & 0.390 &  35.3 &  34.04 & $\!\!\pm\!\!$ &  4.12 & $\!\!\pm\!\!$ &  1.44 \\
0.42--0.50 & 0.459 &  25.0 &  24.09 & $\!\!\pm\!\!$ &  2.90 & $\!\!\pm\!\!$ &  0.96 & 0.459 &  35.4 &  23.40 & $\!\!\pm\!\!$ &  2.93 & $\!\!\pm\!\!$ &  0.96 \\
0.50--0.60 & 0.544 &  25.5 &  25.01 & $\!\!\pm\!\!$ &  2.66 & $\!\!\pm\!\!$ &  0.99 & 0.541 &  35.0 &  23.64 & $\!\!\pm\!\!$ &  2.72 & $\!\!\pm\!\!$ &  0.93 \\
0.60--0.72 & 0.649 &  24.4 &  16.32 & $\!\!\pm\!\!$ &  1.87 & $\!\!\pm\!\!$ &  0.74 & 0.658 &  34.8 &  14.30 & $\!\!\pm\!\!$ &  1.84 & $\!\!\pm\!\!$ &  0.68 \\
0.72--0.90 &  &  &  \multicolumn{5}{c||}{ } & 0.794 &  34.6 &   9.70 & $\!\!\pm\!\!$ &  1.25 & $\!\!\pm\!\!$ &  0.61 \\
 \hline
 \hline
   & \multicolumn{7}{c||}{$40<\theta<50$}
  & \multicolumn{7}{c|}{$50<\theta<60$} \\
 \hline
 $p_{\rm T}$ & $\langle p_{\rm T} \rangle$ & $\langle \theta \rangle$
  & \multicolumn{5}{c||}{${\rm d}^2 \sigma /{\rm d}p{\rm d}\Omega$}
  &$\langle p_{\rm T} \rangle$ & $\langle \theta \rangle$
  & \multicolumn{5}{c|}{${\rm d}^2 \sigma /{\rm d}p{\rm d}\Omega$} \\
 \hline
0.30--0.36 & 0.327 &  45.1 &  24.69 & $\!\!\pm\!\!$ &  3.52 & $\!\!\pm\!\!$ &  1.03 &  &  &  \multicolumn{5}{c|}{ } \\
0.36--0.42 & 0.393 &  45.2 &  27.74 & $\!\!\pm\!\!$ &  3.71 & $\!\!\pm\!\!$ &  1.10 & 0.387 &  55.8 &  20.10 & $\!\!\pm\!\!$ &  3.07 & $\!\!\pm\!\!$ &  0.88 \\
0.42--0.50 & 0.456 &  45.1 &  23.65 & $\!\!\pm\!\!$ &  2.96 & $\!\!\pm\!\!$ &  0.93 & 0.458 &  55.1 &  18.02 & $\!\!\pm\!\!$ &  2.54 & $\!\!\pm\!\!$ &  0.74 \\
0.50--0.60 & 0.545 &  45.0 &  17.32 & $\!\!\pm\!\!$ &  2.30 & $\!\!\pm\!\!$ &  0.72 & 0.550 &  55.2 &  13.05 & $\!\!\pm\!\!$ &  1.96 & $\!\!\pm\!\!$ &  0.63 \\
0.60--0.72 & 0.657 &  44.7 &  10.89 & $\!\!\pm\!\!$ &  1.66 & $\!\!\pm\!\!$ &  0.56 & 0.663 &  54.5 &   9.40 & $\!\!\pm\!\!$ &  1.58 & $\!\!\pm\!\!$ &  0.54 \\
0.72--0.90 & 0.805 &  44.8 &   9.28 & $\!\!\pm\!\!$ &  1.26 & $\!\!\pm\!\!$ &  0.62 & 0.786 &  54.3 &   5.25 & $\!\!\pm\!\!$ &  0.98 & $\!\!\pm\!\!$ &  0.39 \\
0.90--1.25 & 1.024 &  43.7 &   2.02 & $\!\!\pm\!\!$ &  0.41 & $\!\!\pm\!\!$ &  0.21 & 1.010 &  54.7 &   2.14 & $\!\!\pm\!\!$ &  0.45 & $\!\!\pm\!\!$ &  0.25 \\
 \hline
 \hline
   & \multicolumn{7}{c||}{$60<\theta<75$}
  & \multicolumn{7}{c|}{$75<\theta<90$} \\
 \hline
 $p_{\rm T}$ & $\langle p_{\rm T} \rangle$ & $\langle \theta \rangle$
  & \multicolumn{5}{c||}{${\rm d}^2 \sigma /{\rm d}p{\rm d}\Omega$}
  &$\langle p_{\rm T} \rangle$ & $\langle \theta \rangle$
  & \multicolumn{5}{c|}{${\rm d}^2 \sigma /{\rm d}p{\rm d}\Omega$} \\
 \hline
0.36--0.42 & 0.396 &  68.4 &  23.44 & $\!\!\pm\!\!$ &  2.62 & $\!\!\pm\!\!$ &  1.17 &  &  &  \multicolumn{5}{c|}{ } \\
0.42--0.50 & 0.462 &  66.9 &  21.70 & $\!\!\pm\!\!$ &  2.25 & $\!\!\pm\!\!$ &  0.86 & 0.464 &  81.7 &  12.36 & $\!\!\pm\!\!$ &  1.67 & $\!\!\pm\!\!$ &  0.64 \\
0.50--0.60 & 0.554 &  66.9 &  12.09 & $\!\!\pm\!\!$ &  1.55 & $\!\!\pm\!\!$ &  0.56 & 0.561 &  81.1 &   8.45 & $\!\!\pm\!\!$ &  1.30 & $\!\!\pm\!\!$ &  0.50 \\
0.60--0.72 & 0.665 &  67.3 &   8.91 & $\!\!\pm\!\!$ &  1.28 & $\!\!\pm\!\!$ &  0.58 & 0.652 &  81.7 &   3.95 & $\!\!\pm\!\!$ &  0.83 & $\!\!\pm\!\!$ &  0.33 \\
0.72--0.90 & 0.830 &  66.7 &   3.92 & $\!\!\pm\!\!$ &  0.70 & $\!\!\pm\!\!$ &  0.37 & 0.825 &  81.7 &   2.91 & $\!\!\pm\!\!$ &  0.62 & $\!\!\pm\!\!$ &  0.32 \\
0.90--1.25 & 1.063 &  67.2 &   0.90 & $\!\!\pm\!\!$ &  0.24 & $\!\!\pm\!\!$ &  0.13 & 1.031 &  80.7 &   0.58 & $\!\!\pm\!\!$ &  0.20 & $\!\!\pm\!\!$ &  0.10 \\
 \hline
 \hline
  & \multicolumn{7}{c||}{$90<\theta<105$}
  & \multicolumn{7}{c|}{$105<\theta<125$} \\
 \hline
 $p_{\rm T}$ & $\langle p_{\rm T} \rangle$ & $\langle \theta \rangle$
  & \multicolumn{5}{c||}{${\rm d}^2 \sigma /{\rm d}p{\rm d}\Omega$}
  &$\langle p_{\rm T} \rangle$ & $\langle \theta \rangle$
  & \multicolumn{5}{c|}{${\rm d}^2 \sigma /{\rm d}p{\rm d}\Omega$} \\
 \hline
0.36--0.42 &  &  &  \multicolumn{5}{c||}{ } & 0.393 & 112.5 &   7.10 & $\!\!\pm\!\!$ &  1.30 & $\!\!\pm\!\!$ &  0.38 \\
0.42--0.50 & 0.468 &  96.4 &   6.41 & $\!\!\pm\!\!$ &  1.26 & $\!\!\pm\!\!$ &  0.43 & 0.462 & 114.0 &   3.66 & $\!\!\pm\!\!$ &  0.81 & $\!\!\pm\!\!$ &  0.26 \\
0.50--0.60 & 0.552 &  96.5 &   4.78 & $\!\!\pm\!\!$ &  0.97 & $\!\!\pm\!\!$ &  0.39 & 0.562 & 112.9 &   2.41 & $\!\!\pm\!\!$ &  0.64 & $\!\!\pm\!\!$ &  0.26 \\
0.60--0.72 & 0.679 &  96.1 &   2.80 & $\!\!\pm\!\!$ &  0.74 & $\!\!\pm\!\!$ &  0.29 & 0.689 & 109.0 &   0.88 & $\!\!\pm\!\!$ &  0.36 & $\!\!\pm\!\!$ &  0.14 \\
0.72--0.90 & 0.846 &  97.0 &   0.59 & $\!\!\pm\!\!$ &  0.27 & $\!\!\pm\!\!$ &  0.08 & 0.803 & 109.4 &   0.29 & $\!\!\pm\!\!$ &  0.17 & $\!\!\pm\!\!$ &  0.07 \\
0.90--1.25 & 1.063 &  98.3 &   0.20 & $\!\!\pm\!\!$ &  0.13 & $\!\!\pm\!\!$ &  0.04 &  &  &  \multicolumn{5}{c|}{ } \\
 \hline
 \end{tabular}
 \end{center}
 \end{scriptsize}
 \end{table}

%% file: table.pip.pipbe12.tex
  
 \begin{table}[h]
 \begin{scriptsize}
 \caption{Double-differential inclusive
  cross-section ${\rm d}^2 \sigma /{\rm d}p{\rm d}\Omega$
  [mb/(GeV/{\it c} sr)] of the production of $\pi^+$'s
  in $\pi^+$ + Be $\rightarrow$ $\pi^+$ + X interactions
  with $+12.0$~GeV/{\it c} beam momentum;
  the first error is statistical, the second systematic; 
 $p_{\rm T}$ in GeV/{\it c}, polar angle $\theta$ in degrees.}
 \label{pip.pipbe12}
 \begin{center}
 \begin{tabular}{|c||c|c|rcrcr||c|c|rcrcr|}
 \hline
   & \multicolumn{7}{c||}{$20<\theta<30$}
  & \multicolumn{7}{c|}{$30<\theta<40$} \\
 \hline
 $p_{\rm T}$ & $\langle p_{\rm T} \rangle$ & $\langle \theta \rangle$
  & \multicolumn{5}{c||}{${\rm d}^2 \sigma /{\rm d}p{\rm d}\Omega$}
  &$\langle p_{\rm T} \rangle$ & $\langle \theta \rangle$
  & \multicolumn{5}{c|}{${\rm d}^2 \sigma /{\rm d}p{\rm d}\Omega$} \\
 \hline
0.10--0.13 & 0.117 &  24.8 &  35.14 & $\!\!\pm\!\!$ &  7.12 & $\!\!\pm\!\!$ &  2.68 & 0.115 &  36.3 &  40.25 & $\!\!\pm\!\!$ &  7.54 & $\!\!\pm\!\!$ &  3.21 \\
0.13--0.16 & 0.149 &  24.2 &  49.55 & $\!\!\pm\!\!$ &  7.52 & $\!\!\pm\!\!$ &  3.10 & 0.146 &  33.9 &  43.15 & $\!\!\pm\!\!$ &  7.16 & $\!\!\pm\!\!$ &  2.78 \\
0.16--0.20 & 0.180 &  24.6 &  88.52 & $\!\!\pm\!\!$ &  8.54 & $\!\!\pm\!\!$ &  4.63 & 0.182 &  34.7 &  61.79 & $\!\!\pm\!\!$ &  7.47 & $\!\!\pm\!\!$ &  3.49 \\
0.20--0.24 & 0.221 &  24.9 &  76.40 & $\!\!\pm\!\!$ &  8.04 & $\!\!\pm\!\!$ &  3.55 & 0.220 &  35.4 &  66.66 & $\!\!\pm\!\!$ &  7.43 & $\!\!\pm\!\!$ &  3.15 \\
0.24--0.30 & 0.268 &  24.2 &  93.08 & $\!\!\pm\!\!$ &  7.11 & $\!\!\pm\!\!$ &  3.59 & 0.268 &  34.8 &  60.65 & $\!\!\pm\!\!$ &  5.76 & $\!\!\pm\!\!$ &  2.36 \\
0.30--0.36 & 0.328 &  24.9 &  65.62 & $\!\!\pm\!\!$ &  5.87 & $\!\!\pm\!\!$ &  2.21 & 0.327 &  34.4 &  51.40 & $\!\!\pm\!\!$ &  5.12 & $\!\!\pm\!\!$ &  1.83 \\
0.36--0.42 & 0.387 &  24.2 &  62.31 & $\!\!\pm\!\!$ &  5.59 & $\!\!\pm\!\!$ &  2.10 & 0.383 &  34.5 &  38.13 & $\!\!\pm\!\!$ &  4.49 & $\!\!\pm\!\!$ &  1.37 \\
0.42--0.50 & 0.456 &  24.5 &  45.10 & $\!\!\pm\!\!$ &  4.14 & $\!\!\pm\!\!$ &  1.78 & 0.455 &  34.2 &  21.54 & $\!\!\pm\!\!$ &  2.80 & $\!\!\pm\!\!$ &  0.85 \\
0.50--0.60 & 0.541 &  24.8 &  31.91 & $\!\!\pm\!\!$ &  3.08 & $\!\!\pm\!\!$ &  1.72 & 0.540 &  34.6 &  19.08 & $\!\!\pm\!\!$ &  2.42 & $\!\!\pm\!\!$ &  1.00 \\
0.60--0.72 & 0.651 &  24.5 &  21.60 & $\!\!\pm\!\!$ &  2.28 & $\!\!\pm\!\!$ &  1.79 & 0.645 &  34.9 &  14.67 & $\!\!\pm\!\!$ &  1.84 & $\!\!\pm\!\!$ &  1.10 \\
0.72--0.90 &  &  &  \multicolumn{5}{c||}{ } & 0.793 &  34.4 &   5.69 & $\!\!\pm\!\!$ &  0.84 & $\!\!\pm\!\!$ &  0.66 \\
 \hline
 \hline
   & \multicolumn{7}{c||}{$40<\theta<50$}
  & \multicolumn{7}{c|}{$50<\theta<60$} \\
 \hline
 $p_{\rm T}$ & $\langle p_{\rm T} \rangle$ & $\langle \theta \rangle$
  & \multicolumn{5}{c||}{${\rm d}^2 \sigma /{\rm d}p{\rm d}\Omega$}
  &$\langle p_{\rm T} \rangle$ & $\langle \theta \rangle$
  & \multicolumn{5}{c|}{${\rm d}^2 \sigma /{\rm d}p{\rm d}\Omega$} \\
 \hline
0.10--0.13 & 0.114 &  44.7 &  21.46 & $\!\!\pm\!\!$ &  5.40 & $\!\!\pm\!\!$ &  1.80 &  &  &  \multicolumn{5}{c|}{ } \\
0.13--0.16 & 0.144 &  44.7 &  40.74 & $\!\!\pm\!\!$ &  6.75 & $\!\!\pm\!\!$ &  2.81 & 0.145 &  55.9 &  17.62 & $\!\!\pm\!\!$ &  4.41 & $\!\!\pm\!\!$ &  1.37 \\
0.16--0.20 & 0.181 &  45.8 &  35.78 & $\!\!\pm\!\!$ &  5.57 & $\!\!\pm\!\!$ &  2.06 & 0.178 &  54.4 &  19.83 & $\!\!\pm\!\!$ &  4.15 & $\!\!\pm\!\!$ &  1.20 \\
0.20--0.24 & 0.218 &  45.0 &  37.42 & $\!\!\pm\!\!$ &  5.36 & $\!\!\pm\!\!$ &  1.91 & 0.221 &  55.0 &  29.93 & $\!\!\pm\!\!$ &  5.04 & $\!\!\pm\!\!$ &  1.61 \\
0.24--0.30 & 0.268 &  44.5 &  29.80 & $\!\!\pm\!\!$ &  3.99 & $\!\!\pm\!\!$ &  1.24 & 0.268 &  54.2 &  24.21 & $\!\!\pm\!\!$ &  3.56 & $\!\!\pm\!\!$ &  1.06 \\
0.30--0.36 & 0.326 &  44.8 &  28.71 & $\!\!\pm\!\!$ &  3.82 & $\!\!\pm\!\!$ &  1.11 & 0.326 &  54.7 &  17.58 & $\!\!\pm\!\!$ &  3.06 & $\!\!\pm\!\!$ &  0.74 \\
0.36--0.42 & 0.394 &  44.7 &  30.87 & $\!\!\pm\!\!$ &  4.08 & $\!\!\pm\!\!$ &  1.24 & 0.391 &  54.6 &  14.56 & $\!\!\pm\!\!$ &  2.70 & $\!\!\pm\!\!$ &  0.66 \\
0.42--0.50 & 0.455 &  44.1 &  22.43 & $\!\!\pm\!\!$ &  2.98 & $\!\!\pm\!\!$ &  0.99 & 0.456 &  54.7 &  11.25 & $\!\!\pm\!\!$ &  2.00 & $\!\!\pm\!\!$ &  0.59 \\
0.50--0.60 & 0.551 &  44.8 &   9.91 & $\!\!\pm\!\!$ &  1.69 & $\!\!\pm\!\!$ &  0.56 & 0.545 &  53.9 &  11.76 & $\!\!\pm\!\!$ &  1.94 & $\!\!\pm\!\!$ &  0.71 \\
0.60--0.72 & 0.657 &  44.9 &  10.69 & $\!\!\pm\!\!$ &  1.64 & $\!\!\pm\!\!$ &  0.78 & 0.645 &  54.9 &   3.72 & $\!\!\pm\!\!$ &  0.91 & $\!\!\pm\!\!$ &  0.31 \\
0.72--0.90 & 0.791 &  44.1 &   5.08 & $\!\!\pm\!\!$ &  0.87 & $\!\!\pm\!\!$ &  0.56 & 0.795 &  54.5 &   1.95 & $\!\!\pm\!\!$ &  0.52 & $\!\!\pm\!\!$ &  0.22 \\
0.90--1.25 &  &  &  \multicolumn{5}{c||}{ } & 1.080 &  54.3 &   0.35 & $\!\!\pm\!\!$ &  0.14 & $\!\!\pm\!\!$ &  0.07 \\
 \hline
 \hline
   & \multicolumn{7}{c||}{$60<\theta<75$}
  & \multicolumn{7}{c|}{$75<\theta<90$} \\
 \hline
 $p_{\rm T}$ & $\langle p_{\rm T} \rangle$ & $\langle \theta \rangle$
  & \multicolumn{5}{c||}{${\rm d}^2 \sigma /{\rm d}p{\rm d}\Omega$}
  &$\langle p_{\rm T} \rangle$ & $\langle \theta \rangle$
  & \multicolumn{5}{c|}{${\rm d}^2 \sigma /{\rm d}p{\rm d}\Omega$} \\
 \hline
0.13--0.16 & 0.144 &  67.9 &  15.72 & $\!\!\pm\!\!$ &  3.32 & $\!\!\pm\!\!$ &  1.26 & 0.144 &  81.0 &  10.28 & $\!\!\pm\!\!$ &  2.58 & $\!\!\pm\!\!$ &  0.93 \\
0.16--0.20 & 0.181 &  66.7 &  21.22 & $\!\!\pm\!\!$ &  3.35 & $\!\!\pm\!\!$ &  1.26 & 0.182 &  83.3 &  15.35 & $\!\!\pm\!\!$ &  2.88 & $\!\!\pm\!\!$ &  0.94 \\
0.20--0.24 & 0.218 &  67.5 &  19.17 & $\!\!\pm\!\!$ &  3.19 & $\!\!\pm\!\!$ &  0.99 & 0.219 &  83.8 &   9.02 & $\!\!\pm\!\!$ &  2.07 & $\!\!\pm\!\!$ &  0.55 \\
0.24--0.30 & 0.273 &  67.5 &  16.16 & $\!\!\pm\!\!$ &  2.36 & $\!\!\pm\!\!$ &  0.70 & 0.268 &  82.6 &   7.90 & $\!\!\pm\!\!$ &  1.59 & $\!\!\pm\!\!$ &  0.39 \\
0.30--0.36 & 0.329 &  67.9 &  10.81 & $\!\!\pm\!\!$ &  1.99 & $\!\!\pm\!\!$ &  0.46 & 0.328 &  81.7 &   9.06 & $\!\!\pm\!\!$ &  1.77 & $\!\!\pm\!\!$ &  0.46 \\
0.36--0.42 & 0.394 &  66.0 &  11.42 & $\!\!\pm\!\!$ &  1.95 & $\!\!\pm\!\!$ &  0.55 & 0.389 &  83.7 &   6.21 & $\!\!\pm\!\!$ &  1.46 & $\!\!\pm\!\!$ &  0.38 \\
0.42--0.50 & 0.467 &  66.4 &   5.86 & $\!\!\pm\!\!$ &  1.17 & $\!\!\pm\!\!$ &  0.33 & 0.469 &  81.6 &   4.74 & $\!\!\pm\!\!$ &  1.08 & $\!\!\pm\!\!$ &  0.35 \\
0.50--0.60 & 0.548 &  67.5 &   5.42 & $\!\!\pm\!\!$ &  1.02 & $\!\!\pm\!\!$ &  0.39 & 0.556 &  81.8 &   2.88 & $\!\!\pm\!\!$ &  0.76 & $\!\!\pm\!\!$ &  0.28 \\
0.60--0.72 & 0.672 &  67.1 &   2.62 & $\!\!\pm\!\!$ &  0.68 & $\!\!\pm\!\!$ &  0.25 & 0.664 &  81.2 &   0.97 & $\!\!\pm\!\!$ &  0.41 & $\!\!\pm\!\!$ &  0.12 \\
0.72--0.90 & 0.820 &  65.6 &   1.13 & $\!\!\pm\!\!$ &  0.32 & $\!\!\pm\!\!$ &  0.16 & 0.872 &  81.4 &   0.40 & $\!\!\pm\!\!$ &  0.18 & $\!\!\pm\!\!$ &  0.08 \\
0.90--1.25 & 1.069 &  65.8 &   0.09 & $\!\!\pm\!\!$ &  0.06 & $\!\!\pm\!\!$ &  0.02 &  &  &  \multicolumn{5}{c|}{ } \\
 \hline
 \hline
  & \multicolumn{7}{c||}{$90<\theta<105$}
  & \multicolumn{7}{c|}{$105<\theta<125$} \\
 \hline
 $p_{\rm T}$ & $\langle p_{\rm T} \rangle$ & $\langle \theta \rangle$
  & \multicolumn{5}{c||}{${\rm d}^2 \sigma /{\rm d}p{\rm d}\Omega$}
  &$\langle p_{\rm T} \rangle$ & $\langle \theta \rangle$
  & \multicolumn{5}{c|}{${\rm d}^2 \sigma /{\rm d}p{\rm d}\Omega$} \\
 \hline
0.13--0.16 & 0.146 &  96.5 &  12.51 & $\!\!\pm\!\!$ &  2.88 & $\!\!\pm\!\!$ &  1.07 & 0.143 & 114.9 &   9.25 & $\!\!\pm\!\!$ &  2.12 & $\!\!\pm\!\!$ &  0.76 \\
0.16--0.20 & 0.179 &  96.9 &  10.03 & $\!\!\pm\!\!$ &  2.17 & $\!\!\pm\!\!$ &  0.74 & 0.177 & 116.0 &   9.61 & $\!\!\pm\!\!$ &  1.84 & $\!\!\pm\!\!$ &  0.62 \\
0.20--0.24 & 0.218 &  96.1 &  10.93 & $\!\!\pm\!\!$ &  2.32 & $\!\!\pm\!\!$ &  0.70 & 0.221 & 112.4 &   4.74 & $\!\!\pm\!\!$ &  1.43 & $\!\!\pm\!\!$ &  0.31 \\
0.24--0.30 & 0.268 &  96.6 &   9.37 & $\!\!\pm\!\!$ &  1.88 & $\!\!\pm\!\!$ &  0.53 & 0.265 & 113.9 &   3.88 & $\!\!\pm\!\!$ &  1.00 & $\!\!\pm\!\!$ &  0.27 \\
0.30--0.36 & 0.329 &  97.1 &   3.40 & $\!\!\pm\!\!$ &  1.08 & $\!\!\pm\!\!$ &  0.23 & 0.319 & 113.5 &   2.03 & $\!\!\pm\!\!$ &  0.72 & $\!\!\pm\!\!$ &  0.18 \\
0.36--0.42 & 0.398 &  98.5 &   2.04 & $\!\!\pm\!\!$ &  0.83 & $\!\!\pm\!\!$ &  0.18 & 0.391 & 113.8 &   1.59 & $\!\!\pm\!\!$ &  0.58 & $\!\!\pm\!\!$ &  0.20 \\
0.42--0.50 & 0.475 &  96.8 &   1.64 & $\!\!\pm\!\!$ &  0.67 & $\!\!\pm\!\!$ &  0.18 & 0.472 & 110.7 &   0.37 & $\!\!\pm\!\!$ &  0.26 & $\!\!\pm\!\!$ &  0.05 \\
0.50--0.60 & 0.542 &  95.5 &   1.48 & $\!\!\pm\!\!$ &  0.52 & $\!\!\pm\!\!$ &  0.21 &  &  &  \multicolumn{5}{c|}{ } \\
0.60--0.72 &  &  &  \multicolumn{5}{c||}{ } & 0.675 & 112.8 &   0.52 & $\!\!\pm\!\!$ &  0.26 & $\!\!\pm\!\!$ &  0.12 \\
0.72--0.90 & 0.766 &  93.0 &   0.22 & $\!\!\pm\!\!$ &  0.16 & $\!\!\pm\!\!$ &  0.06 &  &  &  \multicolumn{5}{c|}{ } \\
 \hline
 \end{tabular}
 \end{center}
 \end{scriptsize}
 \end{table}

%% file: table.pim.pipbe12.tex
  
 \begin{table}[h]
 \begin{scriptsize}
 \caption{Double-differential inclusive
  cross-section ${\rm d}^2 \sigma /{\rm d}p{\rm d}\Omega$
  [mb/(GeV/{\it c} sr)] of the production of $\pi^-$'s
  in $\pi^+$ + Be $\rightarrow$ $\pi^-$ + X interactions
  with $+12.0$~GeV/{\it c} beam momentum;
  the first error is statistical, the second systematic; 
 $p_{\rm T}$ in GeV/{\it c}, polar angle $\theta$ in degrees.}
 \label{pim.pipbe12}
 \begin{center}
 \begin{tabular}{|c||c|c|rcrcr||c|c|rcrcr|}
 \hline
   & \multicolumn{7}{c||}{$20<\theta<30$}
  & \multicolumn{7}{c|}{$30<\theta<40$} \\
 \hline
 $p_{\rm T}$ & $\langle p_{\rm T} \rangle$ & $\langle \theta \rangle$
  & \multicolumn{5}{c||}{${\rm d}^2 \sigma /{\rm d}p{\rm d}\Omega$}
  &$\langle p_{\rm T} \rangle$ & $\langle \theta \rangle$
  & \multicolumn{5}{c|}{${\rm d}^2 \sigma /{\rm d}p{\rm d}\Omega$} \\
 \hline
0.10--0.13 & 0.116 &  24.6 &  57.74 & $\!\!\pm\!\!$ &  8.98 & $\!\!\pm\!\!$ &  4.49 & 0.114 &  34.7 &  30.36 & $\!\!\pm\!\!$ &  5.92 & $\!\!\pm\!\!$ &  2.47 \\
0.13--0.16 & 0.145 &  24.9 &  65.62 & $\!\!\pm\!\!$ &  9.02 & $\!\!\pm\!\!$ &  4.41 & 0.147 &  35.3 &  48.43 & $\!\!\pm\!\!$ &  7.48 & $\!\!\pm\!\!$ &  3.26 \\
0.16--0.20 & 0.183 &  24.6 &  64.93 & $\!\!\pm\!\!$ &  7.34 & $\!\!\pm\!\!$ &  3.50 & 0.184 &  34.6 &  38.24 & $\!\!\pm\!\!$ &  5.44 & $\!\!\pm\!\!$ &  2.19 \\
0.20--0.24 & 0.221 &  24.7 &  63.89 & $\!\!\pm\!\!$ &  7.29 & $\!\!\pm\!\!$ &  3.00 & 0.220 &  34.9 &  60.38 & $\!\!\pm\!\!$ &  6.97 & $\!\!\pm\!\!$ &  2.98 \\
0.24--0.30 & 0.271 &  25.1 &  67.36 & $\!\!\pm\!\!$ &  5.97 & $\!\!\pm\!\!$ &  2.52 & 0.271 &  34.8 &  44.52 & $\!\!\pm\!\!$ &  4.82 & $\!\!\pm\!\!$ &  1.78 \\
0.30--0.36 & 0.330 &  24.3 &  54.35 & $\!\!\pm\!\!$ &  5.29 & $\!\!\pm\!\!$ &  1.85 & 0.331 &  35.2 &  34.54 & $\!\!\pm\!\!$ &  4.30 & $\!\!\pm\!\!$ &  1.27 \\
0.36--0.42 & 0.389 &  24.2 &  40.93 & $\!\!\pm\!\!$ &  4.61 & $\!\!\pm\!\!$ &  1.47 & 0.391 &  34.8 &  27.82 & $\!\!\pm\!\!$ &  3.80 & $\!\!\pm\!\!$ &  1.09 \\
0.42--0.50 & 0.456 &  24.8 &  31.08 & $\!\!\pm\!\!$ &  3.50 & $\!\!\pm\!\!$ &  1.31 & 0.462 &  35.4 &  15.87 & $\!\!\pm\!\!$ &  2.38 & $\!\!\pm\!\!$ &  0.74 \\
0.50--0.60 & 0.547 &  24.6 &  21.39 & $\!\!\pm\!\!$ &  2.62 & $\!\!\pm\!\!$ &  1.20 & 0.554 &  35.2 &  13.59 & $\!\!\pm\!\!$ &  2.12 & $\!\!\pm\!\!$ &  0.79 \\
0.60--0.72 & 0.653 &  24.4 &  10.69 & $\!\!\pm\!\!$ &  1.65 & $\!\!\pm\!\!$ &  0.83 & 0.660 &  35.0 &   6.45 & $\!\!\pm\!\!$ &  1.20 & $\!\!\pm\!\!$ &  0.54 \\
0.72--0.90 &  &  &  \multicolumn{5}{c||}{ } & 0.800 &  36.2 &   2.58 & $\!\!\pm\!\!$ &  0.65 & $\!\!\pm\!\!$ &  0.29 \\
 \hline
 \hline
   & \multicolumn{7}{c||}{$40<\theta<50$}
  & \multicolumn{7}{c|}{$50<\theta<60$} \\
 \hline
 $p_{\rm T}$ & $\langle p_{\rm T} \rangle$ & $\langle \theta \rangle$
  & \multicolumn{5}{c||}{${\rm d}^2 \sigma /{\rm d}p{\rm d}\Omega$}
  &$\langle p_{\rm T} \rangle$ & $\langle \theta \rangle$
  & \multicolumn{5}{c|}{${\rm d}^2 \sigma /{\rm d}p{\rm d}\Omega$} \\
 \hline
0.10--0.13 & 0.117 &  45.5 &  18.69 & $\!\!\pm\!\!$ &  4.70 & $\!\!\pm\!\!$ &  1.78 &  &  &  \multicolumn{5}{c|}{ } \\
0.13--0.16 & 0.144 &  44.1 &  32.03 & $\!\!\pm\!\!$ &  5.60 & $\!\!\pm\!\!$ &  2.46 & 0.146 &  55.4 &  33.93 & $\!\!\pm\!\!$ &  6.29 & $\!\!\pm\!\!$ &  2.58 \\
0.16--0.20 & 0.178 &  44.3 &  40.68 & $\!\!\pm\!\!$ &  5.98 & $\!\!\pm\!\!$ &  2.54 & 0.180 &  54.9 &  20.99 & $\!\!\pm\!\!$ &  4.06 & $\!\!\pm\!\!$ &  1.33 \\
0.20--0.24 & 0.220 &  44.5 &  29.01 & $\!\!\pm\!\!$ &  4.59 & $\!\!\pm\!\!$ &  1.58 & 0.218 &  54.8 &  24.86 & $\!\!\pm\!\!$ &  4.42 & $\!\!\pm\!\!$ &  1.41 \\
0.24--0.30 & 0.271 &  44.9 &  34.47 & $\!\!\pm\!\!$ &  4.28 & $\!\!\pm\!\!$ &  1.49 & 0.267 &  54.1 &  15.61 & $\!\!\pm\!\!$ &  2.86 & $\!\!\pm\!\!$ &  0.73 \\
0.30--0.36 & 0.332 &  45.5 &  20.65 & $\!\!\pm\!\!$ &  3.19 & $\!\!\pm\!\!$ &  0.85 & 0.333 &  53.9 &  12.89 & $\!\!\pm\!\!$ &  2.40 & $\!\!\pm\!\!$ &  0.69 \\
0.36--0.42 & 0.386 &  44.9 &  20.49 & $\!\!\pm\!\!$ &  3.25 & $\!\!\pm\!\!$ &  0.90 & 0.387 &  55.2 &  12.65 & $\!\!\pm\!\!$ &  2.53 & $\!\!\pm\!\!$ &  0.66 \\
0.42--0.50 & 0.455 &  44.0 &   9.87 & $\!\!\pm\!\!$ &  1.79 & $\!\!\pm\!\!$ &  0.56 & 0.454 &  53.9 &   7.58 & $\!\!\pm\!\!$ &  1.62 & $\!\!\pm\!\!$ &  0.47 \\
0.50--0.60 & 0.548 &  44.8 &   7.65 & $\!\!\pm\!\!$ &  1.42 & $\!\!\pm\!\!$ &  0.55 & 0.540 &  55.1 &   4.94 & $\!\!\pm\!\!$ &  1.24 & $\!\!\pm\!\!$ &  0.35 \\
0.60--0.72 & 0.649 &  44.0 &   5.49 & $\!\!\pm\!\!$ &  1.17 & $\!\!\pm\!\!$ &  0.49 & 0.642 &  53.2 &   2.85 & $\!\!\pm\!\!$ &  0.79 & $\!\!\pm\!\!$ &  0.28 \\
0.72--0.90 & 0.811 &  43.7 &   1.48 & $\!\!\pm\!\!$ &  0.47 & $\!\!\pm\!\!$ &  0.18 & 0.775 &  55.4 &   1.85 & $\!\!\pm\!\!$ &  0.53 & $\!\!\pm\!\!$ &  0.24 \\
 \hline
 \hline
   & \multicolumn{7}{c||}{$60<\theta<75$}
  & \multicolumn{7}{c|}{$75<\theta<90$} \\
 \hline
 $p_{\rm T}$ & $\langle p_{\rm T} \rangle$ & $\langle \theta \rangle$
  & \multicolumn{5}{c||}{${\rm d}^2 \sigma /{\rm d}p{\rm d}\Omega$}
  &$\langle p_{\rm T} \rangle$ & $\langle \theta \rangle$
  & \multicolumn{5}{c|}{${\rm d}^2 \sigma /{\rm d}p{\rm d}\Omega$} \\
 \hline
0.13--0.16 & 0.146 &  66.3 &  11.91 & $\!\!\pm\!\!$ &  2.79 & $\!\!\pm\!\!$ &  1.00 & 0.144 &  83.5 &  12.66 & $\!\!\pm\!\!$ &  2.89 & $\!\!\pm\!\!$ &  1.02 \\
0.16--0.20 & 0.175 &  66.3 &  16.88 & $\!\!\pm\!\!$ &  2.84 & $\!\!\pm\!\!$ &  1.09 & 0.177 &  83.2 &  14.35 & $\!\!\pm\!\!$ &  2.63 & $\!\!\pm\!\!$ &  1.01 \\
0.20--0.24 & 0.216 &  68.0 &  13.44 & $\!\!\pm\!\!$ &  2.60 & $\!\!\pm\!\!$ &  0.75 & 0.214 &  83.0 &  11.01 & $\!\!\pm\!\!$ &  2.30 & $\!\!\pm\!\!$ &  0.72 \\
0.24--0.30 & 0.266 &  66.8 &   8.88 & $\!\!\pm\!\!$ &  1.68 & $\!\!\pm\!\!$ &  0.43 & 0.261 &  81.9 &  10.46 & $\!\!\pm\!\!$ &  1.88 & $\!\!\pm\!\!$ &  0.55 \\
0.30--0.36 & 0.323 &  67.3 &  11.35 & $\!\!\pm\!\!$ &  1.95 & $\!\!\pm\!\!$ &  0.55 & 0.328 &  82.3 &   4.73 & $\!\!\pm\!\!$ &  1.22 & $\!\!\pm\!\!$ &  0.31 \\
0.36--0.42 & 0.377 &  64.5 &   7.16 & $\!\!\pm\!\!$ &  1.47 & $\!\!\pm\!\!$ &  0.44 & 0.385 &  81.0 &   4.66 & $\!\!\pm\!\!$ &  1.25 & $\!\!\pm\!\!$ &  0.35 \\
0.42--0.50 & 0.456 &  65.8 &   5.81 & $\!\!\pm\!\!$ &  1.10 & $\!\!\pm\!\!$ &  0.44 & 0.448 &  81.3 &   4.22 & $\!\!\pm\!\!$ &  1.00 & $\!\!\pm\!\!$ &  0.38 \\
0.50--0.60 & 0.535 &  66.1 &   3.97 & $\!\!\pm\!\!$ &  0.89 & $\!\!\pm\!\!$ &  0.32 & 0.534 &  81.3 &   2.47 & $\!\!\pm\!\!$ &  0.68 & $\!\!\pm\!\!$ &  0.30 \\
0.60--0.72 & 0.629 &  67.1 &   1.17 & $\!\!\pm\!\!$ &  0.41 & $\!\!\pm\!\!$ &  0.13 & 0.647 &  81.0 &   1.00 & $\!\!\pm\!\!$ &  0.41 & $\!\!\pm\!\!$ &  0.15 \\
0.72--0.90 & 0.811 &  65.2 &   0.85 & $\!\!\pm\!\!$ &  0.30 & $\!\!\pm\!\!$ &  0.12 & 0.773 &  88.4 &   0.21 & $\!\!\pm\!\!$ &  0.15 & $\!\!\pm\!\!$ &  0.05 \\
 \hline
 \hline
  & \multicolumn{7}{c||}{$90<\theta<105$}
  & \multicolumn{7}{c|}{$105<\theta<125$} \\
 \hline
 $p_{\rm T}$ & $\langle p_{\rm T} \rangle$ & $\langle \theta \rangle$
  & \multicolumn{5}{c||}{${\rm d}^2 \sigma /{\rm d}p{\rm d}\Omega$}
  &$\langle p_{\rm T} \rangle$ & $\langle \theta \rangle$
  & \multicolumn{5}{c|}{${\rm d}^2 \sigma /{\rm d}p{\rm d}\Omega$} \\
 \hline
0.13--0.16 & 0.145 &  94.3 &   9.09 & $\!\!\pm\!\!$ &  2.43 & $\!\!\pm\!\!$ &  0.84 & 0.144 & 116.4 &   9.99 & $\!\!\pm\!\!$ &  2.08 & $\!\!\pm\!\!$ &  0.95 \\
0.16--0.20 & 0.180 &  97.5 &   8.41 & $\!\!\pm\!\!$ &  1.84 & $\!\!\pm\!\!$ &  0.75 & 0.178 & 113.4 &   8.07 & $\!\!\pm\!\!$ &  1.51 & $\!\!\pm\!\!$ &  0.78 \\
0.20--0.24 & 0.221 &  96.1 &   7.81 & $\!\!\pm\!\!$ &  1.85 & $\!\!\pm\!\!$ &  0.61 & 0.216 & 111.1 &   3.20 & $\!\!\pm\!\!$ &  1.01 & $\!\!\pm\!\!$ &  0.29 \\
0.24--0.30 & 0.261 &  96.8 &   4.65 & $\!\!\pm\!\!$ &  1.13 & $\!\!\pm\!\!$ &  0.39 & 0.261 & 114.1 &   2.86 & $\!\!\pm\!\!$ &  0.83 & $\!\!\pm\!\!$ &  0.24 \\
0.30--0.36 & 0.325 &  96.2 &   5.81 & $\!\!\pm\!\!$ &  1.33 & $\!\!\pm\!\!$ &  0.52 & 0.340 & 113.0 &   1.37 & $\!\!\pm\!\!$ &  0.56 & $\!\!\pm\!\!$ &  0.14 \\
0.36--0.42 & 0.377 &  97.0 &   3.11 & $\!\!\pm\!\!$ &  0.94 & $\!\!\pm\!\!$ &  0.37 & 0.380 & 111.3 &   1.27 & $\!\!\pm\!\!$ &  0.57 & $\!\!\pm\!\!$ &  0.16 \\
0.42--0.50 & 0.453 &  94.7 &   1.49 & $\!\!\pm\!\!$ &  0.53 & $\!\!\pm\!\!$ &  0.23 & 0.444 & 118.4 &   1.36 & $\!\!\pm\!\!$ &  0.56 & $\!\!\pm\!\!$ &  0.21 \\
0.50--0.60 & 0.524 &  96.4 &   0.60 & $\!\!\pm\!\!$ &  0.35 & $\!\!\pm\!\!$ &  0.11 & 0.509 & 112.6 &   0.24 & $\!\!\pm\!\!$ &  0.17 & $\!\!\pm\!\!$ &  0.05 \\
0.60--0.72 & 0.634 & 101.7 &   0.55 & $\!\!\pm\!\!$ &  0.27 & $\!\!\pm\!\!$ &  0.15 &  &  &  \multicolumn{5}{c|}{ } \\
 \hline
 \end{tabular}
 \end{center}
 \end{scriptsize}
 \end{table}

%% file: table.pro.pimbe12.tex
  
 \begin{table}[h]
 \begin{scriptsize}
 \caption{Double-differential inclusive
  cross-section ${\rm d}^2 \sigma /{\rm d}p{\rm d}\Omega$
  [mb/(GeV/{\it c} sr)] of the production of protons
  in $\pi^-$ + Be $\rightarrow$ p + X interactions
  with $-12.0$~GeV/{\it c} beam momentum;
  the first error is statistical, the second systematic; 
 $p_{\rm T}$ in GeV/{\it c}, polar angle $\theta$ in degrees.}
 \label{pro.pimbe12}
 \begin{center}
 \begin{tabular}{|c||c|c|rcrcr||c|c|rcrcr|}
 \hline
   & \multicolumn{7}{c||}{$20<\theta<30$}
  & \multicolumn{7}{c|}{$30<\theta<40$} \\
 \hline
 $p_{\rm T}$ & $\langle p_{\rm T} \rangle$ & $\langle \theta \rangle$
  & \multicolumn{5}{c||}{${\rm d}^2 \sigma /{\rm d}p{\rm d}\Omega$}
  &$\langle p_{\rm T} \rangle$ & $\langle \theta \rangle$
  & \multicolumn{5}{c|}{${\rm d}^2 \sigma /{\rm d}p{\rm d}\Omega$} \\
 \hline
0.20--0.24 & 0.219 &  24.7 &  24.37 & $\!\!\pm\!\!$ &  1.24 & $\!\!\pm\!\!$ &  1.58 &  &  &  \multicolumn{5}{c|}{ } \\
0.24--0.30 & 0.270 &  25.2 &  27.58 & $\!\!\pm\!\!$ &  1.06 & $\!\!\pm\!\!$ &  1.57 & 0.270 &  35.1 &  26.28 & $\!\!\pm\!\!$ &  1.03 & $\!\!\pm\!\!$ &  1.44 \\
0.30--0.36 & 0.330 &  24.9 &  25.60 & $\!\!\pm\!\!$ &  1.02 & $\!\!\pm\!\!$ &  1.26 & 0.329 &  34.8 &  24.47 & $\!\!\pm\!\!$ &  0.98 & $\!\!\pm\!\!$ &  1.19 \\
0.36--0.42 & 0.390 &  25.1 &  23.23 & $\!\!\pm\!\!$ &  0.95 & $\!\!\pm\!\!$ &  1.07 & 0.389 &  34.8 &  21.37 & $\!\!\pm\!\!$ &  0.93 & $\!\!\pm\!\!$ &  0.96 \\
0.42--0.50 & 0.460 &  24.9 &  20.04 & $\!\!\pm\!\!$ &  0.76 & $\!\!\pm\!\!$ &  0.81 & 0.459 &  35.1 &  19.61 & $\!\!\pm\!\!$ &  0.78 & $\!\!\pm\!\!$ &  0.79 \\
0.50--0.60 & 0.548 &  24.9 &  17.86 & $\!\!\pm\!\!$ &  0.64 & $\!\!\pm\!\!$ &  0.71 & 0.549 &  34.8 &  15.55 & $\!\!\pm\!\!$ &  0.61 & $\!\!\pm\!\!$ &  0.64 \\
0.60--0.72 & 0.656 &  25.0 &  14.16 & $\!\!\pm\!\!$ &  0.51 & $\!\!\pm\!\!$ &  0.66 & 0.654 &  34.7 &  11.53 & $\!\!\pm\!\!$ &  0.48 & $\!\!\pm\!\!$ &  0.56 \\
0.72--0.90 &  &  &  \multicolumn{5}{c||}{ } & 0.802 &  34.8 &   7.41 & $\!\!\pm\!\!$ &  0.32 & $\!\!\pm\!\!$ &  0.47 \\
 \hline
 \hline
   & \multicolumn{7}{c||}{$40<\theta<50$}
  & \multicolumn{7}{c|}{$50<\theta<60$} \\
 \hline
 $p_{\rm T}$ & $\langle p_{\rm T} \rangle$ & $\langle \theta \rangle$
  & \multicolumn{5}{c||}{${\rm d}^2 \sigma /{\rm d}p{\rm d}\Omega$}
  &$\langle p_{\rm T} \rangle$ & $\langle \theta \rangle$
  & \multicolumn{5}{c|}{${\rm d}^2 \sigma /{\rm d}p{\rm d}\Omega$} \\
 \hline
0.30--0.36 & 0.329 &  44.9 &  25.24 & $\!\!\pm\!\!$ &  0.99 & $\!\!\pm\!\!$ &  1.27 &  &  &  \multicolumn{5}{c|}{ } \\
0.36--0.42 & 0.389 &  44.9 &  20.63 & $\!\!\pm\!\!$ &  0.89 & $\!\!\pm\!\!$ &  0.97 & 0.391 &  54.9 &  21.24 & $\!\!\pm\!\!$ &  0.89 & $\!\!\pm\!\!$ &  0.95 \\
0.42--0.50 & 0.460 &  44.9 &  17.35 & $\!\!\pm\!\!$ &  0.72 & $\!\!\pm\!\!$ &  0.67 & 0.458 &  55.0 &  16.67 & $\!\!\pm\!\!$ &  0.71 & $\!\!\pm\!\!$ &  0.72 \\
0.50--0.60 & 0.548 &  44.9 &  14.17 & $\!\!\pm\!\!$ &  0.60 & $\!\!\pm\!\!$ &  0.60 & 0.546 &  54.8 &  12.02 & $\!\!\pm\!\!$ &  0.55 & $\!\!\pm\!\!$ &  0.56 \\
0.60--0.72 & 0.657 &  44.8 &   9.43 & $\!\!\pm\!\!$ &  0.44 & $\!\!\pm\!\!$ &  0.48 & 0.656 &  54.9 &   7.89 & $\!\!\pm\!\!$ &  0.42 & $\!\!\pm\!\!$ &  0.44 \\
0.72--0.90 & 0.798 &  44.9 &   6.12 & $\!\!\pm\!\!$ &  0.31 & $\!\!\pm\!\!$ &  0.41 & 0.794 &  54.7 &   4.43 & $\!\!\pm\!\!$ &  0.26 & $\!\!\pm\!\!$ &  0.33 \\
0.90--1.25 & 1.035 &  44.9 &   1.84 & $\!\!\pm\!\!$ &  0.12 & $\!\!\pm\!\!$ &  0.19 & 1.028 &  55.1 &   1.14 & $\!\!\pm\!\!$ &  0.09 & $\!\!\pm\!\!$ &  0.14 \\
 \hline
 \hline
   & \multicolumn{7}{c||}{$60<\theta<75$}
  & \multicolumn{7}{c|}{$75<\theta<90$} \\
 \hline
 $p_{\rm T}$ & $\langle p_{\rm T} \rangle$ & $\langle \theta \rangle$
  & \multicolumn{5}{c||}{${\rm d}^2 \sigma /{\rm d}p{\rm d}\Omega$}
  &$\langle p_{\rm T} \rangle$ & $\langle \theta \rangle$
  & \multicolumn{5}{c|}{${\rm d}^2 \sigma /{\rm d}p{\rm d}\Omega$} \\
 \hline
0.36--0.42 & 0.390 &  67.3 &  17.90 & $\!\!\pm\!\!$ &  0.65 & $\!\!\pm\!\!$ &  0.85 &  &  &  \multicolumn{5}{c|}{ } \\
0.42--0.50 & 0.459 &  67.4 &  14.10 & $\!\!\pm\!\!$ &  0.51 & $\!\!\pm\!\!$ &  0.56 & 0.459 &  82.1 &  11.64 & $\!\!\pm\!\!$ &  0.47 & $\!\!\pm\!\!$ &  0.55 \\
0.50--0.60 & 0.549 &  67.3 &  10.83 & $\!\!\pm\!\!$ &  0.42 & $\!\!\pm\!\!$ &  0.50 & 0.545 &  82.0 &   7.52 & $\!\!\pm\!\!$ &  0.35 & $\!\!\pm\!\!$ &  0.47 \\
0.60--0.72 & 0.658 &  67.3 &   6.07 & $\!\!\pm\!\!$ &  0.30 & $\!\!\pm\!\!$ &  0.40 & 0.650 &  81.8 &   4.11 & $\!\!\pm\!\!$ &  0.25 & $\!\!\pm\!\!$ &  0.35 \\
0.72--0.90 & 0.795 &  66.7 &   3.21 & $\!\!\pm\!\!$ &  0.18 & $\!\!\pm\!\!$ &  0.30 & 0.794 &  81.7 &   1.65 & $\!\!\pm\!\!$ &  0.13 & $\!\!\pm\!\!$ &  0.19 \\
0.90--1.25 & 1.033 &  66.7 &   0.76 & $\!\!\pm\!\!$ &  0.07 & $\!\!\pm\!\!$ &  0.12 & 1.041 &  81.1 &   0.43 & $\!\!\pm\!\!$ &  0.05 & $\!\!\pm\!\!$ &  0.07 \\
 \hline
 \hline
  & \multicolumn{7}{c||}{$90<\theta<105$}
  & \multicolumn{7}{c|}{$105<\theta<125$} \\
 \hline
 $p_{\rm T}$ & $\langle p_{\rm T} \rangle$ & $\langle \theta \rangle$
  & \multicolumn{5}{c||}{${\rm d}^2 \sigma /{\rm d}p{\rm d}\Omega$}
  &$\langle p_{\rm T} \rangle$ & $\langle \theta \rangle$
  & \multicolumn{5}{c|}{${\rm d}^2 \sigma /{\rm d}p{\rm d}\Omega$} \\
 \hline
0.36--0.42 &  &  &  \multicolumn{5}{c||}{ } & 0.389 & 114.0 &   5.02 & $\!\!\pm\!\!$ &  0.31 & $\!\!\pm\!\!$ &  0.28 \\
0.42--0.50 & 0.457 &  96.5 &   6.38 & $\!\!\pm\!\!$ &  0.35 & $\!\!\pm\!\!$ &  0.43 & 0.454 & 113.8 &   3.29 & $\!\!\pm\!\!$ &  0.22 & $\!\!\pm\!\!$ &  0.24 \\
0.50--0.60 & 0.547 &  97.0 &   3.60 & $\!\!\pm\!\!$ &  0.24 & $\!\!\pm\!\!$ &  0.29 & 0.546 & 113.1 &   1.58 & $\!\!\pm\!\!$ &  0.15 & $\!\!\pm\!\!$ &  0.17 \\
0.60--0.72 & 0.655 &  96.8 &   1.95 & $\!\!\pm\!\!$ &  0.18 & $\!\!\pm\!\!$ &  0.21 & 0.650 & 114.3 &   0.77 & $\!\!\pm\!\!$ &  0.10 & $\!\!\pm\!\!$ &  0.13 \\
0.72--0.90 & 0.788 &  96.4 &   0.65 & $\!\!\pm\!\!$ &  0.08 & $\!\!\pm\!\!$ &  0.09 & 0.786 & 111.7 &   0.15 & $\!\!\pm\!\!$ &  0.04 & $\!\!\pm\!\!$ &  0.04 \\
0.90--1.25 & 1.040 &  95.5 &   0.17 & $\!\!\pm\!\!$ &  0.03 & $\!\!\pm\!\!$ &  0.03 & 1.011 & 111.4 &   0.07 & $\!\!\pm\!\!$ &  0.02 & $\!\!\pm\!\!$ &  0.04 \\
 \hline
 \end{tabular}
 \end{center}
 \end{scriptsize}
 \end{table}

%% file: table.pip.pimbe12.tex
  
 \begin{table}[h]
 \begin{scriptsize}
 \caption{Double-differential inclusive
  cross-section ${\rm d}^2 \sigma /{\rm d}p{\rm d}\Omega$
  [mb/(GeV/{\it c} sr)] of the production of $\pi^+$'s
  in $\pi^-$ + Be $\rightarrow$ $\pi^+$ + X interactions
  with $-12.0$~GeV/{\it c} beam momentum;
  the first error is statistical, the second systematic; 
 $p_{\rm T}$ in GeV/{\it c}, polar angle $\theta$ in degrees.}
 \label{pip.pimbe12}
 \begin{center}
 \begin{tabular}{|c||c|c|rcrcr||c|c|rcrcr|}
 \hline
   & \multicolumn{7}{c||}{$20<\theta<30$}
  & \multicolumn{7}{c|}{$30<\theta<40$} \\
 \hline
 $p_{\rm T}$ & $\langle p_{\rm T} \rangle$ & $\langle \theta \rangle$
  & \multicolumn{5}{c||}{${\rm d}^2 \sigma /{\rm d}p{\rm d}\Omega$}
  &$\langle p_{\rm T} \rangle$ & $\langle \theta \rangle$
  & \multicolumn{5}{c|}{${\rm d}^2 \sigma /{\rm d}p{\rm d}\Omega$} \\
 \hline
0.10--0.13 & 0.115 &  24.7 &  37.81 & $\!\!\pm\!\!$ &  2.11 & $\!\!\pm\!\!$ &  3.18 & 0.116 &  34.8 &  27.36 & $\!\!\pm\!\!$ &  1.72 & $\!\!\pm\!\!$ &  2.38 \\
0.13--0.16 & 0.147 &  24.5 &  56.39 & $\!\!\pm\!\!$ &  2.36 & $\!\!\pm\!\!$ &  3.53 & 0.145 &  34.8 &  36.32 & $\!\!\pm\!\!$ &  1.89 & $\!\!\pm\!\!$ &  2.46 \\
0.16--0.20 & 0.181 &  24.3 &  60.11 & $\!\!\pm\!\!$ &  2.04 & $\!\!\pm\!\!$ &  3.17 & 0.180 &  34.7 &  37.98 & $\!\!\pm\!\!$ &  1.59 & $\!\!\pm\!\!$ &  2.07 \\
0.20--0.24 & 0.220 &  24.6 &  66.01 & $\!\!\pm\!\!$ &  2.09 & $\!\!\pm\!\!$ &  3.01 & 0.220 &  34.7 &  45.76 & $\!\!\pm\!\!$ &  1.75 & $\!\!\pm\!\!$ &  2.22 \\
0.24--0.30 & 0.270 &  24.6 &  60.22 & $\!\!\pm\!\!$ &  1.60 & $\!\!\pm\!\!$ &  2.30 & 0.270 &  34.8 &  38.20 & $\!\!\pm\!\!$ &  1.27 & $\!\!\pm\!\!$ &  1.50 \\
0.30--0.36 & 0.329 &  24.4 &  55.53 & $\!\!\pm\!\!$ &  1.53 & $\!\!\pm\!\!$ &  1.89 & 0.329 &  34.6 &  37.06 & $\!\!\pm\!\!$ &  1.26 & $\!\!\pm\!\!$ &  1.32 \\
0.36--0.42 & 0.389 &  24.5 &  42.58 & $\!\!\pm\!\!$ &  1.34 & $\!\!\pm\!\!$ &  1.48 & 0.388 &  34.7 &  29.05 & $\!\!\pm\!\!$ &  1.12 & $\!\!\pm\!\!$ &  1.05 \\
0.42--0.50 & 0.458 &  24.7 &  33.59 & $\!\!\pm\!\!$ &  1.01 & $\!\!\pm\!\!$ &  1.34 & 0.457 &  34.5 &  20.29 & $\!\!\pm\!\!$ &  0.78 & $\!\!\pm\!\!$ &  0.82 \\
0.50--0.60 & 0.547 &  24.5 &  19.53 & $\!\!\pm\!\!$ &  0.67 & $\!\!\pm\!\!$ &  1.08 & 0.546 &  34.5 &  13.46 & $\!\!\pm\!\!$ &  0.55 & $\!\!\pm\!\!$ &  0.72 \\
0.60--0.72 & 0.655 &  24.6 &  12.66 & $\!\!\pm\!\!$ &  0.48 & $\!\!\pm\!\!$ &  1.03 & 0.655 &  34.6 &   7.63 & $\!\!\pm\!\!$ &  0.37 & $\!\!\pm\!\!$ &  0.59 \\
0.72--0.90 &  &  &  \multicolumn{5}{c||}{ } & 0.799 &  34.5 &   3.21 & $\!\!\pm\!\!$ &  0.17 & $\!\!\pm\!\!$ &  0.40 \\
 \hline
 \hline
   & \multicolumn{7}{c||}{$40<\theta<50$}
  & \multicolumn{7}{c|}{$50<\theta<60$} \\
 \hline
 $p_{\rm T}$ & $\langle p_{\rm T} \rangle$ & $\langle \theta \rangle$
  & \multicolumn{5}{c||}{${\rm d}^2 \sigma /{\rm d}p{\rm d}\Omega$}
  &$\langle p_{\rm T} \rangle$ & $\langle \theta \rangle$
  & \multicolumn{5}{c|}{${\rm d}^2 \sigma /{\rm d}p{\rm d}\Omega$} \\
 \hline
0.10--0.13 & 0.115 &  44.8 &  20.24 & $\!\!\pm\!\!$ &  1.52 & $\!\!\pm\!\!$ &  1.83 &  &  &  \multicolumn{5}{c|}{ } \\
0.13--0.16 & 0.145 &  44.8 &  22.63 & $\!\!\pm\!\!$ &  1.47 & $\!\!\pm\!\!$ &  1.59 & 0.146 &  55.2 &  18.96 & $\!\!\pm\!\!$ &  1.36 & $\!\!\pm\!\!$ &  1.44 \\
0.16--0.20 & 0.180 &  44.9 &  27.72 & $\!\!\pm\!\!$ &  1.37 & $\!\!\pm\!\!$ &  1.61 & 0.180 &  54.8 &  20.77 & $\!\!\pm\!\!$ &  1.17 & $\!\!\pm\!\!$ &  1.23 \\
0.20--0.24 & 0.220 &  44.9 &  27.68 & $\!\!\pm\!\!$ &  1.34 & $\!\!\pm\!\!$ &  1.41 & 0.219 &  54.8 &  17.76 & $\!\!\pm\!\!$ &  1.08 & $\!\!\pm\!\!$ &  0.94 \\
0.24--0.30 & 0.270 &  44.8 &  28.76 & $\!\!\pm\!\!$ &  1.13 & $\!\!\pm\!\!$ &  1.19 & 0.270 &  54.4 &  18.53 & $\!\!\pm\!\!$ &  0.88 & $\!\!\pm\!\!$ &  0.81 \\
0.30--0.36 & 0.330 &  44.6 &  22.58 & $\!\!\pm\!\!$ &  0.96 & $\!\!\pm\!\!$ &  0.87 & 0.330 &  54.7 &  14.68 & $\!\!\pm\!\!$ &  0.78 & $\!\!\pm\!\!$ &  0.60 \\
0.36--0.42 & 0.388 &  44.6 &  18.41 & $\!\!\pm\!\!$ &  0.86 & $\!\!\pm\!\!$ &  0.80 & 0.391 &  54.5 &  10.85 & $\!\!\pm\!\!$ &  0.65 & $\!\!\pm\!\!$ &  0.53 \\
0.42--0.50 & 0.458 &  44.5 &  13.56 & $\!\!\pm\!\!$ &  0.65 & $\!\!\pm\!\!$ &  0.60 & 0.458 &  54.4 &   8.72 & $\!\!\pm\!\!$ &  0.49 & $\!\!\pm\!\!$ &  0.44 \\
0.50--0.60 & 0.545 &  44.8 &   7.94 & $\!\!\pm\!\!$ &  0.41 & $\!\!\pm\!\!$ &  0.54 & 0.547 &  54.5 &   6.43 & $\!\!\pm\!\!$ &  0.39 & $\!\!\pm\!\!$ &  0.40 \\
0.60--0.72 & 0.650 &  44.1 &   4.72 & $\!\!\pm\!\!$ &  0.29 & $\!\!\pm\!\!$ &  0.39 & 0.650 &  54.4 &   3.19 & $\!\!\pm\!\!$ &  0.25 & $\!\!\pm\!\!$ &  0.28 \\
0.72--0.90 & 0.790 &  43.9 &   2.51 & $\!\!\pm\!\!$ &  0.18 & $\!\!\pm\!\!$ &  0.28 & 0.792 &  54.7 &   1.55 & $\!\!\pm\!\!$ &  0.13 & $\!\!\pm\!\!$ &  0.18 \\
0.90--1.25 &  &  &  \multicolumn{5}{c||}{ } & 1.020 &  55.1 &   0.29 & $\!\!\pm\!\!$ &  0.03 & $\!\!\pm\!\!$ &  0.06 \\
 \hline
 \hline
   & \multicolumn{7}{c||}{$60<\theta<75$}
  & \multicolumn{7}{c|}{$75<\theta<90$} \\
 \hline
 $p_{\rm T}$ & $\langle p_{\rm T} \rangle$ & $\langle \theta \rangle$
  & \multicolumn{5}{c||}{${\rm d}^2 \sigma /{\rm d}p{\rm d}\Omega$}
  &$\langle p_{\rm T} \rangle$ & $\langle \theta \rangle$
  & \multicolumn{5}{c|}{${\rm d}^2 \sigma /{\rm d}p{\rm d}\Omega$} \\
 \hline
0.13--0.16 & 0.146 &  66.6 &  15.03 & $\!\!\pm\!\!$ &  0.96 & $\!\!\pm\!\!$ &  1.15 & 0.146 &  81.6 &  10.31 & $\!\!\pm\!\!$ &  0.78 & $\!\!\pm\!\!$ &  0.83 \\
0.16--0.20 & 0.180 &  67.0 &  15.09 & $\!\!\pm\!\!$ &  0.79 & $\!\!\pm\!\!$ &  0.90 & 0.180 &  82.5 &  11.99 & $\!\!\pm\!\!$ &  0.70 & $\!\!\pm\!\!$ &  0.79 \\
0.20--0.24 & 0.220 &  67.3 &  14.63 & $\!\!\pm\!\!$ &  0.78 & $\!\!\pm\!\!$ &  0.75 & 0.218 &  81.8 &   9.59 & $\!\!\pm\!\!$ &  0.63 & $\!\!\pm\!\!$ &  0.56 \\
0.24--0.30 & 0.269 &  66.7 &  12.15 & $\!\!\pm\!\!$ &  0.57 & $\!\!\pm\!\!$ &  0.56 & 0.271 &  81.3 &   7.54 & $\!\!\pm\!\!$ &  0.45 & $\!\!\pm\!\!$ &  0.38 \\
0.30--0.36 & 0.329 &  66.9 &   9.86 & $\!\!\pm\!\!$ &  0.52 & $\!\!\pm\!\!$ &  0.42 & 0.327 &  81.7 &   5.51 & $\!\!\pm\!\!$ &  0.38 & $\!\!\pm\!\!$ &  0.29 \\
0.36--0.42 & 0.389 &  67.3 &   7.28 & $\!\!\pm\!\!$ &  0.43 & $\!\!\pm\!\!$ &  0.39 & 0.387 &  82.1 &   4.04 & $\!\!\pm\!\!$ &  0.31 & $\!\!\pm\!\!$ &  0.30 \\
0.42--0.50 & 0.456 &  66.9 &   5.77 & $\!\!\pm\!\!$ &  0.33 & $\!\!\pm\!\!$ &  0.36 & 0.456 &  81.1 &   3.10 & $\!\!\pm\!\!$ &  0.26 & $\!\!\pm\!\!$ &  0.22 \\
0.50--0.60 & 0.548 &  67.1 &   3.53 & $\!\!\pm\!\!$ &  0.23 & $\!\!\pm\!\!$ &  0.27 & 0.544 &  81.6 &   1.74 & $\!\!\pm\!\!$ &  0.16 & $\!\!\pm\!\!$ &  0.17 \\
0.60--0.72 & 0.656 &  66.2 &   1.91 & $\!\!\pm\!\!$ &  0.15 & $\!\!\pm\!\!$ &  0.20 & 0.647 &  81.5 &   0.89 & $\!\!\pm\!\!$ &  0.10 & $\!\!\pm\!\!$ &  0.12 \\
0.72--0.90 & 0.787 &  66.3 &   0.59 & $\!\!\pm\!\!$ &  0.06 & $\!\!\pm\!\!$ &  0.09 & 0.776 &  80.3 &   0.22 & $\!\!\pm\!\!$ &  0.04 & $\!\!\pm\!\!$ &  0.04 \\
0.90--1.25 & 1.016 &  66.1 &   0.09 & $\!\!\pm\!\!$ &  0.02 & $\!\!\pm\!\!$ &  0.02 & 1.021 &  80.9 &   0.06 & $\!\!\pm\!\!$ &  0.02 & $\!\!\pm\!\!$ &  0.02 \\
 \hline
 \hline
  & \multicolumn{7}{c||}{$90<\theta<105$}
  & \multicolumn{7}{c|}{$105<\theta<125$} \\
 \hline
 $p_{\rm T}$ & $\langle p_{\rm T} \rangle$ & $\langle \theta \rangle$
  & \multicolumn{5}{c||}{${\rm d}^2 \sigma /{\rm d}p{\rm d}\Omega$}
  &$\langle p_{\rm T} \rangle$ & $\langle \theta \rangle$
  & \multicolumn{5}{c|}{${\rm d}^2 \sigma /{\rm d}p{\rm d}\Omega$} \\
 \hline
0.13--0.16 & 0.147 &  97.3 &   8.37 & $\!\!\pm\!\!$ &  0.68 & $\!\!\pm\!\!$ &  0.74 & 0.145 & 114.2 &   6.89 & $\!\!\pm\!\!$ &  0.51 & $\!\!\pm\!\!$ &  0.63 \\
0.16--0.20 & 0.179 &  96.9 &   9.48 & $\!\!\pm\!\!$ &  0.59 & $\!\!\pm\!\!$ &  0.76 & 0.179 & 114.2 &   6.06 & $\!\!\pm\!\!$ &  0.39 & $\!\!\pm\!\!$ &  0.49 \\
0.20--0.24 & 0.221 &  97.2 &   6.57 & $\!\!\pm\!\!$ &  0.48 & $\!\!\pm\!\!$ &  0.55 & 0.218 & 113.7 &   4.46 & $\!\!\pm\!\!$ &  0.35 & $\!\!\pm\!\!$ &  0.32 \\
0.24--0.30 & 0.267 &  97.0 &   5.42 & $\!\!\pm\!\!$ &  0.38 & $\!\!\pm\!\!$ &  0.31 & 0.268 & 113.0 &   3.28 & $\!\!\pm\!\!$ &  0.25 & $\!\!\pm\!\!$ &  0.22 \\
0.30--0.36 & 0.329 &  97.6 &   3.18 & $\!\!\pm\!\!$ &  0.28 & $\!\!\pm\!\!$ &  0.21 & 0.329 & 112.8 &   1.69 & $\!\!\pm\!\!$ &  0.18 & $\!\!\pm\!\!$ &  0.17 \\
0.36--0.42 & 0.386 &  96.8 &   2.11 & $\!\!\pm\!\!$ &  0.22 & $\!\!\pm\!\!$ &  0.20 & 0.387 & 113.5 &   1.20 & $\!\!\pm\!\!$ &  0.16 & $\!\!\pm\!\!$ &  0.15 \\
0.42--0.50 & 0.458 &  96.8 &   1.48 & $\!\!\pm\!\!$ &  0.17 & $\!\!\pm\!\!$ &  0.18 & 0.458 & 111.9 &   0.45 & $\!\!\pm\!\!$ &  0.08 & $\!\!\pm\!\!$ &  0.07 \\
0.50--0.60 & 0.544 &  95.3 &   0.92 & $\!\!\pm\!\!$ &  0.12 & $\!\!\pm\!\!$ &  0.14 & 0.541 & 111.5 &   0.27 & $\!\!\pm\!\!$ &  0.06 & $\!\!\pm\!\!$ &  0.06 \\
0.60--0.72 & 0.645 &  95.3 &   0.38 & $\!\!\pm\!\!$ &  0.07 & $\!\!\pm\!\!$ &  0.08 & 0.660 & 110.2 &   0.10 & $\!\!\pm\!\!$ &  0.03 & $\!\!\pm\!\!$ &  0.04 \\
0.72--0.90 & 0.784 &  97.0 &   0.11 & $\!\!\pm\!\!$ &  0.03 & $\!\!\pm\!\!$ &  0.03 &  &  &  \multicolumn{5}{c|}{ } \\
 \hline
 \end{tabular}
 \end{center}
 \end{scriptsize}
 \end{table}

%% file: table.pim.pimbe12.tex
  
 \begin{table}[h]
 \begin{scriptsize}
 \caption{Double-differential inclusive
  cross-section ${\rm d}^2 \sigma /{\rm d}p{\rm d}\Omega$
  [mb/(GeV/{\it c} sr)] of the production of $\pi^-$'s
  in $\pi^-$ + Be $\rightarrow$ $\pi^-$ + X interactions
  with $-12.0$~GeV/{\it c} beam momentum;
  the first error is statistical, the second systematic; 
 $p_{\rm T}$ in GeV/{\it c}, polar angle $\theta$ in degrees.}
 \label{pim.pimbe12}
 \begin{center}
 \begin{tabular}{|c||c|c|rcrcr||c|c|rcrcr|}
 \hline
   & \multicolumn{7}{c||}{$20<\theta<30$}
  & \multicolumn{7}{c|}{$30<\theta<40$} \\
 \hline
 $p_{\rm T}$ & $\langle p_{\rm T} \rangle$ & $\langle \theta \rangle$
  & \multicolumn{5}{c||}{${\rm d}^2 \sigma /{\rm d}p{\rm d}\Omega$}
  &$\langle p_{\rm T} \rangle$ & $\langle \theta \rangle$
  & \multicolumn{5}{c|}{${\rm d}^2 \sigma /{\rm d}p{\rm d}\Omega$} \\
 \hline
0.10--0.13 & 0.115 &  24.7 &  61.40 & $\!\!\pm\!\!$ &  2.64 & $\!\!\pm\!\!$ &  4.67 & 0.116 &  34.9 &  37.58 & $\!\!\pm\!\!$ &  2.00 & $\!\!\pm\!\!$ &  3.13 \\
0.13--0.16 & 0.146 &  24.6 &  74.65 & $\!\!\pm\!\!$ &  2.71 & $\!\!\pm\!\!$ &  4.59 & 0.145 &  34.8 &  51.55 & $\!\!\pm\!\!$ &  2.24 & $\!\!\pm\!\!$ &  3.35 \\
0.16--0.20 & 0.181 &  24.6 &  89.44 & $\!\!\pm\!\!$ &  2.50 & $\!\!\pm\!\!$ &  4.58 & 0.179 &  34.6 &  56.00 & $\!\!\pm\!\!$ &  1.94 & $\!\!\pm\!\!$ &  2.98 \\
0.20--0.24 & 0.220 &  24.7 &  93.92 & $\!\!\pm\!\!$ &  2.50 & $\!\!\pm\!\!$ &  4.10 & 0.220 &  34.8 &  55.62 & $\!\!\pm\!\!$ &  1.92 & $\!\!\pm\!\!$ &  2.54 \\
0.24--0.30 & 0.269 &  24.7 &  89.26 & $\!\!\pm\!\!$ &  1.99 & $\!\!\pm\!\!$ &  3.18 & 0.270 &  34.5 &  56.17 & $\!\!\pm\!\!$ &  1.57 & $\!\!\pm\!\!$ &  2.08 \\
0.30--0.36 & 0.329 &  24.6 &  80.56 & $\!\!\pm\!\!$ &  1.90 & $\!\!\pm\!\!$ &  2.52 & 0.329 &  34.4 &  48.82 & $\!\!\pm\!\!$ &  1.45 & $\!\!\pm\!\!$ &  1.61 \\
0.36--0.42 & 0.389 &  24.8 &  61.66 & $\!\!\pm\!\!$ &  1.66 & $\!\!\pm\!\!$ &  1.99 & 0.388 &  34.7 &  40.31 & $\!\!\pm\!\!$ &  1.31 & $\!\!\pm\!\!$ &  1.37 \\
0.42--0.50 & 0.457 &  24.6 &  49.30 & $\!\!\pm\!\!$ &  1.28 & $\!\!\pm\!\!$ &  1.96 & 0.457 &  34.6 &  31.63 & $\!\!\pm\!\!$ &  1.01 & $\!\!\pm\!\!$ &  1.25 \\
0.50--0.60 & 0.546 &  24.6 &  35.99 & $\!\!\pm\!\!$ &  0.98 & $\!\!\pm\!\!$ &  1.90 & 0.545 &  34.7 &  19.87 & $\!\!\pm\!\!$ &  0.71 & $\!\!\pm\!\!$ &  1.05 \\
0.60--0.72 & 0.654 &  24.7 &  20.76 & $\!\!\pm\!\!$ &  0.67 & $\!\!\pm\!\!$ &  1.53 & 0.655 &  34.9 &  12.42 & $\!\!\pm\!\!$ &  0.52 & $\!\!\pm\!\!$ &  0.90 \\
0.72--0.90 &  &  &  \multicolumn{5}{c||}{ } & 0.798 &  34.5 &   6.06 & $\!\!\pm\!\!$ &  0.29 & $\!\!\pm\!\!$ &  0.62 \\
 \hline
 \hline
   & \multicolumn{7}{c||}{$40<\theta<50$}
  & \multicolumn{7}{c|}{$50<\theta<60$} \\
 \hline
 $p_{\rm T}$ & $\langle p_{\rm T} \rangle$ & $\langle \theta \rangle$
  & \multicolumn{5}{c||}{${\rm d}^2 \sigma /{\rm d}p{\rm d}\Omega$}
  &$\langle p_{\rm T} \rangle$ & $\langle \theta \rangle$
  & \multicolumn{5}{c|}{${\rm d}^2 \sigma /{\rm d}p{\rm d}\Omega$} \\
 \hline
0.10--0.13 & 0.115 &  44.6 &  26.91 & $\!\!\pm\!\!$ &  1.66 & $\!\!\pm\!\!$ &  2.31 &  &  &  \multicolumn{5}{c|}{ } \\
0.13--0.16 & 0.146 &  44.9 &  32.36 & $\!\!\pm\!\!$ &  1.73 & $\!\!\pm\!\!$ &  2.29 & 0.146 &  54.9 &  26.40 & $\!\!\pm\!\!$ &  1.53 & $\!\!\pm\!\!$ &  1.92 \\
0.16--0.20 & 0.181 &  45.0 &  39.25 & $\!\!\pm\!\!$ &  1.64 & $\!\!\pm\!\!$ &  2.18 & 0.180 &  54.8 &  28.89 & $\!\!\pm\!\!$ &  1.38 & $\!\!\pm\!\!$ &  1.71 \\
0.20--0.24 & 0.220 &  44.9 &  40.39 & $\!\!\pm\!\!$ &  1.62 & $\!\!\pm\!\!$ &  1.94 & 0.220 &  54.9 &  34.31 & $\!\!\pm\!\!$ &  1.52 & $\!\!\pm\!\!$ &  1.68 \\
0.24--0.30 & 0.269 &  44.6 &  35.17 & $\!\!\pm\!\!$ &  1.22 & $\!\!\pm\!\!$ &  1.36 & 0.269 &  54.9 &  25.48 & $\!\!\pm\!\!$ &  1.05 & $\!\!\pm\!\!$ &  1.01 \\
0.30--0.36 & 0.329 &  44.7 &  30.35 & $\!\!\pm\!\!$ &  1.14 & $\!\!\pm\!\!$ &  1.07 & 0.330 &  54.6 &  22.27 & $\!\!\pm\!\!$ &  0.95 & $\!\!\pm\!\!$ &  0.88 \\
0.36--0.42 & 0.389 &  44.9 &  25.10 & $\!\!\pm\!\!$ &  1.01 & $\!\!\pm\!\!$ &  1.00 & 0.389 &  54.4 &  18.13 & $\!\!\pm\!\!$ &  0.85 & $\!\!\pm\!\!$ &  0.84 \\
0.42--0.50 & 0.458 &  44.7 &  19.57 & $\!\!\pm\!\!$ &  0.79 & $\!\!\pm\!\!$ &  0.86 & 0.458 &  54.8 &  12.33 & $\!\!\pm\!\!$ &  0.62 & $\!\!\pm\!\!$ &  0.60 \\
0.50--0.60 & 0.548 &  44.6 &  12.89 & $\!\!\pm\!\!$ &  0.57 & $\!\!\pm\!\!$ &  0.77 & 0.545 &  54.7 &   8.63 & $\!\!\pm\!\!$ &  0.46 & $\!\!\pm\!\!$ &  0.56 \\
0.60--0.72 & 0.655 &  44.7 &   7.61 & $\!\!\pm\!\!$ &  0.40 & $\!\!\pm\!\!$ &  0.60 & 0.650 &  55.0 &   5.04 & $\!\!\pm\!\!$ &  0.32 & $\!\!\pm\!\!$ &  0.42 \\
0.72--0.90 & 0.796 &  44.8 &   4.16 & $\!\!\pm\!\!$ &  0.25 & $\!\!\pm\!\!$ &  0.46 & 0.796 &  54.7 &   2.06 & $\!\!\pm\!\!$ &  0.17 & $\!\!\pm\!\!$ &  0.24 \\
0.90--1.25 &  &  &  \multicolumn{5}{c||}{ } & 1.035 &  54.9 &   0.27 & $\!\!\pm\!\!$ &  0.03 & $\!\!\pm\!\!$ &  0.06 \\
 \hline
 \hline
   & \multicolumn{7}{c||}{$60<\theta<75$}
  & \multicolumn{7}{c|}{$75<\theta<90$} \\
 \hline
 $p_{\rm T}$ & $\langle p_{\rm T} \rangle$ & $\langle \theta \rangle$
  & \multicolumn{5}{c||}{${\rm d}^2 \sigma /{\rm d}p{\rm d}\Omega$}
  &$\langle p_{\rm T} \rangle$ & $\langle \theta \rangle$
  & \multicolumn{5}{c|}{${\rm d}^2 \sigma /{\rm d}p{\rm d}\Omega$} \\
 \hline
0.13--0.16 & 0.145 &  67.0 &  20.47 & $\!\!\pm\!\!$ &  1.10 & $\!\!\pm\!\!$ &  1.50 & 0.145 &  82.7 &  14.90 & $\!\!\pm\!\!$ &  0.88 & $\!\!\pm\!\!$ &  1.39 \\
0.16--0.20 & 0.180 &  66.9 &  22.46 & $\!\!\pm\!\!$ &  0.96 & $\!\!\pm\!\!$ &  1.30 & 0.180 &  82.0 &  17.77 & $\!\!\pm\!\!$ &  0.85 & $\!\!\pm\!\!$ &  1.09 \\
0.20--0.24 & 0.219 &  67.2 &  22.17 & $\!\!\pm\!\!$ &  0.98 & $\!\!\pm\!\!$ &  1.01 & 0.219 &  82.4 &  16.30 & $\!\!\pm\!\!$ &  0.82 & $\!\!\pm\!\!$ &  0.83 \\
0.24--0.30 & 0.269 &  67.2 &  18.12 & $\!\!\pm\!\!$ &  0.70 & $\!\!\pm\!\!$ &  0.70 & 0.269 &  81.8 &  11.77 & $\!\!\pm\!\!$ &  0.57 & $\!\!\pm\!\!$ &  0.49 \\
0.30--0.36 & 0.329 &  66.8 &  15.01 & $\!\!\pm\!\!$ &  0.66 & $\!\!\pm\!\!$ &  0.57 & 0.330 &  82.0 &   8.08 & $\!\!\pm\!\!$ &  0.47 & $\!\!\pm\!\!$ &  0.38 \\
0.36--0.42 & 0.388 &  66.8 &  10.98 & $\!\!\pm\!\!$ &  0.54 & $\!\!\pm\!\!$ &  0.51 & 0.390 &  82.0 &   6.17 & $\!\!\pm\!\!$ &  0.41 & $\!\!\pm\!\!$ &  0.37 \\
0.42--0.50 & 0.458 &  66.5 &   8.19 & $\!\!\pm\!\!$ &  0.40 & $\!\!\pm\!\!$ &  0.43 & 0.458 &  82.0 &   5.03 & $\!\!\pm\!\!$ &  0.33 & $\!\!\pm\!\!$ &  0.35 \\
0.50--0.60 & 0.544 &  66.5 &   5.11 & $\!\!\pm\!\!$ &  0.29 & $\!\!\pm\!\!$ &  0.36 & 0.544 &  81.2 &   2.28 & $\!\!\pm\!\!$ &  0.19 & $\!\!\pm\!\!$ &  0.22 \\
0.60--0.72 & 0.656 &  66.5 &   2.65 & $\!\!\pm\!\!$ &  0.19 & $\!\!\pm\!\!$ &  0.25 & 0.654 &  81.9 &   1.13 & $\!\!\pm\!\!$ &  0.12 & $\!\!\pm\!\!$ &  0.14 \\
0.72--0.90 & 0.794 &  66.4 &   1.09 & $\!\!\pm\!\!$ &  0.10 & $\!\!\pm\!\!$ &  0.16 & 0.792 &  80.8 &   0.37 & $\!\!\pm\!\!$ &  0.05 & $\!\!\pm\!\!$ &  0.07 \\
0.90--1.25 & 1.008 &  66.1 &   0.14 & $\!\!\pm\!\!$ &  0.02 & $\!\!\pm\!\!$ &  0.03 & 0.994 &  82.7 &   0.03 & $\!\!\pm\!\!$ &  0.02 & $\!\!\pm\!\!$ &  0.01 \\
 \hline
 \hline
  & \multicolumn{7}{c||}{$90<\theta<105$}
  & \multicolumn{7}{c|}{$105<\theta<125$} \\
 \hline
 $p_{\rm T}$ & $\langle p_{\rm T} \rangle$ & $\langle \theta \rangle$
  & \multicolumn{5}{c||}{${\rm d}^2 \sigma /{\rm d}p{\rm d}\Omega$}
  &$\langle p_{\rm T} \rangle$ & $\langle \theta \rangle$
  & \multicolumn{5}{c|}{${\rm d}^2 \sigma /{\rm d}p{\rm d}\Omega$} \\
 \hline
0.13--0.16 & 0.144 &  97.5 &  14.53 & $\!\!\pm\!\!$ &  0.88 & $\!\!\pm\!\!$ &  1.45 & 0.145 & 114.6 &  11.01 & $\!\!\pm\!\!$ &  0.64 & $\!\!\pm\!\!$ &  0.85 \\
0.16--0.20 & 0.180 &  97.0 &  13.24 & $\!\!\pm\!\!$ &  0.70 & $\!\!\pm\!\!$ &  0.89 & 0.178 & 114.6 &  10.36 & $\!\!\pm\!\!$ &  0.54 & $\!\!\pm\!\!$ &  0.71 \\
0.20--0.24 & 0.219 &  96.3 &  12.73 & $\!\!\pm\!\!$ &  0.70 & $\!\!\pm\!\!$ &  0.72 & 0.219 & 114.3 &   7.09 & $\!\!\pm\!\!$ &  0.46 & $\!\!\pm\!\!$ &  0.43 \\
0.24--0.30 & 0.268 &  96.9 &   7.41 & $\!\!\pm\!\!$ &  0.46 & $\!\!\pm\!\!$ &  0.36 & 0.266 & 113.9 &   5.35 & $\!\!\pm\!\!$ &  0.33 & $\!\!\pm\!\!$ &  0.32 \\
0.30--0.36 & 0.329 &  96.7 &   6.16 & $\!\!\pm\!\!$ &  0.41 & $\!\!\pm\!\!$ &  0.38 & 0.329 & 113.2 &   2.96 & $\!\!\pm\!\!$ &  0.24 & $\!\!\pm\!\!$ &  0.26 \\
0.36--0.42 & 0.387 &  97.3 &   4.18 & $\!\!\pm\!\!$ &  0.33 & $\!\!\pm\!\!$ &  0.35 & 0.385 & 112.0 &   1.76 & $\!\!\pm\!\!$ &  0.19 & $\!\!\pm\!\!$ &  0.21 \\
0.42--0.50 & 0.457 &  97.0 &   2.70 & $\!\!\pm\!\!$ &  0.23 & $\!\!\pm\!\!$ &  0.26 & 0.458 & 112.4 &   0.92 & $\!\!\pm\!\!$ &  0.12 & $\!\!\pm\!\!$ &  0.14 \\
0.50--0.60 & 0.545 &  96.1 &   1.22 & $\!\!\pm\!\!$ &  0.14 & $\!\!\pm\!\!$ &  0.17 & 0.554 & 111.6 &   0.23 & $\!\!\pm\!\!$ &  0.05 & $\!\!\pm\!\!$ &  0.04 \\
0.60--0.72 & 0.662 &  97.0 &   0.44 & $\!\!\pm\!\!$ &  0.08 & $\!\!\pm\!\!$ &  0.08 & 0.633 & 109.6 &   0.09 & $\!\!\pm\!\!$ &  0.03 & $\!\!\pm\!\!$ &  0.02 \\
0.72--0.90 & 0.769 &  96.3 &   0.08 & $\!\!\pm\!\!$ &  0.02 & $\!\!\pm\!\!$ &  0.02 &  &  &  \multicolumn{5}{c|}{ } \\
 \hline
 \end{tabular}
 \end{center}
 \end{scriptsize}
 \end{table}

%% file: table.pro.probe15.tex
  
 \begin{table}[h]
 \begin{scriptsize}
 \caption{Double-differential inclusive
  cross-section ${\rm d}^2 \sigma /{\rm d}p{\rm d}\Omega$
  [mb/(GeV/{\it c} sr)] of the production of protons
  in p + Be $\rightarrow$ p + X interactions
  with $+15.0$~GeV/{\it c} beam momentum;
  the first error is statistical, the second systematic; 
 $p_{\rm T}$ in GeV/{\it c}, polar angle $\theta$ in degrees.}
 \label{pro.probe15}
 \begin{center}
 \begin{tabular}{|c||c|c|rcrcr||c|c|rcrcr|}
 \hline
   & \multicolumn{7}{c||}{$20<\theta<30$}
  & \multicolumn{7}{c|}{$30<\theta<40$} \\
 \hline
 $p_{\rm T}$ & $\langle p_{\rm T} \rangle$ & $\langle \theta \rangle$
  & \multicolumn{5}{c||}{${\rm d}^2 \sigma /{\rm d}p{\rm d}\Omega$}
  &$\langle p_{\rm T} \rangle$ & $\langle \theta \rangle$
  & \multicolumn{5}{c|}{${\rm d}^2 \sigma /{\rm d}p{\rm d}\Omega$} \\
 \hline
0.20--0.24 & 0.220 &  25.0 &  38.48 & $\!\!\pm\!\!$ &  1.80 & $\!\!\pm\!\!$ &  2.13 &  &  &  \multicolumn{5}{c|}{ } \\
0.24--0.30 & 0.269 &  24.8 &  45.17 & $\!\!\pm\!\!$ &  1.57 & $\!\!\pm\!\!$ &  2.24 & 0.270 &  34.9 &  41.34 & $\!\!\pm\!\!$ &  1.48 & $\!\!\pm\!\!$ &  1.96 \\
0.30--0.36 & 0.329 &  24.9 &  42.76 & $\!\!\pm\!\!$ &  1.52 & $\!\!\pm\!\!$ &  1.89 & 0.329 &  34.7 &  40.33 & $\!\!\pm\!\!$ &  1.47 & $\!\!\pm\!\!$ &  1.63 \\
0.36--0.42 & 0.388 &  24.9 &  38.44 & $\!\!\pm\!\!$ &  1.41 & $\!\!\pm\!\!$ &  1.45 & 0.387 &  34.8 &  36.60 & $\!\!\pm\!\!$ &  1.43 & $\!\!\pm\!\!$ &  1.40 \\
0.42--0.50 & 0.458 &  24.8 &  35.42 & $\!\!\pm\!\!$ &  1.17 & $\!\!\pm\!\!$ &  1.17 & 0.456 &  34.9 &  29.87 & $\!\!\pm\!\!$ &  1.10 & $\!\!\pm\!\!$ &  0.96 \\
0.50--0.60 & 0.545 &  24.8 &  30.77 & $\!\!\pm\!\!$ &  0.97 & $\!\!\pm\!\!$ &  1.01 & 0.544 &  34.9 &  27.28 & $\!\!\pm\!\!$ &  0.95 & $\!\!\pm\!\!$ &  0.86 \\
0.60--0.72 & 0.651 &  24.9 &  23.67 & $\!\!\pm\!\!$ &  0.77 & $\!\!\pm\!\!$ &  0.94 & 0.652 &  34.9 &  20.16 & $\!\!\pm\!\!$ &  0.74 & $\!\!\pm\!\!$ &  0.80 \\
0.72--0.90 &  &  &  \multicolumn{5}{c||}{ } & 0.796 &  34.6 &  13.08 & $\!\!\pm\!\!$ &  0.49 & $\!\!\pm\!\!$ &  0.77 \\
 \hline
 \hline
   & \multicolumn{7}{c||}{$40<\theta<50$}
  & \multicolumn{7}{c|}{$50<\theta<60$} \\
 \hline
 $p_{\rm T}$ & $\langle p_{\rm T} \rangle$ & $\langle \theta \rangle$
  & \multicolumn{5}{c||}{${\rm d}^2 \sigma /{\rm d}p{\rm d}\Omega$}
  &$\langle p_{\rm T} \rangle$ & $\langle \theta \rangle$
  & \multicolumn{5}{c|}{${\rm d}^2 \sigma /{\rm d}p{\rm d}\Omega$} \\
 \hline
0.30--0.36 & 0.329 &  44.9 &  35.19 & $\!\!\pm\!\!$ &  1.34 & $\!\!\pm\!\!$ &  1.31 &  &  &  \multicolumn{5}{c|}{ } \\
0.36--0.42 & 0.389 &  44.8 &  35.39 & $\!\!\pm\!\!$ &  1.38 & $\!\!\pm\!\!$ &  1.18 & 0.390 &  54.9 &  30.73 & $\!\!\pm\!\!$ &  1.25 & $\!\!\pm\!\!$ &  1.00 \\
0.42--0.50 & 0.459 &  45.1 &  27.81 & $\!\!\pm\!\!$ &  1.06 & $\!\!\pm\!\!$ &  0.84 & 0.458 &  55.0 &  27.63 & $\!\!\pm\!\!$ &  1.05 & $\!\!\pm\!\!$ &  0.90 \\
0.50--0.60 & 0.549 &  45.0 &  24.39 & $\!\!\pm\!\!$ &  0.93 & $\!\!\pm\!\!$ &  0.80 & 0.548 &  55.1 &  19.33 & $\!\!\pm\!\!$ &  0.82 & $\!\!\pm\!\!$ &  0.75 \\
0.60--0.72 & 0.656 &  44.9 &  15.07 & $\!\!\pm\!\!$ &  0.65 & $\!\!\pm\!\!$ &  0.67 & 0.656 &  55.0 &  13.96 & $\!\!\pm\!\!$ &  0.64 & $\!\!\pm\!\!$ &  0.68 \\
0.72--0.90 & 0.801 &  44.7 &   9.83 & $\!\!\pm\!\!$ &  0.44 & $\!\!\pm\!\!$ &  0.59 & 0.798 &  54.8 &   7.93 & $\!\!\pm\!\!$ &  0.40 & $\!\!\pm\!\!$ &  0.53 \\
0.90--1.25 & 1.042 &  44.7 &   3.58 & $\!\!\pm\!\!$ &  0.19 & $\!\!\pm\!\!$ &  0.35 & 1.033 &  54.5 &   2.36 & $\!\!\pm\!\!$ &  0.16 & $\!\!\pm\!\!$ &  0.25 \\
 \hline
 \hline
   & \multicolumn{7}{c||}{$60<\theta<75$}
  & \multicolumn{7}{c|}{$75<\theta<90$} \\
 \hline
 $p_{\rm T}$ & $\langle p_{\rm T} \rangle$ & $\langle \theta \rangle$
  & \multicolumn{5}{c||}{${\rm d}^2 \sigma /{\rm d}p{\rm d}\Omega$}
  &$\langle p_{\rm T} \rangle$ & $\langle \theta \rangle$
  & \multicolumn{5}{c|}{${\rm d}^2 \sigma /{\rm d}p{\rm d}\Omega$} \\
 \hline
0.36--0.42 & 0.391 &  67.4 &  30.19 & $\!\!\pm\!\!$ &  1.01 & $\!\!\pm\!\!$ &  1.17 &  &  &  \multicolumn{5}{c|}{ } \\
0.42--0.50 & 0.459 &  67.5 &  22.43 & $\!\!\pm\!\!$ &  0.74 & $\!\!\pm\!\!$ &  0.70 & 0.456 &  81.8 &  18.38 & $\!\!\pm\!\!$ &  0.68 & $\!\!\pm\!\!$ &  0.74 \\
0.50--0.60 & 0.547 &  67.2 &  17.44 & $\!\!\pm\!\!$ &  0.64 & $\!\!\pm\!\!$ &  0.72 & 0.546 &  81.5 &  11.95 & $\!\!\pm\!\!$ &  0.51 & $\!\!\pm\!\!$ &  0.64 \\
0.60--0.72 & 0.656 &  67.2 &   9.77 & $\!\!\pm\!\!$ &  0.44 & $\!\!\pm\!\!$ &  0.59 & 0.653 &  82.0 &   5.69 & $\!\!\pm\!\!$ &  0.35 & $\!\!\pm\!\!$ &  0.46 \\
0.72--0.90 & 0.796 &  66.4 &   5.08 & $\!\!\pm\!\!$ &  0.27 & $\!\!\pm\!\!$ &  0.45 & 0.790 &  81.6 &   1.84 & $\!\!\pm\!\!$ &  0.16 & $\!\!\pm\!\!$ &  0.20 \\
0.90--1.25 & 1.027 &  66.5 &   1.30 & $\!\!\pm\!\!$ &  0.10 & $\!\!\pm\!\!$ &  0.19 & 1.047 &  80.8 &   0.51 & $\!\!\pm\!\!$ &  0.07 & $\!\!\pm\!\!$ &  0.08 \\
 \hline
 \hline
  & \multicolumn{7}{c||}{$90<\theta<105$}
  & \multicolumn{7}{c|}{$105<\theta<125$} \\
 \hline
 $p_{\rm T}$ & $\langle p_{\rm T} \rangle$ & $\langle \theta \rangle$
  & \multicolumn{5}{c||}{${\rm d}^2 \sigma /{\rm d}p{\rm d}\Omega$}
  &$\langle p_{\rm T} \rangle$ & $\langle \theta \rangle$
  & \multicolumn{5}{c|}{${\rm d}^2 \sigma /{\rm d}p{\rm d}\Omega$} \\
 \hline
0.36--0.42 &  &  &  \multicolumn{5}{c||}{ } & 0.389 & 113.1 &   6.78 & $\!\!\pm\!\!$ &  0.41 & $\!\!\pm\!\!$ &  0.27 \\
0.42--0.50 & 0.457 &  96.6 &  11.28 & $\!\!\pm\!\!$ &  0.55 & $\!\!\pm\!\!$ &  0.66 & 0.457 & 112.4 &   5.32 & $\!\!\pm\!\!$ &  0.33 & $\!\!\pm\!\!$ &  0.29 \\
0.50--0.60 & 0.545 &  96.7 &   6.14 & $\!\!\pm\!\!$ &  0.38 & $\!\!\pm\!\!$ &  0.44 & 0.544 & 112.6 &   2.08 & $\!\!\pm\!\!$ &  0.20 & $\!\!\pm\!\!$ &  0.20 \\
0.60--0.72 & 0.650 &  96.3 &   2.59 & $\!\!\pm\!\!$ &  0.25 & $\!\!\pm\!\!$ &  0.26 & 0.651 & 111.1 &   0.74 & $\!\!\pm\!\!$ &  0.11 & $\!\!\pm\!\!$ &  0.11 \\
0.72--0.90 & 0.786 &  96.2 &   1.08 & $\!\!\pm\!\!$ &  0.12 & $\!\!\pm\!\!$ &  0.13 & 0.768 & 110.6 &   0.36 & $\!\!\pm\!\!$ &  0.07 & $\!\!\pm\!\!$ &  0.08 \\
0.90--1.25 & 1.028 &  97.3 &   0.23 & $\!\!\pm\!\!$ &  0.05 & $\!\!\pm\!\!$ &  0.04 & 1.000 & 115.3 &   0.03 & $\!\!\pm\!\!$ &  0.02 & $\!\!\pm\!\!$ &  0.02 \\
 \hline
 \end{tabular}
 \end{center}
 \end{scriptsize}
 \end{table}

%% file: table.pip.probe15.tex
  
 \begin{table}[h]
 \begin{scriptsize}
 \caption{Double-differential inclusive
  cross-section ${\rm d}^2 \sigma /{\rm d}p{\rm d}\Omega$
  [mb/(GeV/{\it c} sr)] of the production of $\pi^+$'s
  in p + Be $\rightarrow$ $\pi^+$ + X interactions
  with $+15.0$~GeV/{\it c} beam momentum;
  the first error is statistical, the second systematic; 
 $p_{\rm T}$ in GeV/{\it c}, polar angle $\theta$ in degrees.}
 \label{pip.probe15}
 \begin{center}
 \begin{tabular}{|c||c|c|rcrcr||c|c|rcrcr|}
 \hline
   & \multicolumn{7}{c||}{$20<\theta<30$}
  & \multicolumn{7}{c|}{$30<\theta<40$} \\
 \hline
 $p_{\rm T}$ & $\langle p_{\rm T} \rangle$ & $\langle \theta \rangle$
  & \multicolumn{5}{c||}{${\rm d}^2 \sigma /{\rm d}p{\rm d}\Omega$}
  &$\langle p_{\rm T} \rangle$ & $\langle \theta \rangle$
  & \multicolumn{5}{c|}{${\rm d}^2 \sigma /{\rm d}p{\rm d}\Omega$} \\
 \hline
0.10--0.13 & 0.116 &  24.5 &  62.64 & $\!\!\pm\!\!$ &  3.15 & $\!\!\pm\!\!$ &  4.54 & 0.116 &  35.0 &  39.74 & $\!\!\pm\!\!$ &  2.47 & $\!\!\pm\!\!$ &  2.93 \\
0.13--0.16 & 0.146 &  24.6 &  73.73 & $\!\!\pm\!\!$ &  3.14 & $\!\!\pm\!\!$ &  4.29 & 0.145 &  34.7 &  49.47 & $\!\!\pm\!\!$ &  2.56 & $\!\!\pm\!\!$ &  2.86 \\
0.16--0.20 & 0.181 &  24.4 &  92.11 & $\!\!\pm\!\!$ &  2.93 & $\!\!\pm\!\!$ &  4.44 & 0.179 &  34.6 &  56.14 & $\!\!\pm\!\!$ &  2.30 & $\!\!\pm\!\!$ &  2.71 \\
0.20--0.24 & 0.220 &  24.6 &  98.89 & $\!\!\pm\!\!$ &  2.98 & $\!\!\pm\!\!$ &  4.10 & 0.220 &  34.6 &  62.38 & $\!\!\pm\!\!$ &  2.36 & $\!\!\pm\!\!$ &  2.57 \\
0.24--0.30 & 0.268 &  24.7 &  93.70 & $\!\!\pm\!\!$ &  2.34 & $\!\!\pm\!\!$ &  3.24 & 0.268 &  34.7 &  61.70 & $\!\!\pm\!\!$ &  1.93 & $\!\!\pm\!\!$ &  2.10 \\
0.30--0.36 & 0.329 &  24.5 &  77.37 & $\!\!\pm\!\!$ &  2.10 & $\!\!\pm\!\!$ &  2.29 & 0.327 &  34.6 &  50.29 & $\!\!\pm\!\!$ &  1.70 & $\!\!\pm\!\!$ &  1.47 \\
0.36--0.42 & 0.387 &  24.7 &  66.57 & $\!\!\pm\!\!$ &  1.99 & $\!\!\pm\!\!$ &  1.94 & 0.387 &  34.6 &  37.79 & $\!\!\pm\!\!$ &  1.44 & $\!\!\pm\!\!$ &  1.08 \\
0.42--0.50 & 0.456 &  24.7 &  48.96 & $\!\!\pm\!\!$ &  1.45 & $\!\!\pm\!\!$ &  1.75 & 0.456 &  34.6 &  32.25 & $\!\!\pm\!\!$ &  1.16 & $\!\!\pm\!\!$ &  1.06 \\
0.50--0.60 & 0.544 &  24.4 &  33.06 & $\!\!\pm\!\!$ &  1.01 & $\!\!\pm\!\!$ &  1.70 & 0.545 &  34.7 &  21.86 & $\!\!\pm\!\!$ &  0.84 & $\!\!\pm\!\!$ &  1.02 \\
0.60--0.72 & 0.652 &  24.5 &  20.05 & $\!\!\pm\!\!$ &  0.70 & $\!\!\pm\!\!$ &  1.55 & 0.651 &  34.6 &  12.97 & $\!\!\pm\!\!$ &  0.58 & $\!\!\pm\!\!$ &  0.91 \\
0.72--0.90 &  &  &  \multicolumn{5}{c||}{ } & 0.790 &  34.9 &   6.23 & $\!\!\pm\!\!$ &  0.30 & $\!\!\pm\!\!$ &  0.70 \\
 \hline
 \hline
   & \multicolumn{7}{c||}{$40<\theta<50$}
  & \multicolumn{7}{c|}{$50<\theta<60$} \\
 \hline
 $p_{\rm T}$ & $\langle p_{\rm T} \rangle$ & $\langle \theta \rangle$
  & \multicolumn{5}{c||}{${\rm d}^2 \sigma /{\rm d}p{\rm d}\Omega$}
  &$\langle p_{\rm T} \rangle$ & $\langle \theta \rangle$
  & \multicolumn{5}{c|}{${\rm d}^2 \sigma /{\rm d}p{\rm d}\Omega$} \\
 \hline
0.10--0.13 & 0.117 &  45.0 &  28.29 & $\!\!\pm\!\!$ &  2.11 & $\!\!\pm\!\!$ &  2.15 &  &  &  \multicolumn{5}{c|}{ } \\
0.13--0.16 & 0.146 &  44.9 &  37.70 & $\!\!\pm\!\!$ &  2.25 & $\!\!\pm\!\!$ &  2.23 & 0.146 &  54.9 &  28.62 & $\!\!\pm\!\!$ &  1.83 & $\!\!\pm\!\!$ &  1.77 \\
0.16--0.20 & 0.180 &  44.6 &  45.28 & $\!\!\pm\!\!$ &  2.08 & $\!\!\pm\!\!$ &  2.22 & 0.179 &  54.7 &  32.54 & $\!\!\pm\!\!$ &  1.70 & $\!\!\pm\!\!$ &  1.60 \\
0.20--0.24 & 0.220 &  44.9 &  38.78 & $\!\!\pm\!\!$ &  1.80 & $\!\!\pm\!\!$ &  1.66 & 0.220 &  54.6 &  34.14 & $\!\!\pm\!\!$ &  1.78 & $\!\!\pm\!\!$ &  1.41 \\
0.24--0.30 & 0.270 &  44.8 &  38.39 & $\!\!\pm\!\!$ &  1.49 & $\!\!\pm\!\!$ &  1.32 & 0.270 &  54.7 &  29.37 & $\!\!\pm\!\!$ &  1.31 & $\!\!\pm\!\!$ &  0.99 \\
0.30--0.36 & 0.328 &  44.8 &  35.72 & $\!\!\pm\!\!$ &  1.43 & $\!\!\pm\!\!$ &  1.06 & 0.329 &  54.6 &  24.12 & $\!\!\pm\!\!$ &  1.18 & $\!\!\pm\!\!$ &  0.72 \\
0.36--0.42 & 0.389 &  44.7 &  28.77 & $\!\!\pm\!\!$ &  1.27 & $\!\!\pm\!\!$ &  0.85 & 0.389 &  54.9 &  18.85 & $\!\!\pm\!\!$ &  1.03 & $\!\!\pm\!\!$ &  0.59 \\
0.42--0.50 & 0.458 &  44.8 &  20.46 & $\!\!\pm\!\!$ &  0.93 & $\!\!\pm\!\!$ &  0.70 & 0.460 &  54.6 &  13.10 & $\!\!\pm\!\!$ &  0.73 & $\!\!\pm\!\!$ &  0.50 \\
0.50--0.60 & 0.548 &  44.2 &  13.22 & $\!\!\pm\!\!$ &  0.65 & $\!\!\pm\!\!$ &  0.61 & 0.544 &  54.5 &   8.56 & $\!\!\pm\!\!$ &  0.52 & $\!\!\pm\!\!$ &  0.43 \\
0.60--0.72 & 0.651 &  44.5 &   8.38 & $\!\!\pm\!\!$ &  0.46 & $\!\!\pm\!\!$ &  0.55 & 0.652 &  54.8 &   5.37 & $\!\!\pm\!\!$ &  0.39 & $\!\!\pm\!\!$ &  0.39 \\
0.72--0.90 & 0.799 &  44.6 &   3.70 & $\!\!\pm\!\!$ &  0.24 & $\!\!\pm\!\!$ &  0.38 & 0.791 &  54.6 &   2.28 & $\!\!\pm\!\!$ &  0.18 & $\!\!\pm\!\!$ &  0.24 \\
0.90--1.25 &  &  &  \multicolumn{5}{c||}{ } & 1.031 &  54.2 &   0.46 & $\!\!\pm\!\!$ &  0.05 & $\!\!\pm\!\!$ &  0.08 \\
 \hline
 \hline
   & \multicolumn{7}{c||}{$60<\theta<75$}
  & \multicolumn{7}{c|}{$75<\theta<90$} \\
 \hline
 $p_{\rm T}$ & $\langle p_{\rm T} \rangle$ & $\langle \theta \rangle$
  & \multicolumn{5}{c||}{${\rm d}^2 \sigma /{\rm d}p{\rm d}\Omega$}
  &$\langle p_{\rm T} \rangle$ & $\langle \theta \rangle$
  & \multicolumn{5}{c|}{${\rm d}^2 \sigma /{\rm d}p{\rm d}\Omega$} \\
 \hline
0.13--0.16 & 0.145 &  67.4 &  19.67 & $\!\!\pm\!\!$ &  1.23 & $\!\!\pm\!\!$ &  1.28 & 0.147 &  82.2 &  17.61 & $\!\!\pm\!\!$ &  1.17 & $\!\!\pm\!\!$ &  1.23 \\
0.16--0.20 & 0.181 &  67.4 &  22.41 & $\!\!\pm\!\!$ &  1.15 & $\!\!\pm\!\!$ &  1.12 & 0.179 &  82.1 &  17.64 & $\!\!\pm\!\!$ &  0.98 & $\!\!\pm\!\!$ &  0.92 \\
0.20--0.24 & 0.220 &  67.2 &  25.84 & $\!\!\pm\!\!$ &  1.21 & $\!\!\pm\!\!$ &  1.04 & 0.221 &  82.2 &  15.72 & $\!\!\pm\!\!$ &  0.92 & $\!\!\pm\!\!$ &  0.61 \\
0.24--0.30 & 0.269 &  66.8 &  18.14 & $\!\!\pm\!\!$ &  0.83 & $\!\!\pm\!\!$ &  0.59 & 0.266 &  82.4 &  12.29 & $\!\!\pm\!\!$ &  0.69 & $\!\!\pm\!\!$ &  0.39 \\
0.30--0.36 & 0.329 &  66.8 &  13.63 & $\!\!\pm\!\!$ &  0.70 & $\!\!\pm\!\!$ &  0.41 & 0.327 &  82.3 &   9.85 & $\!\!\pm\!\!$ &  0.63 & $\!\!\pm\!\!$ &  0.34 \\
0.36--0.42 & 0.389 &  67.1 &  11.48 & $\!\!\pm\!\!$ &  0.65 & $\!\!\pm\!\!$ &  0.45 & 0.389 &  82.1 &   5.41 & $\!\!\pm\!\!$ &  0.44 & $\!\!\pm\!\!$ &  0.24 \\
0.42--0.50 & 0.456 &  66.9 &   7.64 & $\!\!\pm\!\!$ &  0.44 & $\!\!\pm\!\!$ &  0.34 & 0.460 &  82.0 &   3.58 & $\!\!\pm\!\!$ &  0.30 & $\!\!\pm\!\!$ &  0.22 \\
0.50--0.60 & 0.544 &  66.4 &   6.23 & $\!\!\pm\!\!$ &  0.38 & $\!\!\pm\!\!$ &  0.38 & 0.548 &  81.5 &   2.26 & $\!\!\pm\!\!$ &  0.21 & $\!\!\pm\!\!$ &  0.18 \\
0.60--0.72 & 0.654 &  66.4 &   2.57 & $\!\!\pm\!\!$ &  0.20 & $\!\!\pm\!\!$ &  0.22 & 0.657 &  81.3 &   1.06 & $\!\!\pm\!\!$ &  0.13 & $\!\!\pm\!\!$ &  0.11 \\
0.72--0.90 & 0.789 &  66.8 &   1.05 & $\!\!\pm\!\!$ &  0.11 & $\!\!\pm\!\!$ &  0.14 & 0.787 &  81.0 &   0.32 & $\!\!\pm\!\!$ &  0.06 & $\!\!\pm\!\!$ &  0.05 \\
0.90--1.25 & 1.008 &  64.8 &   0.16 & $\!\!\pm\!\!$ &  0.02 & $\!\!\pm\!\!$ &  0.03 & 1.052 &  78.8 &   0.05 & $\!\!\pm\!\!$ &  0.02 & $\!\!\pm\!\!$ &  0.02 \\
 \hline
 \hline
  & \multicolumn{7}{c||}{$90<\theta<105$}
  & \multicolumn{7}{c|}{$105<\theta<125$} \\
 \hline
 $p_{\rm T}$ & $\langle p_{\rm T} \rangle$ & $\langle \theta \rangle$
  & \multicolumn{5}{c||}{${\rm d}^2 \sigma /{\rm d}p{\rm d}\Omega$}
  &$\langle p_{\rm T} \rangle$ & $\langle \theta \rangle$
  & \multicolumn{5}{c|}{${\rm d}^2 \sigma /{\rm d}p{\rm d}\Omega$} \\
 \hline
0.13--0.16 & 0.146 &  96.9 &  14.04 & $\!\!\pm\!\!$ &  0.99 & $\!\!\pm\!\!$ &  0.96 & 0.144 & 114.7 &  12.91 & $\!\!\pm\!\!$ &  0.87 & $\!\!\pm\!\!$ &  0.85 \\
0.16--0.20 & 0.179 &  97.0 &  15.15 & $\!\!\pm\!\!$ &  0.88 & $\!\!\pm\!\!$ &  0.80 & 0.180 & 113.8 &  10.22 & $\!\!\pm\!\!$ &  0.62 & $\!\!\pm\!\!$ &  0.56 \\
0.20--0.24 & 0.219 &  97.2 &  10.26 & $\!\!\pm\!\!$ &  0.69 & $\!\!\pm\!\!$ &  0.41 & 0.221 & 112.7 &   7.81 & $\!\!\pm\!\!$ &  0.58 & $\!\!\pm\!\!$ &  0.36 \\
0.24--0.30 & 0.269 &  96.7 &   8.02 & $\!\!\pm\!\!$ &  0.55 & $\!\!\pm\!\!$ &  0.28 & 0.270 & 113.8 &   5.14 & $\!\!\pm\!\!$ &  0.38 & $\!\!\pm\!\!$ &  0.25 \\
0.30--0.36 & 0.327 &  97.4 &   4.92 & $\!\!\pm\!\!$ &  0.39 & $\!\!\pm\!\!$ &  0.24 & 0.329 & 113.7 &   2.55 & $\!\!\pm\!\!$ &  0.26 & $\!\!\pm\!\!$ &  0.17 \\
0.36--0.42 & 0.387 &  96.8 &   3.39 & $\!\!\pm\!\!$ &  0.34 & $\!\!\pm\!\!$ &  0.22 & 0.385 & 111.7 &   1.55 & $\!\!\pm\!\!$ &  0.20 & $\!\!\pm\!\!$ &  0.15 \\
0.42--0.50 & 0.458 &  96.4 &   2.08 & $\!\!\pm\!\!$ &  0.24 & $\!\!\pm\!\!$ &  0.18 & 0.455 & 113.3 &   0.72 & $\!\!\pm\!\!$ &  0.11 & $\!\!\pm\!\!$ &  0.08 \\
0.50--0.60 & 0.545 &  97.4 &   1.09 & $\!\!\pm\!\!$ &  0.15 & $\!\!\pm\!\!$ &  0.13 & 0.539 & 111.1 &   0.26 & $\!\!\pm\!\!$ &  0.06 & $\!\!\pm\!\!$ &  0.04 \\
0.60--0.72 & 0.646 &  96.2 &   0.25 & $\!\!\pm\!\!$ &  0.06 & $\!\!\pm\!\!$ &  0.04 & 0.657 & 111.5 &   0.08 & $\!\!\pm\!\!$ &  0.03 & $\!\!\pm\!\!$ &  0.02 \\
0.72--0.90 & 0.808 &  94.3 &   0.10 & $\!\!\pm\!\!$ &  0.03 & $\!\!\pm\!\!$ &  0.02 &  &  &  \multicolumn{5}{c|}{ } \\
 \hline
 \end{tabular}
 \end{center}
 \end{scriptsize}
 \end{table}

%% file: table.pim.probe15.tex
  
 \begin{table}[h]
 \begin{scriptsize}
 \caption{Double-differential inclusive
  cross-section ${\rm d}^2 \sigma /{\rm d}p{\rm d}\Omega$
  [mb/(GeV/{\it c} sr)] of the production of $\pi^-$'s
  in p + Be $\rightarrow$ $\pi^-$ + X interactions
  with $+15.0$~GeV/{\it c} beam momentum;
  the first error is statistical, the second systematic; 
 $p_{\rm T}$ in GeV/{\it c}, polar angle $\theta$ in degrees.}
 \label{pim.probe15}
 \begin{center}
 \begin{tabular}{|c||c|c|rcrcr||c|c|rcrcr|}
 \hline
   & \multicolumn{7}{c||}{$20<\theta<30$}
  & \multicolumn{7}{c|}{$30<\theta<40$} \\
 \hline
 $p_{\rm T}$ & $\langle p_{\rm T} \rangle$ & $\langle \theta \rangle$
  & \multicolumn{5}{c||}{${\rm d}^2 \sigma /{\rm d}p{\rm d}\Omega$}
  &$\langle p_{\rm T} \rangle$ & $\langle \theta \rangle$
  & \multicolumn{5}{c|}{${\rm d}^2 \sigma /{\rm d}p{\rm d}\Omega$} \\
 \hline
0.10--0.13 & 0.116 &  24.7 &  58.26 & $\!\!\pm\!\!$ &  3.00 & $\!\!\pm\!\!$ &  4.14 & 0.116 &  34.8 &  41.24 & $\!\!\pm\!\!$ &  2.36 & $\!\!\pm\!\!$ &  3.18 \\
0.13--0.16 & 0.146 &  24.6 &  73.61 & $\!\!\pm\!\!$ &  3.09 & $\!\!\pm\!\!$ &  4.36 & 0.146 &  34.8 &  48.14 & $\!\!\pm\!\!$ &  2.37 & $\!\!\pm\!\!$ &  2.87 \\
0.16--0.20 & 0.180 &  24.8 &  84.86 & $\!\!\pm\!\!$ &  2.76 & $\!\!\pm\!\!$ &  4.12 & 0.181 &  34.7 &  58.68 & $\!\!\pm\!\!$ &  2.32 & $\!\!\pm\!\!$ &  2.89 \\
0.20--0.24 & 0.221 &  24.8 &  81.72 & $\!\!\pm\!\!$ &  2.66 & $\!\!\pm\!\!$ &  3.33 & 0.220 &  34.8 &  59.64 & $\!\!\pm\!\!$ &  2.32 & $\!\!\pm\!\!$ &  2.46 \\
0.24--0.30 & 0.270 &  24.8 &  80.92 & $\!\!\pm\!\!$ &  2.18 & $\!\!\pm\!\!$ &  2.65 & 0.270 &  34.8 &  54.98 & $\!\!\pm\!\!$ &  1.79 & $\!\!\pm\!\!$ &  1.83 \\
0.30--0.36 & 0.330 &  24.7 &  72.41 & $\!\!\pm\!\!$ &  2.10 & $\!\!\pm\!\!$ &  2.01 & 0.330 &  34.8 &  46.11 & $\!\!\pm\!\!$ &  1.61 & $\!\!\pm\!\!$ &  1.30 \\
0.36--0.42 & 0.389 &  24.7 &  53.33 & $\!\!\pm\!\!$ &  1.74 & $\!\!\pm\!\!$ &  1.52 & 0.391 &  34.9 &  39.69 & $\!\!\pm\!\!$ &  1.56 & $\!\!\pm\!\!$ &  1.14 \\
0.42--0.50 & 0.458 &  24.6 &  38.90 & $\!\!\pm\!\!$ &  1.28 & $\!\!\pm\!\!$ &  1.39 & 0.460 &  34.8 &  25.49 & $\!\!\pm\!\!$ &  1.00 & $\!\!\pm\!\!$ &  0.90 \\
0.50--0.60 & 0.550 &  24.6 &  27.89 & $\!\!\pm\!\!$ &  1.00 & $\!\!\pm\!\!$ &  1.41 & 0.550 &  34.7 &  17.13 & $\!\!\pm\!\!$ &  0.75 & $\!\!\pm\!\!$ &  0.84 \\
0.60--0.72 & 0.656 &  24.7 &  14.96 & $\!\!\pm\!\!$ &  0.64 & $\!\!\pm\!\!$ &  1.07 & 0.661 &  34.6 &   9.58 & $\!\!\pm\!\!$ &  0.50 & $\!\!\pm\!\!$ &  0.66 \\
0.72--0.90 &  &  &  \multicolumn{5}{c||}{ } & 0.808 &  35.1 &   4.44 & $\!\!\pm\!\!$ &  0.28 & $\!\!\pm\!\!$ &  0.44 \\
 \hline
 \hline
   & \multicolumn{7}{c||}{$40<\theta<50$}
  & \multicolumn{7}{c|}{$50<\theta<60$} \\
 \hline
 $p_{\rm T}$ & $\langle p_{\rm T} \rangle$ & $\langle \theta \rangle$
  & \multicolumn{5}{c||}{${\rm d}^2 \sigma /{\rm d}p{\rm d}\Omega$}
  &$\langle p_{\rm T} \rangle$ & $\langle \theta \rangle$
  & \multicolumn{5}{c|}{${\rm d}^2 \sigma /{\rm d}p{\rm d}\Omega$} \\
 \hline
0.10--0.13 & 0.117 &  44.8 &  29.37 & $\!\!\pm\!\!$ &  2.10 & $\!\!\pm\!\!$ &  2.53 &  &  &  \multicolumn{5}{c|}{ } \\
0.13--0.16 & 0.145 &  44.9 &  34.86 & $\!\!\pm\!\!$ &  2.07 & $\!\!\pm\!\!$ &  2.21 & 0.146 &  54.8 &  27.80 & $\!\!\pm\!\!$ &  1.76 & $\!\!\pm\!\!$ &  1.97 \\
0.16--0.20 & 0.180 &  44.8 &  38.95 & $\!\!\pm\!\!$ &  1.87 & $\!\!\pm\!\!$ &  1.95 & 0.181 &  54.6 &  29.48 & $\!\!\pm\!\!$ &  1.56 & $\!\!\pm\!\!$ &  1.54 \\
0.20--0.24 & 0.220 &  44.7 &  38.21 & $\!\!\pm\!\!$ &  1.77 & $\!\!\pm\!\!$ &  1.61 & 0.221 &  54.7 &  31.13 & $\!\!\pm\!\!$ &  1.65 & $\!\!\pm\!\!$ &  1.30 \\
0.24--0.30 & 0.270 &  44.6 &  35.67 & $\!\!\pm\!\!$ &  1.42 & $\!\!\pm\!\!$ &  1.20 & 0.269 &  54.7 &  28.04 & $\!\!\pm\!\!$ &  1.30 & $\!\!\pm\!\!$ &  0.93 \\
0.30--0.36 & 0.329 &  44.6 &  29.03 & $\!\!\pm\!\!$ &  1.26 & $\!\!\pm\!\!$ &  0.84 & 0.330 &  54.5 &  22.78 & $\!\!\pm\!\!$ &  1.09 & $\!\!\pm\!\!$ &  0.73 \\
0.36--0.42 & 0.389 &  44.8 &  24.70 & $\!\!\pm\!\!$ &  1.19 & $\!\!\pm\!\!$ &  0.77 & 0.388 &  54.9 &  17.83 & $\!\!\pm\!\!$ &  0.93 & $\!\!\pm\!\!$ &  0.60 \\
0.42--0.50 & 0.458 &  44.8 &  18.20 & $\!\!\pm\!\!$ &  0.85 & $\!\!\pm\!\!$ &  0.69 & 0.459 &  54.8 &  10.94 & $\!\!\pm\!\!$ &  0.64 & $\!\!\pm\!\!$ &  0.44 \\
0.50--0.60 & 0.547 &  44.6 &  12.03 & $\!\!\pm\!\!$ &  0.61 & $\!\!\pm\!\!$ &  0.64 & 0.544 &  54.8 &   8.73 & $\!\!\pm\!\!$ &  0.54 & $\!\!\pm\!\!$ &  0.52 \\
0.60--0.72 & 0.655 &  44.7 &   6.42 & $\!\!\pm\!\!$ &  0.39 & $\!\!\pm\!\!$ &  0.48 & 0.652 &  54.6 &   4.39 & $\!\!\pm\!\!$ &  0.35 & $\!\!\pm\!\!$ &  0.35 \\
0.72--0.90 & 0.791 &  44.6 &   3.22 & $\!\!\pm\!\!$ &  0.26 & $\!\!\pm\!\!$ &  0.34 & 0.791 &  54.7 &   1.51 & $\!\!\pm\!\!$ &  0.14 & $\!\!\pm\!\!$ &  0.17 \\
0.90--1.25 &  &  &  \multicolumn{5}{c||}{ } & 1.028 &  54.9 &   0.44 & $\!\!\pm\!\!$ &  0.06 & $\!\!\pm\!\!$ &  0.08 \\
 \hline
 \hline
   & \multicolumn{7}{c||}{$60<\theta<75$}
  & \multicolumn{7}{c|}{$75<\theta<90$} \\
 \hline
 $p_{\rm T}$ & $\langle p_{\rm T} \rangle$ & $\langle \theta \rangle$
  & \multicolumn{5}{c||}{${\rm d}^2 \sigma /{\rm d}p{\rm d}\Omega$}
  &$\langle p_{\rm T} \rangle$ & $\langle \theta \rangle$
  & \multicolumn{5}{c|}{${\rm d}^2 \sigma /{\rm d}p{\rm d}\Omega$} \\
 \hline
0.13--0.16 & 0.145 &  67.2 &  22.57 & $\!\!\pm\!\!$ &  1.34 & $\!\!\pm\!\!$ &  1.67 & 0.146 &  81.7 &  16.54 & $\!\!\pm\!\!$ &  1.11 & $\!\!\pm\!\!$ &  1.23 \\
0.16--0.20 & 0.180 &  67.0 &  21.41 & $\!\!\pm\!\!$ &  1.03 & $\!\!\pm\!\!$ &  1.11 & 0.180 &  82.5 &  16.80 & $\!\!\pm\!\!$ &  0.89 & $\!\!\pm\!\!$ &  0.95 \\
0.20--0.24 & 0.220 &  67.3 &  21.47 & $\!\!\pm\!\!$ &  1.05 & $\!\!\pm\!\!$ &  0.83 & 0.220 &  82.0 &  17.27 & $\!\!\pm\!\!$ &  0.98 & $\!\!\pm\!\!$ &  0.80 \\
0.24--0.30 & 0.269 &  66.9 &  19.54 & $\!\!\pm\!\!$ &  0.87 & $\!\!\pm\!\!$ &  0.61 & 0.268 &  82.1 &  12.08 & $\!\!\pm\!\!$ &  0.65 & $\!\!\pm\!\!$ &  0.38 \\
0.30--0.36 & 0.329 &  66.9 &  12.82 & $\!\!\pm\!\!$ &  0.66 & $\!\!\pm\!\!$ &  0.39 & 0.329 &  82.0 &   7.94 & $\!\!\pm\!\!$ &  0.50 & $\!\!\pm\!\!$ &  0.29 \\
0.36--0.42 & 0.387 &  66.9 &  10.25 & $\!\!\pm\!\!$ &  0.56 & $\!\!\pm\!\!$ &  0.39 & 0.389 &  81.6 &   6.04 & $\!\!\pm\!\!$ &  0.46 & $\!\!\pm\!\!$ &  0.30 \\
0.42--0.50 & 0.457 &  66.9 &   7.75 & $\!\!\pm\!\!$ &  0.45 & $\!\!\pm\!\!$ &  0.39 & 0.457 &  81.6 &   4.24 & $\!\!\pm\!\!$ &  0.34 & $\!\!\pm\!\!$ &  0.29 \\
0.50--0.60 & 0.546 &  66.8 &   5.93 & $\!\!\pm\!\!$ &  0.37 & $\!\!\pm\!\!$ &  0.39 & 0.543 &  82.0 &   2.64 & $\!\!\pm\!\!$ &  0.23 & $\!\!\pm\!\!$ &  0.22 \\
0.60--0.72 & 0.651 &  67.0 &   2.23 & $\!\!\pm\!\!$ &  0.17 & $\!\!\pm\!\!$ &  0.20 & 0.651 &  82.8 &   0.76 & $\!\!\pm\!\!$ &  0.11 & $\!\!\pm\!\!$ &  0.09 \\
0.72--0.90 & 0.798 &  66.6 &   0.84 & $\!\!\pm\!\!$ &  0.09 & $\!\!\pm\!\!$ &  0.11 & 0.788 &  82.4 &   0.32 & $\!\!\pm\!\!$ &  0.06 & $\!\!\pm\!\!$ &  0.05 \\
0.90--1.25 & 1.025 &  66.3 &   0.18 & $\!\!\pm\!\!$ &  0.03 & $\!\!\pm\!\!$ &  0.04 & 1.018 &  80.3 &   0.05 & $\!\!\pm\!\!$ &  0.02 & $\!\!\pm\!\!$ &  0.02 \\
 \hline
 \hline
  & \multicolumn{7}{c||}{$90<\theta<105$}
  & \multicolumn{7}{c|}{$105<\theta<125$} \\
 \hline
 $p_{\rm T}$ & $\langle p_{\rm T} \rangle$ & $\langle \theta \rangle$
  & \multicolumn{5}{c||}{${\rm d}^2 \sigma /{\rm d}p{\rm d}\Omega$}
  &$\langle p_{\rm T} \rangle$ & $\langle \theta \rangle$
  & \multicolumn{5}{c|}{${\rm d}^2 \sigma /{\rm d}p{\rm d}\Omega$} \\
 \hline
0.13--0.16 & 0.146 &  97.6 &  12.82 & $\!\!\pm\!\!$ &  0.90 & $\!\!\pm\!\!$ &  1.01 & 0.146 & 114.7 &  11.03 & $\!\!\pm\!\!$ &  0.68 & $\!\!\pm\!\!$ &  0.85 \\
0.16--0.20 & 0.180 &  96.8 &  13.07 & $\!\!\pm\!\!$ &  0.75 & $\!\!\pm\!\!$ &  0.78 & 0.180 & 114.2 &   8.30 & $\!\!\pm\!\!$ &  0.49 & $\!\!\pm\!\!$ &  0.46 \\
0.20--0.24 & 0.219 &  97.1 &  11.84 & $\!\!\pm\!\!$ &  0.76 & $\!\!\pm\!\!$ &  0.52 & 0.219 & 113.7 &   6.30 & $\!\!\pm\!\!$ &  0.45 & $\!\!\pm\!\!$ &  0.34 \\
0.24--0.30 & 0.268 &  97.0 &   6.88 & $\!\!\pm\!\!$ &  0.45 & $\!\!\pm\!\!$ &  0.31 & 0.267 & 113.6 &   4.91 & $\!\!\pm\!\!$ &  0.36 & $\!\!\pm\!\!$ &  0.28 \\
0.30--0.36 & 0.330 &  97.1 &   4.71 & $\!\!\pm\!\!$ &  0.40 & $\!\!\pm\!\!$ &  0.23 & 0.327 & 113.8 &   2.65 & $\!\!\pm\!\!$ &  0.24 & $\!\!\pm\!\!$ &  0.20 \\
0.36--0.42 & 0.386 &  96.6 &   3.28 & $\!\!\pm\!\!$ &  0.33 & $\!\!\pm\!\!$ &  0.23 & 0.391 & 112.7 &   1.72 & $\!\!\pm\!\!$ &  0.22 & $\!\!\pm\!\!$ &  0.17 \\
0.42--0.50 & 0.457 &  96.9 &   2.06 & $\!\!\pm\!\!$ &  0.22 & $\!\!\pm\!\!$ &  0.19 & 0.452 & 113.2 &   0.95 & $\!\!\pm\!\!$ &  0.15 & $\!\!\pm\!\!$ &  0.12 \\
0.50--0.60 & 0.548 &  96.3 &   0.96 & $\!\!\pm\!\!$ &  0.14 & $\!\!\pm\!\!$ &  0.12 & 0.537 & 113.0 &   0.19 & $\!\!\pm\!\!$ &  0.05 & $\!\!\pm\!\!$ &  0.03 \\
0.60--0.72 & 0.645 &  96.4 &   0.26 & $\!\!\pm\!\!$ &  0.06 & $\!\!\pm\!\!$ &  0.04 & 0.629 & 114.1 &   0.08 & $\!\!\pm\!\!$ &  0.04 & $\!\!\pm\!\!$ &  0.02 \\
0.72--0.90 & 0.786 &  96.0 &   0.11 & $\!\!\pm\!\!$ &  0.03 & $\!\!\pm\!\!$ &  0.02 &  &  &  \multicolumn{5}{c|}{ } \\
 \hline
 \end{tabular}
 \end{center}
 \end{scriptsize}
 \end{table}

%% file: table.pro.pipbe15.tex
  
 \begin{table}[h]
 \begin{scriptsize}
 \caption{Double-differential inclusive
  cross-section ${\rm d}^2 \sigma /{\rm d}p{\rm d}\Omega$
  [mb/(GeV/{\it c} sr)] of the production of protons
  in $\pi^+$ + Be $\rightarrow$ p + X interactions
  with $+15.0$~GeV/{\it c} beam momentum;
  the first error is statistical, the second systematic; 
 $p_{\rm T}$ in GeV/{\it c}, polar angle $\theta$ in degrees.}
 \label{pro.pipbe15}
 \begin{center}
 \begin{tabular}{|c||c|c|rcrcr||c|c|rcrcr|}
 \hline
   & \multicolumn{7}{c||}{$20<\theta<30$}
  & \multicolumn{7}{c|}{$30<\theta<40$} \\
 \hline
 $p_{\rm T}$ & $\langle p_{\rm T} \rangle$ & $\langle \theta \rangle$
  & \multicolumn{5}{c||}{${\rm d}^2 \sigma /{\rm d}p{\rm d}\Omega$}
  &$\langle p_{\rm T} \rangle$ & $\langle \theta \rangle$
  & \multicolumn{5}{c|}{${\rm d}^2 \sigma /{\rm d}p{\rm d}\Omega$} \\
 \hline
0.20--0.24 & 0.218 &  24.2 &  16.22 & $\!\!\pm\!\!$ &  9.16 & $\!\!\pm\!\!$ &  1.13 &  &  &  \multicolumn{5}{c|}{ } \\
0.24--0.30 & 0.273 &  25.0 &  49.57 & $\!\!\pm\!\!$ & 14.27 & $\!\!\pm\!\!$ &  3.06 & 0.262 &  36.1 &  25.30 & $\!\!\pm\!\!$ & 10.35 & $\!\!\pm\!\!$ &  1.53 \\
0.30--0.36 & 0.336 &  26.1 &  20.22 & $\!\!\pm\!\!$ &  9.42 & $\!\!\pm\!\!$ &  1.16 & 0.323 &  35.2 &  44.39 & $\!\!\pm\!\!$ & 13.63 & $\!\!\pm\!\!$ &  2.44 \\
0.36--0.42 & 0.379 &  24.9 &  13.87 & $\!\!\pm\!\!$ &  7.05 & $\!\!\pm\!\!$ &  0.77 & 0.380 &  32.5 &  18.20 & $\!\!\pm\!\!$ &  8.55 & $\!\!\pm\!\!$ &  0.95 \\
0.42--0.50 & 0.461 &  22.3 &  13.35 & $\!\!\pm\!\!$ &  5.87 & $\!\!\pm\!\!$ &  0.66 & 0.457 &  35.4 &  39.84 & $\!\!\pm\!\!$ & 11.22 & $\!\!\pm\!\!$ &  1.99 \\
0.50--0.60 & 0.539 &  25.8 &  11.22 & $\!\!\pm\!\!$ &  5.00 & $\!\!\pm\!\!$ &  0.55 & 0.548 &  34.8 &  18.04 & $\!\!\pm\!\!$ &  6.65 & $\!\!\pm\!\!$ &  0.87 \\
0.60--0.72 & 0.648 &  24.9 &  19.92 & $\!\!\pm\!\!$ &  5.83 & $\!\!\pm\!\!$ &  1.08 & 0.650 &  35.1 &   7.76 & $\!\!\pm\!\!$ &  4.10 & $\!\!\pm\!\!$ &  0.42 \\
0.72--0.90 &  &  &  \multicolumn{5}{c||}{ } & 0.799 &  35.4 &  11.65 & $\!\!\pm\!\!$ &  4.10 & $\!\!\pm\!\!$ &  0.81 \\
 \hline
 \hline
   & \multicolumn{7}{c||}{$40<\theta<50$}
  & \multicolumn{7}{c|}{$50<\theta<60$} \\
 \hline
 $p_{\rm T}$ & $\langle p_{\rm T} \rangle$ & $\langle \theta \rangle$
  & \multicolumn{5}{c||}{${\rm d}^2 \sigma /{\rm d}p{\rm d}\Omega$}
  &$\langle p_{\rm T} \rangle$ & $\langle \theta \rangle$
  & \multicolumn{5}{c|}{${\rm d}^2 \sigma /{\rm d}p{\rm d}\Omega$} \\
 \hline
0.30--0.36 & 0.335 &  43.6 &  22.96 & $\!\!\pm\!\!$ &  9.40 & $\!\!\pm\!\!$ &  1.24 &  &  &  \multicolumn{5}{c|}{ } \\
0.36--0.42 & 0.385 &  44.4 &  31.96 & $\!\!\pm\!\!$ & 11.47 & $\!\!\pm\!\!$ &  1.60 & 0.397 &  54.1 &  38.67 & $\!\!\pm\!\!$ & 12.25 & $\!\!\pm\!\!$ &  2.02 \\
0.42--0.50 & 0.452 &  45.4 &  19.59 & $\!\!\pm\!\!$ &  7.77 & $\!\!\pm\!\!$ &  1.01 & 0.452 &  53.4 &   8.63 & $\!\!\pm\!\!$ &  5.11 & $\!\!\pm\!\!$ &  0.43 \\
0.50--0.60 & 0.554 &  44.5 &  17.12 & $\!\!\pm\!\!$ &  6.80 & $\!\!\pm\!\!$ &  0.85 & 0.545 &  56.4 &  18.53 & $\!\!\pm\!\!$ &  7.05 & $\!\!\pm\!\!$ &  0.99 \\
0.60--0.72 & 0.634 &  42.2 &   8.95 & $\!\!\pm\!\!$ &  4.29 & $\!\!\pm\!\!$ &  0.52 & 0.657 &  57.0 &   9.87 & $\!\!\pm\!\!$ &  4.78 & $\!\!\pm\!\!$ &  0.61 \\
0.72--0.90 & 0.825 &  42.8 &   7.54 & $\!\!\pm\!\!$ &  3.24 & $\!\!\pm\!\!$ &  0.53 & 0.813 &  58.2 &   3.96 & $\!\!\pm\!\!$ &  2.52 & $\!\!\pm\!\!$ &  0.31 \\
0.90--1.25 & 1.063 &  44.2 &   2.79 & $\!\!\pm\!\!$ &  1.41 & $\!\!\pm\!\!$ &  0.29 & 1.029 &  54.2 &   1.76 & $\!\!\pm\!\!$ &  1.21 & $\!\!\pm\!\!$ &  0.20 \\
 \hline
 \hline
   & \multicolumn{7}{c||}{$60<\theta<75$}
  & \multicolumn{7}{c|}{$75<\theta<90$} \\
 \hline
 $p_{\rm T}$ & $\langle p_{\rm T} \rangle$ & $\langle \theta \rangle$
  & \multicolumn{5}{c||}{${\rm d}^2 \sigma /{\rm d}p{\rm d}\Omega$}
  &$\langle p_{\rm T} \rangle$ & $\langle \theta \rangle$
  & \multicolumn{5}{c|}{${\rm d}^2 \sigma /{\rm d}p{\rm d}\Omega$} \\
 \hline
0.36--0.42 & 0.396 &  70.2 &  21.16 & $\!\!\pm\!\!$ &  7.49 & $\!\!\pm\!\!$ &  1.08 &  &  &  \multicolumn{5}{c|}{ } \\
0.42--0.50 & 0.456 &  65.3 &  15.77 & $\!\!\pm\!\!$ &  5.34 & $\!\!\pm\!\!$ &  0.84 & 0.466 &  81.5 &   9.58 & $\!\!\pm\!\!$ &  4.29 & $\!\!\pm\!\!$ &  0.55 \\
0.50--0.60 & 0.540 &  66.1 &  15.79 & $\!\!\pm\!\!$ &  5.21 & $\!\!\pm\!\!$ &  0.85 & 0.536 &  82.8 &   6.25 & $\!\!\pm\!\!$ &  3.25 & $\!\!\pm\!\!$ &  0.41 \\
0.60--0.72 & 0.659 &  70.0 &   9.15 & $\!\!\pm\!\!$ &  3.68 & $\!\!\pm\!\!$ &  0.63 & 0.653 &  82.2 &   7.67 & $\!\!\pm\!\!$ &  3.50 & $\!\!\pm\!\!$ &  0.66 \\
0.72--0.90 & 0.798 &  69.4 &   4.09 & $\!\!\pm\!\!$ &  2.09 & $\!\!\pm\!\!$ &  0.39 & 0.745 &  78.3 &   1.49 & $\!\!\pm\!\!$ &  1.22 & $\!\!\pm\!\!$ &  0.17 \\
0.90--1.25 & 1.076 &  65.8 &   1.63 & $\!\!\pm\!\!$ &  1.00 & $\!\!\pm\!\!$ &  0.24 &  &  &  \multicolumn{5}{c|}{ } \\
 \hline
 \hline
  & \multicolumn{7}{c||}{$90<\theta<105$}
  & \multicolumn{7}{c|}{$105<\theta<125$} \\
 \hline
 $p_{\rm T}$ & $\langle p_{\rm T} \rangle$ & $\langle \theta \rangle$
  & \multicolumn{5}{c||}{${\rm d}^2 \sigma /{\rm d}p{\rm d}\Omega$}
  &$\langle p_{\rm T} \rangle$ & $\langle \theta \rangle$
  & \multicolumn{5}{c|}{${\rm d}^2 \sigma /{\rm d}p{\rm d}\Omega$} \\
 \hline
0.36--0.42 &  &  &  \multicolumn{5}{c||}{ } & 0.388 & 114.7 &   5.72 & $\!\!\pm\!\!$ &  3.31 & $\!\!\pm\!\!$ &  0.34 \\
0.42--0.50 & 0.458 & 100.2 &   8.10 & $\!\!\pm\!\!$ &  4.07 & $\!\!\pm\!\!$ &  0.57 & 0.445 & 109.9 &   7.82 & $\!\!\pm\!\!$ &  3.55 & $\!\!\pm\!\!$ &  0.54 \\
0.50--0.60 & 0.518 &  98.8 &   8.15 & $\!\!\pm\!\!$ &  3.74 & $\!\!\pm\!\!$ &  0.67 &  &  &  \multicolumn{5}{c|}{ } \\
0.60--0.72 & 0.649 &  99.3 &   2.61 & $\!\!\pm\!\!$ &  2.07 & $\!\!\pm\!\!$ &  0.27 &  &  &  \multicolumn{5}{c|}{ } \\
 \hline
 \end{tabular}
 \end{center}
 \end{scriptsize}
 \end{table}

%% file: table.pip.pipbe15.tex
  
 \begin{table}[h]
 \begin{scriptsize}
 \caption{Double-differential inclusive
  cross-section ${\rm d}^2 \sigma /{\rm d}p{\rm d}\Omega$
  [mb/(GeV/{\it c} sr)] of the production of $\pi^+$'s
  in $\pi^+$ + Be $\rightarrow$ $\pi^+$ + X interactions
  with $+15.0$~GeV/{\it c} beam momentum;
  the first error is statistical, the second systematic; 
 $p_{\rm T}$ in GeV/{\it c}, polar angle $\theta$ in degrees.}
 \label{pip.pipbe15}
 \begin{center}
 \begin{tabular}{|c||c|c|rcrcr||c|c|rcrcr|}
 \hline
   & \multicolumn{7}{c||}{$20<\theta<30$}
  & \multicolumn{7}{c|}{$30<\theta<40$} \\
 \hline
 $p_{\rm T}$ & $\langle p_{\rm T} \rangle$ & $\langle \theta \rangle$
  & \multicolumn{5}{c||}{${\rm d}^2 \sigma /{\rm d}p{\rm d}\Omega$}
  &$\langle p_{\rm T} \rangle$ & $\langle \theta \rangle$
  & \multicolumn{5}{c|}{${\rm d}^2 \sigma /{\rm d}p{\rm d}\Omega$} \\
 \hline
0.13--0.16 &  &  &  \multicolumn{5}{c||}{ } & 0.141 &  33.9 &  42.30 & $\!\!\pm\!\!$ & 19.30 & $\!\!\pm\!\!$ &  3.17 \\
0.16--0.20 &  &  &  \multicolumn{5}{c||}{ } & 0.180 &  35.1 &  33.27 & $\!\!\pm\!\!$ & 15.08 & $\!\!\pm\!\!$ &  2.11 \\
0.24--0.30 & 0.264 &  24.6 &  44.67 & $\!\!\pm\!\!$ & 13.69 & $\!\!\pm\!\!$ &  2.14 & 0.267 &  33.6 &  40.55 & $\!\!\pm\!\!$ & 13.55 & $\!\!\pm\!\!$ &  2.02 \\
0.30--0.36 &  &  &  \multicolumn{5}{c||}{ } & 0.330 &  35.2 &  49.89 & $\!\!\pm\!\!$ & 15.04 & $\!\!\pm\!\!$ &  2.37 \\
0.36--0.42 & 0.378 &  24.5 &  69.69 & $\!\!\pm\!\!$ & 17.99 & $\!\!\pm\!\!$ &  3.31 & 0.391 &  35.8 &  21.90 & $\!\!\pm\!\!$ &  9.77 & $\!\!\pm\!\!$ &  1.07 \\
0.42--0.50 & 0.461 &  24.6 &  48.52 & $\!\!\pm\!\!$ & 12.79 & $\!\!\pm\!\!$ &  2.45 & 0.456 &  34.0 &  29.16 & $\!\!\pm\!\!$ &  9.56 & $\!\!\pm\!\!$ &  1.50 \\
0.50--0.60 & 0.555 &  23.4 &  33.71 & $\!\!\pm\!\!$ &  8.82 & $\!\!\pm\!\!$ &  2.12 & 0.527 &  35.4 &   9.19 & $\!\!\pm\!\!$ &  4.60 & $\!\!\pm\!\!$ &  0.57 \\
0.60--0.72 & 0.656 &  25.5 &  24.11 & $\!\!\pm\!\!$ &  6.89 & $\!\!\pm\!\!$ &  2.08 & 0.641 &  36.8 &   5.15 & $\!\!\pm\!\!$ &  3.38 & $\!\!\pm\!\!$ &  0.43 \\
0.72--0.90 &  &  &  \multicolumn{5}{c||}{ } & 0.809 &  32.7 &   4.61 & $\!\!\pm\!\!$ &  2.24 & $\!\!\pm\!\!$ &  0.57 \\
 \hline
 \hline
   & \multicolumn{7}{c||}{$40<\theta<50$}
  & \multicolumn{7}{c|}{$50<\theta<60$} \\
 \hline
 $p_{\rm T}$ & $\langle p_{\rm T} \rangle$ & $\langle \theta \rangle$
  & \multicolumn{5}{c||}{${\rm d}^2 \sigma /{\rm d}p{\rm d}\Omega$}
  &$\langle p_{\rm T} \rangle$ & $\langle \theta \rangle$
  & \multicolumn{5}{c|}{${\rm d}^2 \sigma /{\rm d}p{\rm d}\Omega$} \\
 \hline
0.10--0.13 & 0.120 &  47.2 &  25.06 & $\!\!\pm\!\!$ & 17.75 & $\!\!\pm\!\!$ &  2.46 &  &  &  \multicolumn{5}{c|}{ } \\
0.13--0.16 &  &  &  \multicolumn{5}{c||}{ } & 0.150 &  54.4 &  24.24 & $\!\!\pm\!\!$ & 14.06 & $\!\!\pm\!\!$ &  2.18 \\
0.16--0.20 & 0.185 &  44.4 &  26.34 & $\!\!\pm\!\!$ & 13.40 & $\!\!\pm\!\!$ &  1.79 & 0.174 &  55.7 &  13.90 & $\!\!\pm\!\!$ &  9.85 & $\!\!\pm\!\!$ &  0.99 \\
0.20--0.24 & 0.223 &  45.8 &  38.82 & $\!\!\pm\!\!$ & 15.92 & $\!\!\pm\!\!$ &  2.42 & 0.221 &  53.4 &  26.72 & $\!\!\pm\!\!$ & 13.47 & $\!\!\pm\!\!$ &  1.80 \\
0.24--0.30 & 0.275 &  44.3 &  36.38 & $\!\!\pm\!\!$ & 12.86 & $\!\!\pm\!\!$ &  1.95 & 0.262 &  56.3 &  18.39 & $\!\!\pm\!\!$ &  8.71 & $\!\!\pm\!\!$ &  1.04 \\
0.30--0.36 & 0.332 &  45.0 &  37.02 & $\!\!\pm\!\!$ & 12.44 & $\!\!\pm\!\!$ &  1.93 & 0.319 &  55.4 &  13.61 & $\!\!\pm\!\!$ &  7.84 & $\!\!\pm\!\!$ &  0.75 \\
0.36--0.42 & 0.397 &  45.5 &  21.52 & $\!\!\pm\!\!$ &  9.61 & $\!\!\pm\!\!$ &  1.17 & 0.387 &  54.7 &  26.06 & $\!\!\pm\!\!$ & 10.64 & $\!\!\pm\!\!$ &  1.60 \\
0.42--0.50 & 0.458 &  43.4 &  15.84 & $\!\!\pm\!\!$ &  7.07 & $\!\!\pm\!\!$ &  0.91 & 0.457 &  56.4 &   9.77 & $\!\!\pm\!\!$ &  5.64 & $\!\!\pm\!\!$ &  0.65 \\
0.50--0.60 & 0.545 &  43.7 &   7.56 & $\!\!\pm\!\!$ &  4.36 & $\!\!\pm\!\!$ &  0.50 &  &  &  \multicolumn{5}{c|}{ } \\
0.60--0.72 & 0.628 &  45.0 &   6.45 & $\!\!\pm\!\!$ &  3.68 & $\!\!\pm\!\!$ &  0.55 &  &  &  \multicolumn{5}{c|}{ } \\
0.72--0.90 & 0.777 &  43.9 &   6.20 & $\!\!\pm\!\!$ &  2.86 & $\!\!\pm\!\!$ &  0.73 & 0.744 &  55.6 &   2.54 & $\!\!\pm\!\!$ &  1.79 & $\!\!\pm\!\!$ &  0.34 \\
 \hline
 \hline
   & \multicolumn{7}{c||}{$60<\theta<75$}
  & \multicolumn{7}{c|}{$75<\theta<90$} \\
 \hline
 $p_{\rm T}$ & $\langle p_{\rm T} \rangle$ & $\langle \theta \rangle$
  & \multicolumn{5}{c||}{${\rm d}^2 \sigma /{\rm d}p{\rm d}\Omega$}
  &$\langle p_{\rm T} \rangle$ & $\langle \theta \rangle$
  & \multicolumn{5}{c|}{${\rm d}^2 \sigma /{\rm d}p{\rm d}\Omega$} \\
 \hline
0.13--0.16 & 0.148 &  66.1 &  28.94 & $\!\!\pm\!\!$ & 13.00 & $\!\!\pm\!\!$ &  2.70 & 0.143 &  82.1 &  18.80 & $\!\!\pm\!\!$ & 10.84 & $\!\!\pm\!\!$ &  1.77 \\
0.16--0.20 & 0.182 &  66.1 &  16.43 & $\!\!\pm\!\!$ &  8.31 & $\!\!\pm\!\!$ &  1.18 & 0.182 &  78.6 &  15.65 & $\!\!\pm\!\!$ &  7.91 & $\!\!\pm\!\!$ &  1.23 \\
0.20--0.24 & 0.236 &  63.6 &   8.08 & $\!\!\pm\!\!$ &  5.72 & $\!\!\pm\!\!$ &  0.52 & 0.216 &  78.6 &  12.84 & $\!\!\pm\!\!$ &  7.42 & $\!\!\pm\!\!$ &  0.91 \\
0.24--0.30 & 0.263 &  67.1 &  13.13 & $\!\!\pm\!\!$ &  6.03 & $\!\!\pm\!\!$ &  0.73 & 0.257 &  83.3 &  14.90 & $\!\!\pm\!\!$ &  6.66 & $\!\!\pm\!\!$ &  0.90 \\
0.30--0.36 & 0.332 &  67.4 &  17.84 & $\!\!\pm\!\!$ &  6.92 & $\!\!\pm\!\!$ &  1.01 & 0.326 &  83.9 &  15.90 & $\!\!\pm\!\!$ &  6.95 & $\!\!\pm\!\!$ &  1.08 \\
0.42--0.50 & 0.449 &  71.0 &   5.54 & $\!\!\pm\!\!$ &  3.20 & $\!\!\pm\!\!$ &  0.40 & 0.457 &  84.8 &   5.88 & $\!\!\pm\!\!$ &  3.39 & $\!\!\pm\!\!$ &  0.55 \\
0.50--0.60 &  &  &  \multicolumn{5}{c||}{ } & 0.535 &  80.6 &   4.48 & $\!\!\pm\!\!$ &  2.59 & $\!\!\pm\!\!$ &  0.53 \\
0.60--0.72 & 0.638 &  64.3 &   4.80 & $\!\!\pm\!\!$ &  2.41 & $\!\!\pm\!\!$ &  0.54 &  &  &  \multicolumn{5}{c|}{ } \\
0.72--0.90 & 0.807 &  63.9 &   0.15 & $\!\!\pm\!\!$ &  0.11 & $\!\!\pm\!\!$ &  0.02 &  &  &  \multicolumn{5}{c|}{ } \\
 \hline
 \hline
  & \multicolumn{7}{c||}{$90<\theta<105$}
  & \multicolumn{7}{c|}{$105<\theta<125$} \\
 \hline
 $p_{\rm T}$ & $\langle p_{\rm T} \rangle$ & $\langle \theta \rangle$
  & \multicolumn{5}{c||}{${\rm d}^2 \sigma /{\rm d}p{\rm d}\Omega$}
  &$\langle p_{\rm T} \rangle$ & $\langle \theta \rangle$
  & \multicolumn{5}{c|}{${\rm d}^2 \sigma /{\rm d}p{\rm d}\Omega$} \\
 \hline
0.13--0.16 & 0.141 &  98.0 &  28.25 & $\!\!\pm\!\!$ & 12.65 & $\!\!\pm\!\!$ &  3.10 &  &  &  \multicolumn{5}{c|}{ } \\
0.16--0.20 &  &  &  \multicolumn{5}{c||}{ } & 0.184 & 113.0 &  19.87 & $\!\!\pm\!\!$ &  7.51 & $\!\!\pm\!\!$ &  1.78 \\
0.20--0.24 & 0.215 &  95.3 &   7.34 & $\!\!\pm\!\!$ &  5.19 & $\!\!\pm\!\!$ &  0.70 & 0.227 & 114.3 &  16.88 & $\!\!\pm\!\!$ &  7.55 & $\!\!\pm\!\!$ &  1.39 \\
0.24--0.30 & 0.274 & 100.2 &   5.74 & $\!\!\pm\!\!$ &  4.06 & $\!\!\pm\!\!$ &  0.41 & 0.249 & 120.1 &   4.21 & $\!\!\pm\!\!$ &  2.97 & $\!\!\pm\!\!$ &  0.36 \\
0.30--0.36 & 0.321 &  91.0 &   7.52 & $\!\!\pm\!\!$ &  4.34 & $\!\!\pm\!\!$ &  0.69 &  &  &  \multicolumn{5}{c|}{ } \\
0.36--0.42 & 0.405 &  94.9 &   5.56 & $\!\!\pm\!\!$ &  3.93 & $\!\!\pm\!\!$ &  0.57 &  &  &  \multicolumn{5}{c|}{ } \\
0.42--0.50 & 0.469 &  94.2 &   4.32 & $\!\!\pm\!\!$ &  3.06 & $\!\!\pm\!\!$ &  0.53 &  &  &  \multicolumn{5}{c|}{ } \\
0.50--0.60 &  &  &  \multicolumn{5}{c||}{ } & 0.539 & 112.7 &   2.46 & $\!\!\pm\!\!$ &  1.74 & $\!\!\pm\!\!$ &  0.52 \\
 \hline
 \end{tabular}
 \end{center}
 \end{scriptsize}
 \end{table}

%% file: table.pim.pipbe15.tex
  
 \begin{table}[h]
 \begin{scriptsize}
 \caption{Double-differential inclusive
  cross-section ${\rm d}^2 \sigma /{\rm d}p{\rm d}\Omega$
  [mb/(GeV/{\it c} sr)] of the production of $\pi^-$'s
  in $\pi^+$ + Be $\rightarrow$ $\pi^-$ + X interactions
  with $+15.0$~GeV/{\it c} beam momentum;
  the first error is statistical, the second systematic; 
 $p_{\rm T}$ in GeV/{\it c}, polar angle $\theta$ in degrees.}
 \label{pim.pipbe15}
 \begin{center}
 \begin{tabular}{|c||c|c|rcrcr||c|c|rcrcr|}
 \hline
   & \multicolumn{7}{c||}{$20<\theta<30$}
  & \multicolumn{7}{c|}{$30<\theta<40$} \\
 \hline
 $p_{\rm T}$ & $\langle p_{\rm T} \rangle$ & $\langle \theta \rangle$
  & \multicolumn{5}{c||}{${\rm d}^2 \sigma /{\rm d}p{\rm d}\Omega$}
  &$\langle p_{\rm T} \rangle$ & $\langle \theta \rangle$
  & \multicolumn{5}{c|}{${\rm d}^2 \sigma /{\rm d}p{\rm d}\Omega$} \\
 \hline
0.10--0.13 &  &  &  \multicolumn{5}{c||}{ } & 0.122 &  31.7 &  16.01 & $\!\!\pm\!\!$ & 11.79 & $\!\!\pm\!\!$ &  1.53 \\
0.13--0.16 & 0.156 &  24.6 &  21.23 & $\!\!\pm\!\!$ & 14.83 & $\!\!\pm\!\!$ &  1.52 & 0.146 &  32.6 &  25.48 & $\!\!\pm\!\!$ & 14.08 & $\!\!\pm\!\!$ &  2.05 \\
0.16--0.20 &  &  &  \multicolumn{5}{c||}{ } & 0.184 &  33.9 &  42.39 & $\!\!\pm\!\!$ & 17.35 & $\!\!\pm\!\!$ &  2.84 \\
0.20--0.24 & 0.221 &  24.2 &  43.46 & $\!\!\pm\!\!$ & 16.56 & $\!\!\pm\!\!$ &  2.41 & 0.217 &  35.7 &  13.82 & $\!\!\pm\!\!$ &  9.80 & $\!\!\pm\!\!$ &  0.84 \\
0.24--0.30 & 0.270 &  24.9 &  62.09 & $\!\!\pm\!\!$ & 16.68 & $\!\!\pm\!\!$ &  3.00 & 0.263 &  35.9 &  27.09 & $\!\!\pm\!\!$ & 11.10 & $\!\!\pm\!\!$ &  1.40 \\
0.30--0.36 & 0.329 &  23.9 &  58.34 & $\!\!\pm\!\!$ & 16.19 & $\!\!\pm\!\!$ &  2.72 & 0.335 &  33.4 &  24.25 & $\!\!\pm\!\!$ &  9.91 & $\!\!\pm\!\!$ &  1.22 \\
0.36--0.42 & 0.388 &  23.3 &  32.11 & $\!\!\pm\!\!$ & 11.40 & $\!\!\pm\!\!$ &  1.55 & 0.391 &  33.3 &  39.28 & $\!\!\pm\!\!$ & 13.14 & $\!\!\pm\!\!$ &  2.15 \\
0.42--0.50 & 0.463 &  24.5 &  28.44 & $\!\!\pm\!\!$ &  9.49 & $\!\!\pm\!\!$ &  1.52 & 0.457 &  34.6 &  23.96 & $\!\!\pm\!\!$ &  8.47 & $\!\!\pm\!\!$ &  1.45 \\
0.50--0.60 & 0.533 &  24.9 &  13.74 & $\!\!\pm\!\!$ &  6.15 & $\!\!\pm\!\!$ &  0.90 & 0.542 &  35.4 &  25.50 & $\!\!\pm\!\!$ &  8.07 & $\!\!\pm\!\!$ &  1.84 \\
0.60--0.72 & 0.673 &  23.9 &   6.00 & $\!\!\pm\!\!$ &  3.47 & $\!\!\pm\!\!$ &  0.52 & 0.656 &  33.4 &   5.93 & $\!\!\pm\!\!$ &  3.43 & $\!\!\pm\!\!$ &  0.55 \\
 \hline
 \hline
   & \multicolumn{7}{c||}{$40<\theta<50$}
  & \multicolumn{7}{c|}{$50<\theta<60$} \\
 \hline
 $p_{\rm T}$ & $\langle p_{\rm T} \rangle$ & $\langle \theta \rangle$
  & \multicolumn{5}{c||}{${\rm d}^2 \sigma /{\rm d}p{\rm d}\Omega$}
  &$\langle p_{\rm T} \rangle$ & $\langle \theta \rangle$
  & \multicolumn{5}{c|}{${\rm d}^2 \sigma /{\rm d}p{\rm d}\Omega$} \\
 \hline
0.13--0.16 &  &  &  \multicolumn{5}{c||}{ } & 0.136 &  55.7 &  28.29 & $\!\!\pm\!\!$ & 15.79 & $\!\!\pm\!\!$ &  2.82 \\
0.16--0.20 & 0.180 &  43.8 &  32.13 & $\!\!\pm\!\!$ & 14.49 & $\!\!\pm\!\!$ &  2.36 & 0.174 &  53.3 &  12.63 & $\!\!\pm\!\!$ &  8.93 & $\!\!\pm\!\!$ &  1.01 \\
0.20--0.24 & 0.214 &  46.0 &  31.20 & $\!\!\pm\!\!$ & 14.03 & $\!\!\pm\!\!$ &  2.10 & 0.214 &  52.7 &  25.52 & $\!\!\pm\!\!$ & 12.80 & $\!\!\pm\!\!$ &  1.82 \\
0.24--0.30 & 0.273 &  47.1 &  26.65 & $\!\!\pm\!\!$ & 10.91 & $\!\!\pm\!\!$ &  1.53 & 0.259 &  54.4 &  18.85 & $\!\!\pm\!\!$ &  9.43 & $\!\!\pm\!\!$ &  1.16 \\
0.30--0.36 & 0.309 &  44.4 &   8.35 & $\!\!\pm\!\!$ &  5.91 & $\!\!\pm\!\!$ &  0.48 & 0.328 &  56.3 &  16.48 & $\!\!\pm\!\!$ &  8.24 & $\!\!\pm\!\!$ &  1.09 \\
0.36--0.42 & 0.383 &  43.7 &  30.08 & $\!\!\pm\!\!$ & 11.37 & $\!\!\pm\!\!$ &  1.86 &  &  &  \multicolumn{5}{c|}{ } \\
0.42--0.50 & 0.460 &  44.1 &  27.12 & $\!\!\pm\!\!$ &  9.04 & $\!\!\pm\!\!$ &  1.84 &  &  &  \multicolumn{5}{c|}{ } \\
0.50--0.60 & 0.557 &  42.0 &   6.28 & $\!\!\pm\!\!$ &  3.64 & $\!\!\pm\!\!$ &  0.52 & 0.554 &  51.2 &   7.45 & $\!\!\pm\!\!$ &  4.30 & $\!\!\pm\!\!$ &  0.69 \\
0.60--0.72 &  &  &  \multicolumn{5}{c||}{ } & 0.630 &  56.2 &   4.35 & $\!\!\pm\!\!$ &  3.07 & $\!\!\pm\!\!$ &  0.50 \\
 \hline
 \hline
   & \multicolumn{7}{c||}{$60<\theta<75$}
  & \multicolumn{7}{c|}{$75<\theta<90$} \\
 \hline
 $p_{\rm T}$ & $\langle p_{\rm T} \rangle$ & $\langle \theta \rangle$
  & \multicolumn{5}{c||}{${\rm d}^2 \sigma /{\rm d}p{\rm d}\Omega$}
  &$\langle p_{\rm T} \rangle$ & $\langle \theta \rangle$
  & \multicolumn{5}{c|}{${\rm d}^2 \sigma /{\rm d}p{\rm d}\Omega$} \\
 \hline
0.13--0.16 & 0.146 &  65.7 &  23.65 & $\!\!\pm\!\!$ & 11.86 & $\!\!\pm\!\!$ &  2.22 &  &  &  \multicolumn{5}{c|}{ } \\
0.16--0.20 & 0.182 &  69.8 &   8.01 & $\!\!\pm\!\!$ &  5.66 & $\!\!\pm\!\!$ &  0.69 & 0.187 &  81.6 &   7.04 & $\!\!\pm\!\!$ &  4.98 & $\!\!\pm\!\!$ &  0.68 \\
0.24--0.30 &  &  &  \multicolumn{5}{c||}{ } & 0.263 &  83.3 &   8.15 & $\!\!\pm\!\!$ &  4.71 & $\!\!\pm\!\!$ &  0.61 \\
0.30--0.36 & 0.338 &  71.4 &   7.79 & $\!\!\pm\!\!$ &  4.51 & $\!\!\pm\!\!$ &  0.55 &  &  &  \multicolumn{5}{c|}{ } \\
0.36--0.42 & 0.389 &  69.7 &  16.51 & $\!\!\pm\!\!$ &  6.25 & $\!\!\pm\!\!$ &  1.42 &  &  &  \multicolumn{5}{c|}{ } \\
0.42--0.50 &  &  &  \multicolumn{5}{c||}{ } & 0.456 &  82.2 &  10.25 & $\!\!\pm\!\!$ &  4.59 & $\!\!\pm\!\!$ &  1.11 \\
0.50--0.60 & 0.558 &  68.8 &   5.30 & $\!\!\pm\!\!$ &  3.06 & $\!\!\pm\!\!$ &  0.54 &  &  &  \multicolumn{5}{c|}{ } \\
 \hline
 \hline
  & \multicolumn{7}{c||}{$90<\theta<105$}
  & \multicolumn{7}{c|}{$105<\theta<125$} \\
 \hline
 $p_{\rm T}$ & $\langle p_{\rm T} \rangle$ & $\langle \theta \rangle$
  & \multicolumn{5}{c||}{${\rm d}^2 \sigma /{\rm d}p{\rm d}\Omega$}
  &$\langle p_{\rm T} \rangle$ & $\langle \theta \rangle$
  & \multicolumn{5}{c|}{${\rm d}^2 \sigma /{\rm d}p{\rm d}\Omega$} \\
 \hline
0.13--0.16 &  &  &  \multicolumn{5}{c||}{ } & 0.142 & 116.1 &   6.55 & $\!\!\pm\!\!$ &  4.64 & $\!\!\pm\!\!$ &  0.86 \\
0.20--0.24 & 0.227 &  97.1 &  12.49 & $\!\!\pm\!\!$ &  6.70 & $\!\!\pm\!\!$ &  1.23 &  &  &  \multicolumn{5}{c|}{ } \\
0.24--0.30 & 0.256 &  97.2 &   6.67 & $\!\!\pm\!\!$ &  3.86 & $\!\!\pm\!\!$ &  0.70 &  &  &  \multicolumn{5}{c|}{ } \\
 \hline
 \end{tabular}
 \end{center}
 \end{scriptsize}
 \end{table}

%% file: table.pro.pimbe15.tex
  
 \begin{table}[h]
 \begin{scriptsize}
 \caption{Double-differential inclusive
  cross-section ${\rm d}^2 \sigma /{\rm d}p{\rm d}\Omega$
  [mb/(GeV/{\it c} sr)] of the production of protons
  in $\pi^-$ + Be $\rightarrow$ p + X interactions
  with $-15.0$~GeV/{\it c} beam momentum;
  the first error is statistical, the second systematic; 
 $p_{\rm T}$ in GeV/{\it c}, polar angle $\theta$ in degrees.}
 \label{pro.pimbe15}
 \begin{center}
 \begin{tabular}{|c||c|c|rcrcr||c|c|rcrcr|}
 \hline
   & \multicolumn{7}{c||}{$20<\theta<30$}
  & \multicolumn{7}{c|}{$30<\theta<40$} \\
 \hline
 $p_{\rm T}$ & $\langle p_{\rm T} \rangle$ & $\langle \theta \rangle$
  & \multicolumn{5}{c||}{${\rm d}^2 \sigma /{\rm d}p{\rm d}\Omega$}
  &$\langle p_{\rm T} \rangle$ & $\langle \theta \rangle$
  & \multicolumn{5}{c|}{${\rm d}^2 \sigma /{\rm d}p{\rm d}\Omega$} \\
 \hline
0.20--0.24 & 0.219 &  24.9 &  25.27 & $\!\!\pm\!\!$ &  1.11 & $\!\!\pm\!\!$ &  1.86 &  &  &  \multicolumn{5}{c|}{ } \\
0.24--0.30 & 0.268 &  24.7 &  27.07 & $\!\!\pm\!\!$ &  0.93 & $\!\!\pm\!\!$ &  1.66 & 0.267 &  34.8 &  24.14 & $\!\!\pm\!\!$ &  0.87 & $\!\!\pm\!\!$ &  1.47 \\
0.30--0.36 & 0.327 &  24.8 &  25.87 & $\!\!\pm\!\!$ &  0.90 & $\!\!\pm\!\!$ &  1.51 & 0.326 &  35.0 &  22.96 & $\!\!\pm\!\!$ &  0.84 & $\!\!\pm\!\!$ &  1.41 \\
0.36--0.42 & 0.385 &  25.0 &  23.79 & $\!\!\pm\!\!$ &  0.85 & $\!\!\pm\!\!$ &  1.25 & 0.385 &  34.9 &  21.30 & $\!\!\pm\!\!$ &  0.83 & $\!\!\pm\!\!$ &  1.12 \\
0.42--0.50 & 0.451 &  24.7 &  21.24 & $\!\!\pm\!\!$ &  0.69 & $\!\!\pm\!\!$ &  1.03 & 0.452 &  35.0 &  18.90 & $\!\!\pm\!\!$ &  0.67 & $\!\!\pm\!\!$ &  0.94 \\
0.50--0.60 & 0.538 &  25.0 &  17.99 & $\!\!\pm\!\!$ &  0.56 & $\!\!\pm\!\!$ &  0.86 & 0.538 &  34.9 &  15.67 & $\!\!\pm\!\!$ &  0.55 & $\!\!\pm\!\!$ &  0.76 \\
0.60--0.72 & 0.642 &  24.9 &  12.92 & $\!\!\pm\!\!$ &  0.43 & $\!\!\pm\!\!$ &  0.70 & 0.642 &  34.8 &  11.46 & $\!\!\pm\!\!$ &  0.43 & $\!\!\pm\!\!$ &  0.63 \\
0.72--0.90 &  &  &  \multicolumn{5}{c||}{ } & 0.778 &  35.0 &   6.81 & $\!\!\pm\!\!$ &  0.26 & $\!\!\pm\!\!$ &  0.47 \\
 \hline
 \hline
   & \multicolumn{7}{c||}{$40<\theta<50$}
  & \multicolumn{7}{c|}{$50<\theta<60$} \\
 \hline
 $p_{\rm T}$ & $\langle p_{\rm T} \rangle$ & $\langle \theta \rangle$
  & \multicolumn{5}{c||}{${\rm d}^2 \sigma /{\rm d}p{\rm d}\Omega$}
  &$\langle p_{\rm T} \rangle$ & $\langle \theta \rangle$
  & \multicolumn{5}{c|}{${\rm d}^2 \sigma /{\rm d}p{\rm d}\Omega$} \\
 \hline
0.30--0.36 & 0.331 &  45.1 &  23.42 & $\!\!\pm\!\!$ &  0.85 & $\!\!\pm\!\!$ &  1.19 &  &  &  \multicolumn{5}{c|}{ } \\
0.36--0.42 & 0.389 &  45.0 &  20.03 & $\!\!\pm\!\!$ &  0.79 & $\!\!\pm\!\!$ &  1.06 & 0.389 &  54.9 &  19.23 & $\!\!\pm\!\!$ &  0.76 & $\!\!\pm\!\!$ &  1.04 \\
0.42--0.50 & 0.458 &  44.9 &  17.40 & $\!\!\pm\!\!$ &  0.64 & $\!\!\pm\!\!$ &  0.85 & 0.458 &  54.9 &  15.32 & $\!\!\pm\!\!$ &  0.59 & $\!\!\pm\!\!$ &  0.81 \\
0.50--0.60 & 0.545 &  44.9 &  11.91 & $\!\!\pm\!\!$ &  0.49 & $\!\!\pm\!\!$ &  0.60 & 0.548 &  54.9 &  12.11 & $\!\!\pm\!\!$ &  0.49 & $\!\!\pm\!\!$ &  0.65 \\
0.60--0.72 & 0.653 &  44.8 &   9.01 & $\!\!\pm\!\!$ &  0.38 & $\!\!\pm\!\!$ &  0.52 & 0.654 &  55.0 &   7.68 & $\!\!\pm\!\!$ &  0.37 & $\!\!\pm\!\!$ &  0.48 \\
0.72--0.90 & 0.799 &  45.0 &   5.25 & $\!\!\pm\!\!$ &  0.25 & $\!\!\pm\!\!$ &  0.37 & 0.797 &  55.0 &   4.33 & $\!\!\pm\!\!$ &  0.23 & $\!\!\pm\!\!$ &  0.34 \\
0.90--1.25 & 1.029 &  45.3 &   1.61 & $\!\!\pm\!\!$ &  0.10 & $\!\!\pm\!\!$ &  0.17 & 1.029 &  54.8 &   1.17 & $\!\!\pm\!\!$ &  0.08 & $\!\!\pm\!\!$ &  0.14 \\
 \hline
 \hline
   & \multicolumn{7}{c||}{$60<\theta<75$}
  & \multicolumn{7}{c|}{$75<\theta<90$} \\
 \hline
 $p_{\rm T}$ & $\langle p_{\rm T} \rangle$ & $\langle \theta \rangle$
  & \multicolumn{5}{c||}{${\rm d}^2 \sigma /{\rm d}p{\rm d}\Omega$}
  &$\langle p_{\rm T} \rangle$ & $\langle \theta \rangle$
  & \multicolumn{5}{c|}{${\rm d}^2 \sigma /{\rm d}p{\rm d}\Omega$} \\
 \hline
0.36--0.42 & 0.389 &  67.3 &  17.30 & $\!\!\pm\!\!$ &  0.57 & $\!\!\pm\!\!$ &  0.88 &  &  &  \multicolumn{5}{c|}{ } \\
0.42--0.50 & 0.459 &  67.2 &  14.62 & $\!\!\pm\!\!$ &  0.47 & $\!\!\pm\!\!$ &  0.74 & 0.458 &  81.9 &  11.01 & $\!\!\pm\!\!$ &  0.41 & $\!\!\pm\!\!$ &  0.62 \\
0.50--0.60 & 0.546 &  67.3 &  10.63 & $\!\!\pm\!\!$ &  0.37 & $\!\!\pm\!\!$ &  0.58 & 0.547 &  81.7 &   6.96 & $\!\!\pm\!\!$ &  0.30 & $\!\!\pm\!\!$ &  0.47 \\
0.60--0.72 & 0.652 &  66.9 &   6.64 & $\!\!\pm\!\!$ &  0.28 & $\!\!\pm\!\!$ &  0.46 & 0.652 &  82.1 &   3.63 & $\!\!\pm\!\!$ &  0.21 & $\!\!\pm\!\!$ &  0.32 \\
0.72--0.90 & 0.797 &  67.0 &   2.96 & $\!\!\pm\!\!$ &  0.15 & $\!\!\pm\!\!$ &  0.29 & 0.790 &  81.9 &   1.48 & $\!\!\pm\!\!$ &  0.11 & $\!\!\pm\!\!$ &  0.18 \\
0.90--1.25 & 1.029 &  66.6 &   0.86 & $\!\!\pm\!\!$ &  0.06 & $\!\!\pm\!\!$ &  0.13 & 1.029 &  81.0 &   0.38 & $\!\!\pm\!\!$ &  0.04 & $\!\!\pm\!\!$ &  0.07 \\
 \hline
 \hline
  & \multicolumn{7}{c||}{$90<\theta<105$}
  & \multicolumn{7}{c|}{$105<\theta<125$} \\
 \hline
 $p_{\rm T}$ & $\langle p_{\rm T} \rangle$ & $\langle \theta \rangle$
  & \multicolumn{5}{c||}{${\rm d}^2 \sigma /{\rm d}p{\rm d}\Omega$}
  &$\langle p_{\rm T} \rangle$ & $\langle \theta \rangle$
  & \multicolumn{5}{c|}{${\rm d}^2 \sigma /{\rm d}p{\rm d}\Omega$} \\
 \hline
0.36--0.42 &  &  &  \multicolumn{5}{c||}{ } & 0.387 & 113.5 &   5.43 & $\!\!\pm\!\!$ &  0.28 & $\!\!\pm\!\!$ &  0.35 \\
0.42--0.50 & 0.459 &  97.0 &   6.51 & $\!\!\pm\!\!$ &  0.32 & $\!\!\pm\!\!$ &  0.49 & 0.456 & 113.8 &   3.08 & $\!\!\pm\!\!$ &  0.19 & $\!\!\pm\!\!$ &  0.23 \\
0.50--0.60 & 0.545 &  96.4 &   3.49 & $\!\!\pm\!\!$ &  0.21 & $\!\!\pm\!\!$ &  0.31 & 0.540 & 112.8 &   1.38 & $\!\!\pm\!\!$ &  0.12 & $\!\!\pm\!\!$ &  0.16 \\
0.60--0.72 & 0.650 &  96.4 &   1.56 & $\!\!\pm\!\!$ &  0.14 & $\!\!\pm\!\!$ &  0.18 & 0.653 & 112.6 &   0.52 & $\!\!\pm\!\!$ &  0.07 & $\!\!\pm\!\!$ &  0.09 \\
0.72--0.90 & 0.794 &  95.0 &   0.59 & $\!\!\pm\!\!$ &  0.07 & $\!\!\pm\!\!$ &  0.09 & 0.789 & 112.6 &   0.12 & $\!\!\pm\!\!$ &  0.03 & $\!\!\pm\!\!$ &  0.03 \\
0.90--1.25 & 1.020 &  95.4 &   0.16 & $\!\!\pm\!\!$ &  0.03 & $\!\!\pm\!\!$ &  0.04 &  &  &  \multicolumn{5}{c|}{ } \\
 \hline
 \end{tabular}
 \end{center}
 \end{scriptsize}
 \end{table}

%% file: table.pip.pimbe15.tex
  
 \begin{table}[h]
 \begin{scriptsize}
 \caption{Double-differential inclusive
  cross-section ${\rm d}^2 \sigma /{\rm d}p{\rm d}\Omega$
  [mb/(GeV/{\it c} sr)] of the production of $\pi^+$'s
  in $\pi^-$ + Be $\rightarrow$ $\pi^+$ + X interactions
  with $-15.0$~GeV/{\it c} beam momentum;
  the first error is statistical, the second systematic; 
 $p_{\rm T}$ in GeV/{\it c}, polar angle $\theta$ in degrees.}
 \label{pip.pimbe15}
 \begin{center}
 \begin{tabular}{|c||c|c|rcrcr||c|c|rcrcr|}
 \hline
   & \multicolumn{7}{c||}{$20<\theta<30$}
  & \multicolumn{7}{c|}{$30<\theta<40$} \\
 \hline
 $p_{\rm T}$ & $\langle p_{\rm T} \rangle$ & $\langle \theta \rangle$
  & \multicolumn{5}{c||}{${\rm d}^2 \sigma /{\rm d}p{\rm d}\Omega$}
  &$\langle p_{\rm T} \rangle$ & $\langle \theta \rangle$
  & \multicolumn{5}{c|}{${\rm d}^2 \sigma /{\rm d}p{\rm d}\Omega$} \\
 \hline
0.10--0.13 & 0.115 &  24.6 &  45.98 & $\!\!\pm\!\!$ &  2.08 & $\!\!\pm\!\!$ &  3.81 & 0.115 &  35.0 &  30.15 & $\!\!\pm\!\!$ &  1.63 & $\!\!\pm\!\!$ &  2.64 \\
0.13--0.16 & 0.145 &  24.7 &  55.61 & $\!\!\pm\!\!$ &  2.07 & $\!\!\pm\!\!$ &  3.90 & 0.145 &  35.0 &  39.64 & $\!\!\pm\!\!$ &  1.77 & $\!\!\pm\!\!$ &  2.80 \\
0.16--0.20 & 0.180 &  24.7 &  63.93 & $\!\!\pm\!\!$ &  1.87 & $\!\!\pm\!\!$ &  3.75 & 0.179 &  34.8 &  42.87 & $\!\!\pm\!\!$ &  1.54 & $\!\!\pm\!\!$ &  2.59 \\
0.20--0.24 & 0.219 &  24.6 &  75.96 & $\!\!\pm\!\!$ &  2.01 & $\!\!\pm\!\!$ &  4.00 & 0.219 &  34.6 &  43.58 & $\!\!\pm\!\!$ &  1.50 & $\!\!\pm\!\!$ &  2.36 \\
0.24--0.30 & 0.268 &  24.5 &  64.35 & $\!\!\pm\!\!$ &  1.48 & $\!\!\pm\!\!$ &  3.00 & 0.266 &  34.5 &  42.01 & $\!\!\pm\!\!$ &  1.19 & $\!\!\pm\!\!$ &  2.00 \\
0.30--0.36 & 0.326 &  24.6 &  56.74 & $\!\!\pm\!\!$ &  1.38 & $\!\!\pm\!\!$ &  2.47 & 0.325 &  34.7 &  36.88 & $\!\!\pm\!\!$ &  1.13 & $\!\!\pm\!\!$ &  1.64 \\
0.36--0.42 & 0.384 &  24.6 &  46.39 & $\!\!\pm\!\!$ &  1.24 & $\!\!\pm\!\!$ &  2.01 & 0.385 &  34.4 &  27.42 & $\!\!\pm\!\!$ &  0.94 & $\!\!\pm\!\!$ &  1.23 \\
0.42--0.50 & 0.451 &  24.6 &  34.24 & $\!\!\pm\!\!$ &  0.91 & $\!\!\pm\!\!$ &  1.67 & 0.452 &  34.7 &  20.07 & $\!\!\pm\!\!$ &  0.69 & $\!\!\pm\!\!$ &  0.97 \\
0.50--0.60 & 0.539 &  24.6 &  24.03 & $\!\!\pm\!\!$ &  0.66 & $\!\!\pm\!\!$ &  1.47 & 0.538 &  34.7 &  15.30 & $\!\!\pm\!\!$ &  0.54 & $\!\!\pm\!\!$ &  0.91 \\
0.60--0.72 & 0.643 &  24.5 &  12.77 & $\!\!\pm\!\!$ &  0.42 & $\!\!\pm\!\!$ &  1.09 & 0.642 &  34.5 &   7.79 & $\!\!\pm\!\!$ &  0.32 & $\!\!\pm\!\!$ &  0.64 \\
0.72--0.90 &  &  &  \multicolumn{5}{c||}{ } & 0.781 &  34.6 &   3.80 & $\!\!\pm\!\!$ &  0.17 & $\!\!\pm\!\!$ &  0.45 \\
 \hline
 \hline
   & \multicolumn{7}{c||}{$40<\theta<50$}
  & \multicolumn{7}{c|}{$50<\theta<60$} \\
 \hline
 $p_{\rm T}$ & $\langle p_{\rm T} \rangle$ & $\langle \theta \rangle$
  & \multicolumn{5}{c||}{${\rm d}^2 \sigma /{\rm d}p{\rm d}\Omega$}
  &$\langle p_{\rm T} \rangle$ & $\langle \theta \rangle$
  & \multicolumn{5}{c|}{${\rm d}^2 \sigma /{\rm d}p{\rm d}\Omega$} \\
 \hline
0.10--0.13 & 0.116 &  44.7 &  21.46 & $\!\!\pm\!\!$ &  1.42 & $\!\!\pm\!\!$ &  2.01 &  &  &  \multicolumn{5}{c|}{ } \\
0.13--0.16 & 0.146 &  44.8 &  27.03 & $\!\!\pm\!\!$ &  1.44 & $\!\!\pm\!\!$ &  2.01 & 0.145 &  55.1 &  18.40 & $\!\!\pm\!\!$ &  1.14 & $\!\!\pm\!\!$ &  1.59 \\
0.16--0.20 & 0.181 &  45.0 &  26.95 & $\!\!\pm\!\!$ &  1.21 & $\!\!\pm\!\!$ &  1.69 & 0.179 &  54.8 &  21.02 & $\!\!\pm\!\!$ &  1.04 & $\!\!\pm\!\!$ &  1.36 \\
0.20--0.24 & 0.219 &  44.6 &  28.62 & $\!\!\pm\!\!$ &  1.23 & $\!\!\pm\!\!$ &  1.61 & 0.220 &  54.9 &  19.66 & $\!\!\pm\!\!$ &  0.99 & $\!\!\pm\!\!$ &  1.17 \\
0.24--0.30 & 0.269 &  44.7 &  26.03 & $\!\!\pm\!\!$ &  0.93 & $\!\!\pm\!\!$ &  1.30 & 0.271 &  54.6 &  18.09 & $\!\!\pm\!\!$ &  0.78 & $\!\!\pm\!\!$ &  0.93 \\
0.30--0.36 & 0.330 &  44.8 &  22.51 & $\!\!\pm\!\!$ &  0.87 & $\!\!\pm\!\!$ &  1.05 & 0.328 &  54.6 &  15.71 & $\!\!\pm\!\!$ &  0.71 & $\!\!\pm\!\!$ &  0.77 \\
0.36--0.42 & 0.389 &  44.7 &  18.31 & $\!\!\pm\!\!$ &  0.77 & $\!\!\pm\!\!$ &  0.88 & 0.390 &  54.6 &  12.43 & $\!\!\pm\!\!$ &  0.63 & $\!\!\pm\!\!$ &  0.68 \\
0.42--0.50 & 0.459 &  44.7 &  14.34 & $\!\!\pm\!\!$ &  0.59 & $\!\!\pm\!\!$ &  0.73 & 0.457 &  54.8 &   8.54 & $\!\!\pm\!\!$ &  0.44 & $\!\!\pm\!\!$ &  0.52 \\
0.50--0.60 & 0.547 &  44.7 &   8.60 & $\!\!\pm\!\!$ &  0.39 & $\!\!\pm\!\!$ &  0.55 & 0.546 &  54.5 &   5.77 & $\!\!\pm\!\!$ &  0.32 & $\!\!\pm\!\!$ &  0.43 \\
0.60--0.72 & 0.654 &  44.7 &   5.52 & $\!\!\pm\!\!$ &  0.29 & $\!\!\pm\!\!$ &  0.44 & 0.654 &  54.4 &   3.37 & $\!\!\pm\!\!$ &  0.22 & $\!\!\pm\!\!$ &  0.30 \\
0.72--0.90 & 0.792 &  44.6 &   2.17 & $\!\!\pm\!\!$ &  0.14 & $\!\!\pm\!\!$ &  0.25 & 0.797 &  54.7 &   1.37 & $\!\!\pm\!\!$ &  0.11 & $\!\!\pm\!\!$ &  0.17 \\
0.90--1.25 &  &  &  \multicolumn{5}{c||}{ } & 1.020 &  55.0 &   0.24 & $\!\!\pm\!\!$ &  0.03 & $\!\!\pm\!\!$ &  0.05 \\
 \hline
 \hline
   & \multicolumn{7}{c||}{$60<\theta<75$}
  & \multicolumn{7}{c|}{$75<\theta<90$} \\
 \hline
 $p_{\rm T}$ & $\langle p_{\rm T} \rangle$ & $\langle \theta \rangle$
  & \multicolumn{5}{c||}{${\rm d}^2 \sigma /{\rm d}p{\rm d}\Omega$}
  &$\langle p_{\rm T} \rangle$ & $\langle \theta \rangle$
  & \multicolumn{5}{c|}{${\rm d}^2 \sigma /{\rm d}p{\rm d}\Omega$} \\
 \hline
0.13--0.16 & 0.146 &  67.0 &  13.85 & $\!\!\pm\!\!$ &  0.81 & $\!\!\pm\!\!$ &  1.11 & 0.146 &  82.4 &  10.71 & $\!\!\pm\!\!$ &  0.70 & $\!\!\pm\!\!$ &  0.95 \\
0.16--0.20 & 0.180 &  67.1 &  14.41 & $\!\!\pm\!\!$ &  0.67 & $\!\!\pm\!\!$ &  0.98 & 0.180 &  82.3 &  11.12 & $\!\!\pm\!\!$ &  0.59 & $\!\!\pm\!\!$ &  0.88 \\
0.20--0.24 & 0.220 &  67.2 &  15.85 & $\!\!\pm\!\!$ &  0.73 & $\!\!\pm\!\!$ &  0.90 & 0.219 &  82.1 &  11.09 & $\!\!\pm\!\!$ &  0.60 & $\!\!\pm\!\!$ &  0.67 \\
0.24--0.30 & 0.268 &  67.0 &  12.71 & $\!\!\pm\!\!$ &  0.53 & $\!\!\pm\!\!$ &  0.63 & 0.268 &  82.2 &   7.35 & $\!\!\pm\!\!$ &  0.39 & $\!\!\pm\!\!$ &  0.42 \\
0.30--0.36 & 0.328 &  66.9 &   9.96 & $\!\!\pm\!\!$ &  0.46 & $\!\!\pm\!\!$ &  0.51 & 0.330 &  81.4 &   5.88 & $\!\!\pm\!\!$ &  0.34 & $\!\!\pm\!\!$ &  0.39 \\
0.36--0.42 & 0.389 &  67.0 &   7.93 & $\!\!\pm\!\!$ &  0.41 & $\!\!\pm\!\!$ &  0.45 & 0.389 &  82.0 &   3.82 & $\!\!\pm\!\!$ &  0.27 & $\!\!\pm\!\!$ &  0.30 \\
0.42--0.50 & 0.460 &  67.2 &   5.04 & $\!\!\pm\!\!$ &  0.28 & $\!\!\pm\!\!$ &  0.36 & 0.457 &  82.0 &   2.84 & $\!\!\pm\!\!$ &  0.20 & $\!\!\pm\!\!$ &  0.29 \\
0.50--0.60 & 0.544 &  66.3 &   3.80 & $\!\!\pm\!\!$ &  0.22 & $\!\!\pm\!\!$ &  0.31 & 0.546 &  81.6 &   1.67 & $\!\!\pm\!\!$ &  0.14 & $\!\!\pm\!\!$ &  0.17 \\
0.60--0.72 & 0.654 &  66.4 &   1.72 & $\!\!\pm\!\!$ &  0.12 & $\!\!\pm\!\!$ &  0.21 & 0.652 &  81.9 &   0.81 & $\!\!\pm\!\!$ &  0.08 & $\!\!\pm\!\!$ &  0.12 \\
0.72--0.90 & 0.791 &  65.9 &   0.66 & $\!\!\pm\!\!$ &  0.06 & $\!\!\pm\!\!$ &  0.10 & 0.813 &  81.1 &   0.17 & $\!\!\pm\!\!$ &  0.03 & $\!\!\pm\!\!$ &  0.04 \\
0.90--1.25 & 1.030 &  67.1 &   0.11 & $\!\!\pm\!\!$ &  0.02 & $\!\!\pm\!\!$ &  0.03 & 1.016 &  81.0 &   0.04 & $\!\!\pm\!\!$ &  0.01 & $\!\!\pm\!\!$ &  0.02 \\
 \hline
 \hline
  & \multicolumn{7}{c||}{$90<\theta<105$}
  & \multicolumn{7}{c|}{$105<\theta<125$} \\
 \hline
 $p_{\rm T}$ & $\langle p_{\rm T} \rangle$ & $\langle \theta \rangle$
  & \multicolumn{5}{c||}{${\rm d}^2 \sigma /{\rm d}p{\rm d}\Omega$}
  &$\langle p_{\rm T} \rangle$ & $\langle \theta \rangle$
  & \multicolumn{5}{c|}{${\rm d}^2 \sigma /{\rm d}p{\rm d}\Omega$} \\
 \hline
0.13--0.16 & 0.145 &  97.7 &   7.88 & $\!\!\pm\!\!$ &  0.59 & $\!\!\pm\!\!$ &  0.79 & 0.145 & 114.0 &   7.20 & $\!\!\pm\!\!$ &  0.46 & $\!\!\pm\!\!$ &  0.75 \\
0.16--0.20 & 0.181 &  96.9 &   8.14 & $\!\!\pm\!\!$ &  0.49 & $\!\!\pm\!\!$ &  0.61 & 0.179 & 114.3 &   6.60 & $\!\!\pm\!\!$ &  0.36 & $\!\!\pm\!\!$ &  0.58 \\
0.20--0.24 & 0.219 &  97.3 &   7.33 & $\!\!\pm\!\!$ &  0.46 & $\!\!\pm\!\!$ &  0.57 & 0.217 & 114.6 &   3.84 & $\!\!\pm\!\!$ &  0.29 & $\!\!\pm\!\!$ &  0.33 \\
0.24--0.30 & 0.268 &  97.4 &   5.37 & $\!\!\pm\!\!$ &  0.33 & $\!\!\pm\!\!$ &  0.33 & 0.268 & 113.2 &   3.15 & $\!\!\pm\!\!$ &  0.22 & $\!\!\pm\!\!$ &  0.25 \\
0.30--0.36 & 0.328 &  97.2 &   3.68 & $\!\!\pm\!\!$ &  0.28 & $\!\!\pm\!\!$ &  0.26 & 0.327 & 112.5 &   1.85 & $\!\!\pm\!\!$ &  0.16 & $\!\!\pm\!\!$ &  0.19 \\
0.36--0.42 & 0.387 &  96.5 &   2.27 & $\!\!\pm\!\!$ &  0.20 & $\!\!\pm\!\!$ &  0.24 & 0.386 & 111.8 &   1.03 & $\!\!\pm\!\!$ &  0.12 & $\!\!\pm\!\!$ &  0.14 \\
0.42--0.50 & 0.458 &  95.7 &   1.66 & $\!\!\pm\!\!$ &  0.16 & $\!\!\pm\!\!$ &  0.19 & 0.460 & 112.3 &   0.46 & $\!\!\pm\!\!$ &  0.07 & $\!\!\pm\!\!$ &  0.09 \\
0.50--0.60 & 0.549 &  95.4 &   0.61 & $\!\!\pm\!\!$ &  0.08 & $\!\!\pm\!\!$ &  0.10 & 0.545 & 113.3 &   0.21 & $\!\!\pm\!\!$ &  0.04 & $\!\!\pm\!\!$ &  0.05 \\
0.60--0.72 & 0.647 &  96.9 &   0.26 & $\!\!\pm\!\!$ &  0.05 & $\!\!\pm\!\!$ &  0.05 & 0.646 & 110.9 &   0.07 & $\!\!\pm\!\!$ &  0.02 & $\!\!\pm\!\!$ &  0.02 \\
0.72--0.90 & 0.763 &  95.3 &   0.08 & $\!\!\pm\!\!$ &  0.02 & $\!\!\pm\!\!$ &  0.04 &  &  &  \multicolumn{5}{c|}{ } \\
 \hline
 \end{tabular}
 \end{center}
 \end{scriptsize}
 \end{table}

%% file: table.pim.pimbe15.tex
  
 \begin{table}[h]
 \begin{scriptsize}
 \caption{Double-differential inclusive
  cross-section ${\rm d}^2 \sigma /{\rm d}p{\rm d}\Omega$
  [mb/(GeV/{\it c} sr)] of the production of $\pi^-$'s
  in $\pi^-$ + Be $\rightarrow$ $\pi^-$ + X interactions
  with $-15.0$~GeV/{\it c} beam momentum;
  the first error is statistical, the second systematic; 
 $p_{\rm T}$ in GeV/{\it c}, polar angle $\theta$ in degrees.}
 \label{pim.pimbe15}
 \begin{center}
 \begin{tabular}{|c||c|c|rcrcr||c|c|rcrcr|}
 \hline
   & \multicolumn{7}{c||}{$20<\theta<30$}
  & \multicolumn{7}{c|}{$30<\theta<40$} \\
 \hline
 $p_{\rm T}$ & $\langle p_{\rm T} \rangle$ & $\langle \theta \rangle$
  & \multicolumn{5}{c||}{${\rm d}^2 \sigma /{\rm d}p{\rm d}\Omega$}
  &$\langle p_{\rm T} \rangle$ & $\langle \theta \rangle$
  & \multicolumn{5}{c|}{${\rm d}^2 \sigma /{\rm d}p{\rm d}\Omega$} \\
 \hline
0.10--0.13 & 0.115 &  24.6 &  57.96 & $\!\!\pm\!\!$ &  2.29 & $\!\!\pm\!\!$ &  4.67 & 0.116 &  34.8 &  36.13 & $\!\!\pm\!\!$ &  1.78 & $\!\!\pm\!\!$ &  3.15 \\
0.13--0.16 & 0.146 &  24.7 &  74.78 & $\!\!\pm\!\!$ &  2.42 & $\!\!\pm\!\!$ &  5.04 & 0.146 &  34.6 &  44.08 & $\!\!\pm\!\!$ &  1.83 & $\!\!\pm\!\!$ &  3.12 \\
0.16--0.20 & 0.182 &  24.5 &  83.43 & $\!\!\pm\!\!$ &  2.14 & $\!\!\pm\!\!$ &  4.84 & 0.181 &  34.6 &  49.76 & $\!\!\pm\!\!$ &  1.63 & $\!\!\pm\!\!$ &  2.99 \\
0.20--0.24 & 0.221 &  24.6 &  90.28 & $\!\!\pm\!\!$ &  2.19 & $\!\!\pm\!\!$ &  4.63 & 0.222 &  34.7 &  55.36 & $\!\!\pm\!\!$ &  1.69 & $\!\!\pm\!\!$ &  2.95 \\
0.24--0.30 & 0.272 &  24.5 &  84.12 & $\!\!\pm\!\!$ &  1.71 & $\!\!\pm\!\!$ &  3.75 & 0.271 &  34.6 &  52.07 & $\!\!\pm\!\!$ &  1.34 & $\!\!\pm\!\!$ &  2.40 \\
0.30--0.36 & 0.332 &  24.6 &  74.96 & $\!\!\pm\!\!$ &  1.62 & $\!\!\pm\!\!$ &  3.10 & 0.333 &  34.7 &  42.92 & $\!\!\pm\!\!$ &  1.21 & $\!\!\pm\!\!$ &  1.85 \\
0.36--0.42 & 0.394 &  24.6 &  62.60 & $\!\!\pm\!\!$ &  1.48 & $\!\!\pm\!\!$ &  2.64 & 0.394 &  34.6 &  37.25 & $\!\!\pm\!\!$ &  1.13 & $\!\!\pm\!\!$ &  1.63 \\
0.42--0.50 & 0.464 &  24.6 &  46.78 & $\!\!\pm\!\!$ &  1.10 & $\!\!\pm\!\!$ &  2.21 & 0.464 &  34.6 &  28.82 & $\!\!\pm\!\!$ &  0.86 & $\!\!\pm\!\!$ &  1.39 \\
0.50--0.60 & 0.555 &  24.6 &  30.44 & $\!\!\pm\!\!$ &  0.80 & $\!\!\pm\!\!$ &  1.80 & 0.555 &  34.8 &  21.50 & $\!\!\pm\!\!$ &  0.67 & $\!\!\pm\!\!$ &  1.30 \\
0.60--0.72 & 0.667 &  24.6 &  19.92 & $\!\!\pm\!\!$ &  0.59 & $\!\!\pm\!\!$ &  1.62 & 0.667 &  34.6 &  11.61 & $\!\!\pm\!\!$ &  0.44 & $\!\!\pm\!\!$ &  0.92 \\
0.72--0.90 &  &  &  \multicolumn{5}{c||}{ } & 0.815 &  34.7 &   5.84 & $\!\!\pm\!\!$ &  0.26 & $\!\!\pm\!\!$ &  0.62 \\
 \hline
 \hline
   & \multicolumn{7}{c||}{$40<\theta<50$}
  & \multicolumn{7}{c|}{$50<\theta<60$} \\
 \hline
 $p_{\rm T}$ & $\langle p_{\rm T} \rangle$ & $\langle \theta \rangle$
  & \multicolumn{5}{c||}{${\rm d}^2 \sigma /{\rm d}p{\rm d}\Omega$}
  &$\langle p_{\rm T} \rangle$ & $\langle \theta \rangle$
  & \multicolumn{5}{c|}{${\rm d}^2 \sigma /{\rm d}p{\rm d}\Omega$} \\
 \hline
0.10--0.13 & 0.116 &  44.8 &  28.52 & $\!\!\pm\!\!$ &  1.55 & $\!\!\pm\!\!$ &  2.71 &  &  &  \multicolumn{5}{c|}{ } \\
0.13--0.16 & 0.146 &  44.8 &  31.34 & $\!\!\pm\!\!$ &  1.50 & $\!\!\pm\!\!$ &  2.34 & 0.145 &  54.8 &  24.85 & $\!\!\pm\!\!$ &  1.34 & $\!\!\pm\!\!$ &  1.93 \\
0.16--0.20 & 0.179 &  44.7 &  35.76 & $\!\!\pm\!\!$ &  1.39 & $\!\!\pm\!\!$ &  2.21 & 0.181 &  54.8 &  27.65 & $\!\!\pm\!\!$ &  1.19 & $\!\!\pm\!\!$ &  1.77 \\
0.20--0.24 & 0.220 &  44.8 &  39.50 & $\!\!\pm\!\!$ &  1.44 & $\!\!\pm\!\!$ &  2.20 & 0.220 &  54.7 &  29.04 & $\!\!\pm\!\!$ &  1.23 & $\!\!\pm\!\!$ &  1.63 \\
0.24--0.30 & 0.270 &  44.7 &  32.91 & $\!\!\pm\!\!$ &  1.04 & $\!\!\pm\!\!$ &  1.57 & 0.271 &  54.7 &  22.29 & $\!\!\pm\!\!$ &  0.86 & $\!\!\pm\!\!$ &  1.14 \\
0.30--0.36 & 0.328 &  44.6 &  30.34 & $\!\!\pm\!\!$ &  1.02 & $\!\!\pm\!\!$ &  1.36 & 0.330 &  54.8 &  20.09 & $\!\!\pm\!\!$ &  0.82 & $\!\!\pm\!\!$ &  0.95 \\
0.36--0.42 & 0.388 &  44.8 &  22.96 & $\!\!\pm\!\!$ &  0.87 & $\!\!\pm\!\!$ &  1.07 & 0.389 &  54.6 &  15.70 & $\!\!\pm\!\!$ &  0.71 & $\!\!\pm\!\!$ &  0.78 \\
0.42--0.50 & 0.460 &  44.5 &  17.64 & $\!\!\pm\!\!$ &  0.66 & $\!\!\pm\!\!$ &  0.99 & 0.458 &  54.6 &  13.51 & $\!\!\pm\!\!$ &  0.58 & $\!\!\pm\!\!$ &  0.75 \\
0.50--0.60 & 0.546 &  44.7 &  13.49 & $\!\!\pm\!\!$ &  0.53 & $\!\!\pm\!\!$ &  0.88 & 0.544 &  54.7 &   8.32 & $\!\!\pm\!\!$ &  0.41 & $\!\!\pm\!\!$ &  0.57 \\
0.60--0.72 & 0.654 &  44.5 &   7.21 & $\!\!\pm\!\!$ &  0.35 & $\!\!\pm\!\!$ &  0.60 & 0.649 &  54.6 &   4.74 & $\!\!\pm\!\!$ &  0.28 & $\!\!\pm\!\!$ &  0.42 \\
0.72--0.90 & 0.795 &  44.7 &   3.79 & $\!\!\pm\!\!$ &  0.21 & $\!\!\pm\!\!$ &  0.43 & 0.789 &  54.7 &   1.90 & $\!\!\pm\!\!$ &  0.14 & $\!\!\pm\!\!$ &  0.23 \\
0.90--1.25 &  &  &  \multicolumn{5}{c||}{ } & 1.015 &  54.3 &   0.39 & $\!\!\pm\!\!$ &  0.04 & $\!\!\pm\!\!$ &  0.08 \\
 \hline
 \hline
   & \multicolumn{7}{c||}{$60<\theta<75$}
  & \multicolumn{7}{c|}{$75<\theta<90$} \\
 \hline
 $p_{\rm T}$ & $\langle p_{\rm T} \rangle$ & $\langle \theta \rangle$
  & \multicolumn{5}{c||}{${\rm d}^2 \sigma /{\rm d}p{\rm d}\Omega$}
  &$\langle p_{\rm T} \rangle$ & $\langle \theta \rangle$
  & \multicolumn{5}{c|}{${\rm d}^2 \sigma /{\rm d}p{\rm d}\Omega$} \\
 \hline
0.13--0.16 & 0.145 &  67.6 &  20.41 & $\!\!\pm\!\!$ &  0.97 & $\!\!\pm\!\!$ &  1.67 & 0.146 &  82.5 &  15.64 & $\!\!\pm\!\!$ &  0.82 & $\!\!\pm\!\!$ &  1.46 \\
0.16--0.20 & 0.181 &  67.2 &  21.23 & $\!\!\pm\!\!$ &  0.82 & $\!\!\pm\!\!$ &  1.37 & 0.181 &  82.3 &  16.90 & $\!\!\pm\!\!$ &  0.72 & $\!\!\pm\!\!$ &  1.22 \\
0.20--0.24 & 0.220 &  67.1 &  21.39 & $\!\!\pm\!\!$ &  0.85 & $\!\!\pm\!\!$ &  1.25 & 0.219 &  82.5 &  14.29 & $\!\!\pm\!\!$ &  0.67 & $\!\!\pm\!\!$ &  0.82 \\
0.24--0.30 & 0.268 &  66.8 &  16.96 & $\!\!\pm\!\!$ &  0.61 & $\!\!\pm\!\!$ &  0.80 & 0.267 &  82.0 &  11.35 & $\!\!\pm\!\!$ &  0.49 & $\!\!\pm\!\!$ &  0.57 \\
0.30--0.36 & 0.328 &  66.6 &  14.09 & $\!\!\pm\!\!$ &  0.56 & $\!\!\pm\!\!$ &  0.66 & 0.329 &  82.1 &   7.47 & $\!\!\pm\!\!$ &  0.40 & $\!\!\pm\!\!$ &  0.41 \\
0.36--0.42 & 0.389 &  66.8 &  10.59 & $\!\!\pm\!\!$ &  0.47 & $\!\!\pm\!\!$ &  0.63 & 0.390 &  81.7 &   6.32 & $\!\!\pm\!\!$ &  0.37 & $\!\!\pm\!\!$ &  0.41 \\
0.42--0.50 & 0.456 &  66.9 &   7.76 & $\!\!\pm\!\!$ &  0.35 & $\!\!\pm\!\!$ &  0.48 & 0.459 &  81.5 &   3.68 & $\!\!\pm\!\!$ &  0.24 & $\!\!\pm\!\!$ &  0.30 \\
0.50--0.60 & 0.549 &  66.9 &   5.45 & $\!\!\pm\!\!$ &  0.27 & $\!\!\pm\!\!$ &  0.41 & 0.540 &  82.2 &   2.53 & $\!\!\pm\!\!$ &  0.18 & $\!\!\pm\!\!$ &  0.24 \\
0.60--0.72 & 0.653 &  66.3 &   2.13 & $\!\!\pm\!\!$ &  0.14 & $\!\!\pm\!\!$ &  0.24 & 0.648 &  81.4 &   1.08 & $\!\!\pm\!\!$ &  0.10 & $\!\!\pm\!\!$ &  0.14 \\
0.72--0.90 & 0.795 &  67.0 &   0.87 & $\!\!\pm\!\!$ &  0.08 & $\!\!\pm\!\!$ &  0.12 & 0.787 &  80.0 &   0.40 & $\!\!\pm\!\!$ &  0.05 & $\!\!\pm\!\!$ &  0.07 \\
0.90--1.25 & 1.038 &  65.4 &   0.15 & $\!\!\pm\!\!$ &  0.02 & $\!\!\pm\!\!$ &  0.03 & 0.994 &  80.2 &   0.05 & $\!\!\pm\!\!$ &  0.02 & $\!\!\pm\!\!$ &  0.02 \\
 \hline
 \hline
  & \multicolumn{7}{c||}{$90<\theta<105$}
  & \multicolumn{7}{c|}{$105<\theta<125$} \\
 \hline
 $p_{\rm T}$ & $\langle p_{\rm T} \rangle$ & $\langle \theta \rangle$
  & \multicolumn{5}{c||}{${\rm d}^2 \sigma /{\rm d}p{\rm d}\Omega$}
  &$\langle p_{\rm T} \rangle$ & $\langle \theta \rangle$
  & \multicolumn{5}{c|}{${\rm d}^2 \sigma /{\rm d}p{\rm d}\Omega$} \\
 \hline
0.13--0.16 & 0.146 &  97.0 &  13.15 & $\!\!\pm\!\!$ &  0.74 & $\!\!\pm\!\!$ &  1.20 & 0.144 & 114.4 &  10.34 & $\!\!\pm\!\!$ &  0.55 & $\!\!\pm\!\!$ &  0.95 \\
0.16--0.20 & 0.179 &  97.3 &  14.11 & $\!\!\pm\!\!$ &  0.63 & $\!\!\pm\!\!$ &  1.19 & 0.179 & 114.5 &   9.61 & $\!\!\pm\!\!$ &  0.45 & $\!\!\pm\!\!$ &  0.70 \\
0.20--0.24 & 0.219 &  96.5 &  10.65 & $\!\!\pm\!\!$ &  0.58 & $\!\!\pm\!\!$ &  0.70 & 0.217 & 113.6 &   7.21 & $\!\!\pm\!\!$ &  0.40 & $\!\!\pm\!\!$ &  0.52 \\
0.24--0.30 & 0.268 &  97.2 &   8.52 & $\!\!\pm\!\!$ &  0.43 & $\!\!\pm\!\!$ &  0.47 & 0.269 & 113.8 &   4.65 & $\!\!\pm\!\!$ &  0.27 & $\!\!\pm\!\!$ &  0.34 \\
0.30--0.36 & 0.330 &  96.4 &   5.10 & $\!\!\pm\!\!$ &  0.33 & $\!\!\pm\!\!$ &  0.34 & 0.329 & 113.6 &   2.88 & $\!\!\pm\!\!$ &  0.21 & $\!\!\pm\!\!$ &  0.25 \\
0.36--0.42 & 0.389 &  97.0 &   3.90 & $\!\!\pm\!\!$ &  0.28 & $\!\!\pm\!\!$ &  0.36 & 0.390 & 114.2 &   1.58 & $\!\!\pm\!\!$ &  0.15 & $\!\!\pm\!\!$ &  0.20 \\
0.42--0.50 & 0.460 &  96.5 &   2.41 & $\!\!\pm\!\!$ &  0.19 & $\!\!\pm\!\!$ &  0.28 & 0.454 & 112.7 &   1.00 & $\!\!\pm\!\!$ &  0.10 & $\!\!\pm\!\!$ &  0.15 \\
0.50--0.60 & 0.547 &  96.7 &   0.95 & $\!\!\pm\!\!$ &  0.11 & $\!\!\pm\!\!$ &  0.15 & 0.542 & 112.1 &   0.42 & $\!\!\pm\!\!$ &  0.06 & $\!\!\pm\!\!$ &  0.08 \\
0.60--0.72 & 0.657 &  96.8 &   0.46 & $\!\!\pm\!\!$ &  0.07 & $\!\!\pm\!\!$ &  0.09 & 0.631 & 115.2 &   0.15 & $\!\!\pm\!\!$ &  0.04 & $\!\!\pm\!\!$ &  0.04 \\
0.72--0.90 & 0.796 &  95.1 &   0.14 & $\!\!\pm\!\!$ &  0.03 & $\!\!\pm\!\!$ &  0.04 &  &  &  \multicolumn{5}{c|}{ } \\
 \hline
 \end{tabular}
 \end{center}
 \end{scriptsize}
 \end{table}